\documentclass[11pt]{article}
\usepackage[utf8]{inputenc}
\usepackage{macros}
\graphicspath{{./figures/}}

\begin{document}
\begin{titlepage}
\begin{flushright} 
\
\end{flushright}
\vskip 12mm
\begin{center}
    {\LARGE \bf{5d AGT correspondence of supergroup gauge\\ theories from quantum toroidal $\mathfrak{gl}_{1}$}\par}          
\vskip 2cm    
\begin{center}
{\Large Go Noshita}
\end{center}
\vskip 2cm 
{\it Department of Physics, The University of Tokyo}\\
{\it Hongo 7-3-1, Bunkyo-ku, Tokyo 113-0033, Japan}
\end{center}
\vfill
\begin{abstract}
We discuss the 5d AGT correspondence of supergroup gauge theories with A-type supergroups. We introduce two intertwiners called positive and negative intertwiners to compute the instanton partition function. The positive intertwiner is the ordinary Awata-Feigin-Shiraishi intertwiner while the negative intertwiner is an intertwiner obtained by using central charges with negative levels. We show that composition of them gives the basic Nekrasov factors appearing in supergroup partition functions. We explicitly derive the instanton partition functions of supergroup gauge theories with A and D-type quiver structures. Using the intertwiners, we briefly study the Gaiotto state, $qq$-characters and the relation with quiver W-algebra. Furthermore, we show that the negative intertwiner corresponds to the anti-refined topological vertex recently defined by Kimura and Sugimoto. We also discuss how superquiver theories should appear in our formalism if they exist. The existence of the AGT correspondence of the theories we study in this paper implies that there is a broader 2d/4d (5d/$q$-algebra) correspondence, or more generally the BPS/CFT correspondence, where new non-unitary theories play important roles.
\end{abstract}
\vfill
\end{titlepage}

\tableofcontents

\section{Introduction and summary}\label{sec:Introduction}
\noindent Duality is one of the fundamental issues in the non-perturbative understanding of string theory and gauge theory. When there is suitable amount of supersymmetries, we can compute the instanton partition functions explicitly using localization techniques \cite{Nekrasov:2002qd,Nekrasov:2003rj,Nekrasov2004,Nakajima:2003uh,Nakajima:2005fg,Nakajima2005,Nakajimalecturebook,Pestun:2016zxk}. One of the astonishing properties of supersymmetric gauge theories is the existence of rich geometric and algebraic structures behind them \cite{Nakajima:1994nid,Nakajima:2005fg,schiffmann2012cherednik}. Nekrasov proposed the BPS/CFT correspondence claiming that there is a duality between correlation functions of BPS observables in supersymmetric gauge theories and correlation functions of conformal field theories \cite{Nekrasov:2015wsu,Nekrasov:2016gud,Nekrasov:2016qym,Nekrasov:2016ydq,Nekrasov:2017gzb,Nekrasov:2017rqy}. Various studies have been done to justify this belief. One of the famous correspondence that belongs to this correspondence is the AGT correspondence~\cite{Alday2010,Wyllard2009,Gaiotto:2009we,Gaiotto:2009ma} (see \cite{LeFloch:2020uop} for a review). It is a 2d/4d duality relating Nekrasov partition functions of 4d $\mathcal{N}=2$ theories coming from a certain 6d $\mathcal{N}=(2,0)$ theory compactified on a Riemann surface and conformal blocks of 2d CFTs (e.g. Virasoro, $W_{N}$ algebra). Moreover, we also have a 5d lift up of this correspondence relating Nekrasov partition functions of 5d $\mathcal{N}=1$ gauge theories and symmetries of quantum algebras (e.g. $q$-Virasoro, $q$-$W_{N}$) \cite{awata2010five,Awata_2010,Yanagida_2010,awata2011notes,ohkubo2016crystallization}. Understanding how far the BPS/CFT or AGT correspondence is true is an important task that must be done because new chiral algebras (or quantum algebras) might show the existence of new supersymmetric gauge theories, and vice versa.

An interesting extension in the gauge theory side is the supergroup gauge theory \cite{Vafa:2001qf,Okuda:2006fb,Mikhaylov:2014aoa,Dijkgraaf:2016lym}. Supergroup gauge theory is a theory in which both boson and fermion degrees of freedom appear as local gauge symmetries. Therefore, it breaks the spin-statistics theorem and is inevitably non-unitary, and thus, it has been overlooked as a theory whose physical meaning is not well understood. However, recent studies have shown that even if these theories are non-unitary, they can be analyzed by using methods similar to those used when discussing physical theories. In type~\rom{2}A theory, these gauge theories are constructed using ghost (negative) D-branes\footnote{We call ordinary D-branes as \textit{positive} D-branses and ghost D-branes as \textit{negative} D-branes in this paper.} \cite{Okuda:2006fb,Dijkgraaf:2016lym}. For example, the 4d $\mathcal{N}=2$ $U(N_{+}\,|\,N_{-})$ theory has a Hanany-Witten construction \cite{Hanany:1996ie} realized by $N_{+}$ positive D4-branes and $N_{-}$ negative D4-branes surrounded by two parallel NS5-branes (see (\ref{eq:HW-pureSYM})) \cite{Hanany:1996ie,Dijkgraaf:2016lym}. Seiberg-Witten curves were also determined in the same paper \cite{Dijkgraaf:2016lym}. Taking T-duality, we obtain brane webs in the type~\rom{2}B picture, which give 5d $\mathcal{N}=1$ theories of supergroup gauge theories \cite{Kimura-Sugimoto:antivertex}. Exact formulas for instanton partition functions of these theories were derived from supersymmetric localization in \cite{Kimura-Pestun:supergroup}. A generalization of the refined topological vertex \cite{Iqbal:2007ii,Awata:2005fa,Awata:2008ed,Aganagic:2003db} reproducing instanton partition functions called anti-refined topological vertex (anti-vertex) was as well proposed in \cite{Kimura-Sugimoto:antivertex}. Besides, relations with defects \cite{Kimura-Nieri:defects,Nieri:2021xpe} and integrable systems \cite{Chen:2020rxu} were also discussed previously. All of these previous studies imply the existence of an underlying algebraic structure in the belief of BPS/CFT correspondence. 

The goal of this paper is to determine the algebraic structure of the supergroup gauge theories. In particular, we are interested in the 5d AGT correspondence of these theories. We focus on quiver gauge theories with superunitary gauge groups $U(N_{+}\,|\,N_{-})$ (we still call them A-type for convenience) and A, D-type quiver structures\footnote{Mostly, we only discuss on theories with non-superalgebra quiver structure. For quivers with superalgebra structure (\textit{superquiver} theories), see section~\ref{sec:superquiver}.}. These theories are realized using brane webs of $(p,q)$-branes, positive D5, and negative D5-branes \cite{Aharony:1997bh,Katz:1996fh,Kimura-Sugimoto:antivertex}. To show the correspondence, we use the representation theory of the quantum toroidal $\mathfrak{gl}_{1}$ \cite{Miki2007,Ding:1996mq,Feigin2011,feigin2011quantum,feigin2012quantum} to construct algebraic quantities called the intertwiners. Assigning the intertwiners to trivalent vertices of the brane web, we show that composition of them gives the partition functions of the gauge theory. We call this procedure the intertwiner formalism (see section~\ref{sec:intertwiner} for details). It was first introduced by Awata, Feigin, and Shiraishi \cite{Awata:2011ce} and various extensions were studied in subsequent studies \cite{Awata:2016riz,Mironov:2016yue,Awata:2016bdm,Awata:2016mxc,Awata_2017,Awata_2018,Zenkevich:2018fzl,zenkevich2019mathfrakgln,zenkevich2020mixed,Ghoneim:2020sqi,Bourgine:2017jsi,Bourgine:2017rik,Zhu:2017ysu,Bourgine:2018uod,Bourgine:2019phm}. 

In the original story of AGT correspondence, the 4d $\mathcal{N}=2$ theories come from 6d theories associated with ADE type non-super Lie-algebras (see \cite{LeFloch:2020uop} and the references therein). Theories appearing there do not have structures of superalgebras, and thus, the supergroup gauge theories do not belong to the theories previously studied in the literature. The existence of the AGT correspondence of the theories we study in this paper implies that there is a broader 2d/4d (5d/$q$-algebra) correspondence, or more generally the BPS/CFT correspondence, where new non-unitary theories play important roles.

The main results of this paper are the following.
\begin{itemize}
    \item We introduce a new intertwiner which we call the negative intertwiner to represent negative D5-branes (section~\ref{sec:intertwiner}).
    \item We show that compositions of the ordinary intertwiners and the negative intertwiners give the Nekrasov factors of supergroup gauge theories. We also explicitly derive the total partition functions of A and D-type quiver gauge theories (section~\ref{sec:intertwiner} and \ref{sec:Examples}).
    \item We derive the supergroup analogue of the Gaiotto state, $qq$-character and discuss the relation with the quiver W-algebra (section~\ref{sec:Gaiotto-qq-quiverW}).
    \item We show that the negative intertwiner is related with the anti-vertex. We also complete the procedure to assign framing factors (section~\ref{sec:ref-topvertex}).
    \item We give conjectures regarding superquiver theories where negative NS5-branes are expected to appear (section~\ref{sec:superquiver}).
\end{itemize}

To be concrete, quantum toroidal $\mathfrak{gl}_{1}$ has two classes of representations: vertex operator representations and crystal representations. They are determined by the values of the two central charges of the algebra, which we call level $(\ell_{1},\ell_{2})$. In the 5-brane web, we associate a Fock space to each of the 5-brane. For the NS5-branes, we assign vertex operator representations with level $(1,n),\,(n\in\mathbb{Z})$, while for the D-branes we assign crystal representations with level $(0,\pm1)$. Level $(0,+1)$ (resp. $(0,-1)$) is assigned to positive (resp. negative) D5-branes. Here, we introduced a new crystal representation with level $(0,-1)$ to represent the negative D5-brane. We then derive two type of intertwiners (positive and negative) $\Phi_{\pm}:(0,\pm1)\otimes(1,n)\rightarrow (1,n\pm1)$ and their duals associated with trivalent vertices of the brane web, representing junctions of NS5-branes and positive, negative D5-branes. Compositions of these interwiners give the partition functions of supergroup gauge theories. We show this explicitly for pure supergroup gauge theories, A and D-type quiver gauge theories. In the process, we also generalize the brane webs of D-type quiver in 5d so that they include negative D-branes. The positive intertwiner is just the intertwiner studied in \cite{Awata:2011ce}, while the physical interpretation of the negative intertwiner is new in the literature.

Using the interwiners, we derive supergroup analogues of the Gaiotto state and characterize it by the action of the Drinfeld currents. We also reproduce the $qq$-characters in \cite{Kimura-Pestun:supergroup} using the interwiners. Relation with the quiver W-algebra is discussed too. We will see that all these quantities are derived by a simple modification of the quantities of the non-supergroup case.

After getting used with the interwiner formalism, we then discuss the relation with the refined topological vertex and anti-vertex. We show that the newly introduced negative intertwiner becomes the anti-vertex after taking nontrivial matrix elements. Using the correspondence with the intertwiners, we give a complete procedure how to assign framing factors, which was not explicitly discussed in \cite{Kimura-Sugimoto:antivertex}. The procedure we give is compatible with the gluing rules of the intertwiners. Discussions in the unrefined limit will be also given. Namely, we will see that the positive and negative interwiners will be related with each other in the unrefined limit, which was a property discussed previously in \cite{Kimura-Sugimoto:antivertex}.

S-duality of supergroup gauge theories with non-superalgebra quiver structure implies the existence of non-supergroup gauge theories with superalgebra quiver structure. For 3d supersymmetric gauge theories, a gauge theory with superquiver structure was discussed in \cite{Nekrasov:2018gne,Ishtiaque:2021jan,Orlando:2010uu,Zenkevich:2018fzl}. However, for 5d gauge theories, to the author's knowledge, there is no known construction of \textit{supersymmetric supergroup superquiver gauge theories}. Using the correspondence with the representations of the algebra, we give a discussion on how superquiver gauge theories should appear in our formalism if they exist and give some conjectures we hope to come back in near future. 

The paper is organized as follows. We review the explicit formulas, properties of the instanton partition functions, and D-brane constructions of A-type supergroup with A and D-type quiver gauge theories in section~\ref{sec:Supergroup}. We summarize the representations of quantum toroidal $\mathfrak{gl}_{1}$ in section~\ref{sec:DIMalgebra}. Intertwiners and their relations with the Nekrasov factors are discussed in section~\ref{sec:intertwiner}. In section~\ref{sec:Examples}, we use the positive and negative intertwiners to derive partition functions of pure super Yang-Mills, A-type quiver, and D-type quiver gauge theories. Relations with Gaiotto state, $qq$-character, and quiver W-algebra are discussed in section~\ref{sec:Gaiotto-qq-quiverW}. The correspondences between the intertwiners and the refined-topological vertices are in section~\ref{sec:ref-topvertex}. Finally, we briefly give a discussion on the S-dual of supergroup gauge theories with A-type quiver structure in section~\ref{sec:superquiver}.

\section{Supergroup gauge theory}\label{sec:Supergroup}
In this section, we review the supergroup gauge theory and the instanton partition functions of supergroup quiver gauge theories following \cite{Kimura-Pestun:supergroup}. We give the Lagrangian and show its properties in section~\ref{sec:lagrangian}. We then give the explicit formulas for the partition functions we use in section~\ref{sec:equiv-index}. We also show the properties of the partition functions and the D-brane construction in section~\ref{sec:Nek-property} and section~\ref{sec:supergrp-Dbrane}. 
\subsection{Lagrangian}\label{sec:lagrangian}
Let us first review what a supergroup gauge theory is. Let $G$ be a Lie supergroup. In this paper we only discuss when $G=U(N_{+}\,|\,N_{-})$. Since we are considering elements of supergroups, the vector spaces appearing are supervector spaces or $\mathbb{Z}_{2}$-graded vector spaces (e.g. $\mathbb{C}^{N_{+}\,|\,N_{-}}$):
\begin{align}
    V=V_{+}\oplus V_{-},
\end{align}
where $V$ is a general supervector space and $V_{+},V_{-}$ are vector spaces of the elements with even (bosonic, positive) and odd (fermionic, negative) parities, respectively. Namely, an element $x\in V_{\sigma}$ is called even (resp. odd) when $\sigma=+$ (resp. $\sigma=-$). We will also identify $\sigma=\pm$ with $\sigma=\pm1$ in later calculations. 

We then introduce a dynamical gauge field $A_{\mu}$ taking values in the Lie algebra of $G$. The gauge transformation of it is $A_{\mu}\rightarrow gA_{\mu}g^{-1}+g\partial_{\mu}g^{-1}$, $g\in G$ and the gauge invariant Lagrangian is written as
\begin{align}
    S_{\text{YM}}=-\frac{1}{g^{2}_{\text{YM}}}\int d^{4}x~\Str(F\wedge\ast F),
\end{align}
where $\Str$ is the supertrace. The supertrace over the graded vector space $\mathbb{C}^{N_{+}\,|\,N_{-}}$ is defined as $\Str_{\mathbb{C}^{N_{+}\,|\,N_{-}}}=\Tr_{\mathbb{C}^{N_{+}}}-\Tr_{\mathbb{C}^{N_{-}}}$. Using this supertrace formula, the Lagrangian of supergroup gauge theory with $G=U(N_{+}\,|\,N_{-})$ can be rewritten as 
\begin{align}
     S_{\text{YM}}=-\frac{1}{g^{2}_{\text{YM}}}\int d^{4}x~\Tr_{\mathbb{C}^{N_{+}}} \left(F\wedge*F\right)^{+}-\left(-\frac{1}{g^{2}_{\text{YM}}}\right)\int d^{4}x~\Tr_{\mathbb{C}^{N_{-}}} \left(F\wedge*F\right)^{-},\label{eq:supergroup-decompose}
\end{align}
where the superscript of $\left(F\wedge*F\right)^{\pm}$ denotes the positive/negative contribution. Obviously, this theory is not unitary because the second term has a wrong sign in front of it, and the energy spectrum is not bounded from below. 

We further can introduce a topological term so-called the $\theta$-term
\begin{align}
    S_{\theta}=-\frac{i\theta}{8\pi^{2}}\int d^{4}x~\Str F\wedge F,
\end{align}
and consider non-perturbative configurations called instantons like the ordinary group gauge theory. Anti-self-dual (ASD)/self-dual (SD) YM configurations $\ast F=\mp F$ minimize the YM action for ordinary group gauge theory:
\begin{align}
     S_{\text{YM}}[A]&=-\frac{1}{2g^{2}_{\text{YM}}}\int d^{4}x\,\Tr(F\pm*F)\wedge*(F\pm*F)\pm\frac{1}{g^{2}_{\text{YM}}}\int d^{4}x\, \Tr(F\wedge F)\geq \frac{8\pi^{2}|k|}{g_{\text{YM}}^{2}},
\end{align}
where the minimized Lagrangian is described by the topological number
\begin{align}
     k&=\frac{1}{8\pi^{2}}\int d^{4}x~\Tr(F\wedge F).
\end{align} 
However, for the supergroup gauge theory, the ASD/SD configuration does not minimize the Lagrangian but rather gives a saddle point. This comes from the minus sign in front of the second term in (\ref{eq:supergroup-decompose}). Similar to the ordinary gauge theory case, the ASD/SD configuration is characterized by a topological number
\begin{align}
    k&=\frac{1}{8\pi^{2}}\int d^{4}x~\Str(F\wedge F)=k_{+}-k_{-},\label{eq:super-inst-numb}
\end{align}
where $k_{\pm}$ are interpreted as the positive and negative instanton numbers\footnote{See Table 3.1 of \cite{Kimura:instantoncounting} for the classification of instanton/anti-instanton, positive/negative instanton. The positive and negative instantons obey the ASD YM equation with $k_{+}>0$ and $k_{-}>0$, respectively. The positive anti-instanton and negative anti-instanton obey the SD YM equation with $k_{+}<0$ and $k_{-}<0$, respectively.}. 

We can explicitly compute the partition function with suitable amounts of supersymmetries under $\Omega$ background using the supersymmetric localization \cite{Nekrasov:2002qd,Nekrasov:2003rj,Kimura-Pestun:supergroup} (see \cite{Kimura:instantoncounting} for a nice review). The partition function is schematically evaluated as 
\begin{align}
\begin{split}
 \mathcal{Z}&=\int [\mathcal{D}A]\,e^{-S_{\text{tot}}}=\sum_{k}\mathfrak{q}^{k}\int[\mathcal{D}A^{(k)}_{\text{inst}}]\int[\mathcal{D}\delta A]\, e^{-S_{\text{fluc}}[\delta A]}=\mathcal{Z}_{\text{pert}}\sum_{k\in\mathbb{Z}}\mathfrak{q}^{k}\mathcal{Z}_{k},\\
 \mathcal{Z}_{k}&=\sum_{\substack{k_{+}-k_{-}=k\\ k_{\pm}\geq0}}Z_{k_{+}\,|\,k_{-}},\quad \mathfrak{q}=e^{2\pi i\tau},
\end{split}
\end{align}
where we denoted the complexified gauge constant as $\tau=\frac{\theta}{2\pi}+\frac{4\pi i}{g^{2}_{\text{YM}}}$. The detailed form of this partition function is derived in the next section.

\subsection{Instanton partition function from equivariant index formula}\label{sec:equiv-index}
\begin{figure}[t]
\centering
\tikzset{every picture/.style={line width=0.75pt}} 
\begin{tikzpicture}[x=0.75pt,y=0.75pt,yscale=-1,xscale=1]
\draw   (151.96,55.27) .. controls (151.96,36.04) and (168.14,20.44) .. (188.09,20.44) .. controls (208.05,20.44) and (224.22,36.04) .. (224.22,55.27) .. controls (224.22,74.51) and (208.05,90.11) .. (188.09,90.11) .. controls (168.14,90.11) and (151.96,74.51) .. (151.96,55.27) -- cycle ;
\draw   (304.22,55.29) .. controls (304.22,43.22) and (314.2,33.44) .. (326.52,33.44) .. controls (338.83,33.44) and (348.81,43.22) .. (348.81,55.29) .. controls (348.81,67.36) and (338.83,77.14) .. (326.52,77.14) .. controls (314.2,77.14) and (304.22,67.36) .. (304.22,55.29) -- cycle ;
\draw    (368.5,48.55) -- (401.5,48.55) ;
\draw [shift={(403.5,48.55)}, rotate = 180] [color={rgb, 255:red, 0; green, 0; blue, 0 }  ][line width=0.75]    (10.93,-3.29) .. controls (6.95,-1.4) and (3.31,-0.3) .. (0,0) .. controls (3.31,0.3) and (6.95,1.4) .. (10.93,3.29)   ;
\draw    (368.5,58.55) -- (401.5,58.55) ;
\draw [shift={(366.5,58.55)}, rotate = 0] [color={rgb, 255:red, 0; green, 0; blue, 0 }  ][line width=0.75]    (10.93,-3.29) .. controls (6.95,-1.4) and (3.31,-0.3) .. (0,0) .. controls (3.31,0.3) and (6.95,1.4) .. (10.93,3.29)   ;
\draw    (292.56,58.11) .. controls (243.72,63.72) and (241.92,40.58) .. (291.05,49.17) ;
\draw [shift={(292.56,49.44)}, rotate = 190.44] [color={rgb, 255:red, 0; green, 0; blue, 0 }  ][line width=0.75]    (10.93,-3.29) .. controls (6.95,-1.4) and (3.31,-0.3) .. (0,0) .. controls (3.31,0.3) and (6.95,1.4) .. (10.93,3.29)   ;
\draw    (469.5,60.55) .. controls (518.39,68.69) and (519.54,39.7) .. (470.39,50.11) ;
\draw [shift={(468.89,50.44)}, rotate = 347.39] [color={rgb, 255:red, 0; green, 0; blue, 0 }  ][line width=0.75]    (10.93,-3.29) .. controls (6.95,-1.4) and (3.31,-0.3) .. (0,0) .. controls (3.31,0.3) and (6.95,1.4) .. (10.93,3.29)   ;
\draw   (417.22,55.29) .. controls (417.22,43.22) and (427.2,33.44) .. (439.52,33.44) .. controls (451.83,33.44) and (461.81,43.22) .. (461.81,55.29) .. controls (461.81,67.36) and (451.83,77.14) .. (439.52,77.14) .. controls (427.2,77.14) and (417.22,67.36) .. (417.22,55.29) -- cycle ;
\draw (160.61,48.05) node [anchor=north west][inner sep=0.75pt]  [font=\footnotesize]  {$N_{i,+} |N_{i,-}$};
\draw (315.4,47.74) node [anchor=north west][inner sep=0.75pt]  [font=\footnotesize]  {$N_{i,+}$};
\draw (427.82,48.33) node [anchor=north west][inner sep=0.75pt]  [font=\footnotesize]  {$N_{i,-}$};
\draw (226.24,43.06) node [anchor=north west][inner sep=0.75pt]    {$=$};
\draw (260.79,22.36) node [anchor=north west][inner sep=0.75pt]  [font=\normalsize]  {$N_{++}$};
\draw (474.3,23.49) node [anchor=north west][inner sep=0.75pt]  [font=\normalsize]  {$N_{--}$};
\draw (370.12,66.74) node [anchor=north west][inner sep=0.75pt]  [font=\normalsize]  {$N_{-+}$};
\draw (369.55,22.97) node [anchor=north west][inner sep=0.75pt]  [font=\normalsize]  {$N_{+-}$};
\end{tikzpicture}
\caption{Quiver gauge theory interpretation of supergroup gauge theory}
\label{fig:quiver-supergroup}
\end{figure}
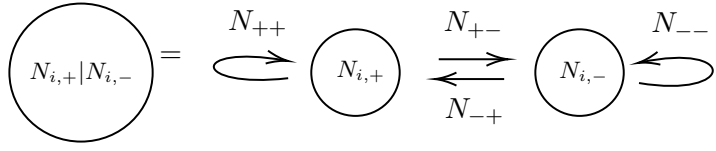

We use the equivariant index formalism to evaluate the instanton partition function of 5d $\mathcal{N}=1$ supersymmetric gauge theories with gauge groups $U(N_{+}\,|\,N_{-})$ following \cite{Kimura-Pestun:supergroup}. We consider generic quiver gauge theories. Let $\Gamma=(\Gamma_{0},\Gamma_{1})$ be a quiver, where $\Gamma_{0}=\{i\}$ and $\Gamma_{1}=\{e:i\rightarrow j\}$ are the set of nodes and edges, respectively. For each node, a supergroup $U(N_{i,+}|N_{i,-})$ is assigned. We denote the instanton and framing bundles over the instanton moduli space at the fixed point as $\mathbf{K}=(\mathbf{K}_{i})_{i\in\Gamma_{0}}$ and $\mathbf{N}=(\mathbf{N}_{i})_{i\in\Gamma_{0}}$. Since we are considering supergroups, the corresponding supercharacters are given by supertraces
\begin{align}
\sch\mathbf{N}_{i}=\ch{\bf{N}}_{i}^{+}-\ch{\bf{N}}_{i}^{-},\quad\sch\mathbf{K}_{i}=\ch{\bf{K}}_{i}^{+}-\ch{\bf{K}}_{i}^{-},\label{eq:schNK}
\end{align}
where each is defined as 
\begin{align}
    \ch\bfN^{\sigma}_{i}=\sum_{\alpha=1}^{N_{i,\sigma}}v_{i,\alpha}^{(\sigma)},\quad \ch\bfK^{\sigma}_{i}=\sum_{x\in\lambda_{i}^{(\sigma)}}\chi_{x}^{(\sigma)},\quad \sigma=\pm,\label{eq:chNK}
\end{align}
where 
\begin{align}
\begin{split}
    \chi_{x}^{(+)}&=v_{i,\alpha}^{\splus}q_{1}^{i-1}q_{2}^{j-1},\quad (\alpha,i,j)\in \lambda_{i,\alpha}^{\splus},\quad \alpha=1,\ldots,N_{i,+},\\
    \chi_{x}^{(-)}&=v_{i,\alpha}^{\sminus}q_{1}^{-i}q_{2}^{-j},\quad (\alpha,i,j)\in\lambda_{i,\alpha}^{\sminus},\quad \alpha=1,\ldots,N_{i,-},
\end{split}\label{eq:boxcoordinate}
\end{align}
and $\lambda^{(\pm)}_{i,\alpha}$ are Young diagrams (see Appendix~\ref{sec:appendix-Young}). The universal sheaf character is obtained as 
\begin{align}
    \sch\bfY_{i}=\sch\bfN_{i}-\ch\hspace{-0.6mm}\land\hspace{-0.5mm}\bfQ\,\sch\bfK_{i},
\end{align}
where 
\begin{align}
    \ch\hspace{-0.6mm}\wedge\hspace{-0.5mm}\bfQ=(1-q_{1})(1-q_{2}).
\end{align}
The dual of a bundle $\bfX$ is denoted as $\bfX^{\vee}$ and the character of it is obtained by taking the inverse as
\begin{align}
    \ch\bfX=\sum_{X}x,\quad \ch \bfX^{\vee}=\sum_{X}x^{-1}.
\end{align}
Using these supercharacters, partition functions are obtained by applying the Dolbeault index,
\begin{align}
    \bI[\bfX]=\prod_{x\in X}(1-x^{-1}).
\end{align}
Note that the $q$ parameters are related to the $\Omega$ background parameters $\epsilon_{1,2}$ as
\begin{align}
    q_{1}=e^{\epsilon_{1}},\quad q_{2}=e^{\epsilon_{2}}.
\end{align}
For later use, we also introduce another parameter $q_{3}$ obeying the condition
\begin{align}
   q_{3}=e^{\epsilon_{3}},\quad q_{1}q_{2}q_{3}=1.
\end{align}
\paragraph{Vector multiplet}
We first consider the vector multiplet contribution. The character formula is written as
\begin{align}
\begin{split}
        \sch\bfV_{i}&=\frac{\sch\bfY^{\vee}_{i}\sch\bfY_{i}}{\ch\swedge\bfQ}=\sch\bfV^{\text{inst}}_{i}+\sch\bfV^{\text{pert}}_{i},\\
        \sch\bfV^{\text{inst}}_{i}&=-\sch\bfN^{\vee}_{i}\sch\bfK_{i}-q_{3}\,\sch\bfK^{\vee}_{i}\sch\bfN_{i}+\ch\swedge\bfQ^{\vee}\sch\bfK^{\vee}_{i}\sch\bfK_{i},\\
        \sch\bfV^{\text{pert}}_{i}&=\frac{\sch \bfN_{i}^{\vee}\sch\bfN_{i}}{\ch\swedge\bfQ},
\end{split}
\end{align}
where we decomposed the character into the perturbative part\footnote{We note that the perturbative part we denote as $\bfX^{\text{pert}}$ for a bundle $\bfX$ in this section is the one-loop contribution and does not include the classical contribution.} and the instanton part. The perturbative part is the part not depending on the supercharacter $\sch\bfK_{i}$. We further can decompose the supercharacter using (\ref{eq:schNK})
\begin{align}
\begin{split}
        \sch\bfV^{\text{inst}}_{i}&=\sum_{\sigma,\sigma'=\pm}\sigma\sigma'\ch\bfV^{\text{inst}}_{i,\sigma\sigma'},\quad \sch\bfV_{i}^{\text{pert}}=\sum_{\sigma,\sigma'=\pm}\sigma\sigma'\ch\bfV_{i,\sigma\sigma'}^{\text{pert}},\\
        \ch\bfV^{\text{inst}}_{i,\sigma\sigma'}&=-\ch\bfN^{\sigma\vee}_{i}\ch\bfK^{\sigma'}_{i}-q_{3}\,\ch\bfK^{\sigma\vee}_{i}\ch\bfN^{\sigma'}_{i}+\ch\swedge\bfQ^{\vee}\ch\bfK^{\sigma\vee}_{i}\ch\bfK^{\sigma'}_{i},\\
        \ch\bfV^{\text{pert}}_{i,\sigma\sigma'}&=\frac{\ch \bfN_{i}^{\sigma\vee}\ch\bfN^{\sigma'}_{i}}{\ch\swedge\bfQ}.
\end{split}
\end{align}
    The instanton partition function is obtained by taking the index which eventually gives
    \begin{align}
    \begin{split}
        \mathcal{Z}^{\text{vec},\text{inst}}_{i}&=\prod_{\sigma,\sigma'=\pm}\mathcal{Z}^{\text{vec},\text{inst}}_{i,\sigma\sigma'},\quad \mathcal{Z}^{\text{vec},\text{inst}}_{i,\sigma\sigma'}=\bI\,[\sigma\sigma'\bfV^{\text{inst}}_{i,\sigma\sigma'}]=N_{\sigma\sigma'}(\vec{v}^{(\sigma)}_{i},\vec{\lambda}^{(\sigma)}_{i}\,|\, \vec{v}^{(\sigma')}_{i},\vec{\lambda}^{(\sigma')}_{i})^{-\sigma\sigma'}.
    \end{split}
    \end{align}
We defined here a generalized Nekrasov factor
\begin{align}
        N_{\sigma\sigma'}(\vec{v}^{(\sigma)}_{1},\vec{\lambda}^{(\sigma)}_{1}\,|\,\vec{v}^{(\sigma')}_{2},\vec{\lambda}^{(\sigma')}_{2})=\prod_{\alpha=1}^{N_{2,\sigma'}}\prod_{x\in\vec{\lambda}_{1}^{\sigma}}\left(1-\frac{\chi^{(\sigma)}_{x}}{q_{3}v^{(\sigma')}_{2,\alpha}}\right)\prod_{\beta=1}^{N_{1,\sigma}}\prod_{x\in\vec{\lambda}^{\sigma'}_{2}}\left(1-\frac{v^{(\sigma)}_{1,\beta}}{\chi^{(\sigma')}_{x}}\right)\prod_{\substack{x\in\vec{\lambda}_{1}^{(\sigma)}\\y\in\vec{\lambda}^{(\sigma')}_{2}}}S\left(\frac{\chi^{(\sigma)}_{x}}{\chi^{(\sigma')}_{y}}\right),\label{eq:generalized-Nekrasov-factor}
\end{align}
where
\begin{align}
        S(z)=\frac{(1-q_{1}z)(1-q_{2}z)}{(1-z)(1-q_{3}^{-1}z)}.
\end{align}
Note that the coordinates of the box $\chi_{x}^{(\sigma)}$ are defined as in (\ref{eq:boxcoordinate}). Explicitly, the total instanton partition function is written as 
    \begin{align}
        \mathcal{Z}^{\text{vec,inst}}_{i}=\frac{N_{\spplus\smminus}(\vec{v}^{\splus}_{i},\vec{\lambda}^{\splus}_{i}\,|\, \vec{v}^{\sminus}_{i},\vec{\lambda}^{\sminus}_{i})N_{\smminus\spplus}(\vec{v}^{\sminus}_{i},\vec{\lambda}^{\sminus}_{i}\,|\, \vec{v}^{\splus}_{i},\vec{\lambda}^{\splus}_{i})}{N_{\spplus\spplus}(\vec{v}^{\splus}_{i},\vec{\lambda}^{\splus}_{i}\,|\, \vec{v}^{\splus}_{i},\vec{\lambda}^{\splus}_{i})N_{\smminus\smminus}(\vec{v}^{\sminus}_{i},\vec{\lambda}^{\sminus}_{i}\,|\, \vec{v}^{\sminus}_{i},\vec{\lambda}^{\sminus}_{i})}.
    \end{align} 
An observation is that Nekrasov factors with the same parities in the subindex (e.g. $N_{++},N_{--}$) are in the denominator while Nekrasov factors with different parities (e.g. $N_{+-},N_{-+}$) are in the numerator. Thus, this contribution can be understood as a quiver gauge theory with two gauge nodes $U(N_{i,+})\times U(N_{i,-})$, where off-diagonal contributions play the role of bifundamental matters connecting the two gauge nodes (see Figure \ref{fig:quiver-supergroup}) \cite{Dijkgraaf:2016lym}. 
    
The perturbative part is read
    \begin{align}
    \begin{split}
        \mathcal{Z}^{\text{vec,pert}}_{i}&=\prod_{\sigma,\sigma'=\pm}\mathcal{Z}^{\text{vec,pert}}_{i,\sigma\sigma'},\\ \mathcal{Z}^{\text{vec,pert}}_{i,\sigma\sigma'}&=\prod_{\alpha=1}^{N_{i,\sigma}}\prod_{\beta=1}^{N_{i,\sigma'}}\prod_{k,k'=1}^{\infty}\left(1-\frac{v_{i,\alpha}^{(\sigma)}}{v_{i,\beta}^{(\sigma')}}q_{1}^{k}q_{2}^{k'}\right)^{\sigma\sigma'}=\prod_{\alpha=1}^{N_{i,\sigma}}\prod_{\beta=1}^{N_{i,\sigma'}}\mathcal{G}\left(q_{1}q_{2}\frac{v_{i,\alpha}^{(\sigma)}}{v^{(\sigma')}_{i,\beta}}\right)^{\sigma\sigma'},
    \end{split}
    \end{align}
    where
    \begin{align}
        \mathcal{G}(z)=\mathcal{G}(z;q_{1},q_{2})=\exp\left(-\sum_{n=1}^{\infty}\frac{z^{n}}{n(1-q_{1}^{n})(1-q_{2}^{n})}\right)   
    \end{align}
    and we assumed $|v_{i,\alpha}^{(\sigma)}/v_{i,\beta}^{(\sigma')}|<1$.
For the diagonal term $\sigma=\sigma'$, this diverge and we have to remove the contribution when $\alpha=\beta$. When some of the Coulomb branch parameters are $|v_{i,\alpha}^{(\sigma)}/v_{i,\beta}^{(\sigma')}|>1$, we do analytic continuation as
\begin{align}
    \mathcal{G}\left(q_{1}q_{2}\frac{v^{(\sigma)}_{i,\alpha}}{v_{i,\beta}^{(\sigma')}}\right)\rightarrow \mathcal{G}\left(\frac{v_{i,\beta}^{(\sigma')}}{v_{i,\alpha}^{(\sigma)}}\right).
\end{align}
See Appendix~\ref{sec:appendix-specialfunc} for details.
\paragraph{ Bifundamental hypermultiplet}
Let us next discuss the contribution from the bifundamental hypermultiplet. The strategy is exactly the same and the character formula is written as 
    \begin{align}
    \begin{split}
        \sch\bfH_{e:i\rightarrow j}&=-\ch\bfM_{e}\frac{\sch\bfY^{\vee}_{i}\sch\bfY_{j}}{\ch\swedge\bfQ}=\sch \bfH_{e:i\rightarrow j}^{\text{inst}}+\sch\bfH_{e:i\rightarrow j}^{\text{pert}},\\
        \sch\bfH_{e:i\rightarrow j}^{\text{inst}}&=\ch\bfM_{e}\left(\sch\bfN^{\vee}_{i}\sch\bfK_{j}+q_{3}\,\sch\bfK^{\vee}_{i}\sch\bfN_{j}-\ch\swedge\bfQ^{\vee}\sch\bfK^{\vee}_{i}\sch\bfK_{j}\right),\\
        \sch\bfH^{\text{pert}}_{e:i\rightarrow j}&=-\ch\bfM_{e}\frac{\sch\bfN_{i}^{\vee}\sch\bfN_{j}}{\ch\swedge\bfQ},
    \end{split}
    \end{align}
where
\begin{align}
    \ch\bfM_{e}=\mu_{e},\quad e\in \Gamma_{1}
\end{align}
is the bifundamental mass. Expanding the supercharacters and taking the index gives
\begin{align}
\begin{split}
    \mathcal{Z}^{\text{bf,inst}}_{e:i\rightarrow j}&=\prod_{\sigma,\sigma'=\pm} \mathcal{Z}^{\text{bf,inst}}_{e:i\rightarrow j,\sigma\sigma'},\quad \mathcal{Z}^{\text{bf,inst}}_{e:i\rightarrow j,\sigma\sigma'}=\bI[\sigma\sigma'\bfH^{\text{bf,inst}}_{e,\sigma\sigma'}]=N_{\sigma\sigma'}(\vec{v}^{(\sigma)}_{i},\vec{\lambda}^{(\sigma)}_{i}\,|\,\mu_{e:i\rightarrow j}\vec{v}^{(\sigma')}_{j},\vec{\lambda}^{(\sigma')}_{j})^{\sigma\sigma'}.
\end{split}
\end{align}
Similarly, the perturbative contribution is given 
    \begin{align}
    \begin{split}
        \mathcal{Z}^{\text{bf,pert}}_{e:i\rightarrow j}&=\prod_{\sigma,\sigma'=\pm}\mathcal{Z}^{\text{bf,pert}}_{e:i\rightarrow j,\sigma\sigma'},\\ \mathcal{Z}^{\text{bf,pert}}_{e:i\rightarrow j,\sigma\sigma'}&=\prod_{\alpha=1}^{N_{i,\sigma}}\prod_{\beta=1}^{N_{j,\sigma'}}\prod_{k,k'=1}^{\infty}\left(1-\mu_{e}^{-1}\frac{v_{i,\alpha}^{(\sigma)}}{v_{j,\beta}^{(\sigma')}}q_{1}^{k}q_{2}^{k'}\right)^{-\sigma\sigma'}=\prod_{\alpha=1}^{N_{i,\sigma}}\prod_{\beta=1}^{N_{j,\sigma'}}\mathcal{G}\left(\frac{q_{1}q_{2}}{\mu_{e}}\frac{v^{(\sigma)}_{i,\alpha}}{v^{(\sigma')}_{j,\beta}}\right)^{-\sigma\sigma'}.
    \end{split}
    \end{align}
Similar to the vector multiplet contribution, we do analytic continuation depending on the ratio of the Coulomb branch parameters.
\paragraph{Fundamental, antifundamental hypermultiplets}
The contributions of the fundamental and antifundamental hypermultiplets come from the following character formulas. For the fundamental hypermultiplets, we have
\begin{align}
\begin{split}
    \sch\bfH^{\text{f}}_{i}&=-\frac{\sch\bfY^{\vee}_{i}\sch\bfM_{i}}{\ch\swedge\bfQ}=\sch\bfH^{\text{f,inst}}_{i}+\sch\bfH^{\text{f,pert}}_{i},\\
    \sch\bfH^{\text{f,inst}}_{i}&=q_{3}\,\sch\bfK^{\vee}_{i}\sch\bfM_{i},\quad \sch\bfH_{i}^{\text{f,pert}}=-\frac{\sch\bfN_{i}^{\vee}\sch\bfM_{i}}{\ch\swedge\bfQ},\end{split}
\end{align}
and for the antifundamental hypermultiplets, we have
\begin{align}
\begin{split}
    \sch\bfH^{\text{af}}_{i}&=-\frac{\sch\wt{\bfM}^{\vee}_{i}\sch\bfY_{i}}{\ch\swedge\bfQ}=\sch\bfH^{\text{af,inst}}_{i}+\sch\bfH^{\text{af,pert}}_{i},\\
    \sch\bfH^{\text{af,inst}}_{i}&=\sch\wt{\bfM}^{\vee}_{i}\sch\bfK_{i},\quad \sch\bfH_{i}^{\text{af,pert}}=-\frac{\sch\wt{\bfM}_{i}^{\vee}\sch\bfN_{i}}{\ch\swedge\bfQ},
\end{split}
\end{align}
where
\begin{align}
    \sch\bfM_{i}=\sum_{f=1}^{N^{\text{f}}_{i,+}}\mu_{i,f}^{\spplus}-\sum_{f=1}^{N^{\text{f}}_{i,-}}\mu^{\smminus}_{i,f},\quad \sch\wt{\bfM}_{i}=\sum_{f=1}^{N^{\text{af}}_{i,\spplus}}\wt{\mu}^{\spplus}_{i,f}-\sum_{f=1}^{N^{\text{af}}_{i,\smminus}}\wt{\mu}^{\smminus}_{i,f}.
\end{align}
Comparing with the bifundamental matters, the difference is that the bifundamental mass is set to be $\mu_{e}=1$ and one of the $\sch\bfY_{i}$ in the numerator is converted to $\sch\bfM_{i}$ or $\sch\wt{\bfM}_{i}$, which are independent of the instanton contribution. Namely, after freezing the gauge degrees of freedom of either the source or target gauge node, we obtain the fundamental and antifundamental contribution. 

Taking the index, the instanton partition function coming from the fundamental/antifundamental multiplet will be 
\begin{align}
\begin{split}
        \mathcal{Z}^{\text{inst}}_{\text{f},i}&=\prod_{\sigma,\sigma'=\pm}\mathcal{Z}^{\text{inst}}_{\text{f},i,\sigma\sigma'},\quad \mathcal{Z}^{\text{inst}}_{\text{af},i}=\prod_{\sigma,\sigma'=\pm}\mathcal{Z}^{\text{inst}}_{\text{af},i,\sigma\sigma'},\\
        \mathcal{Z}^{\text{inst}}_{\text{f},i,\sigma\sigma'}&=\bI[\sigma\sigma'q_{3}\,\ch\bfK^{\sigma\vee}_{i}\ch\bfM^{\sigma'}_{i}]=N_{\sigma\sigma'}(\vec{v}^{(\sigma)}_{i},\vec{\lambda}^{(\sigma)}_{i}\,|\,\vec{\mu}_{i}^{\sigma'},\vec{\emptyset})^{\sigma\sigma'}=\prod_{f=1}^{N^{\text{f}}_{i,\sigma'}}\prod_{x\in\vec{\lambda}_{i}^{\sigma}}\left(1-\frac{\chi^{(\sigma)}_{x}}{q_{3}\mu^{\sigma'}_{i,f}}\right)^{\sigma\sigma'},\\
        \mathcal{Z}^{\text{inst}}_{\text{af},i,\sigma\sigma'}&=\bI[\sigma\sigma'\ch\widetilde{\bfM}^{\vee,\sigma}_{i}\ch\bfK^{\sigma'}_{i}]=N_{\sigma\sigma'}(\vec{\Tilde{\mu}}^{\sigma}_{i},\vec{\emptyset}\,|\,\vec{v}^{(\sigma')}_{i},\vec{\lambda}_{i}^{(\sigma')})^{\sigma\sigma'}=\prod_{\Tilde{f}=1}^{N^{\text{af}}_{i,\sigma}}\prod_{x\in\vec{\lambda}^{\sigma'}_{i}}\left(1-\frac{\Tilde{\mu}^{\sigma}_{i,\Tilde{f}}}{\chi^{(\sigma')}_{x}}\right)^{\sigma\sigma'}.
\end{split}
\end{align}
The perturbative contributions are read as
\begin{align}
    \begin{split}
        \mathcal{Z}^{\text{pert}}_{\text{f},i}&=\prod_{\sigma,\sigma'=\pm}\mathcal{Z}^{\text{pert}}_{\text{f},i,\sigma\sigma'},\quad \mathcal{Z}^{\text{pert}}_{\text{af},i}=\prod_{\sigma,\sigma'=\pm}\mathcal{Z}^{\text{pert}}_{\text{af},i,\sigma\sigma'},\\
        \mathcal{Z}_{\text{f},i,\sigma\sigma'}^{\text{pert}}&=\prod_{\alpha=1}^{N_{i,\sigma}}\prod_{f=1}^{N^{f}_{i,\sigma'}}\prod_{k,k'=1}^{\infty}\left(1-\frac{v^{(\sigma)}_{i,\alpha}}{\mu^{\sigma'}_{i,f}}q_{1}^{k}q_{2}^{k'}\right)^{-\sigma\sigma'}=\prod_{\alpha=1}^{N_{i,\sigma}}\prod_{f=1}^{N^{f}_{i,\sigma'}}\mathcal{G}\left(q_{1}q_{2}\frac{v^{(\sigma)}_{i,\alpha}}{\mu^{\sigma'}_{i,f}}\right)^{-\sigma\sigma'},\\
        \mathcal{Z}_{\text{af},i,\sigma\sigma'}&=\prod_{\alpha=1}^{N_{i,\sigma}}\prod_{f=1}^{N^{\text{af}}_{i,\sigma'}}\prod_{k,k'=1}^{\infty}\left(1-\frac{\Tilde{\mu}^{\sigma'}_{i,f}}{v^{(\sigma)}_{i,\alpha}}q_{1}^{k}q_{2}^{k'}\right)^{-\sigma\sigma'}=\prod_{\alpha=1}^{N_{i,\sigma}}\prod_{f=1}^{N^{\text{af}}_{i,\sigma'}}\mathcal{G}\left(q_{1}q_{2}\frac{\Tilde{\mu}_{i,f}^{\sigma'}}{v^{(\sigma)}_{i,\alpha}}\right)^{-\sigma\sigma'}.
    \end{split}
\end{align}
Similar to other perturbative contributions, analytic continuation will be done if needed.
\paragraph{Topological term}
The topological term is a term counting the number of instantons. As mentioned in (\ref{eq:super-inst-numb}), the topological contribution is characterized by the difference between the positive and negative instanton numbers. When the gauge groups are $U(N_{+}\,|\,N_{-})$, the instantons are characterized by Young diagrams (the bundle $\bfK_{i}$), and thus, the topological term has the following form:
    \begin{align}
        \mathcal{Z}_{i,\text{top}}=\prod_{i=1}^{N}\mathfrak{q}_{i}^{|\vec{\lambda}^{\splus}_{i}|-|\vec{\lambda}^{\sminus}_{i}|},\quad |\vec{\lambda}^{\spm}_{i}|=\sum_{\alpha=1}^{N_{i,\pm}}|\lambda^{\spm}_{i,\alpha}|,
    \end{align}  
where the positive (negative) instanton numbers are identified as $k_{i}^{\pm}=|\vec{\lambda}^{\spm}_{i}|$.
\paragraph{Chern-Simons term}
For 5d gauge theory, we further can add a Chern-Simons term to the Lagrangian as the ordinary group gauge theory \cite{Nakajima:2005fg,Nekrasov:1996cz,Tachikawa:2004ur}. It is labeled by the integer called the Chern-Simons level assigned to each node $(\kappa_{i}^{\sigma})_{\substack{i\in\Gamma_{0},\sigma=\pm}}$. The instanton contribution is written as
\begin{align}
    \mathcal{Z}_{i}^{\text{CS}}&=\prod_{\sigma=\pm}\mathcal{Z}_{i,\sigma}^{\text{CS}},\quad \mathcal{Z}^{\text{CS}}_{i,\sigma}=\prod_{x\in\vec{\lambda}^{(\sigma)}_{i}}\left(\chi^{(\sigma)}_{x}\right)^{\sigma\kappa^{\sigma}_{i}}.
\end{align}
In the original paper \cite{Kimura-Pestun:supergroup}, the authors introduced two integers for each node, giving independent Chern-Simons levels for the positive and negative contributions. However, it seems that when we consider the intertwiner formalism, the Chern-Simons level appearing will be\footnote{This phenomenon itself is nothing special in the context of ABJM theory and supermatrix models \cite{Aharony:2008ug,Marino:2009jd,Drukker:2009hy}, and it actually comes from the gauge invariance of the Chern-Simons term with supergroup. We thank Taro Kimura for pointing this out.} 
\begin{align}
    \kappa_{i}^{+}=\kappa_{i}^{-}\coloneqq\kappa_{i},\quad \forall i\in \Gamma_{0}. \label{eq:CScond}
\end{align}
We will always impose this condition throughout this paper.

\paragraph{Convention}
For later convenience, we introduce a generalized notation. In the original reference \cite{Kimura-Pestun:supergroup}, the authors gathered the Chern roots with even and odd parities. The order of these roots is related to the order of D-branes, so we will not specify the order of the roots but rather use the following notation. Consider a supergroup quiver gauge theory whose gauge nodes are $U(N_{i,+}\,|\,N_{i,-})$ with the condition $N_{i,+}+N_{i,-}=N_{i}$. There are $N_{i}$ Coulomb branch parameters which we denote $v_{i\alpha}\,(\alpha=1,\ldots,N_{i})$. Each of the elements belongs to either $\bfN^{+}_{i}$ or $\bfN^{-}_{i}$ depending on their parities. We denote the parity for $v_{i\alpha}$ as $\sigma_{i\alpha}=\pm$. We further introduce $N_{i}$-tuples of Young diagrams $\lambda_{i\alpha}\,(\alpha=1,\ldots,N_{i})$ which label the fix points of the $U(1)^{2}$ torus actions: 
\begin{align}
    \vec{v}_{i}=(v_{i1},v_{i2},\ldots,v_{iN_{i}}),\quad \vec{\lambda}_{i}=(\lambda_{i1},\lambda_{i2},\ldots,\lambda_{iN_{i}}),\quad \vec{\sigma}_{i}=(\sigma_{i1},\sigma_{i2},\ldots,\sigma_{iN_{i}}).
\end{align}
Then, the basic supercharacters are written as 
\begin{align}
\sch\bfN_{i}&=\sum_{\alpha=1}^{N_{i}}\sigma_{i\alpha}v_{i\alpha},\quad \sch\bfK_{i}=\sum_{\alpha=1}^{N_{i}}\sigma_{i\alpha}\sum_{x\in\lambda_{i\alpha}}\chi_{x}^{(\sigma_{i\alpha})}.
\end{align}
For fundamental and antifundamental contributions, we introduce the mass parameters and their parities as
\begin{align}
\begin{split}
    &\vec{\mu}^{(\text{f})}_{i}=(\mu^{(\text{f})}_{i1},\mu^{(\text{f})}_{i2},\ldots,\mu^{(\text{f})}_{iN^{\text{f}}_{i}}),\quad \vec{\sigma}^{(\text{f})}_{i}=(\sigma^{(\text{f})}_{i1},\sigma^{(\text{f})}_{i2},\ldots,\sigma^{(\text{f})}_{iN^{\text{f}}_{i}}),\\ &\vec{\wt{\mu}}^{(\text{af})}_{i}=(\wt{\mu}^{(\text{af})}_{i1},\wt{\mu}^{(\text{af})}_{i2},\ldots,\wt{\mu}^{(\text{af})}_{iN^{\text{af}}_{i}}),\quad \vec{\wt{\sigma}}_{i}^{(\text{af})}=(\wt{\sigma}^{(\text{af})}_{i1},\wt{\sigma}^{(\text{af})}_{i2},\ldots,\wt{\sigma}^{(\text{af})}_{iN^{\text{af}}_{i}}),
\end{split}
\end{align}
and then, the supercharacters are written as 
\begin{align}
    \sch\bfM_{i}=\sum_{f=1}^{N^{\text{f}}_{i}}\sigma^{(\text{f})}_{if}\mu_{if}^{(\text{f})},\quad \sch\wt{\bfM}_{i}=\sum_{f=1}^{N^{\text{af}}_{i}}\wt{\sigma}^{(\text{af})}_{if}\wt{\mu}_{if}^{(\text{af})}.
\end{align}
Using these conventions, the instanton partition functions of the topological term, Chern-Simons term, bifundamental, vector, and (anti)fundamental contributions are rewritten as 
\begin{align}
\begin{split}
&\mathcal{Z}_{\text{top.}}(\vec{\lambda}_{i},\vec{\sigma}_{i})=\mathfrak{q}_{i}^{\sum\limits_{a=1}^{N_{i}}\sigma_{ia}|\lambda_{i,a}|},\quad\mathcal{Z}_{\text{CS}}(\kappa_{i},\vec{\lambda}_{i},\vec{\sigma}_{i})=\prod_{j=1}^{N_{i}}\prod_{x\in\lambda_{ij}}\left(\chi_{x}^{(\sigma_{ij})}\right)^{\sigma_{ij}\kappa_{i}}, \\
&\mathcal{Z}_{\text{bfd.}}(\vec{v}_{i},\vec{\lambda}_{i},\vec{\sigma}_{i}\,|\,\vec{v}_{j},\vec{\lambda}_{j},\vec{\sigma}_{j}\,|\,\mu_{e:i\rightarrow j})=\prod_{a=1}^{N_{i}}\prod_{b=1}^{N_{j}}N_{\sigma_{ia}\sigma_{jb}}(v_{ia},\lambda_{ia}\,|\,\mu_{e:i\rightarrow j}v_{jb},\lambda_{jb})^{\sigma_{ia}\sigma_{jb}},\\
&\mathcal{Z}_{\text{vec.}}(\vec{v}_{i},\vec{\lambda}_{i},\vec{\sigma}_{i})=\mathcal{Z}_{\text{bfd.}}(\vec{v}_{i},\vec{\lambda}_{i},\vec{\sigma}_{i}\,|\,\vec{v}_{i},\vec{\lambda}_{i},\vec{\sigma}_{i}\,|\,1)^{-1}=\prod_{a,b=1}^{N_{i}}N_{\sigma_{ia}\sigma_{ib}}(v_{ia},\lambda_{ia}\,|\,v_{ib},\lambda_{ib})^{-\sigma_{ia}\sigma_{ib}},\\
&\mathcal{Z}_{\text{f}}(\vec{v}_{i},\vec{\lambda}_{i},\vec{\sigma}_{i}\,|\,\vec{\mu}^{(\text{f})}_{i},\vec{\sigma}^{(\text{f})}_{i})=\mathcal{Z}_{\text{bfd.}}(\vec{v}_{i},\vec{\lambda}_{i},\vec{\sigma}_{i}\,|\,\vec{\mu}^{(\text{f})}_{i},\vec{\emptyset},\vec{\sigma}^{(\text{f})}_{i}\,|\,1),\\
&\mathcal{Z}_{\text{af}}(\vec{v}_{i},\vec{\lambda}_{i},\vec{\sigma}_{i}\,|\,\vec{\wt{\mu}}^{(\text{af})}_{i},\vec{\wt{\sigma}}^{(\text{af})}_{i})=\mathcal{Z}_{\text{bfd.}}(\vec{\wt{\mu}}_{i}^{(\text{af})},\vec{\emptyset},\vec{\wt{\sigma}}^{(\text{af})}_{i}\,|\,\vec{v}_{i},\vec{\lambda}_{i},\vec{\sigma}_{i}\,|\,1),
\end{split}\label{eq:Nekrasov-convention}
\end{align}
where 
\begin{align}
\begin{split}
    N_{\sigma\sigma'}(v_{1},\lambda\,|\,v_{2},\nu)
    &=\prod_{x\in\lambda}\left(1-\frac{\chi_{x}^{(\sigma)}}{q_{3}v_{2}}\right)\prod_{x\in\nu}\left(1-\frac{v_{1}}{\chi_{x}^{(\sigma')}}\right)\prod_{\substack{x\in\lambda\\y\in\nu}}S\left(\frac{\chi_{x}^{(\sigma)}}{\chi_{y}^{(\sigma')}}\right).
\end{split}\label{eq:def-Nekrasov}
\end{align}
The generalized Nekrasov factor of (\ref{eq:generalized-Nekrasov-factor}) is just a composition of this Nekrasov factor (\ref{eq:def-Nekrasov}). Equivalent expressions of these Nekrasov factors are in Appendix~\ref{sec:appendix-nekrasov}.

Using this notation, the total instanton partition function is written as
\begin{align}
\begin{split}
    \mathcal{Z}_{\text{inst.}}[\Gamma]=&\sum_{\{\vec{\lambda}_{i}\}}\prod_{i\in\Gamma_{0}}\mathcal{Z}_{\text{top.}}(\vec{\lambda}_{i},\vec{\sigma}_{i})\mathcal{Z}_{\text{CS}}(\kappa_{i},\vec{\lambda}_{i},\vec{\sigma}_{i})\mathcal{Z}_{\text{vec.}}(\vec{v}_{i},\vec{\lambda}_{i},\vec{\sigma}_{i})\\
    \times&\prod_{i\in\Gamma_{0}}\mathcal{Z}_{\text{f}}(\vec{v}_{i},\vec{\lambda}_{i},\vec{\sigma}_{i}\,|\,\vec{\mu}^{(\text{f})}_{i},\vec{\sigma}^{(\text{f})}_{i})\mathcal{Z}_{\text{af}}(\vec{v}_{i},\vec{\lambda}_{i},\vec{\sigma}_{i}\,|\,\vec{\wt{\mu}}^{(\text{af})}_{i},\vec{\wt{\sigma}}^{(\text{af})}_{i})\\
    \times&\prod_{e:i\rightarrow j\in\Gamma_{1}}\mathcal{Z}_{\text{bfd.}}(\vec{v}_{i},\vec{\lambda}_{i},\vec{\sigma}_{i}\,|\,\vec{v}_{j},\vec{\lambda}_{j},\vec{\sigma}_{j}\,|\,\mu_{e:i\rightarrow j}).
\end{split}\label{eq:supergrp-gen-inst}
\end{align}
\subsection{Properties of the instanton partition functions}\label{sec:Nek-property}
In this section, we briefly study two properties of the instanton partition function in (\ref{eq:supergrp-gen-inst}) which we will use in other sections.
\paragraph{Bifundamental mass}
We can set the bifundamental mass to be $\mu_{e:i\rightarrow j}=\gamma^{-1}$ by using the symmetry of the Nekrasov partition functions. The partition function is invariant under the following transformation as long as we impose the condition (\ref{eq:CScond}) on the Chern Simons levels:
\begin{align}
\begin{split}
    \vec{v}_{i}\rightarrow \alpha_{i}\vec{v}_{i},\quad \vec{\mu}_{i,\text{f}}&\rightarrow \alpha_{i}\vec{\mu}_{i,\text{f}},\quad \vec{\mu}_{i,\text{af}}\rightarrow \alpha_{i}\vec{\mu}_{i,\text{af}},\\
    \mu_{e:i\rightarrow j}\rightarrow \frac{\alpha_{i}}{\alpha_{j}}\mu_{e:i\rightarrow j},&\quad \mathfrak{q}_{i}\rightarrow \alpha_{i}^{-\kappa_{i}}\mathfrak{q}_{i}.
\end{split}\label{eq:Nek-shift-symm}
\end{align}
This is a generalized version of the symmetry used in \cite{Awata:2011ce,Mironov:2016yue,Awata:2016riz,Bourgine:2017jsi} to study the relation with quantum algebras. If we want to set the Chern-Simons levels to be independent with each other $\kappa^{+}_{i}\neq \kappa^{-}_{i}$, to manifest the symmetry (\ref{eq:Nek-shift-symm}), we need to modify the topological term as ${\mathfrak{q}^{\splus}_{i}}^{|\vec{\lambda}^{\splus}_{i}|}{\mathfrak{q}^{\sminus}_{i}}^{-|\vec{\lambda}^{\sminus}_{i}|}$. Then the symmetry transformation of the topological term above is changed to 
\begin{align}
    \mathfrak{q}^{\splus}_{i}\rightarrow \alpha_{i}^{-\kappa^{\spplus}_{i}}\mathfrak{q}^{\splus}_{i},\quad \mathfrak{q}^{\sminus}_{i}\rightarrow \alpha_{i}^{-\kappa^{\smminus}_{i}}\mathfrak{q}^{\sminus}_{i}.
\end{align}
However, since the instanton number is (\ref{eq:super-inst-numb}), it is natural to use the first scenario (\ref{eq:Nek-shift-symm}) instead of introducing two different instanton counting parameters $\mathfrak{q}^{\spm}_{i}$. Actually, only the first scenario seems to appear in the intertwiner formalism we will discuss later. Therefore, we will use only one instanton counting parameter and one Chern-Simons level for each gauge node from now on. Using the symmetry (\ref{eq:Nek-shift-symm}), we can tune the bifundamental mass to be $\gamma^{-1}$. We always set the bifundamental mass to be this value from now on.
\paragraph{Recursive relation of the Nekrasov factor}
From the explicit formula of the Nekrasov factor (\ref{eq:def-Nekrasov}), we obviously have the following recursive relations
\begin{align}
\begin{split}
    \frac{N_{\sigma\sigma'}(v_{1},\lambda_{1}+x\,|\, v_{2},\lambda_{2})}{N_{\sigma\sigma'}(v_{1},\lambda_{1}\,|\,v_{2},\lambda_{2})}&=\left(-\frac{\chi_{x}^{(\sigma)}}{q_{3}v_{2}}\right)\mathcal{Y}^{(\sigma')}_{\lambda_{2}}(q_{3}^{-1}\chi_{x}^{(\sigma)}),\\
    \frac{N_{\sigma\sigma'}(v_{1},\lambda_{1}\,|\,v_{2},\lambda_{2}+x)}{N_{\sigma\sigma'}(v_{1},\lambda_{1}\,|\,v_{2},\lambda_{2})}&=\mathcal{Y}_{\lambda_{1}}^{(\sigma)}(\chi_{x}^{(\sigma')}),
\end{split}
\end{align}
where 
\begin{align}
    \mathcal{Y}_{\lambda}^{(\sigma)}(z)=\left(1-v/z\right)\prod_{x\in\lambda}S\left(\chi_{x}^{(\sigma)}/z\right).\label{eq:Yfunction-def}
\end{align}
These are the $\mathsf{Y}$ functions introduced in \cite{Kimura-Pestun:supergroup} (see also \cite{Poghossian:2010pn,Fucito:2011pn,Nekrasov:2013xda}). They are related to the $qq$ characters and the deformed Seiberg-Witten curve \cite{Poghossian:2010pn,Fucito:2011pn,Nekrasov:2013xda,Nekrasov:2015wsu}. They will appear as a fundamental element in our construction, too. 
\subsection{D-brane construction of supergroup gauge theories}\label{sec:supergrp-Dbrane}
We briefly review D-brane constructions of A-type supergroup gauge theories with A-quiver and D-quiver structures \cite{Hanany:1996ie,Vafa:2001qf,Okuda:2006fb,Mikhaylov:2014aoa,Dijkgraaf:2016lym}.

4d $\mathcal{N}=2$ $SU(N_{+}\,|\,N_{-})$ Yang-Mills theory is realized on the world-volume of positive (ordinary) $\text{D}4^{+}$ branes and negative (ghost) $\text{D}4^{-}$ branes suspended between NS5-branes in type \rom{2}A theory as
\begin{align}
    \begin{tikzpicture}[thick,cross/.style={path picture={ 
  \draw[black]
(path picture bounding box.south east) -- (path picture bounding box.north west) (path picture bounding box.south west) -- (path picture bounding box.north east);
}}]
            \begin{scope}[scale=1.5,xscale=1.2,yscale=0.9]
            \draw[] (-0.9,1.2)--(0.9,1.2);
                \draw[] (-0.9,0.8)--(0.9,0.8);
                \draw[] (-0.9,0.4)--(0.9,0.4);
                \draw[] (-0.9,0)--(0.9,0);
                \draw[dashed] (-0.9,-0.4)--(0.9,-0.4);
                \draw[dashed] (-0.9,-0.8)--(0.9,-0.8);
                \draw[dashed] (-0.9,-1.2)--(0.9,-1.2);
		        \draw[] (0.9,1.5)--(0.9,0)--(0.9,-1.5);
		        \draw[] (-0.9,1.5) -- (-0.9,0)--(-0.9,-1.5);
		        \node[above] at (-0.9,1.5){NS5};
		        \node[above] at (0.9,1.5){NS5};
		        \draw [decorate,decoration={brace,amplitude=5pt,raise=4pt},yshift=0pt] (0.9,1.23) -- (0.9,-0.03) node [xshift=0.8cm ,midway] {};
		        \node[right] at (1.1,0.6) {$N_{+}\hspace{0.2cm} \text{D4}^{+}$};
		        \draw [decorate,decoration={brace,amplitude=5pt,raise=4pt},yshift=0pt] (0.9,-0.37) -- (0.9,-1.23) node [xshift=0.8cm ,midway] {};
		        \node[right] at (1.1,-0.8) {$N_{-}\hspace{0.2cm} \text{D4}^{-}$};
            \end{scope}
\end{tikzpicture} \label{eq:HW-pureSYM}
\end{align}
where we illustrated the $\text{D4}^{+}$ branes in solid horizontal lines, the $\text{D4}^{-}$ branes in dashed horizontal lines, and the NS5 branes in solid vertical lines \cite{Hanany:1996ie,Dijkgraaf:2016lym}. Note also that generally we can order the D-branes in any order but the order above was chosen for simplicity.

\paragraph{$A_{r}$ quiver gauge theory}
The difference with the ordinary group gauge theory is that negative branes are also included in the setup. Similarly to the ordinary group case, we can construct linear quiver supergroup gauge theories by including these negative branes as
\begin{align}
\adjustbox{valign=c}{
    \begin{tikzpicture}[thick,roundnode/.style={circle,fill=blue!50,draw=black},
squarednode/.style={rectangle,draw=black}]
            \begin{scope}[scale=2.3]
		        \draw (-1,0.8)--(-1,0)--(-1,-0.9);
            \draw[] (0,0.5) -- (-1,0.5);
                \draw[] (0,0.3) -- (-1,0.3);
                \draw[] (0,0.1) -- (-1,0.1);
                \draw[] (0,-0.1) -- (-1,-0.1);
                \draw[dashed] (0,-0.3) -- (-1,-0.3);
                \draw[dashed] (0,-0.5) -- (-1,-0.5);
		        \draw[dashed] (0,-0.7) -- (-1,-0.7);
	         \draw[] (1,0.6) -- (0,0.6);
                \draw[] (1,0.4) -- (0,0.4);
                \draw[] (1,0.2) -- (0,0.2);
                \draw[] (1,0) -- (0,0);
                \draw[dashed] (1,-0.2) -- (0,-0.2);
                \draw[dashed] (1,-0.4) -- (0,-0.4);
		        \draw[dashed] (1,-0.6) -- (0,-0.6);
		        \draw (1,0.8)--(1,0)--(1,-0.9);
		        \draw [] (0,0.8) -- (0,0)--(0,-0.9);
		        \draw[] (2,0.5) -- (1,0.5);
                \draw[] (2,0.3) -- (1,0.3);
                \draw[] (2,0.1) -- (1,0.1);
                \draw[] (2,-0.1) -- (1,-0.1);
                \draw[dashed] (2,-0.3) -- (1,-0.3);
                \draw[dashed] (2,-0.5) -- (1,-0.5);
		        \draw[dashed] (2,-0.7) -- (1,-0.7);
		        \draw[] (2,0.8)--(2,0)--(2,-0.9);
		         \draw[] (3,0.6) -- (2,0.6);
                \draw[] (3,0.4) -- (2,0.4);
                \draw[] (3,0.2) -- (2,0.2);
                \draw[] (3,0) -- (2,0);
                \draw[dashed] (3,-0.2) -- (2,-0.2);
                \draw[dashed] (3,-0.4) -- (2,-0.4);
		        \draw[dashed] (3,-0.6) -- (2,-0.6);
		        \draw[] (3,0.8)--(3,0)--(3,-0.9);
		        \node[roundnode,scale=1.3] (A) at (-0.5,-1.5){$4\,|\,3$};
		        \node[roundnode,scale=1.3] (B) at (0.5,-1.5){$4\,|\,3$};
		        \node[roundnode,scale=1.3] (C) at (1.5,-1.5){$4\,|\,3$};
		        \node[roundnode,scale=1.3] (D) at (2.5,-1.5){$4\,|\,3$};
		        \draw[ultra thick] (A)--(B)--(C)--(D);
		        \end{scope}
        \end{tikzpicture}}\label{eq:HW-supergroup-Aquiver}
\end{align}
where for simplicity, we assigned the same supergroup $U(4\,|\,3)$ to all the gauge nodes. The numbers inside the nodes are the ranks of the positive parts and negative parts of the gauge groups. We can also consider the 5d lift-up of this theory by taking T-duality. After this process, we obtain a $(p,q)$ web diagram which is dual to the topological string theory on non-compact toric Calabi-Yau manifold \cite{Aharony:1997bh,Aharony:1997ju,Katz:1996fh,Katz:1997eq,Dijkgraaf:2002fc,Hollowood:2003cv,Leung:1997tw}. For the supergroup case, a web diagram was proposed in \cite{Kimura-Sugimoto:antivertex}. For example, for the pure SYM, we have
\begin{align}
    \begin{tikzpicture}[thick,baseline=(current bounding box.center)]
    \begin{scope}[scale=0.5,yscale=0.5,xscale=0.5]
    \draw [] (-7.5,6)--(7.5,6);
    \draw [] (-3.5,4)--(3.5,4);
    \draw[] (-1.5,2)--(1.5,2);
        \draw[] (-0.5,0)--(0.5,0);
        \draw[dashed] (-1.5,-2)--(1.5,-2);
        \draw [dashed] (-3.5,-4)--(3.5,-4);
        \draw [dashed] (-7.5,-6)--(7.5,-6);
        \draw (-13.5,8)--(-7.5,6)--(-3.5,4)--(-1.5,2)--(-0.5,0)--(-1.5,-2)--(-3.5,-4)--(-7.5,-6)--(-13.5,-8);
        \draw (13.5,8)--(7.5,6)--(3.5,4)--(1.5,2)--(0.5,0)--(1.5,-2)--(3.5,-4)--(7.5,-6)--(13.5,-8);
    \end{scope}
    \node at (0,-2.5) {Type \rom{2}B};
    \node at (8,-2.5) {Type \rom{2}A};
     \draw[red,latex-latex] (3.5,0) -- ++(1,0);
   \node at (4,.5) {T dual};
    \begin{scope}[shift=({8,0}),scale=1.3,xscale=1.2,yscale=0.9]
            \draw[] (-0.9,1.2)--(0.9,1.2);
                \draw[] (-0.9,0.8)--(0.9,0.8);
                \draw[] (-0.9,0.4)--(0.9,0.4);
                \draw[] (-0.9,0)--(0.9,0);
                \draw[dashed] (-0.9,-0.4)--(0.9,-0.4);
                \draw[dashed] (-0.9,-0.8)--(0.9,-0.8);
                \draw[dashed] (-0.9,-1.2)--(0.9,-1.2);
		        \draw[] (0.9,1.5)--(0.9,0)--(0.9,-1.5);
		        \draw[] (-0.9,1.5) -- (-0.9,0)--(-0.9,-1.5);
            \end{scope}
    \end{tikzpicture}
\end{align}
In \cite{Kimura-Sugimoto:antivertex}, the authors introduced a trivalent vertex drawn in dashed lines to incorporate the negative D-branes in the $(p,q)$ brane web. In this paper, we will only draw the horizontal lines, which correspond to D5-branes, in dashed lines. The $(p,q)$ branes with $q\neq 1$ will be always drawn in solid lines. Moreover, to simplify diagrams, instead of using the type \rom{2}B picture, we will always use the type \rom{2}A diagram to draw figures. The slope of the $(p,q)$-brane corresponds to the charge of the brane, and we will simply assign the value of it next to the vertical segment if needed. We often make use of the type \rom{2}A figure, but note that it always means the type \rom{2}B figure because we are considering five-dimensional gauge theories.

\paragraph{$D_{r}$ quiver gauge theory}We can also consider 4d D-type quiver supergroup gauge theories by introducing the $\text{ON}^{0}$ plane \cite{Kapustin:1998fa,Hanany:1999sj,Dijkgraaf:2016lym}. The brane configuration is then given as the following:
\begin{align}
\adjustbox{valign=c}{
    \begin{tikzpicture}[thick,roundnode/.style={circle,fill=blue!50,draw=black},
squarednode/.style={rectangle,draw=black}]
            \begin{scope}[scale=2.3]
		        \draw (-1,0.8)--(-1,0)--(-1,-0.9);
            \draw[] (0,0.5) -- (-1,0.5);
                \draw[] (0,0.3) -- (-1,0.3);
                \draw[] (0,0.1) -- (-1,0.1);
                \draw[] (0,-0.1) -- (-1,-0.1);
                \draw[dashed] (0,-0.3) -- (-1,-0.3);
                \draw[dashed] (0,-0.5) -- (-1,-0.5);
                \draw[] (1,0.6) -- (0,0.6);
                \draw[] (1,0.4) -- (0,0.4);
                \draw[] (1,0.2) -- (0,0.2);
                \draw[] (1,0) -- (0,0);
                \draw[dashed] (1,-0.2) -- (0,-0.2);
                \draw[dashed] (1,-0.4) -- (0,-0.4);
		        \draw (1,0.8)--(1,0)--(1,-0.9);
		        \draw [] (0,0.8) -- (0,0)--(0,-0.9);
		        \draw[] (2,0.5) -- (1,0.5);
                \draw[] (2,0.3) -- (1,0.3);
                \draw[] (2,0.1) -- (1,0.1);
                \draw[] (2,-0.1) -- (1,-0.1);
                \draw[dashed] (2,-0.3) -- (1,-0.3);
                \draw[dashed] (2,-0.5) -- (1,-0.5);
		        \draw[] (2,0.8)--(2,0)--(2,-0.9);
		         \draw[] (3,0.6) -- (2,0.6);
                \draw[] (3,0.4) -- (2,0.4);
                \draw[] (3,0.2) -- (2,0.2);
                \draw[] (3,0) -- (2,0);
                \draw[dashed] (3,-0.2) -- (2,-0.2);
                \draw[dashed] (3,-0.4) -- (2,-0.4);
		        \draw[blue,thick] (3,0.8)--(3,0)--(3,-0.9);
		       \node [blue,above] at (3,0.85) {$\text{ON}^{0}$}; 
		        \node[roundnode,scale=1.3] (A) at (-0.5,-1.5){$4\,|\,2$};
		        \node[roundnode,scale=1.3] (B) at (0.5,-1.5){$4\,|\,2$};
		        \node[roundnode,scale=1.3] (C) at (1.5,-1.5){$4\,|\,2$};
		        \node[roundnode,scale=1.3] (D) at (2.5,-1.9){$2\,|\,1$};
		        \node [roundnode,scale=1.3] (E) at (2.5,-1.1){$2\,|\,1$};
		        \draw[ultra thick] (A)--(B)--(C)--(D);
		        \draw[ultra thick] (C)--(E);
		        \end{scope}
        \end{tikzpicture}}
\end{align}
where we draw the $\text{ON}^{0}$ plane in a blue solid line. 

Taking T-duality of the above system, we can consider 5-brane webs realizing five-dimensional gauge theories. A microscopic description of the $\text{ON}^{0}$ plane as a combination of an NS5-brane and an $\text{ON}^{-}$ plane was proposed in \cite{Hayashi:2015vhy}:
\begin{align}
\begin{tikzpicture}[thick,baseline=(current bounding box.center),scale=1.3]
\begin{scope}
    \draw[] (-0.5,0.5)--(0,0)--(0,-1)--(-0.5,-1.5);
    \draw[] (0,0)--(1,0);
    \draw[] (0,-1)--(1,-1);
    \draw[blue,thick] (1,0.8)--(1,-1.8);
    \node[blue, right] at (1,0.8) {$\text{ON}^{0}$};
    \draw[->,>=latex] (1.5,-0.5)--(2,-0.5);
\end{scope}
\begin{scope}[shift=({3,0})]
    \draw[] (-0.5,0.5)--(0,0)--(0,-1)--(-0.5,-1.5);
    \draw[] (0,0)--(2.5,0);
    \draw[] (0,-1)--(1,-1);
    \draw[red,thick] (1,-1)--(1,-2);
    \node [red,left] at (1,-2){$\text{NS}5$};
    \draw[] (1,-1)--(2,-0.1);
    \draw[red,thick] (2,-0.1)--(2,-0.03);
    \draw[red,thick](2,0.03)--(2,0.8);
    \node [red,left] at (2,0.8){$\text{NS}5$};
    \draw [] (2,-0.1)--(2.5,-0.1);
    \draw [blue,thick] (2.5,0.8)--(2.5,-1.8);
    \node [blue,right] at (2.5,0.8) {$\text{ON}^{-}$};
\end{scope}
\end{tikzpicture}
\end{align}
Using this description, the authors managed to realize D-type quiver gauge theories using 5-brane webs. In this paper, we follow the description in \cite{Kimura:2019gon} and simply draw this web diagram as
\begin{align}
    \begin{tikzpicture}[thick,baseline=(current bounding box.center),scale=1.3]
\begin{scope}
    \draw[] (0,0.5)--(0,-1.5);
    \draw[] (0,0)--(1,0);
    \draw[] (0,-1)--(1,-1);
    \draw[] (1,0.5)--(1,-1.5);
    \draw[very thick] (0.9,-0.1)--(1.1,0.1);
    \draw[very thick] (0.9,0.1)--(1.1,-0.1);
\end{scope}
    \end{tikzpicture}
\end{align}
where we draw the $(p,q)$-brane straight and assigned a vertex $\hspace{-0.15cm}\Dvertex$ to the junction of the branes connected to the $\text{ON}^{-}$ plane. 

We expect that a similar story of \cite{Hayashi:2015vhy} holds also for supergroup gauge theories and that the difference is only the existence of negative D-branes. Namely, dashed horizontal lines will be also included in the web diagram. Thus, in our notation, brane webs for general D-type quiver supergroup gauge theories are drawn as 
\begin{align}
\adjustbox{valign=c}{
     \begin{tikzpicture}[thick]
            \begin{scope}[scale=2.4]
		        \draw (-1,0.7)--(-1,0)--(-1,-0.8);
		        \draw [] (0,0.7) -- (0,0)--(0,-0.8);
		        \draw (1,0.7)--(1,0)--(1,-0.8);
		        \draw[] (2,0.7)--(2,0)--(2,-0.8);
		        \draw[] (3,0.7)--(3,0)--(3,-0.8);
		        \draw[] (4,0.7)--(4,0)--(4,-0.8);
                \draw[] (0,0.4) -- (-1,0.4);
                \draw[dashed] (0,0.2) -- (-1,0.2);
                \draw[] (0,0) -- (-1,0);
                \draw[dashed] (0,-0.2) -- (-1,-0.2);
                \draw[dashed] (0,-0.4) -- (-1,-0.4);
                \draw[] (0,-0.6) -- (-1,-0.6);
                \draw[] (1,0.5) -- (0,0.5);
                \draw[dashed] (1,0.1) -- (0,0.1);
                \draw[] (1,-0.3)--(0,-0.3);
            \draw[dashed] (1,0.3)--(2,0.3);
            \draw[] (1,0)--(2,0);
            \draw[dashed] (1,-0.2)--(2,-0.2);
            \draw[dashed] (1,-0.5)--(2,-0.5);
                \draw[] (3,0.15) -- (2,0.15);
                \draw[] (3,0.45) -- (2,0.45);
                \draw[] (3,-0.1) -- (2,-0.1);
                \draw[dashed] (3,-0.35) -- (2,-0.35);
		        \draw[] (4,0.4) -- (3,0.4);
		        \draw[dashed] (4,0.55) -- (3,0.55);
		        \draw[dashed] (4,0.65) -- (3,0.65);
		        \draw[] (4,0.3) -- (3,0.3);
		        \draw[dashed] (4,0) -- (3,0);
		        \draw[dashed] (4,-0.15) -- (3,-0.15);
		        \draw[dashed] (4,-0.25) -- (3,-0.25);
		        \draw[dashed] (4,-0.45) -- (3,-0.45);
            \end{scope}
            \begin{scope}[scale=2.4,shift=({4,0})]
                \draw [very thick] (-0.040,0.36)--(0.040,0.44);
		        \draw [very thick] (0.040,0.36)--(-0.040,0.44);
		        \draw [very thick] (-0.040,0.51)--(0.040,0.59);
		        \draw [very thick] (0.040,0.51)--(-0.040,0.59);
		        \draw [very thick] (-0.040,0.61)--(0.040,0.69);
		        \draw [very thick] (0.040,0.61)--(-0.040,0.69);
		        \draw [very thick] (-0.040,0.26)--(0.040,0.34);
		        \draw [very thick] (0.040,0.26)--(-0.040,0.34);
            \end{scope}
        \end{tikzpicture}}\label{eq:Dquiverbraneweb1}
\end{align}
and the quiver structure is read as 
\begin{align}
\adjustbox{valign=c}{
     \begin{tikzpicture}[roundnode/.style={circle,draw=black,fill=blue!50}]
            \begin{scope}[scale=2.4]
		        \draw (-1,0.7)--(-1,0)--(-1,-0.8);
		        \draw [] (0,0.7) -- (0,0)--(0,-0.8);
		        \draw (1,0.7)--(1,0)--(1,-0.8);
		        \draw[] (2,0.7)--(2,0)--(2,-0.8);
		        \draw[] (3,0.7)--(3,0)--(3,-0.8);
		        \draw[] (4,0.7)--(4,0)--(4,-0.8);
                \draw[] (0,0.4) -- (-1,0.4);
                \draw[dashed] (0,0.2) -- (-1,0.2);
                \draw[] (0,0) -- (-1,0);
                \draw[dashed] (0,-0.2) -- (-1,-0.2);
                \draw[dashed] (0,-0.4) -- (-1,-0.4);
                \draw[] (0,-0.6) -- (-1,-0.6);
                \draw[] (1,0.5) -- (0,0.5);
                \draw[dashed] (1,0.1) -- (0,0.1);
                \draw[] (1,-0.3)--(0,-0.3);
            \draw[dashed] (1,0.3)--(2,0.3);
            \draw[] (1,0)--(2,0);
            \draw[dashed] (1,-0.2)--(2,-0.2);
            \draw[dashed] (1,-0.5)--(2,-0.5);
                \draw[] (3,0.15) -- (2,0.15);
                \draw[] (3,0.45) -- (2,0.45);
                \draw[] (3,-0.1) -- (2,-0.1);
                \draw[dashed] (3,-0.35) -- (2,-0.35);
		        \draw[] (4,0.4) -- (3,0.4);
		        \draw[dashed] (4,0.55) -- (3,0.55);
		        \draw[dashed] (4,0.65) -- (3,0.65);
		        \draw[] (4,0.3) -- (3,0.3);
		        \draw[dashed] (4,0) -- (3,0);
		        \draw[dashed] (4,-0.15) -- (3,-0.15);
		        \draw[dashed] (4,-0.25) -- (3,-0.25);
		        \draw[dashed] (4,-0.45) -- (3,-0.45);
		        \node[roundnode,scale=1.5] (A) at (-0.5,-0.1){$3\,|\,3$};
		        \node[roundnode,scale=1.5] (B) at (0.5,0.1){$2\,|\,1$};
		        \node[roundnode,scale=1.5] (C) at (1.5,-0.1){$1\,|\,3$};
		        \node[roundnode,scale=1.5] (D) at (2.5,0.025){$3\,|\,1$};
		        \node[roundnode,scale=1.5] (E) at (3.5,0.475){$2\,|\,2$};
		        \node[roundnode,scale=1.5] (F) at (3.5,-0.3){$0\,|\,4$};
		        \draw [ultra thick](A)--(B)--(C)--(D)--(E);
		        \draw [ultra thick](D)--(F);
            \end{scope}
            \begin{scope}[scale=2.4,shift=({4,0})]
                \draw [very thick] (-0.040,0.36)--(0.040,0.44);
		        \draw [very thick] (0.040,0.36)--(-0.040,0.44);
		        \draw [very thick] (-0.040,0.51)--(0.040,0.59);
		        \draw [very thick] (0.040,0.51)--(-0.040,0.59);
		        \draw [very thick] (-0.040,0.61)--(0.040,0.69);
		        \draw [very thick] (0.040,0.61)--(-0.040,0.69);
		        \draw [very thick] (-0.040,0.26)--(0.040,0.34);
		        \draw [very thick] (0.040,0.26)--(-0.040,0.34);
            \end{scope}
        \end{tikzpicture}}\label{eq:Dquiverbraneweb2}
\end{align}

\section{Quantum toroidal \texorpdfstring{$\mathfrak{gl}_{1}$}{gl(1)}}\label{sec:DIMalgebra}
Quantum toroidal $\mathfrak{gl}_{1}$ \cite{Ding:1996mq,Miki2007,feigin2011quantum,Feigin2011,feigin2012quantum} is a quantum algebra whose generators are described in Drinfeld currents 
\begin{align}
    x^{\pm}(z)=\sum_{m\in\mathbb{Z}}x^{\pm}_{m}z^{-m},\quad \psi^{\pm}(z)=\sum_{r\geq 0}\psi^{\pm}_{\pm r}z^{\mp r},\quad \hat{\gamma},\quad \psi_{0}^{+}/\psi_{0}^{-},
\end{align}
where $\hat{\gamma}$ and $\psi_{0}^{+}/\psi_{0}^{-}$ are central elements. We denote this algebra $\mathcal{E}$ for simplicity. There are two independent deformation parameters $q_{1},q_{2},q_{3}$ with the condition $q_{1}q_{2}q_{3}=1$. These $q$-parameters are identified with the parameters introduced in the previous section.

The defining relations are 
\begin{align}
\begin{split}
    [\psi^{\pm}(z),\psi^{\pm}(w)]=0,&\quad
    \psi^{+}(z)\psi^{-}(w)=\frac{g(\hat{\gamma}z/w)}{g(\hat{\gamma}^{-1}z/w)}\psi^{-}(w)\psi^{+}(z),\\
    \psi^{\pm}(z)x^{+}(w)=g\left(\hat{\gamma}^{\pm\frac{1}{2}}z/w\right)x^{+}(w)\psi^{\pm}(z),&\quad
    \psi^{\pm}(z)x^{-}(w)=g\left(\hat{\gamma}^{\mp\frac{1}{2}}z/w\right)^{-1} x^{-}(w)\psi^{\pm}(z),\\
    x^{\pm}(z)x^{\pm}(w)&=g(z/w)^{\pm 1}x^{\pm}(w)x^{\pm}(z),\\
    [x^{+}(z),x^{-}(w)]=\frac{(1-q_{1})(1-q_{2})}{(1-q_{3}^{-1})}&\left(\delta\bl(\hat{\gamma}w/z\br)\psi^{+}\left(\hat{\gamma}^{\frac{1}{2}}w\right) -\delta\bl(\hat{\gamma}^{-1}w/z\br)\psi^{-}\left(\hat{\gamma}^{-\frac{1}{2}}w\right)   \right),\\
    \underset{{z_{1},z_{2},z_{3}}}{\text{Sym}}\frac{z_{2}}{z_{3}}[x^{\pm}(z_{1})&,[x^{\pm}(z_{2}),x^{\pm}(z_{3})]]=0
\end{split}\label{eq:DIMdef}
\end{align}
where the structure function is 
\begin{align}
    g(z)=\frac{(1-q_{1}z)(1-q_{2}z)(1-q_{3}z)}{(1-q_{1}^{-1}z)(1-q_{2}^{-1}z)(1-q_{3}^{-1}z)}.
\end{align}
We note this function satisfy 
\begin{align}
    g(z^{-1})=g(z)^{-1}.
\end{align}
We additionally impose the condition $\psi^{+}_{0}\psi^{-}_{0}=1$.
\paragraph{Hopf algebra structure}
A Hopf algebra $H$ is a bialgebra equipped with a unit $1_{H}$, a counit $\epsilon$, a product $m$, a coproduct $\Delta$ and an antipode satisfying some properties\cite{Miki2007}. In this paper, we only need the coproduct formula, which is given
\begin{align}
\begin{split}
    \Delta(x^{+}(z))&=x^{+}(z)\otimes 1+\psi^{-}(\hat{\gamma}^{1/2}_{(1)}z)\otimes x^{+}(\hat{\gamma}_{(1)}z),\\
    \Delta(x^{-}(z))&=x^{-}(\hat{\gamma}_{(2)}z)\otimes \psi^{+}(\hat{\gamma}^{1/2}_{(2)}z)+1\otimes x^{-}(z),\\
    \Delta(\psi^{+}(z))&=\psi^{+}(\hat{\gamma}_{(2)}^{1/2}z)\otimes\psi^{+}(\hat{\gamma}_{(1)}^{-1/2}z),\\
    \Delta(\psi^{-}(z))&=\psi^{-}(\hat{\gamma}_{(2)}^{-1/2}z)\otimes \psi^{-}(\hat{\gamma}_{(1)}^{1/2}z),\\
    \Delta(\hat{\gamma})&=\hat{\gamma}\otimes \hat{\gamma},
\end{split}\label{eq:coproduct1}
\end{align}
where $\hat{\gamma}_{(1)}=\hat{\gamma}\otimes 1$ and $\hat{\gamma}_{(2)}=1\otimes\hat{\gamma}$. 
\paragraph{Miki automorphism\cite{Miki2007}}
Quantum toroidal $\mathfrak{gl}_{1}$ has an automorphism called Miki-automorphism $\mathcal{S}$ (see also \cite{Miki2007,Bourgine:2018fjy,Sasa:2019rbk}). The explicit action on the currents is complicated but the action of $\mathcal{S}^{2}$ can be written explicitly as
\begin{align}
\begin{split}
    \mathcal{S}^{2}\cdot x^{\pm}(z)=-x^{\mp}(z^{-1}),\quad \mathcal{S}^{2}\cdot \psi^{\pm}(z)=\psi^{\mp}(z^{-1}),\quad
    \mathcal{S}^{2}\cdot (C,\psi_{0}^{-})=(C^{-1},(\psi_{0}^{-})^{-1}).
\end{split}\label{eq:Miki-duality}
\end{align}
\paragraph{Remark}
To consider representations of this algebra, we need to specify the values of the central elements $\hat{\gamma},\psi_{0}^{-}/\psi_{0}^{+}$. In this note, we only use representations with central charges 
\begin{align}
    \bl(\hat{\gamma},\psi^{-}_{0}/\psi^{+}_{0}\br)\mapsto \bl(\gamma^{\ell_{1}},\gamma^{2\ell_{2}}\br),\quad (\ell_{1},\ell_{2})\in\mathbb{Z}^{2}\label{eq:deflevel}
\end{align}
where $\gamma\coloneqq q_{3}^{1/2}$. We call this a level $(\ell_{1},\ell_{2})$ representation. Later, we will frequently use the following function 
\begin{align}
    S\bl(z\br)\coloneqq\frac{(1-q_{1}z)(1-q_{2}z)}{(1-z)(1-q_{3}^{-1}z)} 
\end{align}
satisfying
\begin{align}
    S(q_{3}z)=S(z^{-1}).
\end{align}
This is the function that already appeared in the definition of the Nekrasov factor (\ref{eq:generalized-Nekrasov-factor}) and (\ref{eq:def-Nekrasov}). Using this, the structure function is rewritten as 
\begin{align}
    g(z)=\frac{S(z)}{S(q_{3}z)}.
\end{align}
\subsection{Crystal representations}\label{sec:crystalrep}
When we set the central charge to be $\hat{\gamma}=1$, the Cartan operators $\psi^{\pm}(z)$ commute with each other. Then, we can construct representations using simultaneous eigenstates of these Cartan operators. The eigenstates have a crystal like interpretation and actually they are related to BPS crystals\cite{bershtein2018plane,feigin2013representations,Galakhov:2020vyb,Li:2020rij,Galakhov:2021vbo,Galakhov:2021xum,Yamazaki:2022cdg,Noshita:2021dgj,Noshita:2021ldl}, so we call them crystal representations\footnote{In the literature, these representations are called ``vertical representations" because the vertical subalgebra of the quantum toroidal $\mathfrak{gl}_{1}$ is commutative in this representation. In this paper, we instead call them ``crystal representations" so that we do not get confused with the vertical and horizontal direction of the brane web.}. The basis of the representations we consider is labeled by Young diagrams (see Appendix~\ref{sec:appendix-Young} for the notations). We use the double ket convention like $\dket{\lambda}$ for these representations.
\subsubsection{Level \texorpdfstring{$(0,1)$}{(0,1)} representation}\label{sec:positivecrystal}
\paragraph{Ket representation}
We use the notation similar to the one used in \cite{Bourgine:2017jsi}. The level $(0,1)$ representation is described as 
\begin{align}
\begin{split}
    x^{+}(z)\dket{v,\lambda}&=\sum_{x\in A(\lambda)}\delta\bl(z/\chi^{\splus}_{x}\br)\underset{z=\chi^{\splus}_{x}}{\Res}\frac{1}{z\mathcal{Y}^{\splus}_{\lambda}(z)}\dket{v,\lambda+x},\\
    x^{-}(z)\dket{v,\lambda}&=\gamma^{-1}\sum_{x\in R(\lambda)}\delta\bl(z/\chi^{\splus}_{x}\br)\underset{z= \chi^{\splus}_{x}}{\Res}z^{-1}\mathcal{Y}^{\splus}_{\lambda}(zq_{3}^{-1})\dket{v,\lambda-x},\\
    \psi^{\pm}(z)\dket{v,\lambda}&=\left[\Psi^{\splus}_{\lambda}(z)\right]_{\pm}\dket{v,\lambda},
\end{split}
\end{align}
where $\mathcal{Y}^{\splus}_{\lambda}(z)$ is (\ref{eq:Yfunction-def}), $\chi_{x}^{\splus}=vq_{1}^{i-1}q_{2}^{j-1}$, and
\begin{align}
 \Psi^{\splus}_{\lambda}(z)&=\gamma^{-1}\frac{\mathcal{Y}^{\splus}_{\lambda}(q_{3}^{-1}z)}{\mathcal{Y}^{\splus}_{\lambda}(z)}.\label{eq:positive-Cartan}
\end{align}The symbol $[f(z)]_{\pm}$ means formal expansion in $z^{\mp1}$ and $A(\lambda),R(\lambda)$ are the set of addable boxes and removable boxes (see Appendix~\ref{sec:appendix-Young} for the notation). The operator $x^{+}(z)$ adds a box to the Young diagram while the operator $x^{-}(z)$ removes a box from the Young diagram. 

We claim that this representation physically corresponds to adding and removing \textit{positive} instantons. Each box has an interpretation of an instanton, so we can interpret the operators as operators adding or removing instantons. We say that the operators $x^{+}(z)$ $(\text{resp.\,}x^{-}(z))$ increase (resp.\,decrease) the \textit{total} topological number (\ref{eq:super-inst-numb}) by $\Delta k=\pm1$. For the box adding (removing) process to be related with the adding (removing) process of instantons, we need $k=|\lambda|$, and thus, we conclude this representation corresponds to the positive instanton. 

Note also that the spectral parameter $v$ is related with the Chern roots of the framing bundle~$\bfN$.
\paragraph{Bra representation}
We define the dual (bra) representation by 
\begin{align}
    \dbra{v,\mu}\bl(g(z)\dket{v,\lambda}\br)=\bl(\dbra{v,\mu}g(z)\br)\dket{v,\lambda}.\label{eq:norm-identity}
\end{align}
We choose a normalization 
\begin{align}
\begin{split}
    &\dbraket{\mu|\lambda}={a^{\splus}_{\lambda}}^{-1}\delta_{\lambda,\mu},\quad a^{\splus}_{\lambda}=\frac{(v\gamma)^{-|\lambda|}\prod_{x\in\lambda}\chi^{\splus}_{x}}{N_{++}(v,\lambda|v,\lambda)},\\
    &\mathbbm{1}=\sum_{\lambda}a^{\splus}_{\lambda}\dket{v,\lambda}\dbra{v,\lambda}.
\end{split}
\end{align}
This was chosen so that the dual action act as
\begin{align}
\begin{split}
    \bl(x^{+}(z)\br)^{\dagger}=-x^{-}(z),\quad\bl(x^{-}(z)\br))^{\dagger}=-x^{+}(z),\quad
    \bl(\psi^{\pm}(z)\br)^{\dagger}=\psi^{\pm}(z).
\end{split}
\end{align}
The explicit form is 
\begin{align}
\begin{split}
    \dbra{v,\lambda}\psi^{\pm}(z)&=\left[\Psi_{\lambda}^{\splus}(z)\right]_{\pm}\dbra{v,\lambda},\\
    \dbra{v,\lambda}x^{+}(z)&=-\gamma^{-1}\sum_{x\in R(\lambda)}\dbra{v,\lambda-x}\delta\bl(z/\chi^{\splus}_{x}\br)\underset{z=\chi_{x}^{\splus}}{\Res}z^{-1}\mathcal{Y}^{\splus}_{\lambda}(q_{3}^{-1}z),\\
    \dbra{v,\lambda}x^{-}(z)&=-\sum_{x\in A(\lambda)}\dbra{v,\lambda+x}\delta\bl(z/\chi^{\splus}_{x}\br)\underset{z=\chi^{\splus}_{x}}{\Res}\frac{1}{z\mathcal{Y}^{\splus}_{\lambda}(z)}.
\end{split}
\end{align}
\subsubsection{Level \texorpdfstring{$(0,-1)$}{(0,-1)} representation}\label{sec:negativecrystal}
\paragraph{Ket representation}
Let us introduce a different representation with level $(0,-1)$. See Appendix~\ref{sec:appendix-level(0,-1)rep} for a direct derivation of this representation. The representation is written as 
\begin{align}
\begin{split}
    x^{+}(z)\dket{v,\lambda}&=\gamma\sum_{x\in R(\lambda)}\delta\left(z/\chi_{x}^{\sminus}\right)\underset{z=\chi^{\sminus}_{x}}{\Res}z^{-1}\mathcal{Y}^{\sminus}_{\lambda}(z)\dket{v,\lambda-x},\\
    x^{-}(z)\dket{v,\lambda}&=\sum_{x\in A(\lambda)}\delta\left(z/\chi_{x}^{\sminus}\right)\underset{z=\chi_{x}^{\sminus}}{\Res}z^{-1}\mathcal{Y}^{\sminus}_{\lambda}(q_{3}^{-1}z)^{-1}\dket{v,\lambda+x},\\
    \psi^{\pm}(z)\dket{v,\lambda}&=\left[\Psi^{\sminus}_{\lambda}(z)\right]_{\pm}\dket{v,\lambda},
\end{split}
\end{align}
where the coordinate is $\chi_{x}^{\sminus}=vq_{1}^{-i}q_{2}^{-j}$, the function $\mathcal{Y}^{\sminus}_{\lambda}(z)$ is (\ref{eq:Yfunction-def}), and 
\begin{align}
    &\Psi^{\sminus}_{\lambda}(z)=\gamma\frac{\mathcal{Y}^{\sminus}_{\lambda}(z)}{\mathcal{Y}^{\sminus}_{\lambda}(q_{3}^{-1}z)}.
\end{align}
Obviously, since 
\begin{align}
    \mathcal{Y}^{\sminus}_{\lambda}(z)\rightarrow \begin{dcases}
    1,\quad z\rightarrow \infty\\
    z^{-1},\quad z\rightarrow 0
    \end{dcases},\quad \Psi_{\lambda}^{\sminus}(z)\rightarrow \begin{dcases}
    \gamma=\psi_{0}^{+},\quad z\rightarrow \infty\\
    \gamma^{-1}=\psi_{0}^{-},\quad z\rightarrow 0
    \end{dcases},
\end{align}
we have $\psi_{0}^{-}/\psi_{0}^{+}=\gamma^{-2}$, which gives the level $(0,-1)$. 

Compared to the level $(0,1)$ representation, the operator $x^{+}(z)$ removes a box, while the operator $x^{-}(z)$ adds a box to the Young diagram. Following the discussion in the level $(0,1)$ representation, we say that $x^{\pm}(z)$ changes the \textit{total} instanton number by $\Delta k=\pm1.$ Because $x^{+}(z)$ (resp.\,$x^{-}(z)$) removes (resp.\,adds) boxes from the Young diagram, we need $k=-|\lambda|$. Thus, we should interpret this as a negative instanton contribution.

\paragraph{Bra representation}
We can define the bra representation similarly to the $(0,1)$ representation as 
\begin{align}
    (x^{+}(z))^{\dagger}=-x^{-}(z),\quad (x^{-}(z))^{\dagger}=-x^{+}(z),\quad (\psi^{\pm}(z))^{\dagger}=\psi^{\pm}(z)
\end{align}
and obtain
\begin{align}
\begin{split}
    \dbra{v,\lambda}x^{+}(z)&=-\sum_{x\in A(\lambda)}\delta\left(z/\chi_{x}^{\sminus}\right)\underset{z=\chi_{x}^{\ssminus}}{\Res}z^{-1}\mathcal{Y}_{\lambda}^{\sminus}(q_{3}^{-1}z)^{-1}\dbra{v,\lambda+x},\\
    \dbra{v,\lambda}x^{-}(z)&=-\gamma\sum_{x\in R(\lambda)}\delta\left(z/\chi_{x}^{\sminus}\right)\underset{z=\chi_{x}^{\sminus}}{\Res}z^{-1}\mathcal{Y}_{\lambda}^{\sminus}(z)\dbra{v,\lambda-x},\\
    \dbra{v,\lambda}\psi^{\pm}(z)&=\dbra{v,\lambda}\left[\Psi_{\lambda}^{\sminus}(z)\right]_{\pm}.
\end{split}\label{eq:negative-bra-rep}
\end{align}
The norm of the vectors comes from the identity (\ref{eq:norm-identity}):
\begin{align}
\begin{split}
    \dbraket{v,\mu\,|\,v,\lambda}=(a^{\sminus}_{\lambda})^{-1}\delta_{\lambda,\mu},&\quad a^{\sminus}_{\lambda}=\frac{(q_{3}\gamma v)^{-|\lambda|}\prod_{x\in\lambda}\chi_{x}^{\sminus}}{N_{--}(v,\lambda\,|\,v,\lambda)},\\
    \mathbbm{1}=\sum_{\lambda}a_{\lambda}^{\sminus}\dket{v,\lambda}\dbra{v,\lambda}.
\end{split}
\end{align}
\paragraph{Remark}
The level $(0,-1)$ representation actually can be derived from the level $(0,1)$ representation after using the Miki automorphism in equation (\ref{eq:Miki-duality}). For example, for the Cartan action, 
\begin{align}
\begin{split}
    \rho^{(0,1)}_{v}\circ\mathcal{S}^{2}(\psi^{\pm}(z))\dket{v,\lambda,+}&=\rho^{(0,1)}_{v}(\psi^{\mp}(z^{-1}))\dket{v,\lambda,+}=\left[\Psi_{\lambda}^{\splus}(z^{-1};v)\right]_{\pm}\dket{v,\lambda,+},
\end{split}
\end{align}
where we explicitly wrote the dependence of the spectral parameter $v$ and the representation $\rho^{(0,1)}_{v}$. The plus sign in $\dket{v,\lambda,+}$ represents that it is the bases of the $(0,+1)$ representation. 
Then, 
\begin{align}
    \Psi^{\splus}_{\lambda}(z^{-1};v)=\gamma^{-1}\frac{\mathcal{Y}^{\splus}_{\lambda}(q_{3}^{-1}z^{-1};v)}{\mathcal{Y}^{\splus}_{\lambda}(z^{-1};v)}=\gamma\frac{\left(1-\frac{v^{-1}q_{3}^{-1}}{z}\right)}{\left(1-\frac{v^{-1}}{z}\right)}\prod_{(i,j)\in\lambda}\frac{S(q_{3}^{-1}v^{-1}q_{1}^{-i}q_{2}^{-j}/z)}{S(v^{-1}q_{1}^{-i}q_{2}^{-j}/z)}=\Psi^{\sminus}_{\lambda}(z;q_{3}^{-1}v^{-1}).
\end{align}
For the actions of the Drinfeld currents $x^{\pm}(z)$, we have
\begin{align}
\begin{split}
    \rho^{(0,1)}_{v}\circ\mathcal{S}^{2}(x^{+}(z))\dket{v,\lambda,+}&=-\rho_{v}^{(0,1)}(x^{-}(z^{-1}))\dket{v,\lambda,+}\\&=-\gamma^{-1}\sum_{x\in R(\lambda)}\delta(z\chi_{x,v}^{\splus})\Res_{z=\chi_{x,v}^{\splus}}z^{-1}\mathcal{Y}_{\lambda}^{\splus}(q_{3}^{-1}z;v)\dket{v,\lambda-x,+}\\
    &=-vz\gamma\sum_{x\in R(\lambda)}\delta\left(z/\chi_{x,q_{3}^{-1}v^{-1}}^{\sminus
    }\right)\Res_{z=\chi_{x,q_{3}^{-1}v^{-1}}^{\sminus}}z^{-1}\mathcal{Y}_{\lambda}^{\sminus}(z;q_{3}^{-1}v^{-1})\dket{v,\lambda-x,+},\\
    \rho^{(0,1)}_{v}\circ\mathcal{S}^{2}(x^{-}(z))\dket{v,\lambda,+}&=-\rho_{v}^{(0,1)}(x^{+}(z^{-1}))\dket{v,\lambda,+}\\
    &=-\sum_{x\in A(\lambda)}\delta\left(z\chi^{\splus}_{x,v}\right)\Res_{z=\chi_{x,v}^{\splus}}z^{-1}\mathcal{Y}_{\lambda}(z;v)^{-1}\dket{v,\lambda+x,+}\\
    &=-(vz)^{-1}\sum_{x\in A(\lambda)}\delta\left(z/\chi^{\sminus}_{x,q_{3}^{-1}v^{-1}}\right)\Res_{z=\chi^{\sminus}_{x,q_{3}^{-1}v^{-1}}}z^{-1}\mathcal{Y}_{\lambda}^{\sminus}(q_{3}^{-1}z;q_{3}^{-1}v^{-1})^{-1}\dket{v,\lambda+x,+}
\end{split}
\end{align}
where we denote the spectral parameter dependence of the box content as $\chi^{\spm}_{x,v}$ and used 
\begin{align}
    \mathcal{Y}^{\splus}(z^{-1};v)=(-vz)\mathcal{Y}_{\lambda}^{\sminus}(q_{3}^{-1}z;q_{3}^{-1}v^{-1}),\quad
    \chi_{x,v}^{\splus}=1/\chi_{x,q_{3}^{-1}v^{-1}}^{\sminus}
\end{align}
Obviously, since the rescaling 
\begin{align}
    x^{\pm}(z)\rightarrow (-vz)^{\pm1}x^{\pm}(z),\quad \psi^{\pm}(z)\rightarrow \psi^{\pm}(z)
\end{align}
will not change the algebraic relations, multiplication of the factors $(-vz)^{\pm1}$ is an isomorphism. Thus, the level $(0,-1)$ representation we use is obtained by applying the automorphism $\mathcal{S}^{2}$ and changing the spectral parameters to\footnote{This representation was used in \cite{Bourgine:2018fjy} to study S-duality. The representation is the same but the physical interpretation looks different. In this paper, we relate this representation with negative instantons/ghost branes.} $q_{3}^{-1}v^{-1}$. Namely, in our notation, we have 
\begin{align}
\begin{split}
    \rho^{(0,1)}_{v}\circ\mathcal{S}^{2}&\simeq\rho^{(0,-1)}_{q_{3}^{-1}v^{-1}},\\
    \dket{v,\lambda,+}&\leftrightarrow \dket{q_{3}^{-1}v^{-1},\lambda,-},\\
    \rho^{(0,1)}_{v}\circ\mathcal{S}^{2}(g(z))&\leftrightarrow \rho_{q_{3}^{-1}v^{-1}}^{(0,-1)}(g(z)),\quad g(z)\in \mathcal{E}.
\end{split}
\end{align}

\subsection{Vertex operator representations}\label{sec:vertexoprep}
One of the well-known classes of representations of the quantum toroidal $\mathfrak{gl}_{1}$ is the vertex operator\footnote{In the literature, these representations are called ``horizontal representations". We instead call them ``vertex operator representations". } representation. The Drinfeld currents act as vertex operators in this representation. This representation has a nontrivial central charge $\hat{\gamma}\neq 1$. These representations are related with deformed $\mathcal{W}$-algebras \cite{bershtein2018plane, Shiraishi:1995rp,Feigin:1995sf,Awata:1995zk,Awata:1996dx, FHSSY:2010, Miki2007, Kojima2019, Kojima2021, Harada:2021xnm}.
\subsubsection{Level \texorpdfstring{$(1,n)$}{(1,n)} representation}\label{sec:positivevertexop}
The level $(\ell_{1},\ell_{2})=(1,n),\,n\in\mathbb{Z}$ representations are obtained as
\begin{align}
    x^{+}(z)\mapsto u\gamma^{n}z^{-n}\eta(z),\quad x^{-}(z)\mapsto u^{-1}\gamma^{-n}z^{n}\xi(z),\quad \psi^{\pm}(z)\mapsto \gamma^{\mp n}\varphi^{\pm}(z),\quad \hat{\gamma}\mapsto \gamma, \label{eq:vertex-op-rep-positive}
\end{align}
where 
\begin{align}
\begin{split}
  \eta(z)&=\exp\left(-\sum_{n=1}^{\infty}\frac{z^{n}}{n}q_{1}^{n}(1-q_{2}^{n})a_{-n}\right)\exp\left(-\sum_{n=1}^{\infty}\frac{z^{-n}}{n}(1-q_{1}^{-n})a_{n}\right),\\
  \xi(z)&=\exp\left(\sum_{n=1}^{\infty}\frac{z^{n}}{n}q_{1}^{n}(1-q_{2}^{n})\gamma^{n}a_{-n}\right)\exp\left(\sum_{n=1}^{\infty}\frac{z^{-n}}{n}(1-q_{1}^{-n})\gamma^{n}a_{n}\right),\\
  \varphi^{+}(z)&=\exp\left(\sum_{n>0}\frac{z^{-n}}{n}\gamma^{-\frac{n}{2}}(1-q_{1}^{-n})(q_{3}^{n}-1)a_{n}\right),\\
  \varphi^{-}(z)&=\exp\left(\sum_{n>0}\frac{z^{n}}{n}q_{1}^{n}\gamma^{-\frac{n}{2}}(1-q_{2}^{n})(q_{3}^{n}-1)a_{-n}\right),\\
  [a_{n},a_{m}]&=n\delta_{n+m,0}.
  \end{split}
\end{align}
The contraction formulas of this operators are in Appendix~\ref{sec:appendix-vertexcontr}. The Fock space of this representation is denoted using the normal braket notation such as $\ket{0},\bra{0}$, not the double braket notation used in the crystal representation. In later sections, we use the boson-fermion correspondence to form the basis of the Fock space by Young diagrams $\{\ket{\lambda}\}$ (see Appendix~\ref{sec:appendix-Schurfunc}).

\subsubsection{Level \texorpdfstring{$(-1,n)$}{(-1,n)} representation}\label{sec:negativevertexop}
These representations are obtained by using (\ref{eq:Miki-duality}). In the vertex operator representation, we can introduce a spectral parameter into the coefficients of the vertex operators, so we can remove the negative sign in front of $x^{\mp}(z)$. Then the representation for level $(-1,0)$ is 
\begin{align}
    x^{+}(z)\rightarrow u \xi(z^{-1}),\quad x^{-}(z)\rightarrow u^{-1}\eta(z^{-1}),\quad \psi^{\pm}(z)\rightarrow \varphi^{\mp}(z^{-1}),\quad \hat{\gamma}\rightarrow \gamma^{-1}. 
\end{align}
Namely, in our notation, we have $\rho^{(-1,0)}_{u}\simeq\rho_{-u^{-1}}^{(1,0)}\circ \mathcal{S}^{2}$. 

We can also construct level $(-1,n)$ representations as
\begin{align}
    x^{+}(z)\rightarrow u\gamma^{-n}z^{n}\xi(z^{-1}),\quad x^{-}(z)\rightarrow u^{-1}\gamma^{n}z^{-n}\eta(z^{-1}),\quad \psi^{\pm}(z)\rightarrow \gamma^{\mp n}\varphi^{\mp}(z^{-1}),\quad \hat{\gamma}\rightarrow \gamma^{-1}. 
\end{align}
We will not use this representation in this paper, but we expect this representation is related with \textit{superquiver} theories (see section~\ref{sec:superquiver}).
\section{Intertwiner formalism}\label{sec:intertwiner}
In this section, we study the algebraic objects of quantum toroidal $\mathfrak{gl}_{1}$ called the \textit{intertwiners}. Intertwiner is a homomorphism relating two representations. Ever since the discovery of Awata, Feigin, and Shiraishi \cite{Awata:2011ce}, various studies of 3d/4d/5d/6d AGT correspondence have been done by studying intertwiners of various quantum algebras. The basic strategy is always the same:
\begin{enumerate}
    \item Pick two vertex operator representations $H_{1},H_{2}$ and a crystal representation $V$ with suitable levels and spectral parameters.
    \item Study the intertwiner $\Phi:V\otimes H_{1}\rightarrow H_{2}$ and the dual intertwiner $\Phi^{\ast}:H_{2}\rightarrow H_{1}\otimes V$. They obey a property called the intertwiner property. Sometimes, the property is also called the AFS property, which comes from the names of the authors of \cite{Awata:2011ce}. Intertwiners obeying these properties exist only when suitable levels and spectral parameters are chosen. If they exist, they are unique after choosing the normalizations. S-duality gives other types of intertwiners (e.g. $H_{1}\otimes H_{2}\rightarrow V$) as in \cite{Bourgine:2018fjy}, but we focus on these types of intertwiners.
    \item By studying the composition of these intertwiners, we obtain analogues of Nekrasov partition functions. Generally, they are purely algebraic quantities but sometimes we can relate them with physical quantities.
    \item To study the correspondence with physics, we relate each of the representations with branes. The intertwiners are drawn as trivalent vertices and they are interpreted as junctions of branes (see (\ref{eq:general-AFS-intertwiner})). Composition of these trivalent vertices leads to general brane webs. 
    \begin{align}
        \begin{tikzpicture}
		\begin{scope}[thick,xscale = -1,scale=1.5]
		\node[below] at (1.3,1) {$H_{1}$};
		\node[above] at (0,0.1) {$V$};
		\node[above] at (1.5,-1.1) {$H_{2}$};
		\node[below right] at (1,-0.1) {$\Phi$};
		\draw[postaction={segment={mid arrow}}] (0,0) -- (1,0) -- (1.7,-0.7);
		\draw[postaction={segment={mid arrow}}] (1,1) -- (1,0);
		\end{scope}
		\end{tikzpicture}\hspace{1.5cm}
		\begin{tikzpicture}
		\begin{scope}[thick,xscale = -1,scale=1.5]
		\node[below] at (-0.9,2.1) {$H_{2}$};
		\node[above] at (-0.4,0) {$H_{1}$};
		\node[below] at (1,0.9) {$V$};
		\node[above left] at (0,1.1) {$\Phi^{\ast}$};
		\draw[postaction={segment={mid arrow}}] (0,1) -- (1,1);
		\draw[postaction={segment={mid arrow}}] (-0.7,1.7)--(0,1)--(0,0);
		\end{scope}
		\end{tikzpicture}\label{eq:general-AFS-intertwiner}
    \end{align}
    \item The nontrivial matrix elements of the intertwiners give analogues of refined topological vertices.
\end{enumerate}

In section~\ref{sec:positive-negative-interwiner}, we construct intertwiners with representations $H_{1,2}=(1,n)$, where $n\in\mathbb{Z}$ and \mbox{$V=(0,\pm1)$}. We call the intertwiners with $V=(0,1)$ positive intertwiners, and call those with $V=(0,-1)$ negative interwiners. The negative intertwiners are the new intertwiners we introduce. Contractions and the gluing rules of these intertwiners are given in section~\ref{sec:intertwiner-contractions}. We will see they give Nekrasov factors of the supergroup gauge theory in (\ref{eq:def-Nekrasov}). We also give the correspondence with brane junctions in section~\ref{sec:intertwinercorrespondence}. The positive interwiners are the well known AFS intertwiners of \cite{Awata:2011ce} and are related to junctions of $(p,q)$ branes, where the D5-branes are positive branes. The negative intertwiners will be related to $(p,q)$ brane webs with negative D5-branes.  

\subsection{Positive and negative intertwiners}\label{sec:positive-negative-interwiner}
\paragraph{Intertwiners}
The positive intertwiner $\Phi^{(n)}_{+}[u,v]$ and the negative intertwiner $\Phi^{(n)}_{-}[u,v]$ are maps 
\begin{align}
\begin{split}
\Phi_{\pm}^{(n)}[u,v]:(0,\pm1)_{v}\otimes (1,n)_{u}&\rightarrow (1,n\pm1)_{u'},\\
    \Phi_{\pm}^{(n)}[u,v]=\sum_{\lambda}a_{\lambda}^{\spm}\dbra{v,\lambda}\otimes \Phi_{\pm,\lambda}^{(n)}[u,v],&\quad
    \Phi^{(n)}_{\pm,\lambda}[u,v]=\Phi^{(n)}_{\pm}[u,v]\dket{v,\lambda},\label{eq:intertwiner-map}
\end{split}
\end{align}
that obey the following conditions 
\begin{align}
\begin{split}
    \rho_{u'}^{(1,n+1)}(g(z))\Phi_{+}^{(n)}[u,v]&=\Phi_{+}^{(n)}[u,v](\rho_{v}^{(0,1)}\otimes \rho_{u}^{(1,n)})\Delta(g(z)),\\
    \rho_{u'}^{(1,n-1)}(g(z))\Phi_{-}^{(n)}[u,v]&=\Phi_{-}^{(n)}[u,v](\rho_{v}^{(0,-1)}\otimes \rho_{u}^{(1,n)})\Delta(g(z)),
\end{split}\label{eq:AFSproperty}
\end{align}
where $g(z)\in\mathcal{E}$. We expanded the interwiners in the diagonal basis\footnote{Actually, we should write $\dbra{v,\lambda,\pm},\dket{v,\lambda,\pm}$ to distinguish level $(0,\pm1)$ representations, but we omit the signs $\pm$ for simplicity. Readers should be careful when doing explicit calculations.} $\dket{v,\lambda}$ of $(0,\pm1)_{v}$ in (\ref{eq:intertwiner-map}). Conditions (\ref{eq:AFSproperty}) are the intertwiner relations (AFS properties, AFS relations, etc.). See Appendix~\ref{sec:appendix-AFSproperty} for the diagrammatic interpretation of them.

Solutions of (\ref{eq:AFSproperty}) exist only when $u'=-uv^{\pm1}$, and the answers are
\begin{align}
\begin{split}
\Phi^{(n)}_{\pm,\lambda}[u,v]&=t_{n,\pm}(\lambda,u,v) :\Phi_{\pm,\emptyset}[v]\prod_{x\in\lambda}\eta(\chi_{x}^{\spm})^{\pm1}:,\\
    \Phi_{\pm,\emptyset}[v]=:\Phi_{\emptyset}[v]^{\pm 1}:,&\quad  t_{n,\pm}(\lambda,u,v)=(-u^{\pm1}v)^{|\lambda|}\gamma^{\frac{1\mp 1}{2}|\lambda|}\prod_{x\in\lambda}\left(\frac{\gamma}{\chi_{x}^{\spm}}\right)^{\pm (n\pm1)},
\end{split}
\end{align}
where 
\begin{align}
    \Phi_{\emptyset}[v]=\exp\left(-\sum_{n=1}^{\infty}\frac{v^{n}}{n}\frac{a_{-n}}{1-q_{1}^{-n}}\right)\exp\left(\sum_{n=1}^{\infty}\frac{v^{-n}}{n}\frac{a_{n}}{1-q_{2}^{-n}}\right).
\end{align}
Let us briefly discuss how to solve the relations for the negative interwiner only. Inserting~(\ref{eq:intertwiner-map}) and using (\ref{eq:negative-bra-rep}), the expansion of (\ref{eq:AFSproperty}) is
\begin{align}
\begin{split}
    u'\left(\frac{\gamma}{z}\right)^{n-1}\eta(z)\Phi_{-,\lambda}^{(n)}[u,v]&=\gamma\sum_{x\in R(\lambda)}\delta\left(\frac{z}{\chi_{x}^{\sminus}}\right)\underset{z=\chi_{x}^{\sminus}}{\Res}z^{-1}\mathcal{Y}_{\lambda}^{\sminus}(z)\Phi^{(n)}_{-,\lambda-x}[u,v]\\
    &+u\left(\frac{\gamma}{z}\right)^{n}\left[\Psi^{\sminus}_{\lambda}(z)\right]_{-}\Phi_{-,\lambda}^{(n)}[u,v]\eta(z),\\
    u'^{-1}\left(\frac{z}{\gamma}\right)^{n-1}\xi(z)\Phi^{(n)}_{-,\lambda}[u,v]&=\gamma^{-n}\sum_{x\in A(\lambda)}\delta\left(\frac{\gamma z}{\chi_{x}^{\sminus}}\right)\underset{z=\chi_{x}^{\sminus}}{\Res}z^{-1}\mathcal{Y}_{\lambda}^{\sminus}(q_{3}^{-1}z)^{-1}\Phi^{(n)}_{-,\lambda+x}\varphi^{+}(\gamma^{1/2}z)\\
    &+u^{-1}\left(\frac{z}{\gamma}\right)^{n}\Phi_{-,\lambda}^{(n)}[u,v]\xi(z),\\
    \gamma\varphi^{+}(z)\Phi^{(n)}_{-,\lambda}[u,v]&=\left[\Psi_{\lambda}^{\sminus}(\gamma^{1/2}z)\right]_{+}\Phi^{(n)}_{-,\lambda}[u,v]\varphi^{+}(z),\\
    \gamma^{-1}\varphi^{-}(z)\Phi^{(n)}_{-,\lambda}[u,v]&=\left[\Psi^{\sminus}_{\lambda}(\gamma^{-1/2}z)\right]_{-}\Phi^{(n)}_{-,\lambda}[u,v]\varphi^{-}(z).
\end{split}
\end{align}
The relations with the Cartan part $\varphi^{\pm}(z)$ determine the operator part of the intertwiners. The zero-mode part $t_{n,-}(\lambda,u,v)$ and the condition $u'=-uv^{-1}$ are derived from the relations with $x^{\pm}(z)$ after inserting the operator part. 
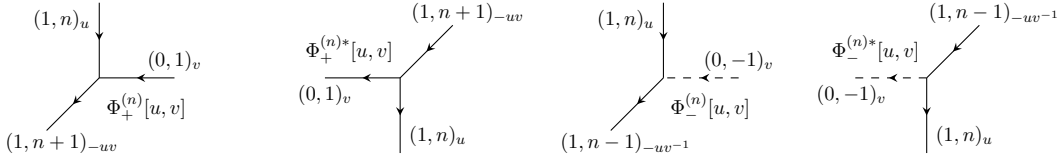
\begin{figure}[t]
	\begin{center}
		\begin{tikzpicture}
		\begin{scope}[xscale = -1]
		\node[below,scale=0.7] at (1.5,1) {$(1,n)_u$};
		\node[above,scale=0.7] at (0,0) {$(0,1)_{ v}$};
		\node[above,scale=0.7] at (1.5,-1.1) {$(1,n+1)_{-uv}$};
		\node[below right,scale=0.7] at (1,-0.1) {$\Phi_{+}^{(n)}[u, v]$};
		\draw[postaction={segment={mid arrow}}] (0,0) -- (1,0) -- (1.7,-0.7);
		\draw[postaction={segment={mid arrow}}] (1,1) -- (1,0);
		\end{scope}
		\begin{scope}[shift=({3,-1}),xscale = -1]
		\node[below,scale=0.7] at (-0.9,2.1) {$(1,n+1)_{-uv}$};
		\node[above,scale=0.7] at (-0.5,0) {$(1,n)_u$};
		\node[below,scale=0.7] at (1,1) {$(0,1)_{v}$};
		\node[above left,scale=0.7] at (0,1.1) {$\Phi_{+}^{(n)\ast}[u,v]$};
		\draw[postaction={segment={mid arrow}}] (0,1) -- (1,1);
		\draw[postaction={segment={mid arrow}}] (-0.7,1.7)--(0,1)--(0,0);
		\end{scope}
		\begin{scope}[shift={(7.5,0)},xscale = -1]
		\node[below,scale=0.7] at (1.5,1) {$(1,n)_u$};
		\node[above,scale=0.7] at (0,0) {$(0,-1)_{ v}$};
		\node[above,scale=0.7] at (1.5,-1.1) {$(1,n-1)_{-uv^{-1}}$};
		\node[below right,scale=0.7] at (1,-0.1) {$\Phi_{-}^{(n)}[u, v]$};
		\draw[dashed,postaction={segment={mid arrow}}] (0,0) -- (1,0);
		\draw[postaction={segment={mid arrow}}] (1,0)-- (1.7,-0.7);
		\draw[postaction={segment={mid arrow}}] (1,1) -- (1,0);
		\end{scope}
		\begin{scope}[shift={(10,-1)},xscale = -1]
		\node[below,scale=0.7] at (-0.9,2.1) {$(1,n-1)_{-uv^{-1}}$};
		\node[above,scale=0.7] at (-0.5,0) {$(1,n)_u$};
		\node[below,scale=0.7] at (1,1) {$(0,-1)_{v}$};
		\node[above left,scale=0.7] at (0,1.1) {$\Phi_{-}^{(n)\ast}[u,v]$};
		\draw[postaction={segment={mid arrow}}] (0,1) -- (0,0);
		\draw[dashed,postaction={segment={mid arrow}}] (0,1)-- (1,1);
		\draw[postaction={segment={mid arrow}}] (-0.7,1.7)--(0,1);
		\end{scope}
		\end{tikzpicture}
	\end{center}
	\caption{$\Phi_{\pm}^{(n)}[u, v]$ and $\Phi_{\pm}^{(n)\ast}[u, v]$. The dashed line represents the representation with negative level $(0,-1)_{v}$. The arrows represent the order of the operator. For example, the intertwiner $\Phi^{(n)}_{+}[u,v]$ is a map $(0,1)_{v}\otimes (1,n)_{u}\rightarrow(1,n+1)_{-uv}$ and thus the arrow is as the figure.}
	\label{fig:vertex}
\end{figure}
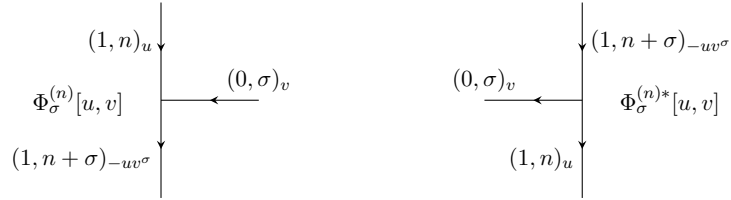
\begin{figure}
    \begin{center}
        \begin{tikzpicture}
            \begin{scope}[scale=1.3]
                \node[above, scale=0.8] at (0,0) {$(0,\sigma)_{v}$};
                \node[left, scale=0.8] at (-1,0.6) {$(1,n)_{u}$};
                \node[left,scale=0.8] at (-1,-0.6) {$(1,n+\sigma)_{-uv^{\sigma}}$};
                \node[left,scale=0.8] at (-1.3,0) {$\Phi^{(n)}_{\sigma}[u,v]$};
    
                \draw[postaction={segment={mid arrow}}] (0,0) -- (-1,0);
		        \draw[postaction={segment={mid arrow}}] (-1,1) -- (-1,0)--(-1,-1);   
            \end{scope}
            \begin{scope}[shift={(3,0)},scale=1.3]
                \node[above, scale=0.8] at (0,0) {$(0,\sigma)_{v}$};
                \node[right, scale=0.8] at (1,0.6) {$(1,n+\sigma)_{-uv^{\sigma}}$};
                \node[left,scale=0.8] at (1,-0.6) {$(1,n)_{u}$};
                \node[right,scale=0.8] at (1.3,0) {$\Phi^{(n)\ast}_{\sigma}[u,v]$};
    
                \draw[postaction={segment={mid arrow}}] (1,0) -- (0,0);
		        \draw[postaction={segment={mid arrow}}] (1,1) -- (1,0)--(1,-1);   
            \end{scope}
        \end{tikzpicture}
    \end{center}
    \caption{Simplified picture of intertwiners. We draw the vertical lines representing NS5-branes as straight lines. The bending of the NS5-branes are represented by the levels written beside them. Note that we have two types of intertwiners and dual intertwiners: the positive one ($\sigma=1$) and the negative one ($\sigma=-1$).}
    \label{fig:simplevertex}
\end{figure}

We illustrate these intertwiners as in Figure \ref{fig:vertex}. Instead of using this pictorial interpretation, we use a rather simplified version of it, where the vertical lines are not bent as in Figure \ref{fig:simplevertex}. We draw the lines associated with representations $(1,n),(0,1)$ in solid lines and representations with $(0,-1)$ in dashed lines. Sometimes, in the simplified diagram, we simply draw the horizontal lines in solid lines and just assign the level $(0,\sigma),\,\sigma=\pm1$ to distinguish whether the level is positive or negative. The arrows represent how the intertwiner maps modules to modules.
\paragraph{Dual intertwiners}
Similar to the intertwiners, we define the dual positive and negative interwiners as maps
\begin{align}
    \begin{split}
       \Phi^{(n)\ast}_{\pm}[u,v]:(1,n\pm1)_{u'}&\rightarrow (1,n)_{u}\otimes (0,\pm1)_{v},\\
    \Phi_{\pm}^{(n)\ast}[u,v]=\sum_{\lambda}a_{\lambda}^{(\spm)}\Phi^{(n)\ast}_{\pm,\lambda}[u,v]\otimes \dket{v,\lambda},&\quad \Phi^{(n)\ast}_{\pm,\lambda}[u,v]=\dbra{v,\lambda}\Phi^{(n)\ast}_{\pm}[u,v]
    \end{split}
\end{align}
obeying the conditions
\begin{align}
    \begin{split}
         (\rho_{u}^{(1,n)}\otimes\rho_{v}^{(0,1)})\Delta(g(z))\Phi_{+}^{(n)\ast}[u,v]&=\Phi_{+}^{(n)\ast}[u,v]\rho_{u'}^{(1,n+1)}(g(z)),\\
         (\rho_{u}^{(1,n)}\otimes\rho_{v}^{(0,-1)})\Delta(g(z))\Phi_{-}^{(n)\ast}[u,v]&=\Phi_{-}^{(n)\ast}[u,v]\rho_{u'}^{(1,n-1)}(g(z)),
    \end{split}\label{eq:dualAFSproperty}
\end{align}
where $g(z)\in\mathcal{E}$. The solutions of these relations exist only when $u'=-uv^{\pm1}$, and they are 
\begin{align}
\begin{split}
    \Phi^{(n)\ast}_{\pm,\lambda}[u,v]&=t^{\ast}_{n,\pm}(\lambda,u,v):\Phi^{\ast}_{\pm,\emptyset}[v]\prod_{x\in\lambda}\xi(\chi_{x}^{\spm})^{\pm1}:,\\
      \Phi^{\ast}_{\pm,\emptyset}[v]=:\Phi^{\ast}[v]^{\pm1}:,&\quad t^{\ast}_{n,\pm}(\lambda,u,v)=u^{\mp|\lambda|}\gamma^{-\frac{1\pm 1}{2}|\lambda|}\prod_{x\in\lambda}\left(\frac{\chi_{x}^{\spm}}{\gamma}\right)^{\pm n},
\end{split}
\end{align}
where
\begin{align}
    \Phi_{\emptyset}^{\ast}[v]=\exp\left(\sum_{n=1}^{\infty}\frac{v^{n}}{n}\frac{\gamma^{n}}{1-q_{1}^{-n}}a_{-n}\right)\exp\left(-\sum_{n=1}^{\infty}\frac{v^{-n}}{n}\frac{\gamma^{n}}{1-q_{2}^{-n}}a_{n}\right).
\end{align}
By expanding the relations (\ref{eq:dualAFSproperty}), we obtain  
\begin{align}
\begin{split}
    u'\left(\frac{\gamma}{z}\right)^{n-1}\Phi^{(n)\ast}_{-,\lambda}[u,v]\eta(z)&=u\left(\frac{\gamma}{z}\right)^{n}\eta(z)\Phi^{(n)\ast}_{-,\lambda}[u,v]\\
    &-\gamma^{n}\sum_{x\in A(\lambda)}\delta\left(\frac{\gamma z}{\chi_{x}^{\sminus}}\right)\underset{z=\chi_{x}^{\sminus}}{\Res}z^{-1}\mathcal{Y}^{\sminus}_{\lambda}(q_{3}^{-1}z)^{-1}\varphi^{-}(\gamma^{1/2}z)\Phi^{(n)\ast}_{-,\lambda+x}[u,v],\\
    u'^{-1}\left(\frac{z}{\gamma}\right)^{n-1}\Phi^{(n)\ast}_{-,\lambda}[u,v]\xi(z)&=u^{-1}\left(\frac{z}{\gamma}\right)^{n}\left[\Psi^{\sminus}_{\lambda}(z)\right]_{+}\xi(z)\Phi^{(n)\ast}_{-,\lambda}[u,v]\\
    &-\gamma\sum_{x\in R(\lambda)}\delta\left(\frac{z}{\chi_{x}^{\sminus}}\right)\underset{z=\chi_{x}^{\sminus}}{\Res}z^{-1}\mathcal{Y}^{\sminus}_{\lambda}(z) \Phi^{(n)\ast}_{-,\lambda-x}[u,v],\\
    \gamma\Phi^{(n)\ast}_{-,\lambda}[u,v]\varphi^{+}(z)&=\left[\Psi^{\sminus}_{\lambda}(\gamma^{-1/2}z)\right]_{+}\varphi^{+}(z)\Phi_{-,\lambda}^{(n)\ast}[u,v],\\
    \gamma^{-1}\Phi^{(n)\ast}_{-,\lambda}[u,v]\varphi^{-}(z)&=\left[\Psi^{\sminus}_{\lambda}(\gamma^{1/2}z)\right]_{-}\varphi^{-}(z)\Phi_{-,\lambda}^{(n)\ast}[u,v].
\end{split}
\end{align}
Similar to the previous intertwiner case, the Cartan part determines the vertex operator part and the other relations determine $u'=-uv^{-1}$ and the zero-mode part. We illustrate this operator as Figure \ref{fig:vertex} and Figure \ref{fig:simplevertex}.

\subsection{Contractions}\label{sec:intertwiner-contractions}
We have two ways to glue the (dual) intertwiners. Gluing horizontal lines give contractions in crystal representations, while gluing vertical lines gives contractions in vertex operator representations.
\paragraph{Contractions in vertex operator representations}
The contraction formulas in the vertex operator representations are summarized as

\begin{itemize}
\item gluing of $\Phi_{\sigma',\mu}^{(n_{2})}[u_{2},v_{2}]\Phi^{(n_{1})}_{\sigma,\lambda}[u_{1},v_{1}]$ with the conditions $ n_{1}+\sigma=n_{2}$, $u_{2}=-u_{1}v_{1}^{\sigma}$:
\begin{align}
  &\adjustbox{valign=c}{\begin{tikzpicture}
            \begin{scope}[scale=1.2]
                \node[above, scale=0.7] at (-0.4,0.2) {$(0,\sigma)_{v_{1}}$};
                \node[right,scale=0.8] at (0,0.2) {$\dket{v_{1},\lambda}$};
                \node[below,scale=0.7] at (-0.4,-1.2) {$(0,\sigma')_{v_{2}}$};
                \node[left, scale=0.7] at (-1,1) {$(1,n_{1})_{u_{1}}$};
                \node[right,scale=0.7] at (-1,-0.2) {$(1,n_{1}+\sigma)_{-u_{1}v_{1}^{\sigma}}$};
                \node[left,scale=0.8] at (-1.3,0.2) {$\Phi^{(n_{1})}_{\sigma}[u_{1},v_{1}]$};
                \node[right, scale=0.7] at (-1,-0.8) {$(1,n_{2})_{u_{2}}$};
                \node[below,scale=0.7] at (-1,-2) {$(1,n_{2}+\sigma')_{-u_{2}v_{2}^{\sigma'}}$};
                \node[right,scale=0.8] at (0,-1.2) {$\dket{v_{2},\mu}$};
                 \node[left,scale=0.8] at (-1.3,-1.2) {$\Phi^{(n_{2})}_{\sigma'}[u_{2},v_{2}]$};
               \draw[postaction={segment={mid arrow}}] (0,0.2) -- (-1,0.2);
              \draw[postaction={segment={mid arrow}}] (0,-1.2)--(-1,-1.2);
		        \draw[postaction={segment={mid arrow}}] (-1,1) -- (-1,0)--(-1,-1)--(-1,-2);
            \end{scope}
    \end{tikzpicture}}=
\begin{array}{cc}
\Phi_{\sigma}^{(n_{1})}[u_{1},v_{1}]&\dket{v_{1},\lambda}\\
\vspace{.1cm}\Phi_{\sigma'}^{(n_{2})}[u_{2},v_{2}]&\dket{v_{2},\mu}\\
\end{array}=\contraction{}{\Phi_{\sigma',\mu}^{(n_{2})}[u_{2},v_{2}]}{}{\Phi^{(n_{1})}_{\sigma,\lambda}[u_{1},v_{1}]}\Phi_{\sigma',\mu}^{(n_{2})}[u_{2},v_{2}]\Phi^{(n_{1})}_{\sigma,\lambda}[u_{1},v_{1}]\label{eq:glue-intertwiner-1}
\end{align}

\item gluing of $\Phi_{\sigma',\mu}^{(n_{2})\ast}[u_{2},v_{2}]\Phi^{(n_{1})\ast}_{\sigma,\lambda}[u_{1},v_{1}]$ with the conditions $n_{1}=n_{2}+\sigma'$, $u_{1}=-u_{2}v_{2}^{\sigma'}$:
\begin{align}
    \adjustbox{valign=c}{\begin{tikzpicture}
            \begin{scope}[scale=1.2]
                \node[above, scale=0.7] at (-1.5,0.2) {$(0,\sigma)_{v_{1}}$};
                \node[left,scale=0.8] at (-2,0.2) {$\dbra{v_{1},\lambda}$};
                \node[below,scale=0.7] at (-1.6,-1.2) {$(0,\sigma')_{v_{2}}$};
                \node[right, scale=0.7] at (-1,1) {$(1,n_{1}+\sigma)_{-u_{1}v_{1}^{\sigma}}$};
                \node[left,scale=0.7] at (-1,-0.2) {$(1,n_{1})_{u_{1}}$};
                \node[left,scale=0.8] at (0.8,0.2) {$\Phi^{(n_{1})\ast}_{\sigma}[u_{1},v_{1}]$};
                \node[left, scale=0.7] at (-1,-0.8) {$(1,n_{2}+\sigma')_{-u_{2}v_{2}^{\sigma'}}$};
                \node[below,scale=0.7] at (-1,-2) {$(1,n_{2})_{u_{2}}$};
                \node[left,scale=0.8] at (-2,-1.2) {$\dbra{v_{2},\mu}$};
                 \node[left,scale=0.8] at (0.8,-1.2) {$\Phi^{(n_{2})\ast}_{\sigma'}[u_{2},v_{2}]$};
               \draw[postaction={segment={mid arrow}}] (-1,0.2) -- (-2,0.2);
              \draw[postaction={segment={mid arrow}}] (-1,-1.2)--(-2,-1.2);
		        \draw[postaction={segment={mid arrow}}] (-1,1) -- (-1,0)--(-1,-1)--(-1,-2);
            \end{scope}
        \end{tikzpicture}}=
\begin{array}{cc}
\dbra{v_{1},\lambda}&\Phi_{\sigma}^{(n_{1})\ast}[u_{1},v_{1}]\\
\dbra{v_{2},\mu}&\vspace{.1cm}\Phi_{\sigma'}^{(n_{2})\ast}[u_{2},v_{2}]
\end{array}=\contraction{}{\Phi_{\sigma',\mu}^{(n_{2})\ast}[u_{2},v_{2}]}{}{\Phi^{(n_{1})\ast}_{\sigma,\lambda}[u_{1},v_{1}]}\Phi_{\sigma',\mu}^{(n_{2})\ast}[u_{2},v_{2}]\Phi^{(n_{1})\ast}_{\sigma,\lambda}[u_{1},v_{1}]\label{eq:glue-intertwiner-2}
\end{align}
\item gluing of $\Phi_{\sigma',\mu}^{(n_{2})\ast}[u_{2},v_{2}]\Phi^{(n_{1})}_{\sigma,\lambda}[u_{1},v_{1}]$ with the conditions $n_{1}+\sigma=n_{2}+\sigma'$, $u_{1}v_{1}^{\sigma}=u_{2}v_{2}^{\sigma'}$:
\begin{align}
\adjustbox{valign=c}{\begin{tikzpicture}
            \begin{scope}[scale=1.2]
                \node[above, scale=0.7] at (-0.5,0.2) {$(0,\sigma)_{v_{1}}$};
                \node[right,scale=0.8] at (0,0.2) {$\dket{v_{1},\lambda}$};
                \node[below,scale=0.7] at (-1.6,-1.2) {$(0,\sigma')_{v_{2}}$};
                \node[right, scale=0.7] at (-1,1) {$(1,n_{1})_{u_{1}}$};
                \node[right,scale=0.7] at (-1,-0.2) {$(1,n_{1}+\sigma)_{-u_{1}v_{1}^{\sigma}}$};
                \node[left,scale=0.8] at (-1.2,0.2) {$\Phi^{(n_{1})}_{\sigma}[u_{1},v_{1}]$};
                \node[left, scale=0.7] at (-1,-0.8) {$(1,n_{2}+\sigma')_{-u_{2}v_{2}^{\sigma'}}$};
                \node[below,scale=0.7] at (-1,-2) {$(1,n_{2})_{u_{2}}$};
                \node[left,scale=0.8] at (-2,-1.2) {$\dbra{v_{2},\mu}$};
                 \node[left,scale=0.8] at (0.8,-1.2) {$\Phi^{(n_{2})\ast}_{\sigma'}[u_{2},v_{2}]$};
               \draw[postaction={segment={mid arrow}}] (0,0.2) -- (-1,0.2);
              \draw[postaction={segment={mid arrow}}] (-1,-1.2)--(-2,-1.2);
		        \draw[postaction={segment={mid arrow}}] (-1,1) -- (-1,0)--(-1,-1)--(-1,-2);
            \end{scope}
        \end{tikzpicture}}=\begin{array}{ccc}
&\Phi_{\sigma}^{(n_{1})}[u_{1},v_{1}]&\dket{v_{1},\lambda}\\
\vspace{.1cm}\dbra{v_{2},\mu}&\Phi_{\sigma'}^{(n_{2})\ast}[u_{2},v_{2}]&
\end{array}=\contraction{}{\Phi_{\sigma',\mu}^{(n_{2})\ast}[u_{2},v_{2}]}{}{\Phi^{(n_{1})}_{\sigma,\lambda}[u_{1},v_{1}]}\Phi_{\sigma',\mu}^{(n_{2})\ast}[u_{2},v_{2}]\Phi^{(n_{1})}_{\sigma,\lambda}[u_{1},v_{1}]\label{eq:glue-intertwiner-3}
\end{align}
\item gluing of $\Phi_{\sigma',\mu}^{(n_{2})}[u_{2},v_{2}]\Phi^{(n_{1})}_{\sigma,\lambda}[u_{1},v_{1}]$ with the conditions $n_{1}=n_{2}$, $u_{1}=u_{2}$:
\begin{align}
\adjustbox{valign=c}{
    \begin{tikzpicture}
            \begin{scope}[scale=1.2]
                \node[above, scale=0.7] at (-1.5,0.2) {$(0,\sigma)_{v_{1}}$};
                \node[left,scale=0.8] at (-2,0.2) {$\dbra{v_{1},\lambda}$};
                \node[below,scale=0.7] at (-0.4,-1.2) {$(0,\sigma')_{v_{2}}$};
                \node[right, scale=0.7] at (-1,1) {$(1,n_{1}+\sigma)_{-u_{1}v_{1}^{\sigma}}$};
                \node[left,scale=0.7] at (-1,-0.2) {$(1,n_{1})_{u_{1}}$};
                \node[left,scale=0.8] at (0.8,0.2) {$\Phi^{(n_{1})\ast}_{\sigma}[u_{1},v_{1}]$};
                \node[right, scale=0.7] at (-1,-0.8) {$(1,n_{2})_{u_{2}}$};
                \node[below,scale=0.7] at (-1,-2) {$(1,n_{2}+\sigma')_{-u_{2}v_{2}^{\sigma'}}$};
                \node[right,scale=0.8] at (0,-1.2) {$\dket{v_{2},\mu}$};
                 \node[left,scale=0.8] at (-1.2,-1.2) {$\Phi^{(n_{2})}_{\sigma'}[u_{2},v_{2}]$};
                 \draw[postaction={segment={mid arrow}}] (-1,0.2) -- (-2,0.2);
              \draw[postaction={segment={mid arrow}}] (0,-1.2)--(-1,-1.2);
		        \draw[postaction={segment={mid arrow}}] (-1,1) -- (-1,0)--(-1,-1)--(-1,-2);
            \end{scope}
        \end{tikzpicture}}=\begin{array}{ccc}
\dbra{v_{1},\lambda}&\Phi_{\sigma}^{(n_{1})\ast}[u_{1},v_{1}]&\\
\vspace{.1cm}&\Phi_{\sigma'}^{(n_{2})}[u_{2},v_{2}]&\dket{v_{2},\mu}
\end{array}=\contraction{}{\Phi_{\sigma',\mu}^{(n_{2})}[u_{2},v_{2}]}{}{\Phi^{(n_{1})\ast}_{\sigma,\lambda}[u_{1},v_{1}]}\Phi_{\sigma',\mu}^{(n_{2})}[u_{2},v_{2}]\Phi^{(n_{1})\ast}_{\sigma,\lambda}[u_{1},v_{1}]\label{eq:glue-intertwiner-4}
\end{align}
\end{itemize}
where
\begin{align}
\begin{split}
    \contraction{}{\Phi_{\sigma',\mu}^{(n_{2})}[u_{2},v_{2}]}{}{\Phi^{(n_{1})}_{\sigma,\lambda}[u_{1},v_{1}]}\Phi_{\sigma',\mu}^{(n_{2})}[u_{2},v_{2}]\Phi^{(n_{1})}_{\sigma,\lambda}[u_{1},v_{1}]&=\bl(\cG(q_{3}^{-1}v_{1}/v_{2})\br)^{\sigma\sigma'}N_{\sigma\sigma'}(v_{1},\lambda\,|\,v_{2},\mu)^{-\sigma\sigma'}:\Phi_{\sigma',\mu}^{(n_{2})}[u_{2},v_{2}]\Phi^{(n_{1})}_{\sigma,\lambda}[u_{1},v_{1}]:,\\
      \contraction{}{\Phi_{\sigma',\mu}^{(n_{2})\ast}[u_{2},v_{2}]}{}{\Phi^{(n_{1})\ast}_{\sigma,\lambda}[u_{1},v_{1}]}\Phi_{\sigma',\mu}^{(n_{2})\ast}[u_{2},v_{2}]\Phi^{(n_{1})\ast}_{\sigma,\lambda}[u_{1},v_{1}]&=\bl(\cG(v_{1}/v_{2})\br)^{\sigma\sigma'}N_{\sigma \sigma'}(q_{3}v_{1},\lambda\,|\,v_{2},\mu)^{-\sigma\sigma'}:\Phi_{\sigma',\mu}^{(n_{2})\ast}[u_{2},v_{2}]\Phi^{(n_{1)\ast}}_{\sigma,\lambda}[u_{1},v_{1}]:,\\
       \contraction{}{\Phi_{\sigma',\mu}^{(n_{2})\ast}[u_{2},v_{2}]}{}{\Phi^{(n_{1})}_{\sigma,\lambda}[u_{1},v_{1}]}\Phi_{\sigma',\mu}^{(n_{2})\ast}[u_{2},v_{2}]\Phi^{(n_{1})}_{\sigma,\lambda}[u_{1},v_{1}]&=\bl(\cG(\gamma^{-1}v_{1}/v_{2})\br)^{-\sigma\sigma'}N_{\sigma\sigma'}(\gamma v_{1},\lambda\,|\,v_{2},\mu)^{\sigma\sigma'}:\Phi_{\sigma',\mu}^{(n_{2})\ast}[u_{2},v_{2}]\Phi^{(n_{1})}_{\sigma,\lambda}[u_{1},v_{1}]:,\\
        \contraction{}{\Phi_{\sigma',\mu}^{(n_{2})}[u_{2},v_{2}]}{}{\Phi^{(n_{1})\ast}_{\sigma,\lambda}[u_{1},v_{1}]}\Phi_{\sigma',\mu}^{(n_{2})}[u_{2},v_{2}]\Phi^{(n_{1})\ast}_{\sigma,\lambda}[u_{1},v_{1}]&=\bl(\cG(\gamma^{-1}v_{1}/v_{2})\br)^{-\sigma\sigma'}N_{\sigma\sigma'}(\gamma v_{1},\lambda\,|\,v_{2},\mu)^{\sigma\sigma'}:\Phi_{\sigma',\mu}^{(n_{2})}[u_{2},v_{2}]\Phi^{(n_{1})\ast}_{\sigma,\lambda}[u_{1},v_{1}]:,
    \end{split}\label{eq:glue-intertwiner-5}
\end{align}
for $\sigma,\sigma'=\pm$. We used the two-dimensional convention representing the vertex operator representations in the vertical direction and the crystal representations in the horizontal direction. Taking contractions in the vertical direction indeed gives the Nekrasov factors of (\ref{eq:def-Nekrasov}). 
\paragraph{Contractions in crystal representations}
Let us consider contractions in the crystal representations which are drawn horizontally:
\begin{align}
\adjustbox{valign=c}{
    \begin{tikzpicture}
            \begin{scope}[scale=1.5]
                \draw[postaction={segment={mid arrow}}] (1,0) -- (-1,0);
                \node [left,scale=0.8] at (-1.1,0) {$\Phi^{(n_{1})}_{\sigma}[u_{1},v]$};
                \node [below,scale=0.8] at (0,0) {$(0,\sigma)_{v}$};
                \node [above, scale=0.8] at (-1,1) {$(1,n_{1})_{u_{1}}$};
                \node [above,scale=0.8] at (1,1) {$(1,n_{2})_{u_{2}}$};
                \node [right,scale=0.8] at (1.1,0) {$\Phi^{(n_{2}-\sigma)\ast}_{\sigma}[u_{2}v^{-\sigma},v]$};
                \node [below, scale=0.8] at (-1,-1) {$(1,n_{1}+\sigma)_{-u_{1}v^{\sigma}}$};
                \node [below, scale=0.8] at (1,-1) {$(1,n_{2}-\sigma)_{-u_{2}v^{-\sigma}}$};
                \node [above,scale=0.7] at (0,0) {$\sum_{\lambda}a_{\lambda}^{(\sigma)}\dket{v,\lambda}\dbra{v,\lambda}$};
		        \draw[postaction={segment={mid arrow}}] (1,1)--(1,0)--(1,-1);
		        \draw[postaction={segment={mid arrow}}] (-1,1) -- (-1,0)--(-1,-1);
            \end{scope}
        \end{tikzpicture}}=\Phi_{\sigma}^{(n_{1})}[u_{1},v]\cdot\Phi_{\sigma}^{(n_{2}-\sigma)*}[u_{2}v^{-\sigma},v]
\end{align}
where the product $\cdot$ means contraction in the crystal representation. Inserting $\mathbbm{1}=\sum_{\lambda}a_{\lambda}^{(\sigma)}\dket{v,\lambda}\dbra{v,\lambda}$, we have
\begin{align}
\begin{split}
&\Phi_{\sigma}^{(n_{1})}[u_{1},v]\cdot\Phi_{\sigma}^{(n_{2}-\sigma)*}[u_{2}v^{-\sigma},v]=\sum_{\lambda}a^{(\sigma)}_{\lambda}\Phi_{\sigma,\lambda}^{(n_{2}-\sigma)*}[u_{2}v^{-\sigma},v]\otimes\Phi_{\sigma,\lambda}^{(n_{1})}[u_{1},v],
\end{split}
\end{align}
where in the second line, the order of the tensor products is opposite compared with the figure. Choosing this order is convenient because the intertwiners act as the following in this order:
\begin{align}
\begin{split}
    \mathcal{F}^{(1,n_{2})}(u_{2})\otimes\mathcal{F}^{(1,n_{1})}(u_{1})\xrightarrow{\Phi^{\ast}\otimes 1} \mathcal{F}^{(1,n_{2})}(u_{2})\otimes\mathcal{F}^{(0,\sigma)}(v)\otimes\mathcal{F}^{(1,n_{1})}(u_{1})\\
    \xrightarrow{1\otimes \Phi}\mathcal{F}^{(1,n_{2}-\sigma)}(-u_{2}v^{-\sigma})\otimes \mathcal{F}^{(1,n_{1}+\sigma)}(-u_{1}v^{\sigma}).
\end{split}
\end{align}

\paragraph{Generalized AFS intertwiner}
For later use, we further introduce a generalized intertwiner as
\begin{align}
\begin{split}
\Phi^{(n)}_{\vec{\sigma}}[u,\vec{v}]&:(0,\vec{\sigma})_{\vec{v}}\otimes (1,n)_{u}\rightarrow (1,n')_{u'},\\
\Phi^{(n^{\ast})*}_{\vec{\sigma}}[u^{\ast},\vec{v}]&:(1,n^{\ast})_{u^{\ast}}\rightarrow(1,{n^{\ast}}')_{{{u^{\ast}}'}}\otimes(0,\vec{\sigma})_{\vec{v}},\\
    \Phi^{(n)}_{\vec{\sigma}}[u,\vec{v}]&=\Phi_{\sigma_{N}}^{(n_{N})}[u_{N},v_{N}]\cdots\Phi_{\sigma_{2}}^{(n_{2})}[u_{2},v_{2}]\Phi_{\sigma_{1}}^{(n_{1})}[u_{1},v_{1}],\\
\Phi^{(n^{\ast})*}_{\vec{\sigma}}[u^{\ast},\vec{v}]&=\Phi_{\sigma_{N}}^{(n_{N}^{*})*}[u_{N}^{*},v_{N}]\cdots\Phi_{\sigma_{2}}^{(n_{2}^{*})*}[u_{2}^{*},v_{2}]
\Phi_{\sigma_{1}}^{(n_{1}^{*})*}[u_{1}^{*},v_{1}],
\end{split}\label{eq:generalizedAFSintertwiner}
\end{align}
where 
\begin{align} 
\begin{split}
&(0,\vec{\sigma})_{\vec{v}}= (0,\sigma_{N})_{v_{N}}\otimes(0,\sigma_{N-1})_{v_{N-1}}\otimes\cdots\otimes(0,\sigma_{1})_{v_{1}},\\
&n'=n+\sum_{i=1}^{N}\sigma_{i},\quad u'=u\prod_{i=1}^{N}(-v_{i})^{\sigma_{i}},\\
&{n^{\ast}}'=n^{\ast}-\sum_{i=1}^{N}\sigma_{i},\quad {u^{\ast}}'=u^{\ast}\prod_{i=1}^{N}(-v_{i})^{-\sigma_{i}}
\end{split}\label{eq:generalizedAFSintertwinerconservation}
\end{align} and the spectral parameters obey the condition (\ref{eq:pureSYMconservation}). We depict these generalized intertwiners using thick arrows as
\begin{align}
    \adjustbox{valign=c}{\begin{tikzpicture}
            \begin{scope}[scale=1.2]
                \draw[postaction={segment={mid arrow}}] (0.4,0.8)--(-0.8,0.8);
               \node [right, scale=0.8]at (0.4,0.8){$(0,\sigma_{1})_{v_{1}}$}; \draw[postaction={segment={mid arrow}}] (0.4,0.4)--(-0.8,0.4);
               \node [right,scale=0.8] at (0.4,0.4){$(0,\sigma_{2})_{v_{2}}$}; 
               \node at (-0.2,0.1) {$\vdots$}; \draw[postaction={segment={mid arrow}}] (0.4,-0.4)--(-0.8,-0.4);
               \node [right,scale=0.8] at (0.4,-0.4){$(0,\sigma_{N-1})_{v_{N-1}}$}; \draw[postaction={segment={mid arrow}}] (0.4,-0.8)--(-0.8,-0.8);
               \node [right,scale=0.8] at (0.4,-0.8){$(0,\sigma_{N})_{v_{N}}$};
		        \draw[] (-0.8,1.1) -- (-0.8,0)--(-0.8,-1.1);
            \end{scope}
\end{tikzpicture}} \rightarrow\quad \adjustbox{valign=c}{ \begin{tikzpicture}
\begin{scope}[scale=1.2]
 \draw[very thick, postaction={segment={mid arrow}}] (0.6,0)--(-0.8,0);
 \node[above, scale=0.8] at (-0.1,0){$(0,\vec{\sigma})_{\vec{v}}$};
 \draw[] (-0.8,1.1) -- (-0.8,0)--(-0.8,-1.1);
\end{scope}
\end{tikzpicture}},\quad \adjustbox{valign=c}{\begin{tikzpicture}
            \begin{scope}[scale=1.2]
                \draw[postaction={segment={mid arrow}}] (0.4,0.8)--(-0.8,0.8);
               \node [left, scale=0.8]at (-0.8,0.8){$(0,\sigma_{1})_{v_{1}}$}; \draw[postaction={segment={mid arrow}}] (0.4,0.4)--(-0.8,0.4);
               \node [left,scale=0.8] at (-0.8,0.4){$(0,\sigma_{2})_{v_{2}}$}; 
               \node at (-0.2,0.1) {$\vdots$}; \draw[postaction={segment={mid arrow}}] (0.4,-0.4)--(-0.8,-0.4);
               \node [left,scale=0.8] at (-0.8,-0.4){$(0,\sigma_{N-1})_{v_{N-1}}$}; \draw[postaction={segment={mid arrow}}] (0.4,-0.8)--(-0.8,-0.8);
               \node [left,scale=0.8] at (-0.8,-0.8){$(0,\sigma_{N})_{v_{N}}$};
		        \draw[] (0.4,1.1)--(0.4,0)--(0.4,-1.1);
            \end{scope}
\end{tikzpicture}}\quad\rightarrow\quad \adjustbox{valign=c}{ \begin{tikzpicture}
\begin{scope}[scale=1.2]
 \draw[very thick, postaction={segment={mid arrow}}] (0.6,0)--(-0.8,0);
 \node[above, scale=0.8] at (-0.1,0){$(0,\vec{\sigma})_{\vec{v}}$};
 \draw[] (0.6,1.1) -- (0.6,0)--(0.6,-1.1);
\end{scope}
\end{tikzpicture}}
\end{align}

The order of the crystal representation was chosen so that the actions of the generators are compatible with the coproduct formula. In this order the maps $\Phi^{(n)}_{\vec{\sigma}}[u,\vec{v}],\Phi^{(n^{\ast})\ast}_{\vec{\sigma}}[u^{\ast},\vec{v}]$ read as 
\begin{align}
\begin{split}
    \Phi^{(n)}_{\vec{\sigma}}[u,\vec{v}]:&\bigotimes_{i=1}^{N}(0,\sigma_{N-i+1})_{v_{N-i+1}}\otimes (1,n_{1})_{u_{1}}\xrightarrow{\Phi_{\sigma_{1}}^{(n_{1})}}\bigotimes_{i=1}^{N-1}(0,\sigma_{N-i+1})_{v_{N-i+1}}\otimes (1,n_{2})_{u_{2}}\\&\qquad\xrightarrow{\Phi^{(n_{2})}_{\sigma_{2}}}\bigotimes_{i=1}^{N-2}(0,\sigma_{N-i+1})_{v_{N-i+1}}\otimes (1,n_{3})_{u_{3}}\xrightarrow{\Phi^{(n_{3})}_{\sigma_{3}}}\cdots\xrightarrow{\Phi^{(n_{N})}_{\sigma_{N}}}(1,n')_{u'},\\
    \Phi^{(n^{\ast})\ast}_{\vec{\sigma}}[u^{\ast},\vec{v}]:&\,(1,n^{\ast})_{u^{\ast}}\xrightarrow{\Phi^{(n_{1}^{\ast})\ast}_{\sigma_{1}}}(1,n_{1}^{\ast})_{u_{1}^{\ast}}\otimes (0,\sigma_{1})_{v_{1}}\xrightarrow{\Phi^{(n_{2}^{\ast})\ast}_{\sigma_{2}}}(1,n_{2}^{\ast})_{u_{2}^{\ast}}\otimes (0,\sigma_{2})_{v_{2}}\otimes (0,\sigma_{1})_{v_{1}}\\
    &\xrightarrow{\Phi_{\sigma_{3}}^{(n_{3}^{\ast})\ast}}(1,n_{3}^{\ast})_{u_{3}^{\ast}}\otimes (0,\sigma_{3})_{v_{3}}\otimes (0,\sigma_{2})_{v_{2}}\otimes (0,\sigma_{1})_{v_{1}}\rightarrow\cdots\rightarrow(1,{n^{\ast}}')_{{u^{\ast}}'}\otimes (0,\vec{\sigma})_{\vec{v}}.
\end{split}
\end{align}
We have the following contraction formulas
\begin{align}
\begin{split}
    \Phi^{(n)}_{\vec{\sigma},\vec{\lambda}}[u,\vec{v}]&=\prod_{i<j}\mathcal{G}\left(\frac{q_{3}^{-1}v_{i}}{v_{j}}\right)^{\sigma_{i}\sigma_{j}}N_{\sigma_{i}\sigma_{j}}(v_{i},\lambda_{i}\,|\,v_{j},\lambda_{j})^{-\sigma_{i}\sigma_{j}}:\Phi^{(n)}_{\vec{\sigma},\vec{\lambda}}[u,\vec{v}]:,\\
    \Phi^{(n^{\ast})*}_{\vec{\sigma},\vec{\lambda}}[u^{\ast},\vec{v}]&=\prod_{i<j}\mathcal{G}\left({\frac{v_{i}}{v_{j}}}\right)^{\sigma_{i}\sigma_{j}}N_{\sigma_{i}\sigma_{j}}(q_{3}v_{i},\lambda{i}\,|\,v_{j},\lambda_{j})^{-\sigma_{i}\sigma_{j}}:\Phi^{(n^{\ast})*}_{\vec{\sigma},\vec{\lambda}}[u^{\ast},\vec{v}]:
\end{split}
\end{align}
and actually these operators are the supergroup generalization of the generalized AFS intertwiners introduced in \cite{Bourgine:2017jsi} up to perturbative factors. 
\subsection{Dictionary of the correspondence}\label{sec:intertwinercorrespondence}
\begin{figure}[t]
    \centering
    \begin{tikzpicture}[thick,scale=1.2]
        \begin{scope}[xscale = -1]
		\node[below,scale=0.9] at (1.5,1) {$(1,n)_u$};
		\node[above,scale=0.9] at (0,0) {$(0,1)_{ v}$};
		\node[above,scale=0.9] at (1.5,-1.1) {$(1,n+1)_{-uv}$};
		\node[below right,scale=0.9] at (1,-0.1) {$\Phi_{+}^{(n)}[u, v]$};
		\draw[postaction={segment={mid arrow}}] (0,0) -- (1,0) -- (1.7,-0.7);
		\draw[postaction={segment={mid arrow}}] (1,1) -- (1,0);
		\end{scope}
	    \begin{scope}[xscale = -1,shift=({0,-3.3})]
		\node[below,scale=0.9] at (1.5,1) {$(1,n)_u$};
		\node[above,scale=0.9] at (0,0) {$(0,-1)_{ v}$};
		\node[above,scale=0.9] at (1.5,-1.1) {$(1,n-1)_{-uv^{-1}}$};
		\node[below right,scale=0.9] at (1,-0.1) {$\Phi_{-}^{(n)}[u, v]$};
		\draw[dashed,postaction={segment={mid arrow}}] (0,0) -- (1,0);
		\draw[postaction={segment={mid arrow}}] (1,0)-- (1.7,-0.7);
		\draw[postaction={segment={mid arrow}}] (1,1) -- (1,0);
		\end{scope}
		\begin{scope}[shift=({3.5,0}),scale=1.2]
                \node[above] at (0,0) {$\text{D4}^{+}$};
                \node[above ] at (-1,0.8) {$\text{NS}5$};
                \draw[] (0,0) -- (-1,0);
		        \draw[] (-1,0.8) -- (-1,0)--(-1,-0.8);   
            \end{scope}
        \begin{scope}[shift=({3.5,-3.3}),scale=1.2]
                \node[above] at (0,0) {$\text{D4}^{-}$};
                \node[above ] at (-1,0.8) {$\text{NS}5$};
                \draw[dashed] (0,0) -- (-1,0);
		        \draw[] (-1,0.8) -- (-1,0)--(-1,-0.8);   
            \end{scope}
        \begin{scope}[shift=({7,-0.2}),xscale = -1]
		\node[above] at (1,1) {$(n,1)$};
		\node[right] at (0,0.2) {$\text{D5}^{+}$};
		\node [right] at (0,-0.2){$(1,0)$};
		\node[below] at (1.7,-0.7){$(n+1,1)$};
	    \draw[] (0,0) -- (1,0) -- (1.7,-0.7);
		\draw[] (1,1) -- (1,0);
		\end{scope}
		 \begin{scope}[shift=({7,-3.5}),xscale = -1]
	    \node [above] at (1,1) {$(n,1)$};
		\node[right] at (0,0.2) {$\text{D5}^{-}$};
		\node [right]at (0.15,-0.2) {$(-1,0)$};
		\node [below] at (1.7,-0.7){$(n-1,1)$};
	    \draw[dashed] (0,0) -- (1,0);
		\draw[] (1,1) -- (1,0) -- (1.7,-0.7);
		\end{scope}
    \end{tikzpicture}
    \caption{Correspondences of the positive and negative intertwiners with brane junctions of type~\rom{2}A and type~\rom{2}B theory. Dual intertwiners have similar physical interpretation.}
    \label{fig:corr-intertwiner}
\end{figure}
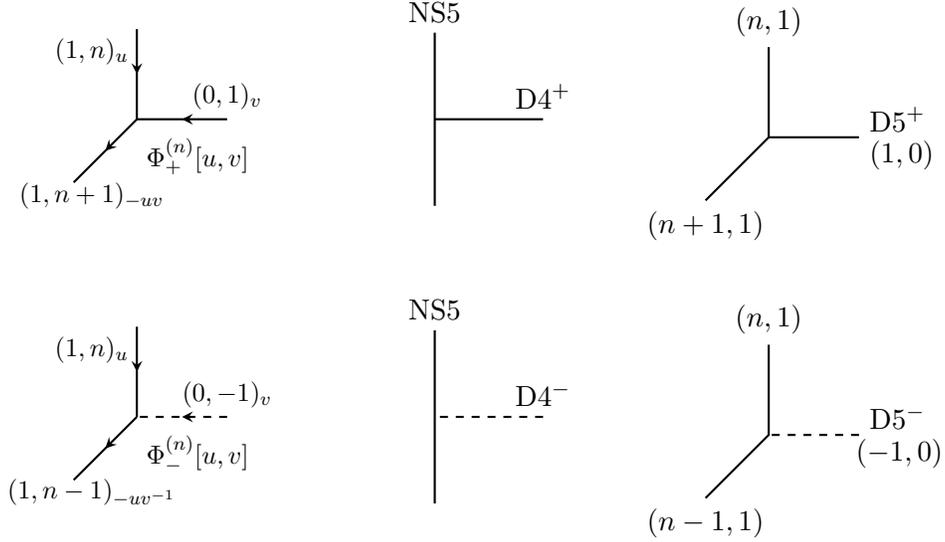
In \cite{Awata:2011ce}, the authors proposed a correspondence between the level $(q,p)$ representation and the $(p,q)$ 5-branes. We would like to extend the dictionary of this correspondence to the supergroup gauge theory case. The difference with the non-supergroup case is the existence of the negative D-branes in the system. 

Looking at the contraction formulas of the (dual) intertwiners in (\ref{eq:glue-intertwiner-5}), we observe that the contractions of $\Phi_{\sigma_{2},\lambda_{2}}\Phi_{\sigma_{1},\lambda_{1}}$ and $\Phi^{\ast}_{\sigma_{2},\lambda_{2}}\Phi_{\sigma_{1},\lambda_{1}}^{\ast}$ give the vector multiplet contributions coming from two D-branes with parities $\sigma_{1}$ and $\sigma_{2}$ (see section~\ref{sec:equiv-index} and equation~(\ref{eq:Nekrasov-convention})). Contractions of intertwiners and dual intertwiners give the contributions of bifundamental multiplets with bifundamental masses $\gamma^{-1}$ (see (\ref{eq:Nekrasov-convention})). Therefore, we propose the following correspondence (see Figure~\ref{fig:corr-intertwiner}):
\begin{align}
\begin{array}{|c|c|c|}
\hline
\text{level of representation}& \text{Type \rom{2}A}&\text{Type \rom{2}B}\\\hline
(0,1) &\text{positive D4-brane}~(\text{D}4^{+})&\text{positive D5-brane}~(\text{D}5^{+})\\\hline
   (0,-1)  & \text{negative D4-brane}~(\text{D}4^{-})&\text{negative D5-brane}~(\text{D}5^{-})\\\hline
   (1,0) & \multirow{2}{*}{\text{ordinary NS5-brane}}&\text{ordinary NS5-brane}\\\cline{1-1}\cline{3-3}
   (1,n) & &\text{$(n,1)$ 5-brane} \\\hline
\end{array}\label{eq:brane-rep-corresp-table}
\end{align}
To be accurate, the correspondence with type \rom{2}A theory should be understood as a correspondence after taking the degenerate limit and considering representations of affine Yangian $\mathfrak{gl}_{1}$ or the Drinfeld double\footnote{Using the algebra introduced in \cite{Bourgine:2018uod}, we can do the intertwiner formalism of 4d gauge theories. One central charge $c$, which corresponds to the level $\ell_{1}$ in (\ref{eq:deflevel}), appears in this algebra. The vertex operator representation has $c=1$, while the crystal representations have $c=0$. The other level $\ell_{2}$ in (\ref{eq:deflevel}) does not appear in this case. This means the level $(1,0)$ and $(1,n)$ representations in the degenerate limit are essentially the same and thus should be understood as the same ordinary NS5-brane.} of it studied in \cite{Bourgine:2018uod}. 

From the above correspondences, the intertwiners are understood as junctions of the positive (negative) D-branes and NS5-branes (see Figure~\ref{fig:corr-intertwiner}). Explicit examples supporting this proposal are given in the section \ref{sec:Examples}.

\begin{figure}[t]
    \centering
    \includegraphics[width=15cm]{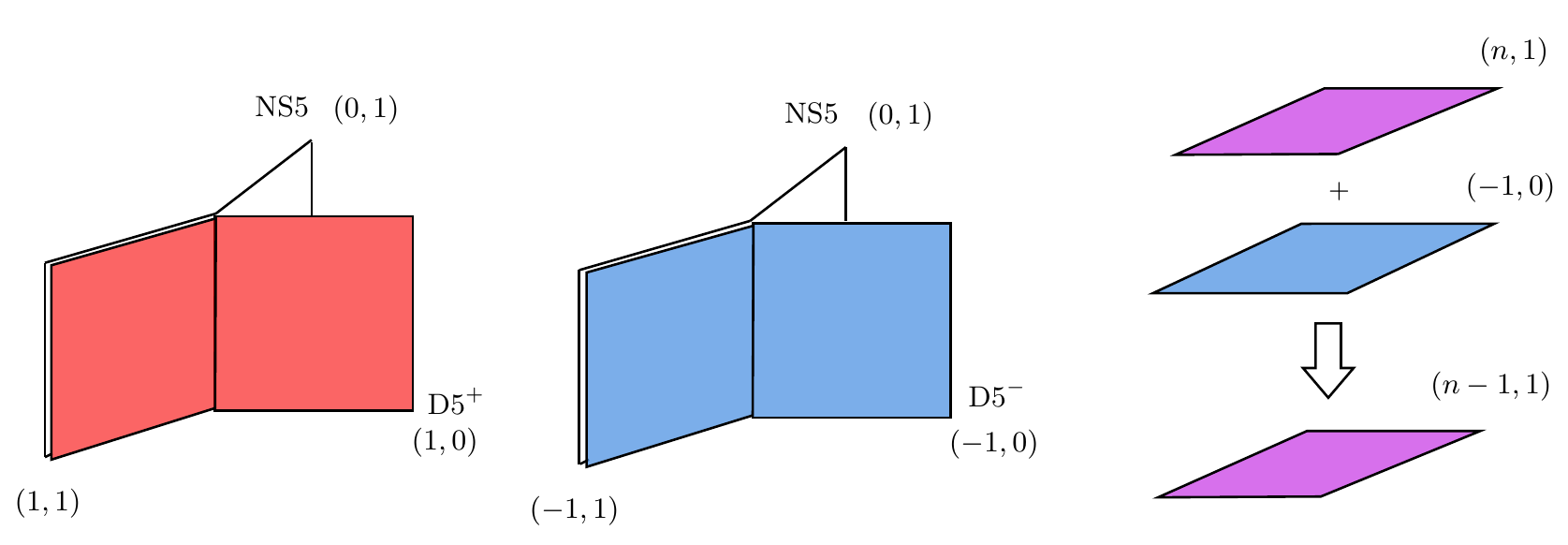}   \caption{Junctions of NS5-brane and positive and negative D5-branes. The charge assigned to negative D5-branes is $(-1,0)$ which correspond to level $(0,-1)$ representation. When NS5-brane intersect with a negative D5-brane, they will form a bound state with charge $(-1,1)$. When general $(n,1)$ 5-brane intersect with a negative D5-brane, we will obtain a $(n-1,1)$ 5-brane.}
    \label{fig:NS5D5junction}
\end{figure}
Negative D-branes have opposite charges compared with the positive D-branes \cite{Okuda:2006fb}. Thus, it is natural to assign the representation $(0,1)$ to the positive D-brane and representation $(0,-1)$ to the negative D-brane. When $n$ D5-branes intersect with a NS5-brane, they will form a bound state $(n+1,1)$ 5-brane. We expect a similar story holds when negative D5-branes intersect with a NS5-brane, but this time since the negative brane has a negative charge, the bound state will be $(n-1,1)$ 5-brane (see Figure~\ref{fig:NS5D5junction}). This conservation law indeed satisfies how the positive and negative intertwiners act on the modules (see Figure~\ref{fig:corr-intertwiner}).

In this paper, we will not give the specific correspondences for general $(p,q)$ 5-branes with representations. Moreover, usually, the charges of the $(p,q)$ 5-branes determine how the brane is bent when they intersect with other branes. A different discussion must be done to determine the angle and how to draw the five-brane webs when negative branes are included. In this paper, we do not specify how the branes are bent nor how the brane web should be drawn to include the information of the charges. We simply assign charges to the branes to do calculations.

\section{Examples}\label{sec:Examples}
We apply the intertwiner formalism introduced in the previous section and derive partition functions of supergroup gauge theories of A and D-type quiver. Using the formalism introduced in \cite{Kimura:2019gon}, we can generalize the computations to other quiver structures. We leave this generalization for future work. In particular, we consider three examples: pure super Yang-Mills (section~\ref{sec:example-pureSYM}), $A_{r}$-quiver (section~\ref{sec:example-Aquiver}), and $D_{r}$-quiver (section~\ref{sec:example-Dquiver}).
\subsection{Pure \texorpdfstring{$U(N\,|\,\sigma_{1},\sigma_{2},\cdots,\sigma_{N})$}{U(N;sigma1,sigma2,...,sigmaN)} gauge theory}\label{sec:example-pureSYM}
Using the intertwiner formalism in deriving partition functions of pure supergroup gauge theories, we need to consider stacks of D5-branes surrounded by NS5-branes with signatures $\sigma_{1},\cdots,\sigma_{N}$, where $\sigma_{i}=+,-$ and $N$ is the total number of D5-branes. Here, we will not determine the signatures because generally we do not have to. However, the ordering of the signatures depend when we use the intertwiner (or topological vertex formalism), so we write it down explicitly. For convenience, we call this theory a $U(N\,|\,\sigma_{1},\sigma_{2},\cdots, \sigma_{N})=U(N\,|\,\vec{\sigma})$ supergroup gauge theory, where $\vec{\sigma}=(\sigma_{1},\sigma_{2},\cdots,\sigma_{N})$. If we do not care of the order of the branes, then this is just a $U(N_{+}\,|\,N_{-})$ theory:
\begin{align}
\begin{split}
    &U(N\,|\,\sigma_{1},\sigma,\cdots,\sigma_{N})=U(N_{+}\,|\,N_{-}),\\
    &N_{+}=|\{i\,|\,\sigma_{i}=+1\}|,\quad N_{-}=|\{i\,|\,\sigma_{i}=-1\}|,\\
    &N_{+}+N_{-}=N.
\end{split}
\end{align}

Using the intertwiner formalism, the partition function is derived as 
\begin{align}
\begin{split}
\mathcal{Z}_{N,\,(\sigma_{1},\cdots,\sigma_{N})}&=
\adjustbox{valign=c,width=6cm}{\tikzset{every picture/.style={line width=0.75pt}} 
\begin{tikzpicture}[x=0.75pt,y=0.75pt,yscale=-1.0,xscale=1.0]
\draw    (232,39) -- (232,249.16) ;
\draw    (392,39) -- (392,249.16) ;
\draw    (232,68) -- (392.8,68.49) ;
\draw    (232,98) -- (392.8,98.49) ;
\draw    (232,128) -- (392.8,128.49) ;
\draw    (232,228) -- (392.8,228.49) ;
\draw [color={rgb, 255:red, 74; green, 144; blue, 226 }  ,draw opacity=1 ]   (232,177) -- (392.8,177.49) ;
\draw    (232,196) -- (392.8,196.49) ;
\draw    (232,153) -- (392.8,153.49) ;
\draw [color={rgb, 255:red, 208; green, 2; blue, 2 }  ,draw opacity=1 ][fill={rgb, 255:red, 208; green, 2; blue, 2 }  ,fill opacity=1 ]   (232,151) -- (232,176) ;
\draw [shift={(232,178)}, rotate = 270] [color={rgb, 255:red, 208; green, 2; blue, 2 }  ,draw opacity=1 ][line width=0.75]    (10.93,-3.29) .. controls (6.95,-1.4) and (3.31,-0.3) .. (0,0) .. controls (3.31,0.3) and (6.95,1.4) .. (10.93,3.29)   ;
\draw [color={rgb, 255:red, 208; green, 2; blue, 2 }  ,draw opacity=1 ][fill={rgb, 255:red, 208; green, 2; blue, 2 }  ,fill opacity=1 ]   (392,174) -- (392,199) ;
\draw [shift={(392,201)}, rotate = 270] [color={rgb, 255:red, 208; green, 2; blue, 2 }  ,draw opacity=1 ][line width=0.75]    (10.93,-3.29) .. controls (6.95,-1.4) and (3.31,-0.3) .. (0,0) .. controls (3.31,0.3) and (6.95,1.4) .. (10.93,3.29)   ;
\draw (300,124.4) node [anchor=north west][inner sep=0.75pt]    {$\vdots $};
\draw (280.2,45.2) node [anchor=north west][inner sep=0.75pt]  [font=\small]  {$( 0,\sigma _{1})_{v_{1}}$};
\draw (280.2,75.2) node [anchor=north west][inner sep=0.75pt]  [font=\small]  {$( 0,\sigma _{2})_{v_{2}}$};
\draw (280.2,105.2) node [anchor=north west][inner sep=0.75pt]  [font=\small]  {$( 0,\sigma _{3})_{v_{3}}$};
\draw (286.2,158.2) node [anchor=north west][inner sep=0.75pt]  [font=\small,color={rgb, 255:red, 74; green, 144; blue, 226 }  ,opacity=1 ]  {$( 0,\sigma _{i})_{v_{i}}$};
\draw (283.2,236.2) node [anchor=north west][inner sep=0.75pt]  [font=\small]  {$( 0,\sigma _{N})_{v_{N}}$};
\draw (201.2,16.2) node [anchor=north west][inner sep=0.75pt]  [font=\small]  {$( 1,n)_{u}$};
\draw (371.2,15.4) node [anchor=north west][inner sep=0.75pt]  [font=\small]  {$\left( 1,n^{*}\right)_{u^{*}}$};
\draw (300,195.4) node [anchor=north west][inner sep=0.75pt]    {$\vdots $};
\draw (172.6,152.4) node [anchor=north west][inner sep=0.75pt]  [font=\small,color={rgb, 255:red, 208; green, 2; blue, 2 }  ,opacity=1 ]  {$( 1,n_{i})_{u_{i}}$};
\draw (398.6,176.4) node [anchor=north west][inner sep=0.75pt]  [font=\small,color={rgb, 255:red, 208; green, 2; blue, 2 }  ,opacity=1 ]  {$\left( 1,n_{i}^{*}\right)_{u_{i}^{*}}$};
\end{tikzpicture}}
=\begin{array}{ccc}
\widehat{\,\underline{0}\,}&&\vspace{1mm}\widehat{\,\underline{0}\,}\\
\Phi_{\sigma_{1}}^{(n_{1})}[u_{1},v_{1}]&\cdot&\Phi_{\sigma_{1}}^{(n_{1}^{*})*}[u_{1}^{*},v_{1}]\\
\Phi_{\sigma_{2}}^{(n_{2})}[u_{2},v_{2}]&\cdot&\Phi_{\sigma_{2}}^{(n_{2}^{*})*}[u_{2}^{*},v_{2}]\\
\vdots&&\vdots\\
\Phi_{\sigma_{N}}^{(n_{N})}[u_{N},v_{N}]&\cdot&\Phi_{\sigma_{N}}^{(n_{N}^{*})*}[u_{N}^{*},v_{N}]\\
\uwidehat{\overline{\,0}\,}&&\uwidehat{\overline{\,0}\,}\\
\end{array}\\
&=\sum_{\lambda_{1}\cdots \lambda_{N}}\prod_{i=1}^{N}a_{\lambda_{i}}^{(\sigma_{i})}\bra{0}\overleftarrow{\prod_{i=1}^{N}}\Phi^{(n_{i})}_{\sigma_{i},\lambda_{i}}[u_{i},v_{i}]\ket{0}\bra{0}\overleftarrow{\prod_{i=1}^{N}}\Phi^{(n^{*}_{i})*}_{\sigma_{i},\lambda_{i}}[u_{i}^{*},v_{i}]\ket{0},
\end{split}
\end{align}
where the product is an ordered product
\begin{align}
    \overleftarrow{\prod_{i=1}^{N}}f_{i}(z)\doteq f_{N}(z)f_{N-1}(z)\cdots f_{2}(z)f_{1}(z).
\end{align}
Note also that from the conservation law, the spectral parameters obey the conditions
\begin{align}
\begin{split}
    &n_{i}=n+\sum_{l=1}^{i-1}\sigma_{l},\quad u_{i}=u\prod_{l=1}^{i-1}(-v_{l}^{\sigma_{l}}),\\
    &n_{i}^{*}=n^{*}-\sum_{l=1}^{i}\sigma_{l},\quad u_{i}^{*}=u^{*}\prod_{l=1}^{i}(-v_{l}^{-\sigma_{l}})
\end{split}\label{eq:pureSYMconservation}
\end{align}
for $i=1,\cdots, N$. Note that using the generalized AFS intertwiners in (\ref{eq:generalizedAFSintertwiner}), the partition function is written in a simpler way
\begin{align}
    \bra{0}\otimes \bra{0} \mathcal{T}_{U(N\,|\,\vec{\sigma})}\ket{0}\otimes\ket{0},\quad \mathcal{T}_{U(N\,|\,\vec{\sigma})}\coloneqq \Phi^{(n)}_{\vec{\sigma}}[u,\vec{v}]\cdot\Phi_{\vec{\sigma}}^{(n^{\ast})\ast}[u^{\ast},\vec{v}],\label{eq:pureSYMoperator}
\end{align}
where the product $\cdot$ is understood as contractions in all the crystal representations. We note this operator itself depends on the sequence of the parities $\vec{\sigma}$.

Computing the contractions and using the properties of the Nekrasov factors in Appendix~\ref{sec:appendix-nekrasov-symm}, the partition function decouples into the perturbative part and instanton part as 
\begin{align}
    \mathcal{Z}_{N,(\sigma_{1},\cdots,\sigma_{N})}&=\mathcal{Z}^{\text{pert}}_{N,(\sigma_{1},\cdots,\sigma_{N})}\mathcal{Z}^{\text{inst}}_{N,(\sigma_{1},\cdots,\sigma_{N})},\\
    \mathcal{Z}^{\text{pert}}_{N,(\sigma_{1},\cdots,\sigma_{N})}&=\prod_{i<j}\mathcal{G}\left(q_{3}^{-1}\frac{v_{i}}{v_{j}}\right)^{\sigma_{i}\sigma_{j}}\mathcal{G}\left(\frac{v_{i}}{v_{j}}\right)^{\sigma_{i}\sigma_{j}},\\
\begin{split}
    \mathcal{Z}^{\text{inst}}_{N,(\sigma_{1},\cdots,\sigma_{N})}&=\sum_{\lambda_{1},\cdots,\lambda_{N}}\prod_{l=1}^{N}\left(a_{\lambda_{l}}^{(\sigma_{l})}t_{n_{l},\sigma_{l}}(\lambda_{l},u_{l},v_{l})t^{\ast}_{n_{l}^{*},\sigma_{l}}(\sigma_{l},u_{l}^{\ast},v_{l})\right)\\
    &\times \prod_{i<j}\left\{\prod_{x\in\lambda_{i}}\left(-\frac{v_{j}}{\chi_{x}^{(\sigma_{i})}}\right)^{\sigma_{i}\sigma_{j}}\prod_{x\in\lambda_{j}}\left(-\frac{\chi_{x}^{(\sigma_{j})}}{q_{3}v_{i}}\right)^{\sigma_{i}\sigma_{j}}\prod_{i\neq j}N_{\sigma_{i}\sigma_{j}}(v_{i},\lambda_{i}\,|\,v_{j},\lambda_{j})^{-\sigma_{i}\sigma_{j}}\right\}.
\end{split}
\end{align}
Using 
\begin{align}
    \prod_{l=1}^{N}a_{\lambda_{l}}^{(\sigma_{l})}=\prod_{l=1}^{N}\left((q_{3}^{\frac{1-\sigma_{l}}{2}}\gamma v_{l})^{-|\lambda_{l}|}\prod_{x\in\lambda_{l}}\chi_{x}^{(\sigma_{l})}\right)\prod_{l=1}^{N}\frac{1}{N_{\sigma_{l}\sigma_{l}}(v_{l},\lambda_{l}\,|\,v_{l},\lambda_{l})}
\end{align}
and computing the zero-modes part in front of the Nekrasov factors, we eventually have 
\begin{align}
    \mathcal{Z}_{N,(\sigma_{1},\ldots,\sigma_{N})}^{\text{inst}}=\sum_{\lambda_{1},\cdots,\lambda_{N}}\mathcal{Z}_{\text{top.}}(\vec{\lambda},\vec{\sigma})\mathcal{Z}_{\text{CS}}(\kappa,\vec{\lambda},\vec{\sigma})\mathcal{Z}_{\text{vec.}}(\vec{v},\vec{\lambda},\vec{\sigma}),
\end{align}
where 
\begin{align}
    \mathfrak{q}=-\frac{u}{u^{\ast}}\gamma^{n-n^{*}}\prod_{i=1}^{N}(-v_{i})^{\sigma_{i}},\quad \kappa=n^{*}-n-\sum_{i=1}^{N}\sigma_{i}.
\end{align}
This matches with the well known formula for gauge theory with ordinary groups when $\sigma_{i}=+1$ for all $i$. Note that the spectral parameters are assumed $|v_{i}/v_{j}|<1,\,i<j$.

Obviously, the partition function is invariant under the ordering since the topological term and the Chern-Simons term is invariant under simultaneous permutation of the Coulomb vev parameters and the parities of the D5-brane. Namely, let $\omega\in\mathfrak{S}_{N}$ be an element of the permutation group and then the partition function obeys the following symmetry:
\begin{align}
    \mathcal{Z}_{N,(\sigma_{1},\ldots,\sigma_{N})}=\mathcal{Z}_{N,(\omega\cdot\sigma_{1},\ldots,\omega\cdot\sigma_{N})},\quad \omega\cdot \sigma_{i}=\sigma_{\omega(i)}\label{eq:super-pureSYM-equiv}
\end{align}
under redefinition of the Coulomb vev parameters
\begin{align}
    v_{i}\rightarrow \omega\cdot v_{i}=v_{\omega(i)}.
\end{align}

Actually, this property is strongly related to the underlying Lie superalgebra structure. For example, let us consider the $U(3\,|\,2)$ case. For this case, we have 3 positive D-branes and 2 negative D-branes. Similarly to the ordinary group gauge theory, we can permute the order of the 3 positive D-branes and the 2 negative D-branes without changing the partition function. Under this permutation, the brane web does not change and it is a trivial symmetry. However, there are 10 different looking brane webs for the $U(3\,|\,2)$ theory:
\begin{align}
    \begin{array}{ccccc}
        \begin{tikzpicture}[cross/.style={path picture={ 
  \draw[black]
(path picture bounding box.south east) -- (path picture bounding box.north west) (path picture bounding box.south west) -- (path picture bounding box.north east);
}}]
            \begin{scope}[scale=1.2]
                \draw[] (-0.8,0.8)--(0.8,0.8);
                \draw[] (-0.8,0.4)--(0.8,0.4);
                \draw[] (-0.8,0)--(0.8,0);
                \draw[dashed] (-0.8,-0.4)--(0.8,-0.4);
                \draw[dashed] (-0.8,-0.8)--(0.8,-0.8);
		        \draw[] (0.8,1.1)--(0.8,0)--(0.8,-1.1);
		        \draw[] (-0.8,1.1) -- (-0.8,0)--(-0.8,-1.1);
		        \draw [] (1.1,0.6) circle (0.12);
		        \draw [] (1.1,0.2) circle (0.12);
		        \draw[cross] (1.1,-0.2) circle (0.12);
		        \draw[] (1.1,-0.6) circle (0.12);
		        \draw (1.1,0.48)--(1.1,0.32);
		        \draw (1.1,-0.08)--(1.1,0.08);
		        \draw (1.1,-0.48)--(1.1,-0.32);
            \end{scope}
\end{tikzpicture} & \quad \begin{tikzpicture}[cross/.style={path picture={ 
  \draw[black]
(path picture bounding box.south east) -- (path picture bounding box.north west) (path picture bounding box.south west) -- (path picture bounding box.north east);
}}]
            \begin{scope}[scale=1.2]
                \draw[] (-0.8,0.8)--(0.8,0.8);
                \draw[] (-0.8,0.4)--(0.8,0.4);
                \draw[dashed] (-0.8,0)--(0.8,0);
                \draw[dashed] (-0.8,-0.4)--(0.8,-0.4);
                \draw[] (-0.8,-0.8)--(0.8,-0.8);
		        \draw[] (0.8,1.1)--(0.8,0)--(0.8,-1.1);
		        \draw[] (-0.8,1.1) -- (-0.8,0)--(-0.8,-1.1);
		        \draw [] (1.1,0.6) circle (0.12);
		        \draw [cross] (1.1,0.2) circle (0.12);
		        \draw[] (1.1,-0.2) circle (0.12);
		        \draw[cross] (1.1,-0.6) circle (0.12);
		        \draw (1.1,0.48)--(1.1,0.32);
		        \draw (1.1,-0.08)--(1.1,0.08);
		        \draw (1.1,-0.48)--(1.1,-0.32);
            \end{scope}
\end{tikzpicture} &\quad\begin{tikzpicture}[cross/.style={path picture={ 
  \draw[black]
(path picture bounding box.south east) -- (path picture bounding box.north west) (path picture bounding box.south west) -- (path picture bounding box.north east);
}}]
            \begin{scope}[scale=1.2]
                \draw[] (-0.8,0.8)--(0.8,0.8);
                \draw[dashed] (-0.8,0.4)--(0.8,0.4);
                \draw[dashed] (-0.8,0)--(0.8,0);
                \draw[] (-0.8,-0.4)--(0.8,-0.4);
                \draw[] (-0.8,-0.8)--(0.8,-0.8);
		        \draw[] (0.8,1.1)--(0.8,0)--(0.8,-1.1);
		        \draw[] (-0.8,1.1) -- (-0.8,0)--(-0.8,-1.1);
		        \draw [cross] (1.1,0.6) circle (0.12);
		        \draw [] (1.1,0.2) circle (0.12);
		        \draw[cross] (1.1,-0.2) circle (0.12);
		        \draw[] (1.1,-0.6) circle (0.12);
		        \draw (1.1,0.48)--(1.1,0.32);
		        \draw (1.1,-0.08)--(1.1,0.08);
		        \draw (1.1,-0.48)--(1.1,-0.32);
            \end{scope}
\end{tikzpicture}&\quad
       \begin{tikzpicture}[cross/.style={path picture={ 
  \draw[black]
(path picture bounding box.south east) -- (path picture bounding box.north west) (path picture bounding box.south west) -- (path picture bounding box.north east);
}}]
            \begin{scope}[scale=1.2]
                \draw[dashed] (-0.8,0.8)--(0.8,0.8);
                \draw[dashed] (-0.8,0.4)--(0.8,0.4);
                \draw[] (-0.8,0)--(0.8,0);
                \draw[] (-0.8,-0.4)--(0.8,-0.4);
                \draw[] (-0.8,-0.8)--(0.8,-0.8);
		        \draw[] (0.8,1.1)--(0.8,0)--(0.8,-1.1);
		        \draw[] (-0.8,1.1) -- (-0.8,0)--(-0.8,-1.1);
		        \draw [] (1.1,0.6) circle (0.12);
		        \draw [cross] (1.1,0.2) circle (0.12);
		        \draw[] (1.1,-0.2) circle (0.12);
		        \draw[] (1.1,-0.6) circle (0.12);
		        \draw (1.1,0.48)--(1.1,0.32);
		        \draw (1.1,-0.08)--(1.1,0.08);
		        \draw (1.1,-0.48)--(1.1,-0.32);
            \end{scope}
\end{tikzpicture}  &\quad\begin{tikzpicture}[cross/.style={path picture={ 
  \draw[black]
(path picture bounding box.south east) -- (path picture bounding box.north west) (path picture bounding box.south west) -- (path picture bounding box.north east);
}}]
            \begin{scope}[scale=1.2]
                \draw[] (-0.8,0.8)--(0.8,0.8);
                \draw[] (-0.8,0.4)--(0.8,0.4);
                \draw[dashed] (-0.8,0)--(0.8,0);
                \draw[] (-0.8,-0.4)--(0.8,-0.4);
                \draw[dashed] (-0.8,-0.8)--(0.8,-0.8);
		        \draw[] (0.8,1.1)--(0.8,0)--(0.8,-1.1);
		        \draw[] (-0.8,1.1) -- (-0.8,0)--(-0.8,-1.1);
		        \draw [] (1.1,0.6) circle (0.12);
		        \draw [cross] (1.1,0.2) circle (0.12);
		        \draw[cross] (1.1,-0.2) circle (0.12);
		        \draw[cross] (1.1,-0.6) circle (0.12);
		        \draw (1.1,0.48)--(1.1,0.32);
		        \draw (1.1,-0.08)--(1.1,0.08);
		        \draw (1.1,-0.48)--(1.1,-0.32);
            \end{scope}
\end{tikzpicture} \vspace{0.5cm}\\\begin{tikzpicture}[cross/.style={path picture={ 
  \draw[black]
(path picture bounding box.south east) -- (path picture bounding box.north west) (path picture bounding box.south west) -- (path picture bounding box.north east);
}}]
            \begin{scope}[scale=1.2]
                \draw[] (-0.8,0.8)--(0.8,0.8);
                \draw[dashed] (-0.8,0.4)--(0.8,0.4);
                \draw[] (-0.8,0)--(0.8,0);
                \draw[dashed] (-0.8,-0.4)--(0.8,-0.4);
                \draw[] (-0.8,-0.8)--(0.8,-0.8);
		        \draw[] (0.8,1.1)--(0.8,0)--(0.8,-1.1);
		        \draw[] (-0.8,1.1) -- (-0.8,0)--(-0.8,-1.1);
		        \draw [cross] (1.1,0.6) circle (0.12);
		        \draw [cross] (1.1,0.2) circle (0.12);
		        \draw[cross] (1.1,-0.2) circle (0.12);
		        \draw[cross] (1.1,-0.6) circle (0.12);
		        \draw (1.1,0.48)--(1.1,0.32);
		        \draw (1.1,-0.08)--(1.1,0.08);
		        \draw (1.1,-0.48)--(1.1,-0.32);
            \end{scope}
\end{tikzpicture}&\quad
      \begin{tikzpicture}[cross/.style={path picture={ 
  \draw[black]
(path picture bounding box.south east) -- (path picture bounding box.north west) (path picture bounding box.south west) -- (path picture bounding box.north east);
}}]
            \begin{scope}[scale=1.2]
                \draw[dashed] (-0.8,0.8)--(0.8,0.8);
                \draw[] (-0.8,0.4)--(0.8,0.4);
                \draw[dashed] (-0.8,0)--(0.8,0);
                \draw[] (-0.8,-0.4)--(0.8,-0.4);
                \draw[] (-0.8,-0.8)--(0.8,-0.8);
		        \draw[] (0.8,1.1)--(0.8,0)--(0.8,-1.1);
		        \draw[] (-0.8,1.1) -- (-0.8,0)--(-0.8,-1.1);
		        \draw [cross] (1.1,0.6) circle (0.12);
		        \draw [cross] (1.1,0.2) circle (0.12);
		        \draw[cross] (1.1,-0.2) circle (0.12);
		        \draw[] (1.1,-0.6) circle (0.12);
		        \draw (1.1,0.48)--(1.1,0.32);
		        \draw (1.1,-0.08)--(1.1,0.08);
		        \draw (1.1,-0.48)--(1.1,-0.32);
            \end{scope}
\end{tikzpicture}&\quad\begin{tikzpicture}[cross/.style={path picture={ 
  \draw[black]
(path picture bounding box.south east) -- (path picture bounding box.north west) (path picture bounding box.south west) -- (path picture bounding box.north east);
}}]
            \begin{scope}[scale=1.2]
                \draw[] (-0.8,0.8)--(0.8,0.8);
                \draw[dashed] (-0.8,0.4)--(0.8,0.4);
                \draw[] (-0.8,0)--(0.8,0);
                \draw[] (-0.8,-0.4)--(0.8,-0.4);
                \draw[dashed] (-0.8,-0.8)--(0.8,-0.8);
		        \draw[] (0.8,1.1)--(0.8,0)--(0.8,-1.1);
		        \draw[] (-0.8,1.1) -- (-0.8,0)--(-0.8,-1.1);
		        \draw [cross] (1.1,0.6) circle (0.12);
		        \draw [cross] (1.1,0.2) circle (0.12);
		        \draw[] (1.1,-0.2) circle (0.12);
		        \draw[cross] (1.1,-0.6) circle (0.12);
		        \draw (1.1,0.48)--(1.1,0.32);
		        \draw (1.1,-0.08)--(1.1,0.08);
		        \draw (1.1,-0.48)--(1.1,-0.32);
            \end{scope}
\end{tikzpicture}   &\quad\begin{tikzpicture}[cross/.style={path picture={ 
  \draw[black]
(path picture bounding box.south east) -- (path picture bounding box.north west) (path picture bounding box.south west) -- (path picture bounding box.north east);
}}]
            \begin{scope}[scale=1.2]
                \draw[dashed] (-0.8,0.8)--(0.8,0.8);
                \draw[] (-0.8,0.4)--(0.8,0.4);
                \draw[] (-0.8,0)--(0.8,0);
                \draw[dashed] (-0.8,-0.4)--(0.8,-0.4);
                \draw[] (-0.8,-0.8)--(0.8,-0.8);
		        \draw[] (0.8,1.1)--(0.8,0)--(0.8,-1.1);
		        \draw[] (-0.8,1.1) -- (-0.8,0)--(-0.8,-1.1);
		        \draw [cross] (1.1,0.6) circle (0.12);
		        \draw [] (1.1,0.2) circle (0.12);
		        \draw[cross] (1.1,-0.2) circle (0.12);
		        \draw[cross] (1.1,-0.6) circle (0.12);
		        \draw (1.1,0.48)--(1.1,0.32);
		        \draw (1.1,-0.08)--(1.1,0.08);
		        \draw (1.1,-0.48)--(1.1,-0.32);
            \end{scope}
\end{tikzpicture} &\quad\begin{tikzpicture}[cross/.style={path picture={ 
  \draw[black]
(path picture bounding box.south east) -- (path picture bounding box.north west) (path picture bounding box.south west) -- (path picture bounding box.north east);
}}]
            \begin{scope}[scale=1.2]
                \draw[dashed] (-0.8,0.8)--(0.8,0.8);
                \draw[] (-0.8,0.4)--(0.8,0.4);
                \draw[] (-0.8,0)--(0.8,0);
                \draw[] (-0.8,-0.4)--(0.8,-0.4);
                \draw[dashed] (-0.8,-0.8)--(0.8,-0.8);
		        \draw[] (0.8,1.1)--(0.8,0)--(0.8,-1.1);
		        \draw[] (-0.8,1.1) -- (-0.8,0)--(-0.8,-1.1);
		         \draw [cross] (1.1,0.6) circle (0.12);
		        \draw [] (1.1,0.2) circle (0.12);
		        \draw[] (1.1,-0.2) circle (0.12);
		        \draw[cross] (1.1,-0.6) circle (0.12);
		        \draw (1.1,0.48)--(1.1,0.32);
		        \draw (1.1,-0.08)--(1.1,0.08);
		        \draw (1.1,-0.48)--(1.1,-0.32);
            \end{scope}
\end{tikzpicture}
    \end{array}
\end{align}
where $\bnode$ is the bosonic root and $\fnode$ is the fermionic root of the superalgebra. When the adjacent branes have the same parities, we assign the bosonic root $\bnode$. When the parities are different, we assign the fermionic root $\fnode$. Equation (\ref{eq:super-pureSYM-equiv}) means that even though there are different D-brane realizations related to the underlying superalgebra structure, the instanton partition function is equivalent. This fact was shown by direct computation for $U(2\,|\,1)$ in \cite{Kimura-Sugimoto:antivertex}. Our result is a straightforward generalization to general configurations.

\subsection{Linear quiver gauge theory}\label{sec:example-Aquiver}
Let us show next an example of linear quiver gauge theory $U(N\,|\,\sigma_{1},\cdots,\sigma_{N})^{\otimes M}$ with $N_{f}=2N$, which was the example studied in \cite{Kimura-Sugimoto:antivertex}:
\begin{align}
\adjustbox{valign=c}{
     \begin{tikzpicture}
            \begin{scope}[scale=2.3]
            \draw[postaction={segment={mid arrow}}] (-1,0.4) -- (-1.7,0.4);
                \draw[postaction={segment={mid arrow}}] (-1,0.6) -- (-1.7,0.6);
                \draw[postaction={segment={mid arrow}}] (-1,0.2) -- (-1.7,0.2);
                \node  at (-1.36, 0.06) {$\vdots$};
                \node  at (-1.36, -0.12) {$\vdots$};
                \draw[postaction={segment={mid arrow}}] (-1,-0.4) -- (-1.7,-0.4);
		        \draw[postaction={segment={mid arrow}}] (-1,-0.6) -- (-1.7,-0.6);
		        \draw (-1,0.8)--(-1,0)--(-1,-0.9);
            \draw[postaction={segment={mid arrow}}] (0,0.3) -- (-1,0.3);
                \draw[postaction={segment={mid arrow}}] (0,0.1) -- (-1,0.1);
                \draw[postaction={segment={mid arrow}}] (0,0.5) -- (-1,0.5);
                \draw[postaction={segment={mid arrow}}] (0,-0.5) -- (-1,-0.5);
		        \draw[postaction={segment={mid arrow}}] (0,-0.7) -- (-1,-0.7);
		        \node  at (-0.5, -0.04) {$\vdots$};
                \node  at (-0.5, -0.22) {$\vdots$};
                \draw[postaction={segment={mid arrow}}] (1,0) -- (0,0);
                \draw[postaction={segment={mid arrow}}] (1,0.4) -- (0,0.4);
                \draw[postaction={segment={mid arrow}}] (1,0.2) -- (0,0.2);
                 \node  at (0.5, -0.14) {$\vdots$};
                \node  at (0.5, -0.32) {$\vdots$};
		        \draw[postaction={segment={mid arrow}}] (1,-0.6) -- (0,-0.6);
		        \draw[postaction={segment={mid arrow}}] (1,-0.8) -- (0,-0.8);
		       \draw (1,0.8)--(1,0)--(1,-0.9);
		        \draw [] (0,0.8) -- (0,0)--(0,-0.9);
		       \draw[] (2,0.8)--(2,0)--(2,-0.9);
		        \node at (1.5, 0.3) {$\cdots$};
		        \node at (1.5, 0.1) {$\cdots$};
		        \node at (1.5, -0.1) {$\cdots$};
		        \node at (1.5, -0.3) {$\cdots$};
		        \node at (1.5, -0.5) {$\cdots$};
		        \draw[postaction={segment={mid arrow}}] (3,0.4) -- (2,0.4);
                \draw[postaction={segment={mid arrow}}] (3,0.2) -- (2,0.2);
                 \node  at (2.5, 0.05) {$\vdots$};
                \node  at (2.5, -0.13) {$\vdots$};
                \draw[postaction={segment={mid arrow}}] (3,-0.4) -- (2,-0.4);
		        \draw[postaction={segment={mid arrow}}] (3,-0.6) -- (2,-0.6);
		        \draw[] (3,0.8)--(3,0)--(3,-0.9);
		         \draw[postaction={segment={mid arrow}}] (3.7,-0.1) -- (3,-0.1);
                \draw[postaction={segment={mid arrow}}] (3.7,0.3) -- (3,0.3);
                \draw[postaction={segment={mid arrow}}] (3.7,0.1) -- (3,0.1);
                 \node  at (3.35, -0.25) {$\vdots$};
                \draw[postaction={segment={mid arrow}}] (3.7,-0.5) -- (3,-0.5);
		        \draw[postaction={segment={mid arrow}}] (3.7,-0.7) -- (3,-0.7);
		         \draw [decorate,decoration={brace,amplitude=5pt,mirror,raise=4pt},yshift=0pt] (-1.7,0.65) -- (-1.7,-0.65) node [xshift=-0.8cm ,midway] {$N$};
                 \draw [decorate,decoration={brace,amplitude=5pt,mirror,raise=4pt},yshift=0pt] (-1,-0.9) -- (3,-0.9) node [midway,xshift=0cm,yshift=-.8cm] {$M+1$};  
            \end{scope}
        \end{tikzpicture}}\label{eq:braneweb-superconformal}
\end{align}

The basic element is the following chain geometry
\begin{align}
    \begin{tikzpicture}
    \begin{scope}[scale=1]
        \draw [->,>=stealth](0,2)--(0,-3.2);
        \draw[postaction={segment={mid arrow}}] (0, 1.6)--(-1.5,1.6);
        \draw[postaction={segment={mid arrow}}] (1.5, 1.2)--(0,1.2);
        \draw[postaction={segment={mid arrow}}] (0, 0.8)--(-1.5,0.8);
        \draw[postaction={segment={mid arrow}}] (0, -0.3)--(-1.5,-0.3);
        \draw[postaction={segment={mid arrow}}] (1.5, -0.8)--(0,-0.8);
        \draw[postaction={segment={mid arrow}}] (0, -1.3)--(-1.5,-1.3);
        \draw[postaction={segment={mid arrow}}] (0, -2.4)--(-1.5,-2.4);
        \draw[postaction={segment={mid arrow}}] (1.5, -2.8)--(0,-2.8);
        \node [scale=0.7,above] at (0, 2) {$(1,n_{i})_{u_{i}}$};
        \node [scale=0.7,left] at (-1.5, 1.6) {$(0,\sigma_{1})_{v_{i-1,1}}$};
        \node [scale=0.7,left] at (-1.5, 0.8) {$(0,\sigma_{2})_{v_{i-1,2}}$};
        \node [scale=0.7,left] at (-1.5, -0.3) {$(0,\sigma_{j})_{v_{i-1,j}}$};
        \node [scale=0.7,right] at (1.5, +1.2) {$(0,\sigma_{1})_{v_{i,1}}$};
        \node [scale=0.7,right] at (1.5, -0.8) {$(0,\sigma_{j})_{v_{i,j}}$};
        \node [scale=0.7,left] at (-1.5, -1.3) {$(0,\sigma_{j+1})_{v_{i-1,j+1}}$};
        \node [scale=0.7,right] at (0,-0.15) {$(1,n_{ij})_{u_{ij}}$};
        \node [scale=0.7,right] at (0,-0.55) {$(1,n'_{ij})_{u'_{ij}}$};
        \node [scale=0.7,right] at (0,-1.1) {$(1,n_{i,j+1})_{u_{i,j+1}}$};
        \node [scale=0.7,left] at (-1.5,-2.4) {$(0,\sigma_{N})_{v_{i-1,N}}$};
        \node [scale=0.7,right] at (1.5,-2.8) {$(0,\sigma_{N})_{v_{iN}}$};
        \node at (0.75, 0.6) {$\vdots$};
        \node at (-0.75, 0.35) {$\vdots$};
        \node at (0.75, -1.6) {$\vdots$};
        \node at (0.75, -2.0) {$\vdots$};
        \node at (-0.75, -1.8) {$\vdots$};
        \end{scope}
    \end{tikzpicture}
\end{align}

The conservation law for the spectral parameters are 
\begin{align}
    \begin{split}
    n_{i1}\coloneqq n_{i},&\quad u_{i1}\coloneqq u_{i},\\
    n_{ij}=n'_{ij}+\sigma_{j},&\quad n_{i,j+1}=n'_{ij}+\sigma_{j},\\ u_{ij}=-u'_{ij}v_{i-1,j}^{\sigma_{j}},&\quad u_{i,j+1}=-u'_{ij}v_{ij}^{\sigma_{j}},
    \end{split}
\end{align}
for $i=1,\ldots, M+1,$ and $j=1,\ldots, N$. The solution is 
\begin{align}
\begin{split}
    n_{ij}=n_{i},&\quad n'_{ij}=n_{i}-\sigma_{j},\\
    u_{ij}=u_{i}\prod_{l=1}^{j-1}\left(\frac{v_{il}}{v_{i-1,l}}\right)^{\sigma_{l}},&\quad u'_{i,j}=-u_{i}v_{i-1,j}^{-\sigma_{j}}\prod_{l=1}^{j-1}\left(\frac{v_{il}}{v_{i-1,l}}\right)^{\sigma_{l}}.
\end{split}
\end{align}
The matrix element of this chain geometry is
\begin{align}
\begin{split}
    \mathcal{T}(u_{i}\,|\, \vec{v}_{i-1},\vec{\lambda}_{i-1}\,|\,\vec{v}_{i},\vec{\lambda}_{i})&\coloneqq\begin{array}{ccc}
&\widehat{\,\underline{0}\,}&\\
\dbra{\lambda_{i-1,1}}&\Phi^{(n'_{i1})*}_{\sigma_{1}}[u'_{i1},v_{i-1,1}]\\
&\Phi^{(n'_{i,1})}_{\sigma_{1}}[u'_{i1},v_{i1}]&\dket{\lambda_{i1}}\\
\dbra{\lambda_{i-1,2}}&\Phi^{(n'_{i2})*}_{\sigma_{2}}[u'_{i2},v_{i-1,2}]\\
&\Phi^{(n'_{i2})}_{\sigma_{2}}[u'_{i2},v_{i2}]&\dket{\lambda_{i2}}\\
&\vdots&\\
\dbra{\lambda_{i-1,N}}&\Phi^{(n'_{iN})*}_{\sigma_{N}}[u'_{iN},v_{i-1,N}]\\
&\Phi^{(n'_{iN})}_{\sigma_{N}}[u'_{iN},v_{iN}]&\dket{\lambda_{iN}}\vspace{0.2cm}\\
&\uwidehat{\overline{\,0}\,}&
\end{array}\\
&=\bra{0}\overleftarrow{\prod_{l=1}^{N}}\left(\Phi^{(n'_{il})}_{\sigma_{l},\,\lambda_{il}}[u'_{il},v_{il}]\Phi^{(n'_{il})*}_{\sigma_{l},\lambda_{i-1,l}}[u'_{il},v_{i-1,l}]\right)\ket{0},
\end{split}
\end{align}
where $\vec{v}_{i}=(v_{i1},\ldots,v_{iN})$ and $\vec{\lambda}_{i}=(\lambda_{i1},\ldots,\lambda_{iN})$. We also impose the condition $\vec{\lambda}_{0}=\vec{\lambda}_{M+1}=\vec{\emptyset}$. Computing the contractions, we have 
\begin{align}
\begin{split}
     &\mathcal{T}(u_{i}\,|\, \vec{v}_{i-1},\vec{\lambda}_{i-1}\,|\,\vec{v}_{i},\vec{\lambda}_{i})\\=&\prod_{l=1}^{N}t_{n'_{il},\sigma_{l}}(\lambda_{il},u'_{il},v_{il})t^{*}_{n'_{il},\sigma_{l}}(\lambda_{i-1,l},u'_{il},v_{i-1,l})\\
     &\times \textcolor{red}{\prod_{k>j}\mathcal{G}\left(\frac{v_{i-1,j}}{v_{i-1,k}}\right)^{\sigma_{j}\sigma_{k}}\mathcal{G}\left(q_{3}^{-1}\frac{v_{ij}}{v_{ik}}\right)^{\sigma_{j}\sigma_{k}}} \textcolor{blue}{\prod_{k\geq j}\mathcal{G}\left(\gamma^{-1}\frac{v_{i-1,j}}{v_{ik}}\right)^{-\sigma_{j}\sigma_{k}}\prod_{k>j}\mathcal{G}\left(\gamma^{-1}\frac{v_{ij}}{v_{i-1,k}}\right)^{-\sigma_{j}\sigma_{k}}}\\
     &\times\textcolor{red}{\prod_{k>j}N_{\sigma_{j}\sigma_{k}}(q_{3}v_{i-1,j},\lambda_{i-1,j}\,|\,v_{i-1,k},\lambda_{i-1,k})^{-\sigma_{j}\sigma_{k}}N_{\sigma_{j}\sigma_{k}}(v_{ij},\lambda_{ij}\,|\,v_{ik},\lambda_{ik})^{-\sigma_{j}\sigma_{k}}}\\
     &\times\textcolor{blue}{\prod_{k\geq j}N_{\sigma_{j}\sigma_{k}}(\gamma v_{i-1,j},\lambda_{i-1,j}\,|\, v_{ik},\lambda_{ik})^{\sigma_{j}\sigma_{k}}\prod_{k>j}N_{\sigma_{j}\sigma_{k}}(\gamma v_{ij},\lambda_{ij}\,|\,v_{i-1,k},\lambda_{i-1,k})^{\sigma_{j}\sigma_{k}}}.
\end{split}
\end{align}
The red terms give half of the contribution of the vector multiplets of the D5-branes attached to both sides of the NS5-brane, while the blue terms give the full bifundamental contribution. Note that to compare with the localization formula, we need use the property of the Nekrasov factor in (\ref{eq:Nekrasovsymm1}).  

After using the formula in (\ref{eq:Nekrasovsymm1}) and dealing with the zero-modes properly, the total partition function will be  
\begin{align}
\begin{split}
    \mathcal{Z}\,[U(N\,|\,\sigma_{1},\cdots,\sigma_{N})^{\otimes M}]&=\sum_{\vec{\lambda}_{1},\vec{\lambda}_{2}\cdots,\vec{\lambda}_{M}}\prod_{i=1}^{M}\prod_{j=1}^{N}a_{\lambda_{ij}}^{(\sigma_{j})}\prod_{i=1}^{M+1}\mathcal{T}(u_{i}\,|\, \vec{v}_{i-1},\vec{\lambda}_{i-1}\,|\,\vec{v}_{i},\vec{\lambda}_{i})\\
    &=\mathcal{Z}_{\text{pert}}\mathcal{Z}_{\text{inst}},
\end{split}
\end{align}
where 
\begin{align}
    \begin{split}
        \mathcal{Z}_{\text{pert}}&=\textcolor{red}{\prod_{k>j}\mathcal{G}\left(\frac{v_{0j}}{v_{0k}}\right)^{\sigma_{j}\sigma_{k}}\prod_{k> j}\mathcal{G}\left(q_{3}^{-1}\frac{v_{M+1,j}}{v_{M+1,k}}\right)^{\sigma_{j}\sigma_{k}}}\textcolor{blue}{\prod_{i=1}^{M}\prod_{k>j}\mathcal{G}\left(\frac{v_{i,j}}{v_{i,k}}\right)^{\sigma_{j}\sigma_{k}}\mathcal{G}\left(q_{3}^{-1}\frac{v_{ij}}{v_{ik}}\right)^{\sigma_{j}\sigma_{k}}}\\
        &\times\textcolor{darkgreen}{\prod_{k\geq j}\mathcal{G}\left(\gamma^{-1}\frac{v_{0,j}}{v_{1k}}\right)^{-\sigma_{j}\sigma_{k}}\prod_{k>j}\mathcal{G}\left(\gamma^{-1}\frac{v_{1j}}{v_{0k}}\right)^{-\sigma_{j}\sigma_{k}} \prod_{k\geq j}\mathcal{G}\left(\gamma^{-1}\frac{v_{M,j}}{v_{M+1,k}}\right)^{-\sigma_{j}\sigma_{k}}\prod_{k>j}\mathcal{G}\left(\gamma^{-1}\frac{v_{M+1,j}}{v_{M,k}}\right)^{-\sigma_{j}\sigma_{k}}} \\
        &\times \textcolor{magenta}{\prod_{i=2}^{M} \prod_{k\geq j}\mathcal{G}\left(\gamma^{-1}\frac{v_{i-1,j}}{v_{ik}}\right)^{-\sigma_{j}\sigma_{k}}\prod_{k>j}\mathcal{G}\left(\gamma^{-1}\frac{v_{ij}}{v_{i-1,k}}\right)^{-\sigma_{j}\sigma_{k}}}.
    \end{split}
\end{align}
The red part is the part independent of the Coulomb vev of the D5-branes, so we can ignore it. The blue part is the perturbative part of the vector multiplets. The green part is the contribution from the fundamental and antifundamental hypermultiplet. The magenta part is the contribution from the bifundamental part.

The instanton part is 
\begin{align}
    \begin{split}
        \mathcal{Z}_{\text{inst.}}&=\sum_{\vec{\lambda}_{1},\ldots,\vec{\lambda}_{M}}\prod_{i=1}^{M}\mathcal{Z}_{\text{top.}}(\vec{\lambda}_{i},\vec{\sigma})\mathcal{Z}_{\text{CS}}(\kappa_{i},\vec{\lambda}_{i},\vec{\sigma})\mathcal{Z}_{\text{vec.}}(\vec{v}_{i},\vec{\lambda}_{i},\vec{\sigma})\\
        &\times \mathcal{Z}_{\text{af}}(\vec{v}_{1},\vec{\lambda}_{1},\vec{\sigma}\,|\,\vec{v}_{0},\vec{\sigma})\left(\prod_{i=1}^{M}\mathcal{Z}_{\text{bfd.}}(\vec{v}_{i-1},\vec{\lambda}_{i-1},\vec{\sigma}\,|\,\gamma^{-1})\right) \mathcal{Z}_{\text{f}}(\vec{v}_{M},\vec{\lambda}_{M},\vec{\sigma}\,|\,\vec{v}_{M+1},\vec{\sigma}),
    \end{split}
\end{align}
where the topological term and the Chern-Simons level are
\begin{align}
\begin{split}
    \kappa_{i}&=n_{i+1}-n_{i},\quad \mathfrak{q}_{i}=-\frac{u_{i}}{u_{i+1}}\prod_{l=1}^{N}\left(\frac{v_{il}}{v_{i-1,l}}\right)^{\sigma_{l}}\gamma^{n_{i}-n_{i+1}-\sum_{k=1}^{N}\sigma_{k}}.
\end{split}
\end{align}
Therefore, we managed to derive the partition function of (\ref{eq:braneweb-superconformal}). As mentioned in equation (\ref{eq:CScond}), we only get one Chern-Simons level for each node.
\paragraph{General linear quiver gauge theory}
To derive general linear quiver gauge theories, instead of using the postive and negative intertwiner one by one, it is easier to use the generalized AFS intertwiner in (\ref{eq:generalizedAFSintertwiner}):
\begin{align}
\begin{split}
\adjustbox{valign=c}{
     \begin{tikzpicture}
            \begin{scope}[scale=2.3]
                \node [above] at (-1,0.7) {$(1,n_{1})_{u_{1}}$};
                \node [above] at (0,0.7) {$(1,n_{2})_{u_{2}}$};
                \node [above] at (1,0.7) {$(1,n_{3})_{u_{3}}$};
                \node [above] at (2,0.7) {$(1,n_{r})_{u_{r}}$};
                \node [above] at (3,0.7) {$(1,n_{r+1})_{u_{r+1}}$};
		        \draw (-1,0.7)--(-1,0)--(-1,-0.7);
                \draw[very thick,postaction={segment={mid arrow}}] (0,0.3) -- (-1,0.3);
                \node [above] at (-0.5,0.3) {$(0,\vec{\sigma}_{1})_{\vec{v}_{1}}$};
                \draw[very thick,postaction={segment={mid arrow}}] (1,0) -- (0,0);
                \node [above] at (0.5,0) {$(0,\vec{\sigma}_{2})_{\vec{v}_{2}}$};
		        \draw (1,0.7)--(1,0)--(1,-0.7);
		        \draw [] (0,0.7) -- (0,0)--(0,-0.7);
		        \draw[] (2,0.7)--(2,0)--(2,-0.7);
		        \node at (1.5, -0.2) {$\cdots$};
                \draw[very thick,postaction={segment={mid arrow}}] (3,-0.4) -- (2,-0.4);
                \node [above] at (2.5,-0.4) {$(0,\vec{\sigma}_{r})_{\vec{v}_{r}}$};
		        \draw[] (3,0.7)--(3,0)--(3,-0.7);
                 \draw [decorate,decoration={brace,amplitude=5pt,mirror,raise=4pt},yshift=0pt] (-1,-0.9) -- (3,-0.9) node [midway,xshift=0cm,yshift=-.8cm] {$r+1$};  
            \end{scope}
        \end{tikzpicture}}
\end{split}\label{eq:general_linear_quiver}
\end{align}
\begin{align}
    \begin{split}
\begin{array}{cccccccccc}
\widehat{\,\underline{0}\,}&&\vspace{1mm}\widehat{\,\underline{0}\,}&&\widehat{\,\underline{0}\,}&&\widehat{\,\underline{0}\,}\\
\Phi_{\vec{\sigma}_{1}}^{(n_{1})}[u_{1},\vec{v}_{1}]&\cdot&\Phi_{\vec{\sigma}_{1}}^{(n_{2})*}[u_{2},\vec{v}_{1}]&&&&\\
&&\Phi_{\vec{\sigma}_{2}}^{(n'_{2})}[u'_{2},\vec{v}_{2}]&\ddots&&&\\
&&&\ddots&&&\\
&&&&&&\\
&&&\ddots&\Phi^{(n_{r})\ast}_{\vec{\sigma}_{r-1}}[u_{r},\vec{v}_{r-1}]&&\\
&&&&\Phi^{(n'_{r})}_{\vec{\sigma}_{r}}[u'_{r},\vec{v}_{r}]&\cdot&\Phi^{(n_{r+1})\ast}_{\vec{\sigma}_{r}}[u_{r+1},\vec{v}_{r}]\vspace{1.3mm}\\
\uwidehat{\overline{\,0}\,}&&\uwidehat{\overline{\,0}\,}&&\uwidehat{\overline{\,0}\,}&&\uwidehat{\overline{\,0}\,}
\end{array}
\end{split}
\end{align}
where 
\begin{align}
    \vec{v}_{i}=(v_{i1},v_{i2},\ldots,v_{iN_{i}}),\quad \vec{\sigma}_{i}=(\sigma_{i1},\sigma_{i2},\ldots,\sigma_{iN_{i}}),\quad  \vec{\lambda}_{i}=(\lambda_{1},\lambda_{2},\ldots,\lambda_{N_{i}}),
\end{align}
for $i=1,\ldots, r$. This gives a supergroup gauge theory $\prod_{i=1}^{r}U(N_{i}\,|\,\vec{\sigma}_{i}\,)$ with linear quiver structure. Actually this quiver is a part of the D-type quiver we will discuss in section~\ref{sec:example-Dquiver}, so see it for the details of the calculation. 

Generally, we can assign matter hypermultiplets to each node of the quiver. We can do the similar process done in the first example by attaching infinite D5-branes to introduce hypermultiplets to the gauge nodes at both ends of the linear quiver. However, to introduce hypermultiplets for each node not only the both ends, this is not enough. Recently, shifted quantum algebras were utilized to introduce these matter hypermultiplets \cite{Bourgine:2021nyw,Bourgine:2022scz}. A similar analysis can be done using the formalism we introduced in this paper.
\subsection{\texorpdfstring{D}{D}-type quiver gauge theory}\label{sec:example-Dquiver}
There are two ways to study D-quiver gauge theories using the intertwiner formalism: either \cite{Bourgine:2017rik} or \cite{Kimura:2019gon}. Inspired by the brane web in (\ref{eq:Dquiverbraneweb1}) and (\ref{eq:Dquiverbraneweb2}), we introduce a new vertex operator\footnote{We note that even for the non-supergroup case, the algebraic property of this vertex operator is still not so clear. We expect by assigning the orientifold plane a module on which quantum toroidal $\mathfrak{gl}_{1}$ acts we can study the algebraic properties of these vertex operators. However, it is beyond the scope of this paper, so we will just introduce a vertex operator reproducing the partition functions.   } representing the brane junction with the vertex $\hspace{-0.15cm}\Dvertex$, which is a supergroup generalization of \cite{Kimura:2019gon}. 
\subsubsection{\texorpdfstring{$D_{1}$}{D1} quiver}\label{sec:example-Dtype-pureSYM}
To study D-type quiver supergroup gauge theories, we can introduce the following vertex operator
\begin{align}
\begin{split}
\bar{\Phi}_{\sigma}^{(n)\ast}[u,v]&:(1,n)_{u}\rightarrow (1,n+\sigma)_{-u(\gamma v)^{\sigma}}\otimes (0,\sigma)_{v},\\
\bar{\Phi}_{\sigma}^{(n)\ast}[u,v]&=\sum_{\lambda}a_{\lambda}^{(\sigma)}\bar{\Phi}_{\sigma,\lambda}^{(n)\ast}[u,v]\otimes\dket{v,\lambda},\\
    \bar{\Phi}_{\sigma,\lambda}^{(n)\ast}[u,v]&=\bar{t}_{n,\sigma}(\lambda,u,v):\Phi^{\ast}_{\emptyset}[q_{3}^{-1}v]^{-\sigma}\prod_{x\in\lambda}\xi(q_{3}^{-1}\chi_{x}^{(\sigma)})^{-\sigma}:,\\
    \bar{t}_{n,\sigma}(\lambda,u,v)&=(-u^{\sigma}v\gamma^{\frac{1+\sigma}{2}})^{|\lambda|}\prod_{x\in\lambda}\left(\chi_{x}^{(\sigma)}\right)^{-\sigma n-1},
\end{split}\label{eq:generalizedAFSintertwiner-Dtype}
\end{align}
which is the supergroup version of the boundary intertwiner introduced in \cite{Kimura:2019gon}. We illustrate this vertex operator as the following diagram
\begin{align}
     \begin{tikzpicture}
            \begin{scope}[scale=1.3]
                \node[above, scale=0.8] at (0,0) {$(0,\sigma)_{v}$};
                \node[right, scale=0.8] at (1,0.6) {$(1,n)_{u}$};
                \node[right,scale=0.8] at (1,-0.6) {$(1,n+\sigma)_{-u(\gamma v)^{\sigma}}$};
                \node[right,scale=0.8] at (1.3,0) {$\bar{\Phi}^{(n)\ast}_{\sigma}[u,v]$};
                \draw [ultra thick] (0.9,-0.1)--(1.1,0.1);
                \draw [ultra thick] (0.9,0.1)--(1.1,-0.1);
                \draw[postaction={segment={mid arrow}}] (1,0) -- (0,0);
		        \draw[postaction={segment={mid arrow}}] (1,1) -- (1,0)--(1,-1);   
            \end{scope}
        \end{tikzpicture}
\end{align}
Obviously, the central charges are not conserved because of the existence of the orientifold, which is the similar situation to the vertex operator introduced in \cite{Kimura:2019gon}. We denote the generalized vertex as
\begin{align}
\begin{split}
    &\bar{\Phi}^{(n)*}_{\vec{\sigma}}[u,\vec{v}]=\bar{\Phi}_{\sigma_{N}}^{(n_{N})}[u_{N},v_{N}]\cdots\bar{\Phi}_{\sigma_{2}}^{(n_{2})}[u_{2},v_{2}]\bar{\Phi}_{\sigma_{1}}^{(n_{1})}[u_{1},v_{1}],\\
    &u_{i}=u\prod_{l=1}^{i-1}(-\gamma v_{l})^{\sigma_{l}},\quad n_{i}=n+\sum_{l=1}^{i-1}\sigma_{l}.
\end{split}
\end{align}
Note that since the vertex operator part of the intertwiner only differs from the dual intertwiner by simultaneous shifts of the spectral parameter $\vec{v}\rightarrow q_{3}^{-1}\vec{v}$, the contraction formulas of $\bar{\Phi}^{*}\bar{\Phi}^{*}$ are exactly the same as that of the dual intertwiners.

Let us consider the simplest case, when there is only one gauge node, which is just the pure SYM:
\begin{align}
\begin{split}
\mathcal{Z}[D_{1}]&=
\adjustbox{valign=c,width=3cm}{
    \begin{tikzpicture}
            \begin{scope}[scale=2]
                \draw [thick] (0.95,-0.05)--(1.05,0.05);
                \draw [thick] (0.95,0.05)--(1.05,-0.05);
                \draw [thick] (0.95,0.35)--(1.05,0.45);
                \draw [thick] (0.95,0.45)--(1.05,0.35);
                \draw [thick] (0.95,0.55)--(1.05,0.65);
                \draw [thick] (0.95,0.65)--(1.05,0.55);
                \draw [thick] (0.95,-0.35)--(1.05,-0.45);
                \draw [thick] (0.95,-0.45)--(1.05,-0.35);
                \draw [thick] (0.95,-0.55)--(1.05,-0.65);
                \draw [thick] (0.95,-0.65)--(1.05,-0.55);
                \draw[postaction={segment={mid arrow}}] (1,0) -- (0,0);
                \node[scale=0.8] at (0.5, 0.25) {$\vdots$};
                \node[scale=0.8] at (0.5, -0.15) {$\vdots$};
                \draw[postaction={segment={mid arrow}}] (1,0.4) -- (0,0.4);
                \draw[postaction={segment={mid arrow}}] (1,0.6) -- (0,0.6);
                \draw[postaction={segment={mid arrow}}] (1,-0.4) -- (0,-0.4);
		        \draw[postaction={segment={mid arrow}}] (1,-0.6) -- (0,-0.6);
		        \draw (1,0.8)--(1,0)--(1,-0.8);
		        \draw (0,0.8) -- (0,0)--(0,-0.8);
            \end{scope}
        \end{tikzpicture}}=\begin{array}{ccc}
\widehat{\,\underline{0}\,}&&\vspace{1mm}\widehat{\,\underline{0}\,}\\
\Phi_{\vec{\sigma}}^{(n)}[u,\vec{v}]&\cdot&\bar{\Phi}_{\vec{\sigma}}^{(n^{*})*}[u^{*},\vec{v}]\vspace{1mm}\\
\uwidehat{\,\overline{0}\,}&&\uwidehat{\,\overline{0}\,}\\
\end{array}\\
&=\sum_{\lambda_{1},\ldots,\lambda_{N}}\prod_{i=1}^{N}a^{(\sigma_{i})}_{\lambda_{i}}t_{n_{i},\sigma_{i}}(\lambda_{i},u_{i},v_{i})\bar{t}_{n^{\ast}_{i},\sigma_{i}}(\lambda_{i},u^{\ast}_{i},v_{i})\\
&\qquad\times\prod_{i<j}N_{\sigma_{i}\sigma_{j}}(v_{i},\lambda_{i}\,|\, v_{j},\lambda_{j})^{-\sigma_{i}\sigma_{j}}N_{\sigma_{i}\sigma_{j}}(q_{3}v_{i},\lambda_{i}\,|\,v_{j},\lambda_{j})^{-\sigma_{i}\sigma_{j}}\\
&=\mathcal{Z}_{\text{pert.}}[D_{1}]\mathcal{Z}_{\text{inst.}}[D_{1}],\\
Z_{\text{inst.}}[D_{1}]&=\sum_{\lambda_{1},\ldots,\lambda_{N}}\mathcal{Z}_{\text{top.}}(\vec{\lambda},\vec{\sigma})\mathcal{Z}_{\text{CS}}(\kappa,\vec{\lambda},\vec{\sigma})\mathcal{Z}_{\text{vec.}}(\vec{v},\vec{\lambda},\vec{\sigma}),
\end{split}
\end{align}
where the topological term and the Chern-Simons level are
\begin{align}
    \mathfrak{q}=-uu^{\ast}\prod_{i=1}^{N}(-v_{i})^{\sigma_{i}}\gamma^{n+1},\quad \kappa=-n-n^{\ast}-\sum_{i=1}^{N}\sigma_{i}.
\end{align}

\subsubsection{\texorpdfstring{$D_{r}$ $(r>1)$}{Dr(r>1)} quiver}\label{sec:example-generalDquiver}
Next, we consider a D-type quiver gauge theory with gauge group $\prod_{i=1}^{r}U(N_{i}\,|\,\sigma_{i1},\ldots,\sigma_{iN_{i}})$:
\begin{align}
\adjustbox{valign=c}{
     \begin{tikzpicture}
            \begin{scope}[scale=2.4]
                \node [above] at (-1,0.7) {$(1,n_{1})_{u_{1}}$};
                \node [above] at (0,0.7) {$(1,n_{2})_{u_{2}}$};
                \node [above] at (1,0.7) {$(1,n_{3})_{u_{3}}$};
                \node [above] at (3,0.7) {$(1,n_{r-1})_{u_{r-1}}$};
                \node[above] at (4,0.7) {$(1,n_{r})_{u_{r}}$};
		        \draw (-1,0.7)--(-1,0)--(-1,-0.8);
		        \draw [] (0,0.7) -- (0,0)--(0,-0.8);
		        \draw (1,0.7)--(1,0)--(1,-0.8);
		        \draw[] (2,0.7)--(2,0)--(2,-0.8);
		        \draw[] (3,0.7)--(3,0)--(3,-0.8);
		        \draw[] (4,0.7)--(4,0)--(4,-0.8);
                \draw[very thick,postaction={segment={mid arrow}}] (0,0.4) -- (-1,0.4);
                \node [above] at (-0.5,0.4) {$(0,\vec{\sigma}_{1})_{\vec{v}_{1}}$};
                \draw[very thick,postaction={segment={mid arrow}}] (1,0) -- (0,0);
                \node [above] at (0.5,0) {$(0,\vec{\sigma}_{2})_{\vec{v}_{2}}$};
		        \node at (1.5, 0) {$\cdots$};
                \draw[very thick,postaction={segment={mid arrow}}] (3,0) -- (2,0);
                \node [above] at (2.5,0) {$(0,\vec{\sigma}_{r-2})_{\vec{v}_{r-2}}$};
		        \draw[very thick,postaction={segment={mid arrow}}] (4,0.35) -- (3,0.35);
		        \draw[very thick,postaction={segment={mid arrow}}] (4,-0.3) -- (3,-0.3);
		        \node [above] at (3.5,0.35){$(0,\vec{\sigma}_{r})_{\vec{v}_{r}}$};
		        \node [above] at (3.5,-0.3){$(0,\vec{\sigma}_{r-1})_{\vec{v}_{r-1}}$};
		        \draw [very thick] (3.95,0.3)--(4.05,0.4);
		        \draw [very thick] (4.05,0.3)--(3.95,0.4);
            \end{scope}
        \end{tikzpicture}}
\end{align}
where we used the generalized intertwiners in (\ref{eq:generalizedAFSintertwiner}) and (\ref{eq:generalizedAFSintertwiner-Dtype}).
For each stack of D5-branes, we assign the Coulomb vev parameters, signatures of branes, and Young diagrams as
\begin{align}
    \vec{v}_{i}=(v_{i1},v_{i2},\ldots,v_{iN_{i}}),\quad \vec{\sigma}_{i}=(\sigma_{i1},\sigma_{i2},\ldots,\sigma_{iN_{i}}),\quad  \vec{\lambda}_{i}=(\lambda_{i1},\lambda_{i2},\ldots,\lambda_{iN_{i}}),
\end{align}
for $i=1,\ldots,r$. The partition function is written as
\begin{align}
\begin{split}
    Z[D_{r}]&=\begin{array}{cccccccccc}
\widehat{\,\underline{0}\,}&&\vspace{1mm}\widehat{\,\underline{0}\,}&&\widehat{\,\underline{0}\,}&&\widehat{\,\underline{0}\,}\\
\Phi_{\vec{\sigma}_{1}}^{(n_{1})}[u_{1},\vec{v}_{1}]&\cdot&\Phi_{\vec{\sigma}_{1}}^{(n_{2})*}[u_{2},\vec{v}_{1}]&&&&\\
&&\Phi_{\vec{\sigma}_{2}}^{(n'_{2})}[u'_{2},\vec{v}_{2}]&\ddots&&&\\
&&&\ddots&&&\\
&&&&\Phi^{(n_{r-1})}_{\vec{\sigma}_{r}}[u_{r-1},\vec{v}_{r}]&\cdot&\bar{\Phi}^{(n_{r})\ast}_{\vec{\sigma}_{r}}[u_{r},\vec{v}_{r}]\\
&&&\ddots&\Phi^{(n'_{r-1})\ast}_{\vec{\sigma}_{r-1}}[u'_{r-1},\vec{v}_{r-2}]&&\\
&&&&\Phi^{(n''_{r-1})}_{\vec{\sigma}_{r-1}}[u''_{r-1},\vec{v}_{r-1}]&\cdot&\Phi^{(n'_{r})\ast}_{\vec{\sigma}_{r}}[u'_{r},\vec{v}_{r-1}]\vspace{1.3mm}\\
\uwidehat{\overline{\,0}\,}&&\uwidehat{\overline{\,0}\,}&&\uwidehat{\overline{\,0}\,}&&\uwidehat{\overline{\,0}\,}
\end{array}\\\\
&=\sum_{\vec{\lambda}_{1},\ldots,\vec{\lambda}_{r-1}}\prod_{i=1}^{r}\prod_{j=1}^{N_{i}}a^{(\sigma_{ij})}_{\lambda_{ij}}\begin{array}{c}
\widehat{\,\underline{0}\,}\\
\Phi^{(n_{1})}_{\vec{\sigma}_{1},\vec{\lambda}_{1}}[u_{1},\vec{v}_{1}]\vspace{1mm}\\
\uwidehat{\overline{\,0\,}}  
\end{array}\prod_{i=2}^{r-2}
\begin{array}{c}
\widehat{\,\underline{0}\,}\\
\Phi^{(n_{i})\ast}_{\vec{\sigma}_{i-1},\vec{\lambda}_{i-1}}[u_{i},\vec{v}_{i-1}]\vspace{1mm}\\
\Phi^{(n'_{i})}_{\vec{\sigma}_{i},\vec{\lambda}_{i}}[u'_{i},\vec{v}_{i}]\vspace{1mm}\\
 \uwidehat{\overline{\,0\,}}
\end{array}\\
&\qquad\qquad\times
\begin{array}{c}
\widehat{\,\underline{0}\,}\\
    \Phi^{(n_{r-1})}_{\vec{\sigma}_{r},\vec{\lambda}_{r}}[u_{r-1},\vec{v}_{r}]  \\
    \Phi^{(n'_{r-1})\ast}_{\vec{\sigma}_{r-1},\vec{\lambda}_{r-2}}[u'_{r-1},\vec{v}_{r-2}]  \vspace{1mm}\\
    \Phi^{(n''_{r-1})}_{\vec{\sigma}_{r-1},\vec{\lambda}_{r-1}}[u''_{r-1},\vec{v}_{r-1}]\vspace{1mm}\\
 \uwidehat{\overline{\,0\,}}
\end{array}
\begin{array}{c}
   \widehat{\,\underline{0}\,}  \\
    \bar{\Phi}^{(n_{r})\ast}_{\vec{\sigma}_{r},\vec{\lambda}_{r}}[u_{r},\vec{v}_{r}] \\
     \Phi^{(n'_{r})\ast}_{\vec{\sigma}_{r},\vec{\lambda}_{r-1}}[u'_{r},\vec{v}_{r-1}]\vspace{1mm}\\
     \uwidehat{\overline{\,0\,}}
\end{array},
\end{split}
\end{align}
where the spectral parameters obey the following conservation laws
\begin{align}
\begin{split}
    &n'_{i}=n_{i}-\sum_{j=1}^{N_{i-1}}\sigma_{i-1,j},\quad u'_{i}=u_{i}\prod_{j=1}^{N_{i-1}}(-v_{i-1,j})^{-\sigma_{i-1,j}},\quad i\leq r-2,\\
    &n'_{r-1}=n_{r-1}+\sum_{j=1}^{N_{r}}\sigma_{rj},\quad n''_{r-1}=n_{r-1}+\sum_{j=1}^{N_{r}}\sigma_{rj}-\sum_{j=1}^{N_{r-2}}\sigma_{r-2,j},\\
    &u'_{r-1}=u_{r-1}\prod_{j=1}^{N_{r}}(-v_{rj})^{\sigma_{rj}},\quad u''_{r-1}=u_{r-1}\prod_{j=1}^{N_{r}}(-v_{rj})^{\sigma_{rj}}\prod_{j=1}^{N_{r-2}}(-v_{r-2,j})^{-\sigma_{r-2,j}},\\
    &n'_{r}=n_{r}+\sum_{j=1}^{N_{r}}\sigma_{rj},\quad u'_{r}=u_{r}\prod_{j=1}^{N_{r}}(-\gamma v_{rj})^{\sigma_{rj}}.
  \end{split} 
\end{align}
The local A-quiver part for $i\leq r-2$ gives
\begin{align}
    \frac{\bra{0}\Phi^{(n_{i}')}_{\vec{\sigma}_{i},\vec{\lambda}_{i}}[u_{i}',\vec{v}_{i}]\Phi^{(n_{i})\ast}_{\vec{\sigma}_{i-1},\vec{\lambda}_{i-1}}[u_{i},\vec{v}_{i-1}]\ket{0}}{\bra{0}\Phi^{n_{i}'}_{\vec{\sigma}_{i},\vec{\lambda}_{i}}[u_{i}',\vec{v}_{i}]\ket{0}\bra{0}\Phi^{(n_{i})\ast}_{\vec{\sigma}_{i-1},\vec{\lambda}_{i-1}}[u_{i},\vec{v}_{i-1}]\ket{0}}&\sim\mathcal{Z}_{\text{bfd.}}(\vec{v}_{i-1},\vec{\sigma}_{i-1},\vec{\lambda}_{i-1}\,|\,\vec{v}_{i},\vec{\lambda}_{i},\vec{\sigma}_{i}\,|\,\gamma^{-1}),\\
    a^{(\vec{\sigma}_{i})}_{\vec{\lambda}_{i}}\bra{0}\Phi^{(n_{i}')}_{\vec{\sigma}_{i},\vec{\lambda}_{i}}[u_{i}',\vec{v}_{i}]\ket{0}\bra{0}\Phi^{(n_{i+1})\ast}_{\vec{\sigma}_{i},\vec{\lambda}_{i}}[u_{i+1},\vec{v}_{i}]\ket{0}&\sim \mathcal{Z}_{\text{top.}}(\vec{\lambda}_{i},\vec{\sigma}_{i})\mathcal{Z}_{\text{CS}}(\kappa_{i},\vec{\lambda}_{i},\vec{\sigma}_{i})\mathcal{Z}_{\text{vec.}}(\vec{v}_{i},\vec{\lambda}_{i},\vec{\sigma}_{i}),
\end{align}
where $\sim$ means we extracted the instanton contribution and omit the perturbative part for simplicity.

The nontrivial part is the contribution coming from the last two vacuum expectation values:
\begin{align}
\begin{split}
 &\frac{\bra{0}\Phi^{(n''_{r-1})}_{\vec{\sigma}_{r-1},\vec{\lambda}_{r-1}}[u''_{r-1},\vec{v}_{r-1}]\Phi^{(n'_{r-1})\ast}_{\vec{\sigma}_{r-1},\vec{\lambda}_{r-2}}[u'_{r-1},\vec{v}_{r-2}]\Phi^{(n_{r-1})}_{\vec{\sigma}_{r},\vec{\lambda}_{r}}[u_{r-1},\vec{v}_{r}]\ket{0}}{\bra{0}\Phi^{(n''_{r-1})}_{\vec{\sigma}_{r-1},\vec{\lambda}_{r-1}}\ket{0}\bra{0}\Phi^{(n'_{r-1})\ast}_{\vec{\sigma}_{r-1},\vec{\lambda}_{r-2}}\ket{0}\bra{0}\Phi^{(n_{r-1})}_{\vec{\sigma}_{r},\vec{\lambda}_{r}}\ket{0}}\\
 \sim &\mathcal{Z}_{\text{bfd.}}(\vec{v}_{r},\vec{\lambda}_{r},\vec{\sigma}_{r}\,|\,\vec{v}_{r-2},\vec{\lambda}_{r-2},\vec{\sigma}_{r-2}\,|\,\gamma^{-1})\mathcal{Z}_{\text{bfd.}}(\vec{v}_{r-2},\vec{\lambda}_{r-2},\vec{\sigma}_{r-2}\,|\,\vec{v}_{r-1},\vec{\lambda}_{r-1},\vec{\sigma}_{r-1}\,|\,\gamma^{-1})\\
&\times \mathcal{Z}_{\text{bfd.}}(\vec{v}_{r},\vec{\lambda}_{r},\vec{\sigma}_{r}\,|\,\vec{v}_{r-1},\vec{\lambda}_{r-1},\vec{\sigma}_{r-1}\,|\,1)^{-1},
\end{split}
\end{align}
and
\begin{align}
\begin{split}
   &\frac{\bra{0}\Phi^{(n'_{r})\ast}_{\vec{\sigma}_{r},\vec{\lambda}_{r-1}}[u'_{r},\vec{v}_{r-1}]\bar{\Phi}^{(n_{r})\ast}_{\vec{\sigma}_{r},\vec{\lambda}_{r}}[u_{r},\vec{v}_{r}]\ket{0}}{\bra{0}\Phi^{(n'_{r})\ast}_{\vec{\sigma}_{r},\vec{\lambda}_{r-1}}\ket{0}\bra{0}\bar{\Phi}^{(n_{r})\ast}_{\vec{\sigma}_{r},\vec{\lambda}_{r}}\ket{0}}
   \sim\mathcal{Z}_{\text{bfd.}}(\vec{v}_{r},\vec{\lambda}_{r},\vec{\sigma}_{r}\,|\,\vec{v}_{r-1},\vec{\lambda}_{r-1},\vec{\sigma}_{r-1}\,|\,1).
\end{split}
\end{align}
The bifundamental contribution between $U(N_{r-1}\,|\,\vec{\sigma}_{r-1})$ and $U(N_{r}\,|\,\vec{\sigma}_{r})$ cancels out with each other and using the result of the $D_{1}$ quiver case, the total partition function is 
\begin{align}
\begin{split}
    \mathcal{Z}[D_{r}]&=\mathcal{Z}_{\text{pert.}}[D_{r}]\mathcal{Z}_{\text{inst.}}[D_{r}],\\
    \mathcal{Z}_{\text{inst.}}[D_{r}]&=\sum_{\vec{\lambda}_{1},\ldots,\vec{\lambda}_{r}}\prod_{i=1}^{r}\mathcal{Z}_{\text{top.}}(\vec{\lambda}_{i},\vec{\sigma}_{i})\mathcal{Z}_{\text{CS}}(\kappa_{i},\vec{\lambda}_{i},\vec{\sigma}_{i})\mathcal{Z}_{\text{vec.}}(\vec{v}_{i},\vec{\lambda}_{i},\vec{\sigma}_{i})\\
   & \times \prod_{i=1}^{r-1}\mathcal{Z}_{\text{bfd.}}(\vec{v}_{i-1},\vec{\lambda}_{i-1},\vec{\sigma}_{i-1}\,|\,\vec{v}_{i},\vec{\lambda}_{i},\vec{\sigma}_{i}\,|\,\gamma^{-1})\\
   &\times \mathcal{Z}_{\text{bfd.}}(\vec{v}_{r},\vec{\lambda}_{r},\vec{\sigma}_{r}\,|\,\vec{v}_{r-2},\vec{\lambda}_{r-2},\vec{\sigma}_{r-2}\,|\,\gamma^{-1}),
\end{split}
\end{align}
where the topological term and Chern-simons level are identified as 
\begin{align}
    \begin{split}
    \mathfrak{q}_{i}&=\begin{dcases}-\frac{u'_{i}}{u_{i+1}}\gamma^{n'_{i}-n_{i+1}}\prod_{l=1}^{N_{i}}(-v_{il})^{\sigma_{il}},\quad 1\leq i\leq r-3\\
    -\frac{u'_{r-2}}{u'_{r-1}}\gamma^{n'_{r-2}-n'_{r-1}}\prod_{l=1}^{N_{r-2}}(-v_{r-2,l})^{\sigma_{r-2,l}},\quad i=r-2,\\
    -\frac{u''_{r-1}}{u'_{r}}\gamma^{n''_{r-1}-n'_{r}}\prod_{l=1}^{N_{r-1}}(-v_{r-1,l})^{\sigma_{r-1,l}},\quad i=r-1,\\
    -u_{r-1}u_{r}\gamma^{n_{r}+1}\prod_{l=1}^{N_{r}}(-v_{rl})^{\sigma_{rl}}\quad i=r,
    \end{dcases}
    \end{split}\\
    \begin{split}
        \kappa_{i}&=\begin{dcases}
        n_{i+1}-n_{i}'-\sum_{l=1}^{N_{i}}\sigma_{il},\quad 1\leq i\leq r-3,\\
        n'_{r-1}-n'_{r-2}-\sum_{l=1}^{N_{r-2}}\sigma_{r-2,l},\quad i=r-2,\\
        n'_{r}-n''_{r-1}-\sum_{l=1}^{N_{r-1}}\sigma_{r-1,l},\quad i=r-1,\\
        -n_{r-1}-n_{r}-\sum_{l=1}^{N_{r}}\sigma_{rl}\quad i=r.
        \end{dcases}
    \end{split}
\end{align}

\section{Gaiotto state, \texorpdfstring{$qq$}{qq}-character, and quiver W-algebra}\label{sec:Gaiotto-qq-quiverW}
Using the intertwiners introduced in section~\ref{sec:intertwiner}, we study the Gaiotto state in section~\ref{sec:Gaiottostate}, the $qq$-characters in section~\ref{sec:qqcharacter}, and the relation with quiver W-algebra in section~\ref{sec:quiverW}.
\subsection{Gaiotto state}\label{sec:Gaiottostate}
We define the Gaiotto state as 
\begin{align}
    \begin{split}
        \dket{G,\vec{v},\vec{\sigma}}&=\frac{\bra{0}\Phi_{\vec{\sigma}}^{(n^{\ast})\ast}[u^{\ast},\vec{v}]\ket{0}}{\bra{0}\Phi^{(n^{\ast})\ast}_{\vec{\sigma},\vec{\emptyset}}[u^{\ast},\vec{v}]\ket{0}}=\sum_{\lambda_{1},\ldots,\lambda_{N}}\prod_{i=1}^{N}a^{(\sigma_{i})}_{\lambda_{i}}\frac{\bra{0}\Phi_{\vec{\sigma},\vec{\lambda}}^{(n^{\ast})\ast}[u^{\ast},\vec{v}]\ket{0}}{\bra{0}\Phi^{(n^{\ast})\ast}_{\vec{\sigma},\vec{\emptyset}}[u^{\ast},\vec{v}]\ket{0}}\dket{\vec{v},\vec{\lambda},\vec{\sigma}}\\
        &=\sum_{\lambda_{1},\ldots,\lambda_{N}}\prod_{i=1}^{N}a_{\lambda_{i}}^{(\sigma_{i})}\prod_{i=1}^{N}t^{\ast}_{n_{i}^{\ast},\sigma_{i}}(\lambda_{i},u_{i}^{\ast},v_{i})\prod_{k<l}N_{\sigma_{k}\sigma_{l}}(q_{3}v_{k},\sigma_{k}\,|\,v_{l},\sigma_{l})^{-\sigma_{k}\sigma_{l}}\dket{\vec{v},\vec{\lambda},\vec{\sigma}}
    \end{split}\label{eq:defGaiottost}
\end{align}
where we used the generalized AFS intertwiner convention in (\ref{eq:generalizedAFSintertwiner}) and 
\begin{align}
    \dket{\vec{v},\vec{\lambda},\vec{\sigma}}=\dket{v_{N},\lambda_{N},\sigma_{N}}\otimes\cdots\dket{v_{2},\lambda_{2},\sigma_{2}}\otimes\dket{v_{1},\lambda_{1},\sigma_{1}}.
\end{align}
We normalized the Gaiotto state so that it is expanded as $\dket{G,\vec{v},\vec{\sigma}}=\dket{\vec{\emptyset}}+\cdots$ by dividing with the empty partition contribution that gives the perturbative part. We note that the explicit form of this Gaiotto state itself depends on the ordering of the parities $\vec{\sigma}$.

Using the intertwiner relations in (\ref{eq:AFSproperty}) (see also Appendix \ref{sec:appendix-AFSproperty}) recursively, one can determine the actions of the generators on the Gaiotto state as
\begin{align}
\begin{split}
    x^{+}(z)\dket{G,\vec{v},\vec{\sigma}}=&-\frac{u^{\ast}}{\prod_{j=1}^{N}(-v_{j})^{\sigma_{j}}}\left(\frac{\gamma}{z}\right)^{n^{\ast}-\sum_{j=1}^{N}\sigma_{j}}\left(\mathcal{Y}^{+}(q_{3}^{-1}z)-\mathcal{Y}^{-}(q_{3}^{-1}z)\right)\dket{G,\vec{v},\vec{\sigma}},\\
    x^{-}(z)\dket{G,\vec{v},\vec{\sigma}}=&-\frac{\prod_{j=1}^{N}(-v_{j}\gamma^{-1})^{\sigma_{j}}}{u^{\ast}}\left(\frac{z}{\gamma}\right)^{n^{\ast}-\sum_{j=1}^{N}\sigma_{j}}\left(\mathcal{Y}^{+}(z)^{-1}-\mathcal{Y}^{-}(z)^{-1}\right)\dket{G,\vec{v},\vec{\sigma}},\\
    \psi^{\pm}(z)\dket{G,\vec{v},\vec{\sigma}}=&\gamma^{-\sum_{j=1}^{N}{\sigma_{j}}}\frac{\mathcal{Y}^{\pm}(q_{3}^{-1}z)}{\mathcal{Y}^{\pm}(z)}\dket{G,\vec{v},\vec{\sigma}},
\end{split}
\end{align}
where the operators $\mathcal{Y}^{\pm}(z)$ are defined as operators acting diagonally on the crystal representation basis as 
\begin{align}
    \mathcal{Y}^{\pm}(z)\dket{\vec{v},\vec{\lambda},\vec{\sigma}}=\left[\mathcal{Y}^{(\vec{\sigma})}_{\vec{\lambda}}(z)\right]_{\pm}\dket{\vec{v},\vec{\lambda},\vec{\sigma}},\quad \mathcal{Y}_{\vec{\lambda}}^{(\vec{\sigma})}(z)\coloneqq\prod_{i=1}^{N}\left(\mathcal{Y}_{\lambda_{i}}^{(\sigma_{i})}(z)\right)^{\sigma_{i}}.
\end{align}
Note that although the Gaiotto state depends on the ordering of the parities, the actions of the Drinfeld currents will not depend on them. This comes from the fact that the coproduct formula~(\ref{eq:coproduct1}) is non-symmetric in the two tensor products and after combining with the Gaiotto states, they will eventually cancel the dependence of the ordering of the parities.

We similarly define the dual Gaiotto state as
\begin{align}
\begin{split}
    \dbra{G,\vec{v},\vec{\sigma}}&=\frac{\bra{0}\Phi^{(n)}_{\vec{\sigma}}[u,\vec{v}]\ket{0}}{\bra{0}\Phi^{(n)}_{\vec{\sigma},\vec{\emptyset}}[u,\vec{v}]\ket{0}}=\sum_{\lambda_{1},\ldots,\lambda_{N}}\prod_{i=1}^{N}a_{\lambda_{i}}^{(\sigma_{i})}\frac{\bra{0}\Phi^{(n)}_{\vec{\sigma},\vec{\lambda}}[u,\vec{v}]\ket{0}}{\bra{0}\Phi^{(n)}_{\vec{\sigma},\vec{\emptyset}}[u,\vec{v}]\ket{0}}\dbra{\vec{v},\vec{\lambda},\vec{\sigma}}.
    \end{split}\label{eq:defdualGaiottost}
\end{align}
Then, the action of the generators on this dual Gaiotto state is written as 
\begin{align}
\begin{split}
\dbra{G,\vec{v},\vec{\sigma}}x^{+}(z)&=u\prod_{j=1}^{N}(-v_{j})^{\sigma_{j}}\left(\frac{\gamma}{z}\right)^{n+\sum\limits_{j=1}^{N}\sigma_{j}}\dbra{G,\vec{v},\vec{\sigma}} \left(\mathcal{Y}^{+}(z)^{-1}-\mathcal{Y}^{-}(z)^{-1}\right),\\
\dbra{G,\vec{v},\vec{\sigma}}x^{-}(z)&=u^{-1}\prod\limits_{j=1}^{N}(-\gamma v_{j})^{-\sigma_{j}}\left(\frac{z}{\gamma}\right)^{n+\sum\limits_{j=1}^{N}\sigma_{j}}\dbra{G,\vec{v},\vec{\sigma}}\left(\mathcal{Y}^{+}(q_{3}^{-1}z)-\mathcal{Y}^{-}(q_{3}^{-1}z)\right),\\
\dbra{G,\vec{v},\vec{\sigma}}\psi^{\pm}(z)&=\gamma^{-\sum_{j=1}^{N}{\sigma_{j}}}\dbra{G,\vec{v},\vec{\sigma}}\frac{\mathcal{Y}^{\pm}(q_{3}^{-1}z)}{\mathcal{Y}^{\pm}(z)},
\end{split}
\end{align}
where the actions of the Drinfeld currents are rewritten as diagonal operators on the Gaiotto state. Moreover, using (\ref{eq:pureSYMoperator}),(\ref{eq:defGaiottost}), and (\ref{eq:defdualGaiottost}), the instanton part of the partition function is rewritten as
\begin{align}
    \mathcal{Z}^{\text{inst}}_{N,(\sigma_{1},\ldots,\sigma_{N})}=\dbraket{G,\vec{v},\vec{\sigma}\,|\,{G,\vec{v},\vec{\sigma}}}.
\end{align}
These are generalizations of the non-supergroup case of the Gaiotto state \cite{Gaiotto:2009ma} (see also \cite{Matsuo:2014rba,Bourgine:2015szm,Bourgine:2016vsq,Bourgine:2017jsi}). The difference is that the diagonal operators of the non-supergroup case are modified in a way that when factors coming from the positive branes such as $\mathcal{Y}^{\splus}_{\lambda}(z)$ appears, factors coming from the negative branes will appear as $\mathcal{Y}^{\sminus}_{\lambda}(z)^{-1}$. Namely, the negative branes' factors always appear in the inverse power of the positive branes' factors. This is a universal property\footnote{Actually, this property is already obvious in the crystal representations in section~\ref{sec:crystalrep}. All the factors appearing in the level $(0,-1)$ representation are obtained from level $(0,1)$ representation by modifying $\mathcal{Y}^{\splus}_{\lambda}(z)\rightarrow \mathcal{Y}^{\sminus}_{\lambda}(z)^{-1}$. } of the supergroup gauge theory and we will see it again in section~\ref{sec:qqcharacter}. Thus, any quantity written as $\mathcal{Y}_{\vec{\lambda}}(z)$ in the non-supergroup case is simply modified to $\mathcal{Y}^{(\vec{\sigma})}_{\vec{\lambda}}(z)$ in general supergoup theories.

\subsection{\texorpdfstring{$qq$}{qq}-character}\label{sec:qqcharacter}
Let us derive the fundamental $qq$-character of $A_{1}$-quiver with supergroup $U(N\,|\,\vec{\sigma})$ as an application of the intertwiner formalism. The fundamental operator in deriving the partition function is (\ref{eq:pureSYMoperator}) (see section~\ref{sec:example-pureSYM}). This operator satisfies the following property
\begin{align}
    \left(\rho_{{u^{\ast}}'}^{(1,{n^{\ast}}')}\otimes \rho_{u'}^{(1,n')}\right)\Delta(g(z))\,\mathcal{T}_{U(N\,|\,\vec{\sigma})}=\mathcal{T}_{U(N\,|\,\vec{\sigma})}\,\left(\rho_{u^{\ast}}^{(1,n^{\ast})}\otimes\rho_{u}^{(1,n)}\right)\Delta(g(z)),\label{eq:pureSYMcommuting}
\end{align}
where $g(z)\in\mathcal{E}$ and the spectral parameters obey the conservation laws (\ref{eq:generalizedAFSintertwinerconservation}). The identity above can be proved easily using the intertwiner properties in (\ref{eq:AFSproperty}) and (\ref{eq:dualAFSproperty}). Let us consider only when $g(z)=x^{+}(z)$ and $N=1$. Extensions to general $N>1$ is obtained easily. Using the figure description in the Appendix~\ref{sec:appendix-AFSproperty}, the left hand side is transformed into the right hand side as follows:
\begin{align}
\begin{split}
    \adjustbox{valign=c}{
        \begin{tikzpicture}[xscale=0.7]
            \begin{scope}[scale=1.5]
                \node [above, scale=0.8] at (-1,0.7){$ $};
                \node [below,scale=0.8] at (-1,-0.7){$1$};
                \node [right,scale=0.8] at (-1,0){$ $};
                \node [right,scale=0.8] at (1,0){$ $};
                \node [above,scale=0.8] at (1,0.7){$ $};
                \node [below,scale=0.8] at (1,-0.7){$x^{+}(z)$};
                \draw[postaction={segment={mid arrow}}] (1,0) -- (-1,0);
		        \draw[postaction={segment={mid arrow}}] (-1,0.7) -- (-1,0)--(-1,-0.7);
		        \draw[postaction={segment={mid arrow}}] (1,0.7) -- (1,0)--(1,-0.7);
            \end{scope}
        \end{tikzpicture}}&+\adjustbox{valign=c}{
        \begin{tikzpicture}[xscale=0.7]
            \begin{scope}[scale=1.5]
                \node [above, scale=0.8] at (-1,0.7){$ $};
                \node [below,scale=0.8] at (-1,-0.7){$x^{+}(\gamma z)$};
                \node [right,scale=0.8] at (-1,0){$ $};
                \node [right,scale=0.8] at (1,0){$ $};
                \node [above,scale=0.8] at (1,0.7){$ $};
                \node [below,scale=0.8] at (1,-0.7){$\psi^{-}(\gamma^{1/2}z)$};
                \draw[postaction={segment={mid arrow}}] (1,0) -- (-1,0);
		        \draw[postaction={segment={mid arrow}}] (-1,0.7) -- (-1,0)--(-1,-0.7);
		        \draw[postaction={segment={mid arrow}}] (1,0.7) -- (1,0)--(1,-0.7);
            \end{scope}
        \end{tikzpicture}}\\
        =\adjustbox{valign=c}{
        \begin{tikzpicture}[xscale=0.7]
            \begin{scope}[scale=1.5]
                \node [above, scale=0.8] at (-1,0.7){$ $};
                \node [below,scale=0.8] at (-1,-0.7){$1$};
                \node [right,scale=0.8] at (-1,0){$ $};
                \node [right,scale=0.8] at (1,0){$ $};
                \node [above,scale=0.8] at (1,0.7){$x^{+}(z)$};
                \node [below,scale=0.8] at (1,-0.7){$ $};
                \draw[postaction={segment={mid arrow}}] (1,0) -- (-1,0);
		        \draw[postaction={segment={mid arrow}}] (-1,0.7) -- (-1,0)--(-1,-0.7);
		        \draw[postaction={segment={mid arrow}}] (1,0.7) -- (1,0)--(1,-0.7);
            \end{scope}
        \end{tikzpicture}}-\adjustbox{valign=c}{
        \begin{tikzpicture}[xscale=0.7]
            \begin{scope}[scale=1.5]
                \node [above, scale=0.8] at (-1,0.7){$ $};
                \node [below,scale=0.8] at (-1,-0.7){$1$};
                \node [right,scale=0.8] at (-1,0){$ $};
                \node [above left,scale=0.8] at (1,0){$x^{+}(\gamma z)$};
                \node [above,scale=0.8] at (1,0.7){$ $};
                \node [below,scale=0.8] at (1,-0.7){$\psi^{-}(\gamma^{1/2}z)$};
                \draw[postaction={segment={mid arrow}}] (1,0) -- (-1,0);
		        \draw[postaction={segment={mid arrow}}] (-1,0.7) -- (-1,0)--(-1,-0.7);
		        \draw[postaction={segment={mid arrow}}] (1,0.7) -- (1,0)--(1,-0.7);
            \end{scope}
        \end{tikzpicture}}
        &+\adjustbox{valign=c}{
        \begin{tikzpicture}[xscale=0.7]
            \begin{scope}[scale=1.5]
                \node [above, scale=0.8] at (-1,0.7){$ $};
                \node [below,scale=0.8] at (-1,-0.7){$ $};
                \node [above right,scale=0.8] at (-1,0){$x^{+}(\gamma z)$};
                \node [above left,scale=0.8] at (1,0){$ $};
                \node [above,scale=0.8] at (1,0.7){$ $};
                \node [below,scale=0.8] at (1,-0.7){$\psi^{-}(\gamma^{1/2}z)$};
                \draw[postaction={segment={mid arrow}}] (1,0) -- (-1,0);
		        \draw[postaction={segment={mid arrow}}] (-1,0.7) -- (-1,0)--(-1,-0.7);
		        \draw[postaction={segment={mid arrow}}] (1,0.7) -- (1,0)--(1,-0.7);
            \end{scope}
        \end{tikzpicture}}+\adjustbox{valign=c}{
        \begin{tikzpicture}[xscale=0.7]
            \begin{scope}[scale=1.5]
                \node [above, scale=0.8] at (-1,0.7){$x^{+}(\gamma z)$};
                \node [below,scale=0.8] at (-1,-0.7){$ $};
                \node [right,scale=0.8] at (-1,0){$ $};
                \node [above left,scale=0.8] at (1,0){$\psi^{-}(\gamma z)$};
                \node [above,scale=0.8] at (1,0.7){$ $};
                \node [below,scale=0.8] at (1,-0.7){$\psi^{-}(\gamma^{1/2}z)$};
                \draw[postaction={segment={mid arrow}}] (1,0) -- (-1,0);
		        \draw[postaction={segment={mid arrow}}] (-1,0.7) -- (-1,0)--(-1,-0.7);
		        \draw[postaction={segment={mid arrow}}] (1,0.7) -- (1,0)--(1,-0.7);
            \end{scope}
        \end{tikzpicture}}\\
        =\adjustbox{valign=c}{
        \begin{tikzpicture}[xscale=0.7]
            \begin{scope}[scale=1.5]
                \node [above, scale=0.8] at (-1,0.7){$ $};
                \node [below,scale=0.8] at (-1,-0.7){$1$};
                \node [right,scale=0.8] at (-1,0){$ $};
                \node [right,scale=0.8] at (1,0){$ $};
                \node [above,scale=0.8] at (1,0.7){$x^{+}(z)$};
                \node [below,scale=0.8] at (1,-0.7){$ $};
                \draw[postaction={segment={mid arrow}}] (1,0) -- (-1,0);
		        \draw[postaction={segment={mid arrow}}] (-1,0.7) -- (-1,0)--(-1,-0.7);
		        \draw[postaction={segment={mid arrow}}] (1,0.7) -- (1,0)--(1,-0.7);
            \end{scope}
        \end{tikzpicture}}&+\adjustbox{valign=c}{
        \begin{tikzpicture}[xscale=0.7]
            \begin{scope}[scale=1.5]
                \node [above, scale=0.8] at (-1,0.7){$x^{+}(\gamma z)$};
                \node [below,scale=0.8] at (-1,-0.7){$ $};
                \node [right,scale=0.8] at (-1,0){$ $};
                \node [above left,scale=0.8] at (1,0){$ $};
                \node [above,scale=0.8] at (1,0.7){$\psi^{-}(\gamma^{1/2}z)$};
                \node [below,scale=0.8] at (1,-0.7){$ $};
                \draw[postaction={segment={mid arrow}}] (1,0) -- (-1,0);
		        \draw[postaction={segment={mid arrow}}] (-1,0.7) -- (-1,0)--(-1,-0.7);
		        \draw[postaction={segment={mid arrow}}] (1,0.7) -- (1,0)--(1,-0.7);
            \end{scope}
        \end{tikzpicture}}
\end{split}
\end{align}
Thus, the operator $\mathcal{T}_{U(N\,|\,\vec{\sigma})}$ is interpreted as a screening operator similar to the non-supergroup case. 

Inserting the operator $\Delta(x^{\pm}(z))$ and taking the expectation value in the vertex operator representation, we can derive the $qq$-characters. In the non-supergroup theory, the $qq$-character turns out to be a polynomial, but for the supergroup case, it will be a rational function\footnote{This property is related to the Seiberg-Witten curve \cite{Vafa:2001qf}. For the pure $SU(N)$ gauge theory, the Seiberg-Witten curve is $z+1/z+\det(x-\Phi)=0$, where $\Phi$ is the scalar matrix of the Coulomb branch parameters, while for the supergroup $SU(N_{+}\,|\,N_{-})$ it is $z+1/z+\sdet(x-\Phi)=0$. If we diagonalize it as $\Phi=(a_{1},\ldots,a_{N_{+}}\,|\,b_{1},\ldots,b_{N_{-}})$, we have $z+1/z+\frac{\prod_{i=1}^{N_{+}}(x-a_{i})}{\prod_{j=1}^{N_{-}}(x-b_{j})}=0$. The super-determinant part is related to why the $qq$-character is a rational function.} \cite{Kimura-Pestun:supergroup}. The authors in \cite{Bourgine:2016vsq,Bourgine:2017jsi} studied this property using the condition (\ref{eq:pureSYMcommuting}) and claimed that the Chern-Simons level will be restricted so that the external branes do not intersect with each other as in \cite{Aharony:1997ju}. We expect that a similar story holds for the supergroup case and that the properties of the $qq$-characters give a restriction to the possible brane webs. We leave a detailed analysis of the $qq$-character for future work. In this paper, we simply apply the derivation of \cite{Bourgine:2017jsi} and only derive the algebraic quantity for the simplest case and give few observations. 

We define the fundamental $qq$-character of the $A_{1}$-quiver as
\begin{align}
\begin{split}
    \chi_{\sAbox}^{+}(z)&\coloneqq\frac{\nu z^{{n^{\ast}}'+\sum_{i=1}^{N}\sigma_{i}}}{{u^{\ast}}'q_{3}^{{n^{\ast}}'}}\frac{\langle\Delta (x^{+}(\gamma^{-1}z))\mathcal{T}_{U(N\,|\,\vec{\sigma})}\rangle}{\langle\mathcal{T}_{U(N\,|\,\vec{\sigma})}\rangle},\quad\nu=-\prod_{l=1}^{N}(-q_{3}v_{l})^{-\sigma_{l}}.
\end{split}
\end{align}
Using the formulas in section~\ref{sec:Gaiottostate} and Appendix~\ref{sec:appendix-contraction-intertwiner}, the $qq$-character is determined as
\begin{align}
\begin{split}
    \chi_{\sAbox}^{+}(z)&=\left\langle\nu z^{\sum_{i=1}^{N}\sigma_{i}}\mathcal{Y}_{\vec{\lambda}}^{(\vec{\sigma})}(q_{3}^{-1}z)+\mathfrak{q}\frac{z^{\kappa}}{\mathcal{Y}_{\vec{\lambda}}^{(\vec{\sigma})}(z)}\right\rangle_{\text{inst.}},
\end{split}
\end{align}
where the expectation value $\langle\mathcal{O}_{\vec{\lambda}}(z)\rangle_{\text{inst.}}$ is
\begin{align}
    \langle\mathcal{O}_{\vec{\lambda}}(z)\rangle_{\text{inst.}}\coloneqq \frac{1}{\mathcal{Z}^{\text{inst}}_{N,(\sigma_{1},\ldots,\sigma_{N})}}\sum_{\lambda_{1},\ldots,\lambda_{N}}\mathcal{O}_{\vec{\lambda}}(z)\mathcal{Z}_{\text{top.}}(\vec{\lambda},\vec{\sigma})\mathcal{Z}_{\text{CS}}(\kappa,\vec{\lambda},\vec{\sigma})\mathcal{Z}_{\text{vec.}}(\vec{v},\vec{\lambda},\vec{\sigma}).
\end{align}
Up to few factors, this is indeed the same $qq$-character derived in \cite{Kimura-Pestun:supergroup}. Note that the qq-character does not depend on the ordering of the parities. As mentioned in section~\ref{sec:Gaiottostate}, the main difference with the non-supergroup case is the existence of negative branes and the corresponding operator $\mathcal{Y}^{\sminus}_{\lambda}(z)$. The $\mathsf{Y}$ function of $N$-stack of D-branes will transform from $\mathcal{Y}_{\vec{\lambda}}(z)$ to $\mathcal{Y}^{(\vec{\sigma})}_{\vec{\lambda}}(z)$, where parities of the D-branes are introduced as additional degrees of freedom. Higher $qq$-characters and $qq$-characters associated with $A_{r}$-quiver can be derived using the results in \cite{Bourgine:2017jsi} and doing this manipulation.

\subsection{Quiver W-algebra}\label{sec:quiverW}
We briefly study the relation with Kimura-Pestun's quiver W-algebra. Obviously, from the brane web of the linear quiver gauge theory in section~\ref{sec:example-pureSYM} and \ref{sec:example-Aquiver} or equation~(\ref{eq:general_linear_quiver}), we can interpret the brane system as a map from tensor products of vertex operators to tensor products of vertex operators. The vertex operators appearing are all of the same type\footnote{From triality, we have three types of vertex operator representations usually denoted $\mathcal{F}_{i}\,(i=1,2,3)$. Since we are using only $\mathcal{F}_{3}$ dressed with some zero mode factors, the arising W-algebra is $\mathcal{F}_{3}\otimes \mathcal{F}_{3}\otimes\cdots\otimes\mathcal{F}_{3}$, which is just the deformed $W_{N}$ algebra (with an additional Heisenberg algebra). As mentioned in section~\ref{sec:negativevertexop}, we also have a different vertex operator representations. Including these representations should give new quiver W-algebras \cite{Kimura:2015rgi}.}, so the arising deformed W-algebra should be the deformed $W_{N}$ algebra ($+$ Heisenberg). If this is so, the partition function itself should be rewritten in products of the screening currents of the W-algebra as in \cite{Kimura:2015rgi}. Let us see this for the simplest pure super Yang-Mills case. Generalizations to other quivers are straightforward.  

For the pure SYM case, the system acts on tensor products of two vertex operator representations. Thus, it should be related with the $q$-Virasoro algebra. There are two screening currents for the $q$-Virasoro algebra: 
\begin{align}
\begin{split}
    \mathbf{S}_{1}(z)&=\exp\left(-\sum_{n=1}^{\infty}\frac{z^{n}}{n}q_{1}^{n}(\gamma^{n}a_{-n}^{(1)}-a_{-n}^{(2)})\right)\exp\left(-\sum_{n=1}^{\infty}\frac{z^{-n}}{n}\frac{1-q_{1}^{-n}}{1-q_{2}^{-n}}(\gamma^{n}a_{n}^{(1)}-a_{n}^{(2)})\right),\\
    \mathbf{S}_{2}(z)&=\exp\left(\sum_{n=1}^{\infty}\frac{z^{n}}{n}\frac{1-q_{2}^{n}}{1-q_{1}^{-n}}(\gamma^{n}a_{-n}^{(1)}-a_{-n}^{(2)})\right)\exp\left(-\sum_{n=1}^{\infty}\frac{z^{-n}}{n}(\gamma^{n}a_{n}^{(1)}-a_{n}^{(2)})\right),
\end{split}
\end{align}
where we omit the zero modes and $a_{n}^{(1)}=a_{n}\otimes 1,\,a_{n}^{(2)}=1\otimes a_{n}$. The screening currents obey the condition
\begin{align}
\begin{split}
    \xi(z)\otimes \eta(z)&=:\mathbf{S}_{1}(z)^{-1}\mathbf{S}_{1}(q_{2}z):=:\mathbf{S}_{2}(z)^{-1}\mathbf{S}_{2}(q_{1}z):.
\end{split}
\end{align}
Using this, they are rewritten as 
\begin{align}
    \mathbf{S}_{1}(z)&=:\prod_{i=1}^{\infty}\xi(q_{2}^{-i}z)\otimes\eta(q_{2}^{-i}z) :,\quad \mathbf{S}_{2}(z)=:\prod_{i=1}^{\infty}\xi(q_{1}^{-i}z)\otimes \eta(q_{1}^{-i}z):.
\end{align}
Then, the intertwiners are rewritten by the above screening currents as
\footnote{Another way to convert the intertwiners to products of screening currents is to use the equivariant index formula in Appendix~\ref{sec:appendix-equiv-nekrasov}. By direct computation, we have
\begin{align}
    \Phi^{\ast}_{\sigma,\lambda}\otimes \Phi_{\sigma,\lambda}&\sim \exp\left(\sum_{n=1}^{\infty}\frac{1}{n}\frac{\gamma^{n}a_{-n}^{(1)}-a_{-n}^{(2)}}{1-q_{1}^{-n}}\sigma\ch\bfY^{\sigma}_{[n]}\right)\exp\left(-\sum_{n=1}^{\infty}\frac{1}{n}\frac{\gamma^{n}a_{n}^{(1)}-a_{n}^{(2)}}{1-q_{2}^{-n}}\sigma\ch\bfY^{\sigma}_{[-n]}\right),
\end{align}
where 
\begin{align}
    \ch\bfY^{\sigma}_{[n]}=v^{n}-(1-q_{1}^{n})(1-q_{2}^{n})\sum_{x\in\lambda}(\chi_{x}^{(\sigma)})^{n}.
\end{align}
Inserting the relation of the Y bundles and X bundles in (\ref{eq:appendix-XYrelation}), we have 
\begin{align}
\begin{split}
    \Phi^{\ast}_{\sigma,\lambda}\otimes\Phi_{\sigma,\lambda}&\simeq:\prod_{x\in\mathcal{X}_{\lambda}^{\sigma}}\mathbf{S}_{1}(q_{1}^{\frac{\sigma-1}{2}}x):=:\prod_{x\in\check{\mathcal{X}}_{\lambda}^{\sigma}}\mathbf{S}_{2}(q_{2}^{\frac{\sigma-1}{2}}x):.
    \end{split}
\end{align}}
\begin{align}
\begin{split}
    \Phi^{\ast}_{+,\lambda}\otimes \Phi_{+,\lambda}&\simeq :\prod_{x\in\lambda_{\infty}}\frac{\mathbf{S}_{1}(\chi_{x}^{\splus})}{\mathbf{S}_{1}(q_{2}\chi_{x}^{\splus})}\prod_{x\in\lambda}\frac{\mathbf{S}_{1}(q_{2}\chi_{x}^{\splus})}{\mathbf{S}_{1}(\chi_{x}^{\splus})}:=:\prod_{x\in\mathcal{X}^{+}_{\lambda}}\mathbf{S}_{1}(x):\\
    &\simeq :\prod_{x\in\lambda_{\infty}}\frac{\mathbf{S}_{2}(\chi_{x}^{\splus})}{\mathbf{S}_{2}(q_{1}\chi_{x}^{\splus})}\prod_{x\in\lambda}\frac{\mathbf{S}_{2}(q_{1}\chi_{x}^{\splus})}{\mathbf{S}_{2}(\chi_{x}^{\splus})}:=:\prod_{x\in\check{\mathcal{X}}^{+}_{\lambda}}\mathbf{S}_{2}(x):,
\end{split}\label{eq:positive-intertwiner-screening}\\
\begin{split}
    \Phi^{\ast}_{-,\lambda}\otimes\Phi_{-,\lambda}&\simeq: \prod_{x\in\lambda_{\infty}}\frac{\mathbf{S}_{1}(q_{2}\chi_{x}^{\sminus})}{\mathbf{S}_{1}(\chi_{x}^{\sminus})}\prod_{x\in\lambda}\frac{\mathbf{S}_{1}(\chi_{x}^{\sminus})}{\mathbf{S}_{1}(q_{2}\chi_{x}^{\sminus})}:=:\prod_{x\in\mathcal{X}^{-}_{\lambda}}\mathbf{S}_{1}(q_{1}^{-1}x):\\
    &\simeq : \prod_{x\in\lambda_{\infty}}\frac{\mathbf{S}_{2}(q_{1}\chi_{x}^{\sminus})}{\mathbf{S}_{2}(\chi_{x}^{\sminus})}\prod_{x\in\lambda}\frac{\mathbf{S}_{2}(\chi_{x}^{\sminus})}{\mathbf{S}_{2}(q_{1}\chi_{x}^{\sminus})}:=:\prod_{x\in\check{\mathcal{X}}^{-}_{\lambda}}\mathbf{S}_{2}(q_{2}^{-1}x):,
\end{split}\label{eq:negative-intertwiner-screening}
\end{align}
where we used 
\begin{align}
\begin{split}
    \Phi_{+,\emptyset}[v]&=:\prod_{i,j=1}^{\infty}\eta(vq_{1}^{i-1}q_{2}^{j-1})^{-1}:,\quad \Phi^{\ast}_{+,\emptyset}[v]=:\prod_{i,j=1}^{\infty}\xi(vq_{1}^{i-1}q_{2}^{j-1})^{-1}:,\\
    \Phi_{-,\emptyset}[v]&=:\prod_{i,j=1}^{\infty}\eta(vq_{1}^{-i}q_{2}^{-j})^{+1}:,\quad \Phi_{-,\emptyset}^{\ast}[v]=:\prod_{i,j=1}^{\infty}\xi(vq_{1}^{-i}q_{2}^{-j})^{+1}:,
\end{split}
\end{align}
and $\lambda_{\infty}=\{(i,j)\,|\,i,j=1,2,\ldots\}$. The symbol $\simeq$ means we omitted the zero mode part. See (\ref{eq:appendix-xvariables}) for the definition of the $x$-variables.

The operator $\mathcal{T}_{U(N\,|\,\vec{\sigma})}$ is now rewritten for example as
\begin{align}
\begin{split}
    \mathcal{T}_{U(N\,|\,\vec{\sigma})}&=\sum_{\vec{\lambda}}\prod_{i=1}^{N}a_{\lambda_{i}}^{(\sigma_{i})}t_{n_{i},\sigma_{i}}(\lambda_{i},u_{i},v_{i})t^{\ast}_{n_{i}^{\ast},\sigma_{i}}(\lambda_{i},u^{\ast}_{i},v_{i})\\
    &\times :\prod_{x_{N}\in\mathcal{X}_{\lambda_{N}}^{\sigma_{N}}}\mathbf{S}_{1}(q_{1}^{\frac{\sigma_{N}-1}{2}}x_{N}):\cdots:\prod_{x_{2}\in\mathcal{X}_{\lambda_{2}}^{\sigma_{2}}}\mathbf{S}_{1}(q_{1}^{\frac{\sigma_{2}-1}{2}}x_{2})::\prod_{x_{1}\in\mathcal{X}_{\lambda_{1}}^{\sigma_{1}}}\mathbf{S}_{1}(q_{1}^{\frac{\sigma_{1}-1}{2}}x_{1}):.
\end{split}
\end{align}
One can also use the other three ways in (\ref{eq:positive-intertwiner-screening}) and (\ref{eq:negative-intertwiner-screening}) to rewrite the intertwiners to screening currents\footnote{See also \cite{Kimura-Nieri:defects,Nieri:2021xpe}, where a similar formula using two types of screening currents to reproduce partition functions of supergroup gauge theories was discussed.}. Adding extra zero modes to the screening currents and dealing with the extra factors in front of the products of screening currents, one can obtain the supergroup analogue of the Z-state defined in \cite{Kimura:2015rgi}:
\begin{align}
\begin{split}
    &\ket{Z_{U(N\,|\,\vec{\sigma})}}=\mathcal{T}_{U(N\,|\,\vec{\sigma})}\ket{0}\otimes \ket{0},\\
    &\mathcal{Z}_{N,(\sigma_{1},\ldots,\sigma_{N})}=\langle0\,|\,Z_{U(N\,|\,\vec{\sigma})}\rangle=\bra{0}\otimes \bra{0}\mathcal{T}_{U(N\,|\,\vec{\sigma})}\ket{0}\otimes \ket{0}.
\end{split}
\end{align}

\section{Comparison with refined topological vertex and anti-vertex}\label{sec:ref-topvertex}
As discussed in detail in section~\ref{sec:Examples}, using the positive and negative intertwiners, we can compute the Nekrasov partition function of supergroup gauge theories. For the non-supergroup case, another traditional way to compute Nekrasov partition functions is to use the (refined) topological vertex \cite{Aganagic:2003db,Iqbal:2007ii,Awata:2005fa,Awata:2008ed,Taki:2007dh}. Since both the intertwiners and the refined topological vertex reproduce the same instanton partition function, it is natural to expect they are related to each other. In the seminal work \cite{Awata:2011ce}, the identification of the refined topological vertex of \cite{Iqbal:2007ii,Awata:2008ed} and the positive intertwiner were shown. 

Actually, an analogue of the refined topological vertex for the supergroup case called anti-refined topological vertex (shortly anti-vertex) was introduced in \cite{Kimura-Sugimoto:antivertex}. A similar identification for this anti-vertex and negative intertwiner should hold because they reproduce the same partition function. 

We review the anti-vertex and give the complete rule for the framing factors in section~\ref{sec:topvertexrule}. We then show that the matrix elements of the interwiners correspond to the topological vertices under the identification $(q_{1},q_{2})=(q,t^{-1})$ in section~\ref{sec:intertwiner-topvertex}. We will see that the positive and negative intertwiners are related with each other in the unrefined limit in section~\ref{sec:unrefinedlimit}, which reproduces the result in \cite{Kimura-Sugimoto:antivertex}. Finally, in section~\ref{sec:gluingtopvertex}, we confirm that the gluing rules for the anti-vertex we propose in section~\ref{sec:topvertexrule} match with the gluing rules of the interwiners.
\subsection{Refined topological vertex and anti-vertex}\label{sec:topvertexrule}
We briefly review the (anti) refined topological vertex in this section (see for example section 4 of \cite{Bao:2013pwa}). We use the Iqbal-Kozcaz-Vafa (IKV) \cite{Iqbal:2007ii} and Kimura-Sugimoto form \cite{Kimura-Sugimoto:antivertex}, which use the Schur functions, instead of the Awata-Kanno form \cite{Awata:2005fa,Awata:2008ed} using Macdonald polynomials. We note that both forms are known to be essentially equivalent \cite{Awata:2008ed,Awata:2011ce}.

The definition of the refined topological vertex is 
\begin{align}
\begin{split}
    C_{\lambda\mu\nu}(t,q)&=\left(\frac{q}{t}\right)^{\frac{||\mu||^{2}+||\nu||^{2}}{2}}t^{\frac{\kappa(\mu)}{2}}t^{\frac{||\nu||^{2}}{2}}\wt{Z}_{\nu}(t,q)\sum_{\eta}\left(\frac{q}{t}\right)^{\frac{|\eta|+|\lambda|-|\mu|}{2}}s_{\lambda^{\rmT}/\eta}(t^{-\rho}q^{-\nu})s_{\mu/\eta}(t^{-\nu^{\rmT}}q^{-\rho})\\
    &=q^{\frac{||\mu||^{2}+||\nu||^{2}}{2}}t^{-\frac{||\mu^{\rmT}||^{2}}{2}}\wt{Z}_{\nu}(t,q)\sum_{\eta}\left(\frac{q}{t}\right)^{\frac{|\eta|+|\lambda|-|\mu|}{2}}s_{\lambda^{\rmT}/\eta}(t^{-\rho}q^{-\nu})s_{\mu/\eta}(t^{-\nu^{\rmT}}q^{-\rho})
\end{split}
\end{align}
where 
\begin{align}
\begin{split}
     \rho=(-\frac{1}{2},-\frac{3}{2},-\frac{5}{2},\cdots),\quad
    \wt{Z}_{\nu}(t,q)&=\prod_{(i,j)\in\nu}(1-t^{l_{\nu}(i,j)+1}q^{a_{\nu}(i,j)})^{-1},
\end{split}
\end{align}
and $s_{\lambda/\eta}(x),\,x=(x_{1},x_{2},\ldots)$ is the skew Schur function. See Appendix~\ref{sec:appendix-Young} and \ref{sec:appendix-Schurfunc} for the notations. The anti refined topological vertex \cite{Kimura-Sugimoto:antivertex} is similarly defined as 
\begin{align}
    \bar{C}_{\lambda\mu\nu}(t,q)=t^{-\frac{1}{2}||\mu^{\rmT}||^{2}}q^{\frac{1}{2}(||\mu||^{2}-||\nu||^{2})}\wt{Z}_{\nu}(t^{-1},q^{-1})\sum_{\eta}\left(\frac{q}{t}\right)^{\frac{1}{2}(|\eta|+|\lambda|-|\mu|)}s_{\lambda^{\rmT}/\eta}(t^{\rho}q^{\nu})s_{\mu/\eta}(q^{\rho}t^{\nu^{\rmT}}).
\end{align}
We depict these vertices as 
\begin{align}
\begin{split}
    &\begin{tikzpicture}[thick,scale=1.2]
		\begin{scope}[scale=1.2]
		\node [below right] at (-1,1){$q,\mu$};
		\node [below right] at (-1.35,-0.35){$t,\lambda$};
		\node [above] at (0,0) {$\nu$};
		\node [left] at (-2,0.2){$C_{\lambda\mu\nu}(t,q)$};
		\node [left] at (-1.5,0.2){$=$};
		\draw[postaction={segment={mid arrow}}] (-1,0)--(-1.7,-0.7);
		\draw[postaction={segment={mid arrow}}] (-1,0)--(0,0);
		\draw[postaction={segment={mid arrow}}] (-1,0) --(-1,1);
		\draw[thick,red] (-0.3,0.1)--(-0.3,-0.1);
		\draw[thick,red] (-0.2,0.1)--(-0.2,-0.1);
		\end{scope}
		\begin{scope}[shift=({5.5,0}),scale=1.2]
		\node [below right] at (-1,1){$q,\mu$};
		\node [below right] at (-1.35,-0.35){$t,\lambda$};
		\node [above] at (0,0) {$\nu$};
		\node [left] at (-2,0.2){$\bar{C}_{\lambda\mu\nu}(t,q)$};
		\node [left] at (-1.5,0.2){$=$};
		\draw[postaction={segment={mid arrow}}] (-1,0)--(-1.7,-0.7);
		\draw[dashed,postaction={segment={mid arrow}}] (-1,0)--(0,0);
		\draw[postaction={segment={mid arrow}}] (-1,0) --(-1,1);
		\draw[thick,red] (-0.3,0.1)--(-0.3,-0.1);
		\draw[thick,red] (-0.2,0.1)--(-0.2,-0.1);
		\end{scope}
	\end{tikzpicture}
 \end{split} \label{eq:top-vertices-def}
\end{align}
where partitions $\lambda,\mu,\nu$ are assigned to all of the legs, while parameters $q,t$ are assigned to two of the three legs. The labels of the vertices are ordered in a clockwise direction. The leg marked with two red strips has no parameter associated with it and is called the preferred direction. In this paper, we will fix the preferred direction to be in the horizontal direction. For the anti-vertex, we use dashed lines only for the preferred direction\footnote{In the original paper \cite{Kimura-Sugimoto:antivertex}, the authors draw all the legs in dashed lines. Since this vertex will be related to the negative intertwiner, we follow the drawings of the intertwiner and draw only the preferred direction in dashed line.}. The arrows are drawn outgoing from the vertex and if the arrow is incoming towards the vertex, the associated partition gets transposed. We also denote these vertices as
\begin{align}
    C^{(\sigma)}_{\mu\nu\lambda}(t,q)&=\begin{dcases}C_{\mu\nu\lambda}(t,q)\quad \sigma=+\\
    \bar{C}_{\mu\nu\lambda}(t,q)\quad \sigma=-
    \end{dcases},
\end{align}
where $\sigma$ determines whether the vertex is the normal vertex or the anti-vertex.

Let us explain how to assign the topological vertices given a brane web. Unfortunately, the gluing rules were not explicitly given in \cite{Kimura-Sugimoto:antivertex} when the framing factors play a role. It seems that the framing factors do not directly come from the charges of the brane web (see the remark at the end of this subsection). We propose the following procedures to compute partition functions using the anti-vertex. In section~\ref{sec:intertwiner-topvertex} and \ref{sec:gluingtopvertex}, we will see that these rules are compatible with the interwiner formalism.

\begin{enumerate}
    \item We start from a brane web satisfying the following conditions.
    \begin{itemize}
        \item Each vertex of the diagram is a trivalent vertex with three edges connected to it.
        \item One of the three edges is always parallel to a fixed direction which we call the \textit{preferred direction}.
        \item Edges non-parallel to the preferred direction are drawn in solid lines.
        \item Edges parallel to the preferred direction are assigned two red strips for convenience.
        \item Edges parallel to the preferred direction are drawn in either solid lines or dashed lines.
    \end{itemize} 
    In this paper, we choose the preferred direction to be the horizontal direction. For the non-supergroup case, the brane web comes from dual toric diagrams \cite{Aharony:1997bh,Aharony:1997ju,Katz:1996fh,Katz:1997eq}. For the supergoup case, whether we have a similar story is non-trivial for the moment, so we just say that the brane web is a general diagram satisfying the above conditions. 
    \item We further assign parameters $q,t$ to each end of the edges of the non-preferred direction as the following:
\begin{align}
\adjustbox{valign=c}{
    \begin{tikzpicture}[thick]
        \begin{scope}
        \draw[] (0,0)--(1.4,0);
        \draw[red] (0.65,0.1)--(0.65,-0.1);
        \draw[red] (0.75,0.1)--(0.75,-0.1);
        \draw (0,0)--(-0.8,-0.8);
        \draw (0,0)--(0,1);
        \draw (1.4,0)--(2.2,-0.8);
        \draw (1.4,0)--(1.4,1);
        \node [above left] at (0,0.2){$t$};
        \node [above right] at (1.4,0.2){$t$};
        \node [left] at (-0.2,-0.15){$q$};
        \node [right] at (1.6,-0.15){$q$};
        \end{scope}
    \end{tikzpicture}}\hspace{1cm}
    \adjustbox{valign=c}{\begin{tikzpicture}[thick]
        \begin{scope}
        \draw[dashed] (0,0)--(1.4,0);
        \draw[red] (0.65,0.1)--(0.65,-0.1);
        \draw[red] (0.75,0.1)--(0.75,-0.1);
        \draw (0,0)--(-0.8,-0.8);
        \draw (0,0)--(0,1);
        \draw (1.4,0)--(2.2,-0.8);
        \draw (1.4,0)--(1.4,1);
        \node [above left] at (0,0.2){$t$};
        \node [above right] at (1.4,0.2){$t$};
        \node [left] at (-0.2,-0.15){$q$};
        \node [right] at (1.6,-0.15){$q$};
        \end{scope}
    \end{tikzpicture}}\hspace{1cm}
    \adjustbox{valign=c}{
    \begin{tikzpicture}[thick]
        \begin{scope}
        \draw (0,0)--(0.7,0);
        \draw[red] (0.35,0.1)--(0.35,-0.1);
        \draw[red] (0.45,0.1)--(0.45,-0.1);
        \draw (0,0)--(-0.8,-0.8);
        \draw (0,0)--(0,1.3);
        \draw[red] (0.35,1.4)--(0.35,1.2);
        \draw[red] (0.45,1.4)--(0.45,1.2);
        \draw (0,1.3)--(0.7,1.3);
        \draw (0,1.3)--(-0.8,2.1);
        \node [left] at (0,0.3){$t$};
        \node [left] at (0,1.0){$q$};
        \node [below] at (-0.2,-0.25){$q$};
        \node [above] at (-0.2,1.55) {$t$};
        \end{scope}
    \end{tikzpicture}}\hspace{1cm}
    \adjustbox{valign=c}{
    \begin{tikzpicture}[thick]
        \begin{scope}
        \draw[dashed] (0,0)--(0.7,0);
        \draw[red] (0.35,0.1)--(0.35,-0.1);
        \draw[red] (0.45,0.1)--(0.45,-0.1);
        \draw (0,0)--(-0.8,-0.8);
        \draw (0,0)--(0,1.3);
        \draw[red] (0.35,1.4)--(0.35,1.2);
        \draw[red] (0.45,1.4)--(0.45,1.2);
        \draw[] (0,1.3)--(0.7,1.3);
        \draw (0,1.3)--(-0.8,2.1);
        \node [left] at (0,0.3){$t$};
        \node [left] at (0,1.0){$q$};
        \node [below] at (-0.2,-0.25){$q$};
        \node [above] at (-0.2,1.55) {$t$};
        \end{scope}
    \end{tikzpicture}}\hspace{1cm} \text{etc.}
\end{align}
Namely, an edge with parameter $q$ (resp.\,$t$) glues with an edge with parameter $t$ (resp.\,$q$). It is exactly the same as the non-supergroup case and does not depend whether the preferred direction is in dashed lines or not.  

We can switch the roles of $q$ and $t$, and the result does not change. After determining the parameter of one of the edges of the brane web, all the other parameters will be determined automatically following the above rule. For convenience, we choose an ordering such that the parameter $t$ is always above compared to $q$. 
\item We assign each edge a Young diagram. For edges extending semi-infinitely, an empty partition is assigned. We also make the edges oriented and assign an arrow pointing to one of the two vertices. The arrows of the three edges surrounding a vertex is chosen to be either incoming or all outgoing from the vertex. 
\item We choose a positive direction of the preferred direction. The opposite direction is called the negative direction. In our situation that the preferred direction is in the horizontal direction, we can choose left or right to be the positive direction. We choose the direction pointing right to be the positive direction. 
\begin{figure}[t]
\centering
\includegraphics[width=10cm]{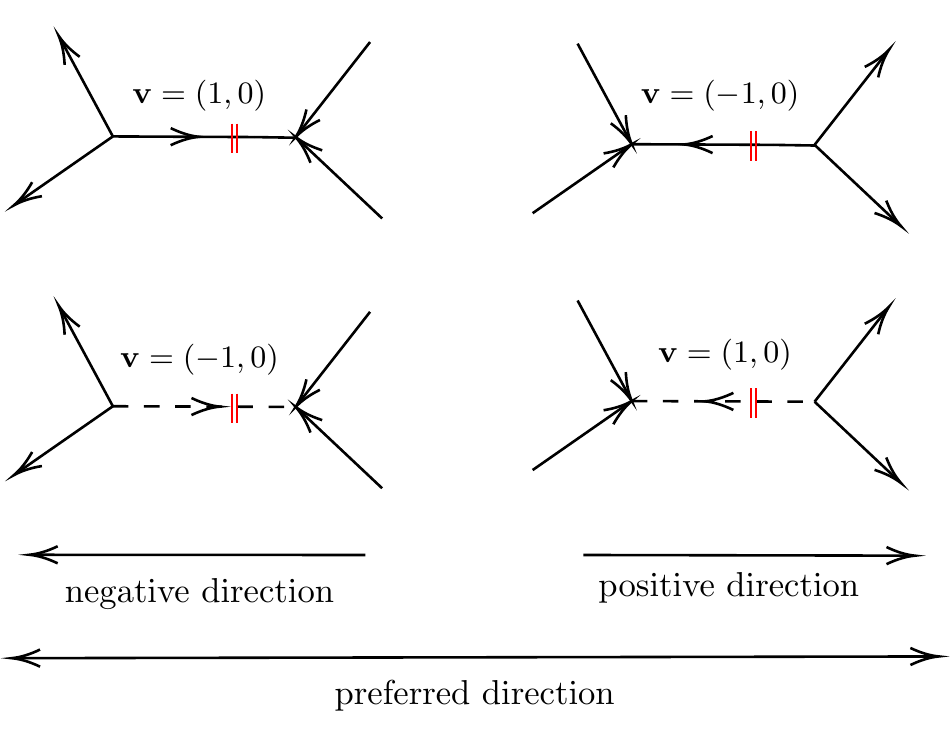}
\caption{Positive and negative direction of the preferred direction. When the edge parallel to the preferred direction is a solid line and points at the positive direction, we assign vector $\mathbf{v}=(1,0)$, when it is a solid line and points at the negative direction, we assign vector $\mathbf{v}=(-1,0)$. Oppositely, when the preferred edge is a dashed line and points at the positive direction, we assign vector $\mathbf{v}=(-1,0)$, when it is a dashed line and points at the negative direction, we assign vector $\mathbf{v}=(1,0)$. }
\label{fig:positive-negative-direction}
\end{figure}

We then, assign two-dimensional integer valued vectors $\mathbf{v}_{i}\in\mathbb{Z}^{2}\,(i=1,2,3)$ to each of the edges obeying the following conditions:
\begin{itemize}
    \item The assigned vectors obey the conservation law:
    \begin{align}
    \sum_{i=1}^{3}\mathbf{v}_{i}=0.
\end{align}
\item For the edges in the preferred direction, either $(1,0)$ or $(-1,0)$ is assigned. If an edge with solid line points at the positive direction of the preferred direction, we assign $(1,0)$, if it points at the negative direction we assign $(-1,0)$. For the edges with dashed lines, we oppositely assign $(-1,0)$ for the positive direction and $(1,0)$ for the negative direction (see Figure~\ref{fig:positive-negative-direction}).
\end{itemize}

\item To each edge, we further associate an integer $\eta$. The integer $\eta$ depends on how four edges are connected to the edge. There are five patterns in our setup and they are drawn in Figure~\ref{fig:frame-factor}. For the non-supergroup case, this integer $\eta$ is derived from the vectors assigned to the four edges. For general supergroup case, we do not have such understanding, so we just assign the integer $\eta$ following Figure~\ref{fig:frame-factor} (see the remark at the end of this section).

We introduce two framing factors as in \cite{Iqbal:2007ii,Taki:2007dh}:
\begin{align}
\begin{split}
    f_{\mu}(t,q)&=(-1)^{|\mu|}t^{\frac{||\mu^{\rmT}||^{2}}{2}}q^{-\frac{||\mu||^{2}}{2}},\\
    \tilde{f}_{\mu}(t,q)&=(-1)^{\mu}\left(\frac{t}{q}\right)^{\frac{|\mu|}{2}}t^{\frac{||\mu^{\rmT}||^{2}}{2}}q^{-\frac{||\mu||^{2}}{2}}=\left(\frac{t}{q}\right)^{\frac{|\mu|}{2}}f_{\mu}(t,q).
\end{split}
\end{align}
Note that these factors have the properties
\begin{align}
    f_{\mu^{\rmT}}(t,q)=f_{\mu}(q,t)^{-1},\quad \tilde{f}_{\mu^{\rmT}}(t,q)=\tilde{f}_{\mu}(q,t)^{-1}.
\end{align}
Using these, we associate a framing factor $f_{\lambda}^{\eta},\tilde{f}_{\lambda}^{\eta}$ to an edge $\mathbf{v}$ with a partition $\lambda$ and an integer~$\eta$ (see for example Figure 12 of \cite{Bao:2013pwa} for how to determine the framing factors of non-supergroup cases). Explicitly, the five patterns in Figure \ref{fig:frame-factor} only appear in our paper. Additionally, we assign a K$\ddot{\text{a}}$hler moduli $(-Q)^{|\lambda|}$ to each edge.  
\begin{figure}[t]
    \centering
    \includegraphics[width=15cm]{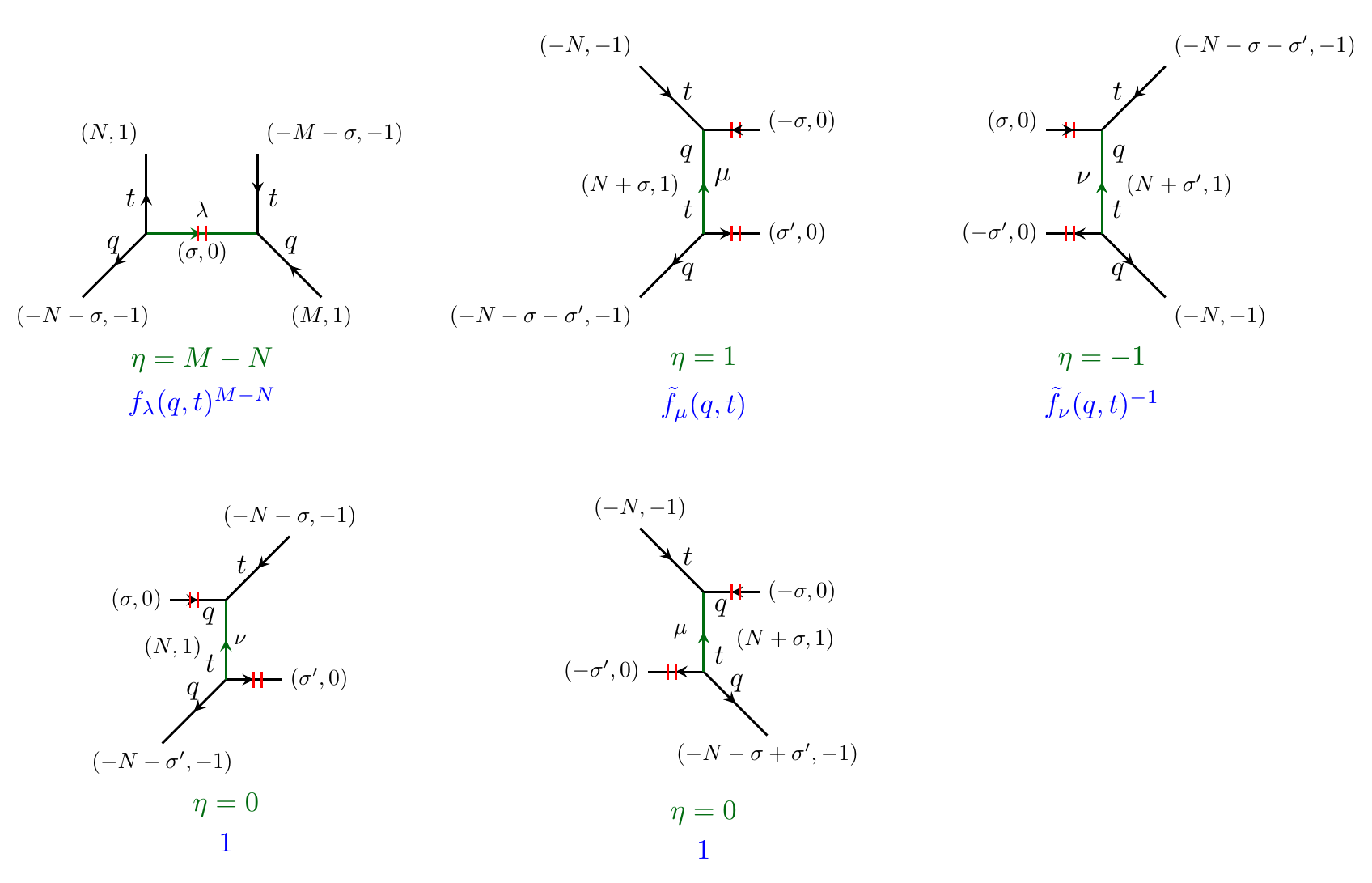}
    \caption{The five patterns of framing factors that appear in this paper. The integer $\eta$ associated with the green edge is written in green. The framing factor for each pattern is written in blue. The parity $\sigma,\sigma=\pm1$ depends on whether the edges in the preferred direction are in dashed lines or not. When $\sigma=+1$, they are positive D-branes and when $\sigma=-1$, they are negative D-branes.  }
    \label{fig:frame-factor}
\end{figure}

\item For each vertex, when there is no dashed line connecting to it, we assign $C_{\lambda\mu\nu}(t,q)$. When there is a dashed line, we assign $\bar{C}_{\lambda\mu\nu}(t,q)$. They are assigned following the rules in (\ref{eq:top-vertices-def}).

\item The final partition function is obtained by doing the above process for each vertex of the brane web. The result for gluings of two trivalent vertices looks like
\begin{align}
    \sum_{\lambda}(-Q)^{|\lambda|}f_{\lambda}^{\eta}C^{(\sigma)}_{\mu_{1}\nu_{1}\lambda}(q,t)C^{(\sigma)}_{\mu_{2}\nu_{2}\lambda^{\rmT}}(t,q)
\end{align}
for example.

\end{enumerate}

\paragraph{Remark} In the context of refined topological vertex (for the non-supergoup case), the framing factors and the integers $\eta$ of an edge are related with the vectors assigned to the four legs connected to the edge. To be concrete, the integer $\eta$ assigned for an edge with vector $\mathbf{v}$ is
\begin{align}
\adjustbox{valign=c}{
    \begin{tikzpicture}[thick]
        \begin{scope}
        \draw[postaction={segment={mid arrow}}] (-1,1)--(0,0);
        \draw[postaction={segment={mid arrow}}] (-1,-0.5)--(0,0);
        \draw [postaction={segment={mid arrow}}](1,0)--(0,0);
        \draw [postaction={segment={mid arrow}}] (1,0)--(1.5,1);
        \draw [postaction={segment={mid arrow}}] (1,0)--(1.8,-0.8);
        \node [above] at (0.5,0){$\mathbf{v}$};
        \node [above right] at (-0.5,0.5){$\mathbf{v}_{1}$};
        \node [below right] at (-0.5,-0.3){$\mathbf{v}_{2}$};
        \node [right] at (1.4,-0.45){$\mathbf{v}_{3}$};
        \node [right] at (1.25,0.5){$\mathbf{v}_{4}$};
        \end{scope}
    \end{tikzpicture}}\qquad \eta=\mathbf{v}_{1}\wedge \mathbf{v}_{3}=\mathbf{v}_{2}\wedge\mathbf{v}_{4}
\end{align}
 where the antisymmetric operation is defined as $\mathbf{v}\wedge \mathbf{w}=v_{1}w_{2}-w_{1}v_{2}$. However, when negative D-branes enter the setup, as one can see in Figure~\ref{fig:frame-factor}, this antisymmetric operation is only true for the case when the preferred direction is glued:
 \begin{align}
 \begin{split}
     \eta=(M,1)\wedge(N,1)=(-M-\sigma,-1)\wedge(-N-\sigma,-1)=M-N,\\
\end{split}
 \end{align}
 For other cases, they do not match with the conventional way:
 \begin{align}
 \begin{split}
     \eta&=(-\sigma,0)\wedge(-N-\sigma-\sigma',-1)=\sigma\neq(-N,-1)\wedge(\sigma',0)=\sigma',\\
     \eta&=(-N-\sigma-\sigma',-1)\wedge(-\sigma',0)=-\sigma'\neq (\sigma,0)\wedge(-N,-1)=-\sigma,\\
     \eta&=(\sigma,0)\wedge(\sigma',0)=0\neq (-N-\sigma,-1)\wedge(-N-\sigma',-1)=\sigma-\sigma',\\
     \eta&=(-\sigma,0)\wedge(-\sigma',0)=0\neq(-N,-1)\wedge(-N-\sigma+\sigma',-1)=\sigma-\sigma'.
\end{split}
\end{align}
and the equality only holds when $\sigma=\sigma'$. One solution for this mismatch may be to redefine the anti-vertex and framing factors so that the conventional way holds for the supergroup case. In this paper, we do not search this direction. Even there is a mismatch, we still can do calculations using the above rules. Moreover, as long as we use the intertwiner formalism, we do not even have to think of framing factors because they are already incorporated in the intertwiners.

\subsection{Correspondence between intertwiners and refined topological vertices}\label{sec:intertwiner-topvertex}
The correspondence in (\ref{eq:brane-rep-corresp-table}) implies that the matrix elements of the intertwiners should correspond with refined topological vertices. It was shown in \cite{Awata:2011ce} that the matrix elements of the positive (dual) intertwiner in Schur basis corresponds to the IKV refined topological vertex. Thus, it is natural for the negative interwiner introduced in this paper to correspond with the anti-refined topological vertex recently introduced by Kimura and Sugimoto \cite{Kimura-Sugimoto:antivertex}. Let us show this correspondence.

After using $q_{1}=q,\,q_{2}=t^{-1}$, the operator part of the negative intertwiner is written as 
\begin{align}
\begin{split}
    :\Phi_{-,\emptyset}[v]\prod_{x\in\lambda}\eta(\chi_{x}^{\sminus})^{-1}:&=\exp\left(\sum_{n=1}^{\infty}\frac{v^{n}}{n}a_{-n}\left\{\frac{1}{1-q^{-n}}+\sum_{x\in\lambda}\left(\frac{t}{q}\right)^{n}q_{x}^{-n}q^{n}(1-t^{-n})\right\}\right)\\
    &\times\exp\left(\sum_{n=1}^{\infty}\frac{v^{-n}}{n}a_{n}\left\{-\frac{1}{1-t^{n}}+\sum_{x\in\lambda}\left(\frac{q}{t}\right)^{n}q_{x}^{n}(1-q^{-n})\right\}\right)
\end{split}
\end{align}
where we shortly wrote $q_{x}=q^{i-1}t^{-j+1},\, x=(i,j)\in\lambda$. Using the following identities 
\begin{align}
\begin{split}
    -\frac{1}{1-t^{n}}+\sum_{x\in\lambda}\left(\frac{q}{t}\right)^{n}q_{x}^{n}(1-q^{-n})&=\sum_{j=1}^{\infty}t^{-nj}q^{n\lambda_{j}^{\rmT}},\\
    \frac{1}{1-q^{-n}}+\sum_{x\in\lambda}\left(\frac{t}{q}\right)^{n}q^{n}(1-t^{-n})q_{x}^{-n}&=\sum_{i=1}^{\infty}q^{-n(i-1)}t^{n\lambda_{i}},
\end{split}
\end{align}
the operator part is rewritten as
\begin{align}
    :\Phi_{-,\emptyset}[v]\prod_{x\in\lambda}\eta(\chi_{x}^{\sminus})^{-1}:&=V_{-}(vq^{\rho+\frac{1}{2}}t^{\lambda})V_{+}(v^{-1}t^{\rho-\frac{1}{2}}q^{\lambda^{\rmT}}),
\end{align}
where we used the vertex operators in (\ref{eq:appendix-Schurvertexop-def}) (see Appendix~\ref{sec:appendix-Schurfunc}). Taking the matrix elements in the Fock basis (\ref{eq:appendix-Fockbasis}) and using the free field realization of the Schur functions in (\ref{eq:appendix-Schurfreefield1}), we have 
\begin{align}
\begin{split}
    \bra{\mu}:\Phi_{-,\emptyset}[v]\prod_{x\in\lambda}\eta(\chi_{x}^{\sminus})^{-1}:\ket{\nu}&=\bra{\mu}V_{-}(vq^{\rho+\frac{1}{2}}t^{\lambda})V_{+}(v^{-1}t^{\rho-\frac{1}{2}}q^{\lambda^{\rmT}})\ket{\nu}\\
    &=(qt^{-\frac{1}{2}}v)^{|\mu|}(q^{\frac{1}{2}}v)^{-|\nu|}\sum_{\eta}\left(\frac{t}{q}\right)^{\frac{1}{2}(|\eta|+|\mu|-|\nu|)}s_{\mu/\eta}(q^{\rho}t^{\lambda})s_{\nu/\eta}(t^{\rho}q^{\lambda^{\rmT}})\\
\end{split}
\end{align}
where we used the property of the Schur functions in (\ref{eq:appendix-Schurproperty}). The zero mode part is rewritten as 
\begin{align}
\begin{split}
    t_{n,-}(\lambda,u,v)&=(u^{-1}\gamma)^{|\lambda|}(-v)^{n|\lambda|}f_{\lambda}(q,t)^{-n+1}
\end{split}
\end{align}
using the framing factor. Then, we obtain the correspondence
\begin{align}
    \begin{split}
        \bra{\mu}\Phi_{-,\lambda}^{(n)}[u,v]\ket{\nu}&=(u^{-1}\gamma)^{|\lambda|}(-v)^{n|\lambda|}(qt^{-\frac{1}{2}}v)^{|\mu|}(-q^{\frac{1}{2}}v)^{-|\nu|}\frac{f_{\nu}(q,t)}{f_{\lambda}(q,t)^{n-1}}t^{\frac{1}{2}||\lambda||^{2}}\wt{Z}_{\lambda}(q^{-1},t^{-1})^{-1}\bar{C}_{\mu^{\rmT}\nu\lambda}(q,t).
    \end{split}
\end{align}

We can do a similar computation for the dual negative intertwiner. The operator part is written as 
\begin{align}
\begin{split}
    :\Phi^{\ast}_{-,\emptyset}[v]\prod_{x\in\lambda}\xi(\chi_{x}^{\sminus})^{-1}:
    &=\wt{V}_{-}(-\gamma vq^{\rho+\frac{1}{2}}t^{\lambda})\wt{V}_{+}(-v^{-1}\gamma q^{\lambda^{\rmT}}t^{\rho-\frac{1}{2}}),
\end{split}
\end{align}
where the operators are defined in (\ref{eq:appendix-Schurvertexop-def}). After taking the matrix element with the Fock basis, using (\ref{eq:appendix-Schurfreefield2}), and then rewriting the zero-mode part, we eventually have
\begin{align}
    \bra{\nu}\Phi^{(n)*}_{-,\lambda}[u,v]\ket{\mu}&=u^{|\lambda|}(-v)^{-n|\lambda|}(-qt^{-\frac{1}{2}}v)^{-|\mu|}(q^{\frac{1}{2}}v)^{|\nu|}\frac{f_{\lambda}(q,t)^{n}}{f_{\nu}(q,t)}q^{\frac{1}{2}||\lambda^{\rmT}||^{2}}\wt{Z}_{\lambda^{\rmT}}(t^{-1},q^{-1})^{-1}\bar{C}_{\mu\nu^{\rmT}\lambda^{\rmT}}(t,q).
\end{align}

Rewriting the positive intertwiners in our notation, we can write the correspondence in a unified way: 
\begin{align}
\begin{split}
    \bra{\mu}\Phi^{(n)}_{\sigma,\lambda}[u,v]\ket{\nu}&=\left(\frac{u^{\sigma}\gamma^{\frac{1-\sigma}{2}}}{(-v)^{n\sigma}}\right)^{|\lambda|}(qt^{-\frac{1}{2}}v)^{|\mu|}(-q^{\frac{1}{2}}v)^{-|\nu|}\frac{f_{\nu}(q,t)}{f_{\lambda}(q,t)^{n+\sigma}}t^{-\frac{\sigma}{2}||\lambda||^{2}}\wt{Z}^{-1}_{\lambda}(q^{\sigma},t^{\sigma})C^{(\sigma)}_{\mu^{\rmT}\nu\lambda}(q,t),\\
    \bra{\nu}\Phi^{(n)*}_{\sigma,\lambda}[u,v]\ket{\mu}&=\left(\frac{(-v)^{n\sigma}}{u^{\sigma}\gamma^{\frac{1+\sigma}{2}}}\right)^{|\lambda|}(-qt^{-\frac{1}{2}}v)^{-|\mu|}(q^{\frac{1}{2}}v)^{|\nu|}\frac{f_{\lambda}(q,t)^{n}}{f_{\nu}(q,t)}q^{-\frac{\sigma}{2}||\lambda^{\rmT}||^{2}}\wt{Z}^{-1}_{\lambda^{\rmT}}(t^{\sigma},q^{\sigma})C^{(\sigma)}_{\mu\nu^{\rmT}\lambda^{\rmT}}(t,q).
\end{split}
\end{align}
Therefore, the interwiners are essentially equivalent to the IKV refined topological vertex and the Kimura-Sugimoto anti-vertex. 

Using diagrams, we have the following identifications
\begin{align}
\begin{split}
    &\begin{tikzpicture}[thick,scale=1.2]
        \begin{scope}[]
		\node [right] at (0,0) {$\dket{v,\lambda}$};
		\node [below] at (-1,-1) {$\uwidehat{\,\overline{\mu}\,}$};
		\node [above] at (-1,1){$\widehat{\,\underline{\nu}\,}$};
		\node[above left,scale=0.9] at (-1,-0.1) {$\Phi_{\sigma}^{(n)}[u, v]$};
		\draw[postaction={segment={mid arrow}}] (0,0) -- (-1,0) -- (-1,-1);
		\draw[postaction={segment={mid arrow}}] (-1,1) -- (-1,0);
		\end{scope}
		\begin{scope}
		\node at (2.5,0){$\Longleftrightarrow$};
		\end{scope}
		\begin{scope}[shift={(6.7,0)},scale=1.2]
		\node [below right] at (-1,1){$t,\nu$};
		\node [below right] at (-1.35,-0.35){$q,\mu$};
		\node [above] at (-0.2,0.05) {$\lambda$};
		\node [right] at (0,0){$(\sigma,0)$};
		\node [above] at (-1,1){$(n,1)$};
		\node [below left] at (-1.7,-0.7){$(n+\sigma,1)$};
		\node [above left] at (-1.1,0){$C^{(\sigma)}_{\mu^{\rmT}\nu\lambda}(q,t)$};
		\draw[postaction={segment={mid arrow}}] (-1.7,-0.7)--(-1,0)--(0,0);
		\draw[postaction={segment={mid arrow}}] (-1,0) --(-1,1);
		\draw[thick,red] (-0.3,0.1)--(-0.3,-0.1);
		\draw[thick,red] (-0.2,0.1)--(-0.2,-0.1);
		\end{scope}
    \end{tikzpicture}\\
    &\begin{tikzpicture}[thick,scale=1.2]
            \begin{scope}[]
                \node[left] at (-1,0) {$\dbra{v,\lambda}$};
                \node [above] at (0,1) {$\widehat{\,\underline{\mu}\,}$};
                \node [below] at (0,-1) {$\uwidehat{\,\overline{\nu}\,}$};
                \node[right] at (0,0) {$\Phi^{(n)\ast}_{\sigma}[u,v]$};
                \draw[postaction={segment={mid arrow}}] (0,0) -- (-1,0);
		        \draw[postaction={segment={mid arrow}}] (0,1) -- (0,0)--(0,-1);   
            \end{scope}
        \begin{scope}
		\node at (2.8,0){$\Longleftrightarrow$};
		\end{scope}
		\begin{scope}[shift={(4.7,0)},scale=1.2]
		\node [above left] at (1,-1){$q,\nu$};
		\node[below]at (1,-1){$(n,1)$};
		\node [above left] at (1.35,0.35){$t,\mu$};
		\node [above] at (0.3,0.1) {$\lambda$};
		\node[left] at (0,0){$(\sigma,0)$};
		\node[above right] at (1.7,0.7){$(n+\sigma,1)$};
		\node [below right] at (1.1,0){$C^{(\sigma)}_{\mu\nu^{\rmT}\lambda^{\rmT}}(t,q)$};
		\draw[postaction={segment={mid arrow}}] (0,0)--(1,0)--(1.7,0.7);
		\draw[postaction={segment={mid arrow}}] (1,-1)--(1,0);
		\draw[thick,red] (0.3,-0.1)--(0.3,0.1);
		\draw[thick,red] (0.2,-0.1)--(0.2,0.1);
		\end{scope}
		\end{tikzpicture}
 \end{split} 
\end{align}
The procedure to obtain the refined topological vertices from the intertwiner is as follows:
\begin{enumerate}
\item We first reverse all the arrows of the interwiners. 
\item Each edge of the intertwiner is assigned a module with levels such as $(0,\sigma),(1,n),(1,n+\sigma)$. The vectors associated with the edges are obtained by switching the components of these levels such as $(\sigma,0),(n,1),(n+\sigma,1)$, respectively. This corresponds to (\ref{eq:brane-rep-corresp-table}).
\item Following the pictorial description in (\ref{eq:top-vertices-def}), we obtain the refined topological vertices. Note that when the arrows are going into the vertex, we use the transpose of the attached Young diagram.
\end{enumerate}

\subsection{Unrefined limit}\label{sec:unrefinedlimit}
Actually the positive and negative intertwiners are directly related with each other in the unrefined limit. A similar discussion for the topological vertices was done in \cite{Kimura-Sugimoto:antivertex}. Let us see what will happen for the intertwiner case. 

We set $q_{1}=q,\,q_{2}=q^{-1},\,q_{3}=1$ in the unrefined limit. We introduce two vertex operators 
\begin{align}
\begin{split}
    \Upsilon[v]&=\exp\left(-\sum_{n=1}^{\infty}\frac{v^{n}}{n}\frac{a_{-n}}{1-q^{-n}}\right)\exp\left(\sum_{n=1}^{\infty}\frac{v^{-n}}{n}\frac{a_{n}}{1-q^{n}}\right),\\
    \phi(z)&=\exp\left(\sum_{n=1}^{\infty}\frac{z^{n}}{n}(1-q^{n})a_{-n}\right)\exp\left(-\sum_{n=1}^{\infty}\frac{z^{-n}}{n}(1-q^{-n})a_{n}\right).
\end{split}
\end{align}
In the unrefined limit, the vertex operators $\eta(z),\xi(z),\Phi_{\emptyset}[v],\Phi_{\emptyset}^{\ast}[v]$ will transform as
\begin{align}
\eta(z)\longrightarrow\phi(z),\quad \xi(z)\longrightarrow :\phi(z)^{-1}:,\quad\Phi_{\emptyset}[v]\longrightarrow \Upsilon[v],\quad \Phi^{\ast}_{\emptyset}[v]\longrightarrow :\Upsilon[v]^{-1}:,\quad q_{3}\rightarrow 1.
\end{align}
The box contents of the Young diagrams $\chi_{x}^{\spm}$ have a nice property in the unrefined limit. For $x=(i,j)\in\lambda$,
\begin{align}
    \chi_{x}^{\splus}\longrightarrow \Tilde{\chi}_{x}^{\splus}\coloneqq vq^{-c_{\lambda}(x)}=vq^{i-j},\quad
    \chi_{x}^{\sminus}\longrightarrow \Tilde{\chi}_{x}^{\sminus}\coloneqq vq^{c_{\lambda}(x)}=vq^{j-i},
\end{align}
where we used $c_{\lambda}(i,j)=j-i$ (see Appendix~\ref{sec:appendix-Young}). Using the property $c_{\lambda^{\rmT}}(x)=-c_{\lambda}(x)$, we have  
\begin{align}
   \prod_{x\in\lambda^{\rmT}}f(\Tilde{\chi}_{x}^{\splus})=\prod_{x\in\lambda}f(\Tilde{\chi}_{x}^{\sminus})
\end{align}
for any function $f(z)$. In the unrefined limit, the vertex operator part of the intertwiners transform as 
\begin{align}
\begin{split}
    :\Phi_{\pm,\emptyset}[v]\prod_{x\in\lambda}\eta(\chi_{x}^{\spm})^{\pm1}:&\longrightarrow :\Upsilon[v]^{\pm1}\prod_{x\in\lambda}\phi(\Tilde{\chi}_{x}^{\spm})^{\pm1}:,\\
    :\Phi^{\ast}_{\pm,\emptyset}[v]\prod_{x\in\lambda}\xi(\chi_{x}^{\spm})^{\pm1}:&\longrightarrow :\Upsilon[v]^{\mp1}\prod_{x\in\lambda}\phi(\Tilde{\chi}_{x}^{\spm})^{\mp1}:.
\end{split}
\end{align}
\begin{figure}[t]
\centering
    \begin{tikzpicture}[thick,scale=1.1]
        \begin{scope}[]
		\node [right] at (0,0) {$\dket{v,\lambda}$};
		\node [below,scale=0.8] at (-1,-1) {$(1,n+1)_{-uv}$};
		\node [above,scale=0.8] at (-1,1){$(1,n)_{u}$};
		\node[left,scale=0.9] at (-1,0) {$\left.\Phi_{+}^{(n)}[u, v]\right|_{\text{unref}}$};
		\draw[postaction={segment={mid arrow}}] (-1,0) -- (-1,-1);
		\draw[postaction={segment={mid arrow}}] (0,0) -- (-1,0);
		\node [above,scale=0.8] at (-0.5,0){$(0,1)_{v}$};
		\draw[postaction={segment={mid arrow}}] (-1,1) -- (-1,0);
		\end{scope}
            \begin{scope}[shift={(4,0)}]
            \node [left] at (-2.35,0){$=$};
                \node[left] at (-1,0) {$\dbra{v,\lambda^{\rmT}}$};
                \node [above,scale=0.8] at (0,1) {$(1,n)_{u}$};
                \node [below,scale=0.8] at (0,-1) {$(1,n+1)_{-uv}$};
                \node[right] at (0,0) {$\left.\Phi^{(n+1)\ast}_{-}[-uv,v]\right|_{\text{unref}}$};
                \draw[dashed,postaction={segment={mid arrow}}] (0,0) -- (-1,0);
		       \node [above, scale=0.8] at (-0.5,0){$(0,-1)_{v}$}; \draw[postaction={segment={mid arrow}}] (0,1) -- (0,0)--(0,-1);   
            \end{scope}
        \begin{scope}[shift={(0,-3.5)}]
		\node [right] at (0,0) {$\dket{v,\lambda}$};
		\node [below,scale=0.8] at (-1,-1) {$(1,n)_{u}$};
		\node [above,scale=0.8] at (-1,1){$(1,n+1)_{-uv}$};
		\node[left,scale=0.9] at (-1,0) {$\left.\Phi_{-}^{(n+1)}[-uv, v]\right|_{\text{unref}}$};
		\draw[postaction={segment={mid arrow}}] (-1,0) -- (-1,-1);
		\draw[dashed,postaction={segment={mid arrow}}] (0,0) -- (-1,0) ;
		\node [above,scale=0.8] at (-0.5,0){$(0,-1)_{v}$};
		\draw[postaction={segment={mid arrow}}] (-1,1) -- (-1,0);
		\end{scope}
            \begin{scope}[shift={(4,-3.5)}]
            \node [left] at (-2.35,0){$=$};
                \node[left] at (-1,0) {$\dbra{v,\lambda^{\rmT}}$};
                \node [above,scale=0.8] at (0,1) {$(1,n+1)_{-uv}$};
                \node [below,scale=0.8] at (0,-1) {$(1,n)_{u}$};
                \node[right] at (0,0) {$\left.\Phi^{(n)\ast}_{+}[u,v]\right|_{\text{unref}}$};
                \draw[postaction={segment={mid arrow}}] (0,0) -- (-1,0);
		       \node [above, scale=0.8] at (-0.5,0){$(0,1)_{v}$}; \draw[postaction={segment={mid arrow}}] (0,1) -- (0,0)--(0,-1);  
            \end{scope}
	\end{tikzpicture}
\caption{Identities of intertwiners in the unrefined limit. Negative branes on one side of the NS5-brane are identical to positive branes on the other side of the NS5-brane in the unrefined limit.}
\label{fig:unrefined-limit}
\end{figure}
The zero modes part transform as 
\begin{align}
\begin{split}
    t_{n,\sigma}(\lambda,u,v)&\longrightarrow (-u^{\sigma}v)^{|\lambda|}\prod_{x\in\lambda}\left(\Tilde{\chi}_{x}^{(\sigma)}\right)^{(-1-n\sigma)},\\
    t^{\ast}_{n,\sigma}(\lambda,u,v)&\longrightarrow u^{-\sigma|\lambda|}\prod_{x\in\lambda}\left(\Tilde{\chi}_{x}^{\sigma}\right)^{\sigma n}.
\end{split}
\end{align}
Using the above limits and properties, the intertwiners obey the following identities in the unrefined limit (see Figure \ref{fig:unrefined-limit}):
\begin{align}
    \begin{split}
        \left.\Phi^{(n)}_{+,\lambda}[u,v]\right|_{\text{unref}}&=\left.\Phi^{(n+1)\ast}_{-,\lambda^{\rmT}}[-uv,v]\right|_{\text{unref}},\\
        \left.\Phi^{(n)\ast}_{+,\lambda^{\rmT}}[u,v]\right|_{\text{unref}}&=\left.\Phi^{(n+1)}_{-,\lambda}[-uv,v]\right|_{\text{unref}}.
    \end{split}\label{eq:unrefined-identity}
\end{align}
To convert to the topological vertices, we take the matrix elements in the Fock representation as
\begin{align}
    \begin{split}
        \bra{\mu^{\rmT}}\left.\Phi^{(n)}_{+,\lambda}[u,v]\right|_{\text{unref}}\ket{\nu}&=\bra{\mu^{\rmT}}\left.\Phi^{(n+1)\ast}_{-,\lambda^{\rmT}}[-uv,v]\right|_{\text{unref}}\ket{\nu},\\
        \bra{\nu^{\rmT}}\left.\Phi^{(n)\ast}_{+,\lambda^{\rmT}}[u,v]\right|_{\text{unref}}\ket{\mu}&=\bra{\nu^{\rmT}}\left.\Phi^{(n+1)}_{-,\lambda}[-uv,v]\right|_{\text{unref}}\ket{\mu}.
    \end{split}
\end{align}
Eventually, we have the relation
\begin{align}
    C_{\mu\nu\lambda}(q,q)=f_{\mu}f_{\nu}^{-1}f_{\lambda}^{-1}\bar{C}_{\nu\mu\lambda}(q,q),\quad f_{\mu}\coloneqq f_{\mu}(q,q)
\end{align}
which matches with the result in \cite{Kimura-Sugimoto:antivertex}.

The identities in (\ref{eq:unrefined-identity}) mean that in the unrefined limit, negative branes extending on the right (resp. left) side of the NS5-brane can be pulled to the left (resp. right) side of the NS5-brane and converted to the positive branes.

\subsection{Gluings of intertwiners and topological vertices}\label{sec:gluingtopvertex}
We managed to show that the matrix elements of the intertwiners correspond with the topological vertices in the previous section. Let us check if the gluing rules are compatible with the ones proposed in section~\ref{sec:topvertexrule}.

There are five ways to glue the intertwiners as shown in section~\ref{sec:intertwiner-contractions}. Taking the matrix element with the Fock basis in the vertex operator representation (see section~\ref{sec:positivevertexop} and Appendix~\ref{sec:appendix-Schurfunc}) and using the correspondence with topological vertices explained in the previous subsection, we can translate the gluing rules of the intertwiners to the gluing rules of the topological vertices. The results are summarized as follows:
\begin{itemize}
\item Gluing of D5-branes (situation (\ref{eq:glue-intertwiner-5})):
    \begin{align}
\begin{split}
\begin{array}{ccc}
\vspace{1mm}\widehat{\,\underline{\nu_{1}}\,}&&\vspace{1mm}\widehat{\,\underline{\mu_{2}}\,}\\
\Phi_{\sigma}^{(N)}[u_{1},v]&\cdot&\Phi_{\sigma}^{(M)*}[u_{2},v]\\
\uwidehat{\overline{\,\mu_{1}}\,}&&\uwidehat{\overline{\,\nu_{2}}\,}\\
\end{array}
&=\sum_{\lambda}a_{\lambda}^{(\sigma)}
\begin{array}{cc}
\vspace{1mm}\widehat{\,\underline{\nu_{1}}\,}&\vspace{1mm}\widehat{\,\underline{\mu_{2}}\,}\\
\Phi_{\sigma,\lambda}^{(N)}[u_{1},v]&\Phi_{\sigma,\lambda}^{(M)*}[u_{2},v]\\
\uwidehat{\overline{\,\mu_{1}}\,}&\uwidehat{\overline{\,\nu_{2}}\,}\\
\end{array}\\
&=(-1)^{|\mu_{2}|+|\nu_{1}|}(qt^{-\frac{1}{2}}v)^{|\mu_{1}|-|\mu_{2}|}(q^{\frac{1}{2}}v)^{|\nu_{2}|-|\nu_{1}|}\frac{f_{\nu_{1}}(q,t)}{f_{\nu_{2}}(q,t)}\\
&\times\sum_{\lambda}\left((-v)^{M-N}\frac{u_{1}}{u_{2}}\right)^{\sigma|\lambda|}f_{\lambda}(q,t)^{M-N}C^{(\sigma)}_{\mu_{1}^{\rmT}\nu_{1}\lambda}(q,t)C^{(\sigma)}_{\mu_{2}\nu_{2}^{\rmT}\lambda^{\rmT}}(t,q)
\end{split}\label{eq:glue-topver-5}
    \end{align}

\item  Gluing of NS5-branes (situation (\ref{eq:glue-intertwiner-1})):
\begin{align}
    \begin{split}
        \begin{array}{cc}
            \vspace{1mm}\widehat{\,\underline{\nu_{1}}\,}&\\
            \Phi_{\sigma}^{(N)}[u,v_{1}]&\dket{\lambda_{1}}\\
            \Phi_{\sigma'}^{(N+\sigma)}[-uv_{1}^{\sigma},v_2]&\dket{\lambda_{2}}\\
            \uwidehat{\,\overline{\nu_{2}}\,}&\\
        \end{array}
        &=\sum_{\mu}\begin{array}{c}
            \vspace{1mm}\widehat{\,\underline{\nu_{1}}\,}\\
            \Phi_{\sigma,\lambda_{1}}^{(N)}[u,v_{1}]\\
             \uwidehat{\overline{\,\mu\,}}\\
             \widehat{\,\underline{\mu}\,}\\
            \Phi_{\sigma',\lambda_{2}}^{(N+\sigma)}[-uv_{1}^{\sigma},v_2]\\
            \uwidehat{\,\overline{\nu_{2}}\,}
        \end{array}\\
        &=\left(\frac{u^{\sigma}\gamma^{\frac{1-\sigma}{2}}}{(-v_{1})^{N\sigma}}\right)^{|\lambda_{1}|}\left(-\frac{u^{\sigma'}v_{1}^{\sigma\sigma'}\gamma^{\frac{1-\sigma'}{2}}}{(-v_{2})^{(N+\sigma)\sigma'}}\right)^{|\lambda_{2}|}(-q^{\frac{1}{2}}v_{1})^{-|\nu_{1}|}(qt^{-\frac{1}{2}}v_{2})^{|\nu_{2}|}\\
        &\times \frac{f_{\nu_{1}}(q,t)}{f_{\lambda_{1}}(q,t)^{N+\sigma}f_{\lambda_{2}}(q,t)^{N+\sigma+\sigma'}}\frac{t^{-\frac{\sigma||\lambda_{1}||^{2}}{2}-\frac{\sigma'||\lambda_{2}||^{2}}{2}}}{\wt{Z}_{\lambda_{1}}(q^{\sigma},t^{\sigma})\wt{Z}_{\lambda_{2}}(q^{\sigma'},t^{\sigma'})}\\
        &\times \sum_{\mu}\left(-\frac{v_{1}}{v_{2}}\right)^{|\mu|}\left(\frac{q}{t}\right)^{\frac{|\mu|}{2}}f_{\mu}(q,t)C^{(\sigma)}_{\mu^{\rmT}\nu_{1}\lambda_{1}}(q,t)C^{(\sigma')}_{\nu_{2}^{\rmT}\mu\lambda_{2}}(q,t)
    \end{split}\label{eq:glue-topver-1}
\end{align}
\item  Gluing of NS5-branes (situation (\ref{eq:glue-intertwiner-2})):
\begin{align}
    \begin{split}
        \begin{array}{cc}
            &\vspace{1mm}\widehat{\,\underline{\mu_{1}}\,}\\
            \dbra{\lambda_{1}}&\Phi_{\sigma}^{(N+\sigma')*}[-uv_{2}^{\sigma'},v_{1}]\\
            \dbra{\lambda_{2}}&\Phi_{\sigma'}^{(N)*}[u,v_2]\\
            &\uwidehat{\,\overline{\mu_{2}}\,}\\
        \end{array}
        &=\sum_{\nu}\begin{array}{c}
            \vspace{1mm}\widehat{\,\underline{\mu_{1}}\,}\\
            \Phi_{\sigma,\lambda_{1}}^{(N+\sigma')*}[-uv_{2}^{\sigma'},v_{1}]\\
             \uwidehat{\overline{\,\nu\,}}\\
             \widehat{\,\underline{\nu}\,}\\
            \Phi_{\sigma',\lambda_{2}}^{(N)*}[u,v_2]\\
            \uwidehat{\,\overline{\mu_{2}}\,}
        \end{array}\\
        &=\left(-\frac{(-v_{1})^{(N+\sigma')\sigma}}{u^{\sigma}v_{2}^{\sigma\sigma'}\gamma^{\frac{1+\sigma}{2}}}\right)^{|\lambda_{1}|}\left(-\frac{(-v_{2})^{N\sigma'}}{u^{\sigma'}\gamma^{\frac{1+\sigma'}{2}}}\right)^{|\lambda_{2}|}(-qt^{-\frac{1}{2}}v_{1})^{-|\mu_{1}|}(q^{\frac{1}{2}}v_{2})^{|\mu_{2}|}\\
        &\times \frac{f_{\lambda_{1}}(q,t)^{N+\sigma'}f_{\lambda_{2}}(q,t)^{N}}{f_{\mu_{2}}(q,t)}\frac{q^{-\frac{\sigma||\lambda^{\rmT}_{1}||^{2}}{2}-\frac{\sigma'||\lambda^{\rmT}_{2}||^{2}}{2}}}{\wt{Z}_{\lambda^{\rmT}_{1}}(t^{\sigma},q^{\sigma})\wt{Z}_{\lambda_{2}^{\rmT}}(t^{\sigma'},q^{\sigma'})}\\
        &\times \sum_{\nu}\left(-\frac{v_{1}}{v_{2}}\right)^{|\nu|}\left(\frac{t}{q}\right)^{\frac{|\nu|}{2}}f_{\nu}(q,t)^{-1}C^{(\sigma)}_{\mu_{1}\nu^{\rmT}\lambda_{1}^{\rmT}}(t,q)C^{(\sigma')}_{\nu\mu_{2}^{\rmT}\lambda_{2}^{\rmT}}(t,q)
    \end{split}\label{eq:glue-topver-2}
\end{align}
\item Gluing of NS5-branes (situation (\ref{eq:glue-intertwiner-3})):
\begin{align}
    \begin{split}
\begin{array}{ccc}
            &\vspace{1mm}\widehat{\,\underline{\nu_{1}}\,}&\\
            &\Phi_{\sigma}^{(N)}[u,v_{1}]&\dket{\lambda_{1}}\\
            \dbra{\lambda_{2}}&\Phi_{\sigma'}^{(N+\sigma-\sigma')*}[uv_{1}^{\sigma}v_{2}^{-\sigma'},v_2]&\\
            &\uwidehat{\overline{\,\nu_{2}\,}}&\\
        \end{array}
        &=\sum_{\mu}\begin{array}{c}
            \vspace{1mm}\widehat{\,\underline{\nu_{1}}\,}\\
            \Phi_{\sigma,\lambda_{1}}^{(N)}[u,v_{1}]\\
             \uwidehat{\overline{\,\mu\,}}\\
             \widehat{\,\underline{\mu}\,}\\
            \Phi_{\sigma',\lambda_{2}}^{(N+\sigma-\sigma')*}[uv_{1}^{\sigma}v_{2}^{-\sigma'},v_2]\\
            \uwidehat{\overline{\,\nu_{2}\,}}
        \end{array}\\
        &=\left(\frac{u^{\sigma}\gamma^{\frac{1-\sigma}{2}}}{(-v_{1})^{N\sigma}}\right)^{|\lambda_{1}|}\left(-\frac{(-v_{2})^{(N+\sigma)\sigma'}}{u^{\sigma'}v_{1}^{\sigma\sigma'}\gamma^{\frac{1+\sigma'}{2}}}\right)^{|\lambda_{2}|}(-q^{\frac{1}{2}}v_{1})^{-|\nu_{1}|}(q^{\frac{1}{2}}v_{2})^{|\nu_{2}|}\\
        &\times \frac{f_{\nu_{1}}(q,t)f_{\lambda_{2}}(q,t)^{N+\sigma-\sigma'}}{f_{\nu_{2}}(q,t)f_{\lambda_{1}}(q,t)^{N+\sigma}}\frac{t^{-\frac{\sigma||\lambda_{1}||^{2}}{2}}q^{-\frac{\sigma'||\lambda_{2}^{\rmT}||^{2}}{2}}}{\wt{Z}_{\lambda_{1}}(q^{\sigma},t^{\sigma})\wt{Z}_{\lambda_{2}^{\rmT}}(t^{\sigma'},q^{\sigma'})}\\
        &\times \sum_{\mu}\left(-\frac{v_{1}}{v_{2}}\right)^{|\mu|}C^{(\sigma)}_{\mu^{\rmT}\nu_{1}\lambda_{1}}(q,t)C^{(\sigma')}_{\mu\nu_{2}^{\rmT}\lambda_{2}^{\rmT}}(t,q)
    \end{split}\label{eq:glue-topver-3}
\end{align}
\item Gluing of NS5-branes (situation (\ref{eq:glue-intertwiner-4})):
\begin{align}
    \begin{split}
        \begin{array}{ccc}
            &\vspace{1mm}\widehat{\,\underline{\mu_{1}}\,}&\\
            \dbra{\lambda_{1}}&\Phi_{\sigma}^{(N)*}[u,v_{1}]&\\
            &\Phi_{\sigma'}^{(N)}[u,v_2]&\dket{\lambda_{2}}\\
            &\uwidehat{\overline{\,\mu_{2}\,}}&\\
        \end{array}
        &=\sum_{\nu}\begin{array}{c}
            \vspace{1mm}\widehat{\,\underline{\mu_{1}}\,}\\
            \Phi_{\sigma,\lambda_{1}}^{(N)*}[u,v_{1}]\\
             \uwidehat{\overline{\,\nu\,}}\\
             \widehat{\,\underline{\nu}\,}\\
            \Phi_{\sigma',\lambda_{2}}^{(N)}[u,v_2]\\
            \uwidehat{\overline{\,\mu_{2}\,}}
        \end{array}\\
        &=\left(\frac{(-v_{1})^{N\sigma}}{u^{\sigma}\gamma^{\frac{1+\sigma}{2}}}\right)^{|\lambda_{1}|}\left(\frac{u^{\sigma'}\gamma^{\frac{1-\sigma'}{2}}}{(-v_{2})^{N\sigma'}}\right)^{|\lambda_{2}|}(-qt^{-\frac{1}{2}}v_{1})^{-|\mu_{1}|}(qt^{-\frac{1}{2}}v_{2})^{-|\mu_{2}|}\\
        &\times \frac{f_{\lambda_{1}}(q,t)^{N}}{f_{\lambda_{2}}(q,t)^{N+\sigma'}}\frac{q^{-\frac{\sigma||\lambda_{1}^{\rmT}||^{2}}{2}}t^{-\frac{\sigma'||\lambda_{2}||^{2}}{2}}}{\wt{Z}_{\lambda_{1}^{\rmT}}(t^{\sigma},q^{\sigma})\wt{Z}_{\lambda_{2}}(q^{\sigma'},t^{\sigma'})}\\
        &\times\sum_{\nu}\left(-\frac{v_{1}}{v_{2}}\right)^{|\nu|}C^{(\sigma)}_{\mu_{1}\nu^{\rmT}\lambda_{1}^{\rmT}}(t,q)C^{(\sigma')}_{\mu_{2}^{\rmT}\nu\lambda_{2}}(q,t)
    \end{split}\label{eq:glue-topver-4}
\end{align} 

\end{itemize}

All of the formulas indeed match with the gluing rules of the topological vertices in Figure \ref{fig:frame-factor}, and thus, we conclude that the procedure we propose in section~\ref{sec:topvertexrule} is correct.
\clearpage
\section{Superquiver theory}\label{sec:superquiver}
In this section, we discuss how superquiver theories should appear in our formalism if they exist. This section is rather a sketch of how to construct these type of theories. A detailed analysis is left for future work. A different construction of superquiver theories for three-dimensional supersymmetric gauge theories and their relations with Bethe/Gauge correspondence and supergroup gauge theories were studied in \cite{Nekrasov:2018gne,Ishtiaque:2021jan,Orlando:2010uu,Zenkevich:2018fzl,Kimura-Nieri:defects}. Studying the relation with these papers is also left to be done. 

Up to the previous section, we managed to derive the correspondence between the negative intertwiners and negative branes of the supergroup gauge theory. The essential point of the negative intertwiner was that it intertwines modules $(0,-1)_{v}\otimes (1,n)_{u}$ and $(1,n-1)_{u'}$. The new part was that representations with negative levels of central charges appear. From the S-dual view point, one would like to consider other intertwiners 
\begin{align}
    &(0,1)_{v}\otimes (-1,n)_{u}\rightarrow (-1,n+1)_{u'},\\
    &(0,-1)_{v}\otimes (-1,n)_{u}\rightarrow (-1,n-1)_{u'},
\end{align}
where vertex operator representations with negative central charges appear. We can illustrate these intertwiners and the dual of them as 
\begin{align}
\begin{split}
\adjustbox{valign=c}{
        \begin{tikzpicture}
            \begin{scope}[scale=1.3]
                \node[above, scale=0.8] at (0,0) {$(0,1)_{v}$};
                \node[left, scale=0.8] at (-1,0.6) {$(-1,n)_{u}$};
                \node[left,scale=0.8] at (-1,-0.6) {$(-1,n+1)_{u'}$};
                \draw[postaction={segment={mid arrow}}] (0,0) -- (-1,0);
		        \draw[dashed,postaction={segment={mid arrow}}] (-1,1) -- (-1,0)--(-1,-1);   
            \end{scope}
        \end{tikzpicture}}
        \hspace{0.2cm}
        \adjustbox{valign=c}{
        \begin{tikzpicture}
            \begin{scope}[scale=1.3]
                \node[above, scale=0.8] at (0,0) {$(0,-1)_{v}$};
                \node[left, scale=0.8] at (-1,0.6) {$(-1,n)_{u}$};
                \node[left,scale=0.8] at (-1,-0.6) {$(-1,n-1)_{u'}$};
                \draw[dashed,postaction={segment={mid arrow}}] (0,0) -- (-1,0);
		        \draw[dashed,postaction={segment={mid arrow}}] (-1,1) -- (-1,0)--(-1,-1);   
            \end{scope}
        \end{tikzpicture}}\hspace{0.2cm}
        \adjustbox{valign=c}{
        \begin{tikzpicture}
            \begin{scope}[scale=1.3]
                \node[above, scale=0.8] at (0,0) {$(0,1)_{v}$};
                \node[right, scale=0.8] at (1,0.6) {$(-1,n+1)_{u'}$};
                \node[left,scale=0.8] at (1,-0.6) {$(-1,n)_{u}$};
                \draw[postaction={segment={mid arrow}}] (1,0) -- (0,0);
		        \draw[dashed,postaction={segment={mid arrow}}] (1,1) -- (1,0)--(1,-1);   
            \end{scope}
        \end{tikzpicture}}\hspace{0.2cm}
        \adjustbox{valign=c}{
         \begin{tikzpicture}
            \begin{scope}[scale=1.3]
                \node[above, scale=0.8] at (0,0) {$(0,-1)_{v}$};
                \node[right, scale=0.8] at (1,0.6) {$(-1,n-1)_{u'}$};
                \node[left,scale=0.8] at (1,-0.6) {$(-1,n)_{u}$};
                \draw[dashed,postaction={segment={mid arrow}}] (1,0) -- (0,0);
		        \draw[dashed,postaction={segment={mid arrow}}] (1,1) -- (1,0)--(1,-1);   
            \end{scope}
        \end{tikzpicture}}
\end{split}\label{eq:superquiverintertwiner}
\end{align}
Combination of these intertwiners with the positive and negative intertwiners leads to the following brane web:
\begin{align}
\adjustbox{valign=c}{
    \begin{tikzpicture}
            \begin{scope}[scale=2.3]
            \draw[] (-1,0) -- (-1.7,0);
                \draw[dashed,] (-1,0.4) -- (-1.7,0.4);
                \draw[] (-1,0.6) -- (-1.7,0.6);
                \draw[] (-1,0.2) -- (-1.7,0.2);
                \draw[dashed,] (-1,-0.2) -- (-1.7,-0.2);
                \draw[dashed,] (-1,-0.4) -- (-1.7,-0.4);
		        \draw[] (-1,-0.6) -- (-1.7,-0.6);
		        \draw (-1,0.8)--(-1,0)--(-1,-0.9);
            \draw[] (0,-0.1) -- (-1,-0.1);
                \draw[dashed] (0,0.3) -- (-1,0.3);
                \draw[dashed] (0,0.1) -- (-1,0.1);
                \draw[] (0,0.5) -- (-1,0.5);
                \draw[dashed] (0,-0.3) -- (-1,-0.3);
                \draw[] (0,-0.5) -- (-1,-0.5);
		        \draw[] (0,-0.7) -- (-1,-0.7);
                \draw[] (1,0) -- (0,0);
                \draw[] (1,0.4) -- (0,0.4);
                \draw[dashed] (1,0.6) -- (0,0.6);
                \draw[dashed] (1,0.2) -- (0,0.2);
                \draw[] (1,-0.2) -- (0,-0.2);
                \draw[dashed] (1,-0.4) -- (0,-0.4);
		        \draw[] (1,-0.6) -- (0,-0.6);
		        \draw (1,0.8)--(1,0)--(1,-0.9);
		        \draw [dashed] (0,0.8) -- (0,0)--(0,-0.9);
		        \draw[] (2,-0.1) -- (1,-0.1);
                \draw[dashed] (2,0.3) -- (1,0.3);
                \draw[] (2,0.1) -- (1,0.1);
                \draw[] (2,0.5) -- (1,0.5);
                \draw[dashed,] (2,-0.3) -- (1,-0.3);
                \draw[] (2,-0.5) -- (1,-0.5);
		        \draw[dashed] (2,-0.7) -- (1,-0.7);
		        \draw[dashed] (2,0.8)--(2,0)--(2,-0.9);
		         \draw[] (3,0) -- (2,0);
                \draw[dashed] (3,0.4) -- (2,0.4);
                \draw[] (3,0.6) -- (2,0.6);
                \draw[dashed] (3,0.2) -- (2,0.2);
                \draw[] (3,-0.2) -- (2,-0.2);
                \draw[dashed,] (3,-0.4) -- (2,-0.4);
		        \draw[] (3,-0.6) -- (2,-0.6);
		        \draw[dashed] (3,0.8)--(3,0)--(3,-0.9);
		         \draw[] (3.7,-0.1) -- (3,-0.1);
                \draw[dashed] (3.7,0.3) -- (3,0.3);
                \draw[] (3.7,0.1) -- (3,0.1);
                \draw[dashed] (3.7,0.5) -- (3,0.5);
                \draw[] (3.7,-0.3) -- (3,-0.3);
                \draw[] (3.7,-0.5) -- (3,-0.5);
		        \draw[dashed] (3.7,-0.7) -- (3,-0.7);
            \end{scope}
        \end{tikzpicture}}
\end{align}
The dashed horizontal lines correspond to the representations like $(0,-1)_{v}$ while the dashed vertical lines correspond to the representations like $(-1,n)_{u}$.The module with level $(-1,0)$ should correspond to negative NS5-branes and level $(-1,n)$ should correspond to a bound state of negative NS5-brane and D5-branes with charge $(0,n)$.

Looking at the diagram vertically, we obtain a supergroup structure for each stack of D4 (D5)-branes. On the other hand, looking at the brane web horizontally, we obtain a superquiver structure. This diagram strongly implies that the gauge theory we obtain here is a \textit{supersymmetric supergroup superquiver gauge theory}. Let us list down few conjectures for linear quiver gauge theories.
\begin{conj}
\label{conj:conjecture1}
The partition function of the pure supergroup gauge theory does not depend on the signatures of the NS5-branes surrounding the stack of D4 (or D5)-branes:
\begin{align}
\adjustbox{valign=c}{
    \begin{tikzpicture}
            \begin{scope}[scale=2]
                \draw[dashed] (1,0) --(0,0);
                \draw[dashed] (1,0.2) -- (0,0.2);
                \draw[] (1,0.4) -- (0,0.4);
                \draw[dashed] (1,0.6) -- (0,0.6);
                \draw[] (1,-0.4) -- (0,-0.4);
		        \draw[dashed] (1,-0.6) -- (0,-0.6);
		        \draw[] (1,-0.2) -- (0,-0.2);
		        \draw (1,0.8)--(1,0)--(1,-0.8);
		        \draw (0,0.8) -- (0,0)--(0,-0.8);
            \end{scope}
        \end{tikzpicture}}\qquad\quad
        \adjustbox{valign=c}{\begin{tikzpicture}
        \begin{scope}[scale=2]
         \draw[dashed] (1,0) --(0,0);
                \draw[dashed] (1,0.2) -- (0,0.2);
                \draw[] (1,0.4) -- (0,0.4);
                \draw[dashed] (1,0.6) -- (0,0.6);
                \draw[] (1,-0.4) -- (0,-0.4);
		        \draw[dashed] (1,-0.6) -- (0,-0.6);
		        \draw[] (1,-0.2) -- (0,-0.2);
		        \draw[dashed] (1,0.8)--(1,0)--(1,-0.8);
		        \draw (0,0.8) -- (0,0)--(0,-0.8);
		  \end{scope}
        \end{tikzpicture}}\qquad\quad  \adjustbox{valign=c}{\begin{tikzpicture}
        \begin{scope}[scale=2]
         \draw[dashed] (1,0) --(0,0);
                \draw[dashed] (1,0.2) -- (0,0.2);
                \draw[] (1,0.4) -- (0,0.4);
                \draw[dashed] (1,0.6) -- (0,0.6);
                \draw[] (1,-0.4) -- (0,-0.4);
		        \draw[dashed] (1,-0.6) -- (0,-0.6);
		        \draw[] (1,-0.2) -- (0,-0.2);
		        \draw (1,0.8)--(1,0)--(1,-0.8);
		        \draw[dashed] (0,0.8) -- (0,0)--(0,-0.8);
		  \end{scope}
        \end{tikzpicture}}\qquad\quad 
        \adjustbox{valign=c}{\begin{tikzpicture}
        \begin{scope}[scale=2]
         \draw[dashed] (1,0) --(0,0);
                \draw[dashed] (1,0.2) -- (0,0.2);
                \draw[] (1,0.4) -- (0,0.4);
                \draw[dashed] (1,0.6) -- (0,0.6);
                \draw[] (1,-0.4) -- (0,-0.4);
		        \draw[dashed] (1,-0.6) -- (0,-0.6);
		        \draw[] (1,-0.2) -- (0,-0.2);
		        \draw[dashed] (1,0.8)--(1,0)--(1,-0.8);
		        \draw[dashed] (0,0.8) -- (0,0)--(0,-0.8);
		  \end{scope}
        \end{tikzpicture}}
\end{align}
Moreover, the partition function does not depend on the order of the D-branes as the pure SYM case explained in sections~\ref{sec:example-pureSYM} and \ref{sec:example-Aquiver}. 
\end{conj}

\begin{conj}
\label{conj:conjecture2}
The S-duality relates supergroup superquiver gauge theories by rotating the brane web diagram by 90 degrees:
\begin{align}
    \adjustbox{valign=c}{\includegraphics[width=15cm]{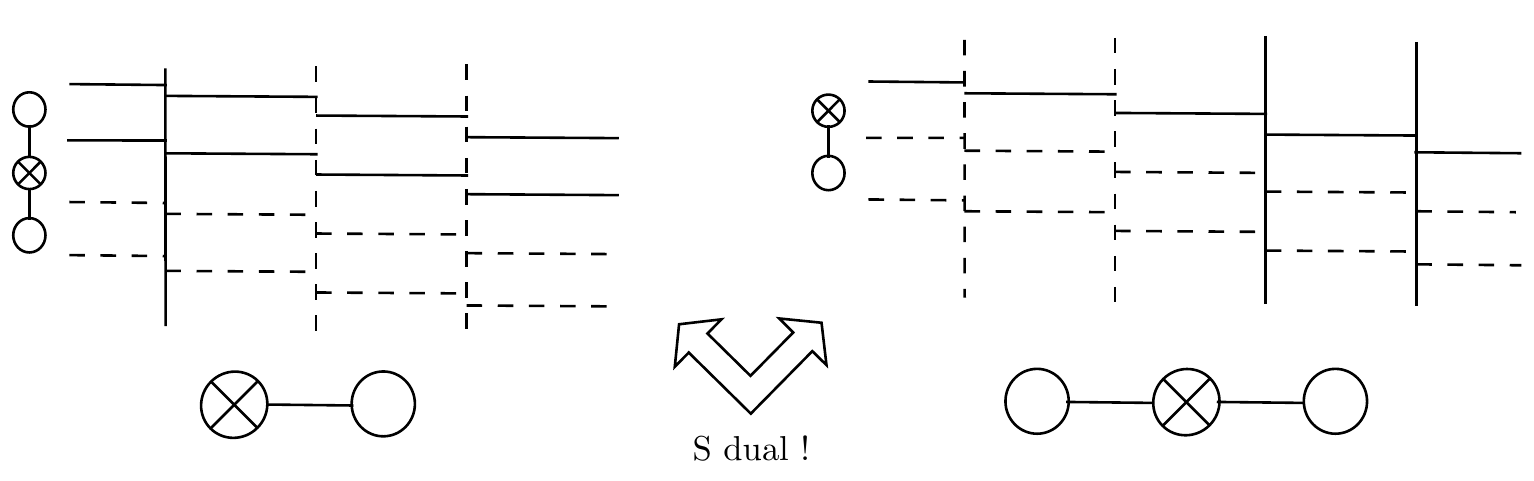}}
\end{align}
In particular, there is a duality between supergroup ordinary gauge theory and ordinary group superquiver gauge theory:
\begin{align}
    \adjustbox{valign=c}{\includegraphics[width=15cm]{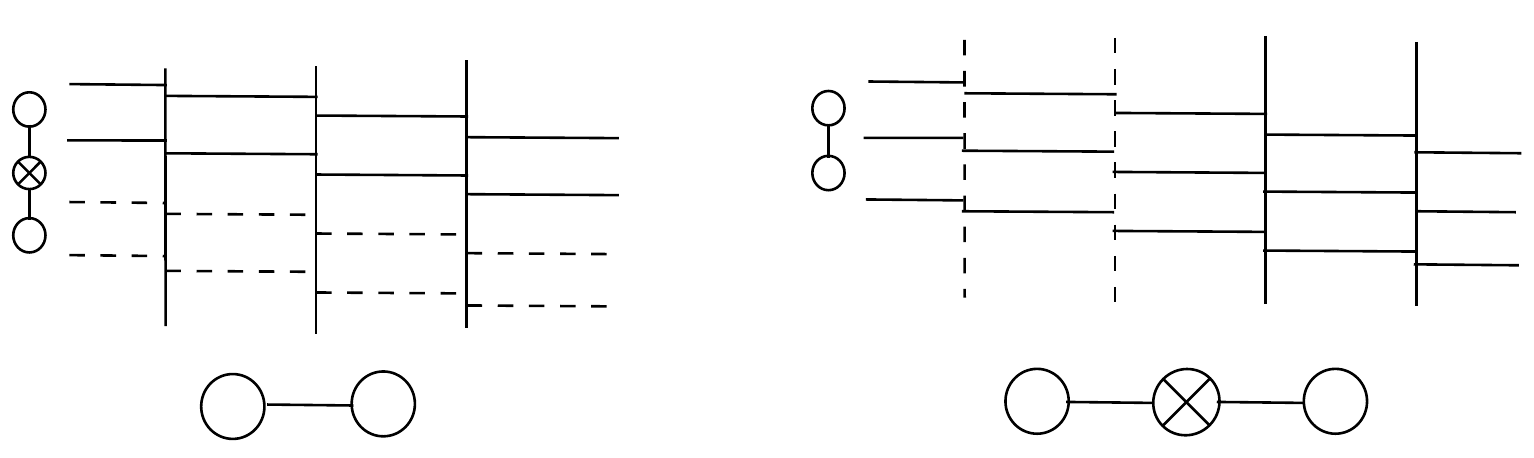}}
\end{align}

\end{conj}

\begin{conj}
\label{conj:conjecture3}
When the rank of the supergroups of the gauge nodes are all the same, then the partition function of the superquiver theory does not depend on the order of the NS5-branes. Namely, after fixing the signatures of the NS5-branes, the partition function does not depend on the order of it. In particular, they are related to the underlying superalgebra structure of the quiver.
\end{conj}
Conjecture \ref{conj:conjecture3} comes from conjecture \ref{conj:conjecture1} and conjecture \ref{conj:conjecture2}. Since the rank of the supergroups are all the same, we can use conjecture~\ref{conj:conjecture1} and reorder the D-branes so that each stack have the same ordering of parities. We then take the S-duality and obtain a superquiver theory with the same ranks of supergroups for each stack of D-branes. Using conjecture \ref{conj:conjecture2}, the brane web is a 90 degrees rotated version of the original one and the NS5-branes are transformed into D-branes now. We can then use again conjecture \ref{conj:conjecture1} and reorder the D-branes. Finally, we use again conjecture \ref{conj:conjecture2} and take the S-duality and rotate back the brane web. During the process, the partition function will not change and we obtain conjecture \ref{conj:conjecture3}.

\section{Conclusion and discussions}\label{sec:conclusion}
We introduced new intertwiners which we call negative intertwiners to reproduce partition functions of supergroup gauge theories and show the 5d AGT correspondence of supergroup gauge theories. Negative intertwiners are obtained by using crystal representations with negative levels of central charges. The new crystal representation is directly related to the character of the negative part of the instanton bundle $\bfK^{-}$. Composition of these intertwiners give the fundamental Nekrasov factors appearing in the partition function of A-type supergroup gauge theories. We also reproduced the partition functions of A and D-type quiver gauge theories using the intertwiner formalism. The negative intertwiners also give the supergroup analogues of the Gaiotto state, $qq$-character, and quiver W-algebra. After taking nontrivial matrix elements of the negative intertwiners, we managed to reproduce the anti-refined topological vertex introduced by Kimura and Sugimoto \cite{Kimura-Sugimoto:antivertex}. Using the correspondence, we gave rules including the framing factor which were not discussed in detail in \cite{Kimura-Sugimoto:antivertex} to compute the partition functions, given a brane web. Finally, we discussed how superquiver gauge theories should appear in our formalism, which is still a conjecture to be confirmed in future. 

Let us list down possible directions for future work.
\begin{itemize}
    \item Although we focused on the 5d AGT correspondence, it is possible to study 2d/3d \cite{Zenkevich:2018fzl,Bourgine:2021nyw}, 4d \cite{Bourgine:2018uod}, and 6d generalizations \cite{Zhu:2017ysu,Foda:2018sce,Saito2013EllipticDA} following previous studies of the intertwiner formalism. Moreover, we can change the space time which the theory is defined on or introduce defects to the system by changing the base quantum toroidal algebra to other quantum toroidal algebras such as quantum toroidal $\mathfrak{gl}_{n}$ \cite{feigin2013representations,Bourgine:2019phm}, $\mathfrak{gl}_{m\,|\,n}$ \cite{bezerra2019quantum,bezerra2021representations}, $D(2,1;\alpha)$ \cite{feigin2021combinatorics, Noshita:2021ldl}, and toroidal quiver BPS algebras \cite{Galakhov:2020vyb,Galakhov:2021vbo,Galakhov:2021xum,Noshita:2021dgj,Noshita:2021ldl}.
\item Understanding the six-dimensional origin of the 4d/2d AGT correspondence for supergroup gauge theories is also interesting \cite{Alday2010}. The existence of the supergroup analogue of the Gaiotto state explained in section~\ref{sec:Gaiottostate} supports this expectation (of course we need to take the degenerate limit for the 4d/2d correspondence). Usually, in the story of AGT correspondence, we start from a certain 6d $\mathcal{N}=(2,0)$ SCFT, compactify it on a punctured Riemann surface, and then we obtain 4d $\mathcal{N}=2$ gauge theories which are dubbed as class $\mathcal{S}$ theories \cite{Gaiotto:2009we,Gaiotto:2009ma} (see \cite{LeFloch:2020uop} for a nice review). The six-dimensional theories are believed to have ADE classifications, but then the theories arising will not have supergroups as gauge groups. Thus, we expect there should be six-dimensional theories associated with superalgebras (they may be non-unitary or non-physical) that lead to this correspondence. M-theoretically, negative M5-branes and M2-branes should also appear in the story, just as how negative D-branes appeared in constructions of supergroup gauge theories \cite{Okuda:2006fb,Dijkgraaf:2016lym,Kimura-Pestun:supergroup,Kimura-Sugimoto:antivertex}. Studying supergroup analogues of the Seiberg-Witten curves appearing in the class $\mathcal{S}$ constructions should help the analysis (see \cite{Dijkgraaf:2016lym}). All these studies should lead us to new non-unitary theories, including non-Lagrangian theories and broader 2d/4d (5d/$q$-algebra) correspondences, or generally, BPS/CFT correspondences.

\item Not only the gauge theoretic aspects as mentioned in the previous paragraph, but also the 2d CFT/quantum algebra side must be studied. One of the reason we used the intertwiner formalism is because we do not have to determine explicitly the corresponding CFT or $q$-deformed algebra. The algebraic structure appears as tensor products of Fock representations, and as long as the spectral parameters are independent with each other, we do not have to take care of singular states and extra truncations, and so the analysis becomes simple. 

In the context of quantum toroidal algebras, tensor product representations of Fock representations give deformed W-algebras. If we denote the Fock spaces with central charges $(\hat{\gamma},\psi_{0}^{-})=(q_{i}^{\sigma/2},1)$ $(i=1,2,3,\,\sigma=\pm1)$ as $\mathcal{F}^{(\sigma)}_{i}$, tensor products of them should lead to new deformed W-algebras.\footnote{We only consider the vertex operator representations for simplicity. There are three types of Fock representations because quantum toroidal $\mathfrak{gl}_{1}$ has triality. Generally, we can place Fock spaces on quiver nodes and obtain quiver W-algebras \cite{Kimura:2015rgi}, but we focus on deformed W-algebras associated with linear quivers.} The known algebras are
\begin{align*}
    \begin{tabular}{|c|c|c|}\hline
        Deformed W-algebra &  Representation &References\\\hline \hline
        $q$-Virasoro $\oplus$ $q$-Heisenberg&  $\mathcal{F}_{3}^{\splus}\otimes\mathcal{F}_{3}^{\splus}$ &\cite{Shiraishi:1995rp}\\\hline
         \multirow{2}{*}{$q$-$W_{N}\oplus q-\text{Heisenberg}$}& $\underbrace{\mathcal{F}^{\splus}_{3}\otimes\cdots\otimes \mathcal{F}^{\splus}_{3}}_{N}$ &\multirow{2}{*}{\cite{Awata:1995zk,Awata:1996dx,Feigin:1995sf,FHSSY:2010}}\\\hline
        \multirow{2}{*}{$q$-$Y_{L,M,N}$}& \multirow{2}{*}{${\mathcal{F}^{\splus}_{1}}^{\otimes L}\otimes {\mathcal{F}_{2}^{\splus}}^{\otimes M}\otimes {\mathcal{F}_{3}^{\splus}}^{\otimes N}$}&\multirow{2}{*}{\cite{Litvinov:2016mgi,bershtein2018plane,Kojima2019,Kojima2021,Harada:2021xnm}}\\
        &&\\\hline
    \end{tabular}
    \label{tab:deformedWalgebra}
\end{align*}
where $Y_{L,M,N}$ in the last row is the $q$-deformed version of the corner vertex operator algebra\footnote{Note that due to the existence of the R-matrix, we can reorder the Fock spaces into any orders.} originally introduced in \cite{Gaiotto:2017euk}. Physically, it was introduced as a CFT obtained at the junction of D3-D5-NS5 branes:
\begin{align*}
    \adjustbox{valign=c}{\includegraphics[width=8cm]{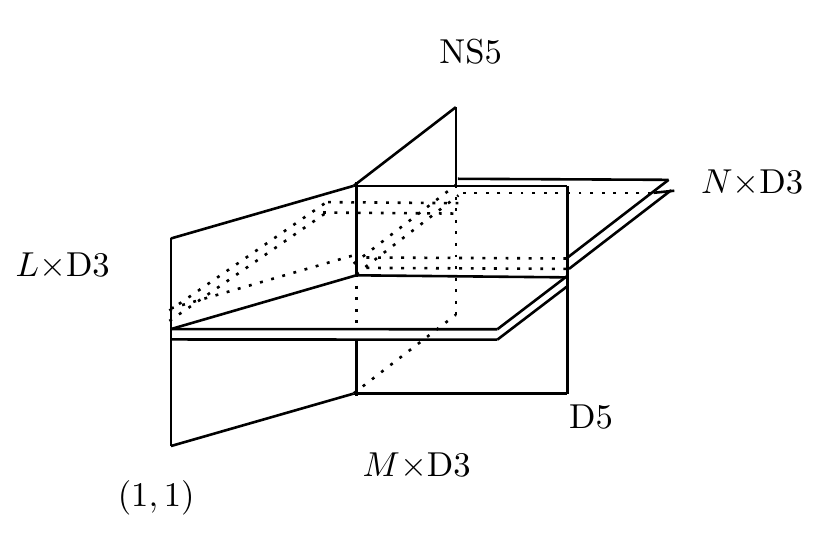}}
\end{align*}
The D3-branes appearing here are positive D3-branes, but generally, we could insert negative D3-branes. This should lead to a larger class of algebras $Y_{\vec{L},\vec{M},\vec{N}}$, where the subindices are changed to two-dimensional vectors $\vec{L}=(L_{+},L_{-}),\,\vec{M}=(M_{+},M_{-}),\,\vec{N}=(N_{+},N_{-})$. The integers $L_{\pm},M_{\pm},N_{\pm}$ are numbers of the positive and negative D3-branes for each stack of D3-branes. In the $q$-deformed language, these algebras are understood as tensor products of both positive and negative Fock representations:
\begin{align}
    {\mathcal{F}^{\splus}_{1}}^{\otimes L_{+}}\otimes{\mathcal{F}^{\sminus}_{1}}^{\otimes L_{-}} \otimes{\mathcal{F}_{2}^{\splus}}^{\otimes M_{+}}\otimes{\mathcal{F}_{2}^{\sminus}}^{\otimes M_{-}} \otimes{\mathcal{F}_{3}^{\splus}}^{\otimes N_{+}}\otimes {\mathcal{F}_{3}^{\sminus}}^{\otimes N_{-}}.
\end{align}
Moreover, since negative branes of one side of the 5-branes can be converted to positive branes of the other side of the 5-branes, we expect there are dualities among these algebras. It is also interesting to glue these algebras and obtain an analogue of the web of W-algebras \cite{Prochazka:2017qum,Prochazka:2018tlo}. These algebra should be dual to the supergroup analogue of the spiked instantons \cite{Rapcak_2019,rapcak2020cohomological,Nekrasov:2015wsu,Nekrasov:2016gud,Nekrasov:2016qym,Nekrasov:2016ydq,Nekrasov:2017gzb,Nekrasov:2017rqy}.

\item The negative level crystal representations have a straightforward generalizations to general BPS crystals associated with toric Calabi-Yau threefolds \cite{Rapcak_2019,rapcak2020cohomological,Galakhov:2020vyb,Galakhov:2021vbo,Galakhov:2021xum,Noshita:2021dgj,Noshita:2021ldl}. We call these crystals \textit{negative crystals} and the crystal representation studied in \cite{Galakhov:2020vyb,Galakhov:2021vbo,Galakhov:2021xum,Noshita:2021dgj,Noshita:2021ldl} as \textit{positive crystals} for convenience. Studying the physical aspects of these positive/negative crystals is also important. For general toroidal quiver BPS algebras, we have Drinfeld currents $x^{\pm}_{i}(z),\psi^{\pm}_{i}(z)$ for each node $i$ of the quiver. Currents $x^{+}_{i}(z)$ add atoms to the crystal and $x^{-}_{i}(z)$ removes atoms from the crystal. To construct negative crystal representations, we need to change the roles of the currents $x^{\pm}_{i}(z)$. Namely, the current $x^{+}_{i}(z)$ removes an atom from the crystal, while $x^{-}_{i}(z)$ adds an atom to the crystal. These negative BPS crystals are expected to correspond to BPS states coming from negative D-branes. As long as we consider these negative crystal representations independently, they are isomorphic to the positive crystal representations, but once we consider tensor product representations mixing both of them, we would get new representations. Understanding what will happen when the spectral parameters are tuned is also important.

\item Completing the discussion in section~\ref{sec:superquiver} is also one future work we hope to come back. The intertwiners in (\ref{eq:superquiverintertwiner}) seem to not obey the intertwiner relations using the coproduct in (\ref{eq:coproduct1}). One of the reason is that the vertex operators with negative level of central charges in section~\ref{sec:negativevertexop} is obtained by transforming the variables as $z\rightarrow z^{-1}$ and thus, when studying contractions of these vertex operators, the radial ordering will change. 

Studying the relation with \cite{Bourgine:2018fjy} might help. Actually, the negative intertwiner studied here is exactly the same intertwiner studied there but the physical motivation is different. The author there studied it for S-duality, while we studied it for supergroup gauge theories. 

\end{itemize}

\section*{Acknowledgements}
 We thank Taro Kimura, Yutaka Matsuo, Tomas Proch\'azka, Yuji Sugimoto, Masahito Yamazaki, and Rui-Dong Zhu for discussions and comments on the draft of this paper. The author also thanks the Yukawa Institute for Theoretical Physics at Kyoto University, where a part of this work was presented during the YITP workshop YITP-W-22-09 on ``Strings and Fields 2022" \cite{NoshitaYITP}. 
The author is supported by JSPS KAKENHI Grant-in-Aid for JSPS fellows Grant No. JP22J20944, JSR Fellowship, and FoPM (WINGS Program), the University of Tokyo.

\appendix
\section{Notation and formulas of Young diagrams}\label{sec:appendix-Young}
\subsection{Young diagrams and formulas}
\begin{figure}[t]\centering
  \begin{tikzpicture}[scale=0.8]
  \draw[-Stealth] (-1,1)--(-1,-1);
  \draw[-Stealth] (-1,1)--(1,1);
  \node [left, scale=0.8] at (-1.3,0) {$i$};
  \node [above, scale=0.8] at (0.5,1.3) {$j$};
\draw (0,0)--(0,-5)--(1,-5)--(1,-4)--(3,-4)--(3,-2)--(5,-2)--(5,-1)--(6,-1)--(6,0)--(0,0);
\draw (1.9,-1.2)--(1.9,-1.8)--(2.5,-1.8)--(2.5,-1.2)--(1.9,-1.2);
\draw [Stealth-Stealth] (2.5,-1.5)--(5,-1.5);
\draw[Stealth-Stealth](1.9,-1.5)--(0,-1.5);
\node[scale=0.8] at (0.9, -1.2) {$a'_{\lambda}(i,j)$};
\node [scale=0.8] at (2.2,-1.5)  {$i,j$};
\draw [Stealth-Stealth] (2.2,-1.8)--(2.2,-4);
\draw [Stealth-Stealth] (2.2,-1.2)--(2.2,0);
\node [right, scale=0.8] at (2.2,-0.6){$l'_{\lambda}(i,j)$};
\draw (3.8,-1.2) node [scale=0.8] {$a_{\lambda}(i,j)$};
\draw (2.2,-2.8) node [left,scale=0.8] {$l_{\lambda}(i,j)$};
  \end{tikzpicture}\label{fig:appendix-Young}
  \caption{Arm and leg (co)length of a box $(i,j)$ in a Young diagram.}
\end{figure}
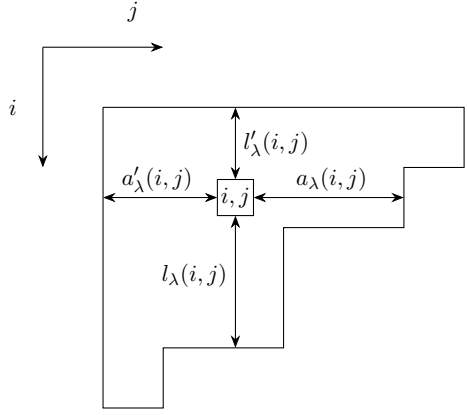
See chapter 1 of \cite{macdonald1998symmetric} for a good reference for formulas of Young diagrams.

The Young diagram (partition) is a sequence of decreasing non-negative integers 
\begin{align}
    \lambda=\{\lambda_{i}\in\mathbb{Z}_{\geq 0}\,|\,\lambda_{1}\geq \lambda_{2}\geq \cdots\}.
\end{align}
We denote the transpose of $\lambda$ as $\lambda^{\rmT}$. The size and norm are defined as 
\begin{align}
    |\lambda|=\sum_{i=1}^{\ell(\lambda)}\lambda_{i},\quad ||\lambda||^{2}=\sum_{i=1}^{\ell(\lambda)}\lambda_{i}^{2},
\end{align}
where $\ell(\lambda)$ is the length of the Young diagram. Note that 
\begin{align}
    |\lambda^{\rmT}|=|\lambda|.
\end{align}
We also define the following quantities (arm length, leg length, arm co-length, leg co-length, and content ($c_{\lambda}(i,j)$)) for $(i,j)\in\lambda$ (see Figure \ref{fig:appendix-Young}),
\begin{align}
    a_{\lambda}(i,j)=\lambda_{i}-j,&\quad l_{\lambda}(i,j)=\lambda_{j}^{\rmT}-i,\\
    a'_{\lambda}(i,j)=j-1,&\quad l'_{\lambda}(i,j)=i-1,\\
    c_{\lambda}(i,j)=j-i.&
\end{align}
The hook length is defined as 
\begin{align}
    h_{\lambda}(i,j)=\lambda_{i}-j+\lambda_{j}^{\rmT}-i+1=a_{\lambda}(i,j)+l_{\lambda}(i,j)+1.
\end{align}

In the context of topological vertex, we frequently use the following quantities
\begin{align}
    n(\lambda)=\sum_{i=1}^{\ell(\lambda)}(i-1)\lambda_{i},&\quad \kappa(\lambda)=2\sum_{x\in\lambda}c(x)=2\sum_{(i,j)\in\lambda}(j-i),\\
    a'_{\lambda}(x)-l'_{\lambda}(x)=c_{\lambda}(x).&
\end{align}
They satisfy the following properties.
\begin{align}
    \kappa(\lambda)&=\sum_{i}\lambda_{i}(\lambda_{i}+1-2i)=2(n(\lambda^{\rmT})-n(\lambda))=||\lambda||^{2}-||\lambda^{\rmT}||^{2},\\
    n(\lambda)&=\frac{1}{2}\sum_{j=1}^{\lambda_{1}}\lambda_{j}^{\rmT}(\lambda_{j}^{\rmT}-1)=\sum_{x\in\lambda}l'_{\lambda}(x)=\sum_{x\in\lambda}l_{\lambda}(x),\\
    n(\lambda^{\rmT})&=\frac{1}{2}\sum_{i=1}^{\ell(\lambda)}\lambda_{i}(\lambda_{i}-1)=\sum_{x\in\lambda}a'_{\lambda}(x)=\sum_{x\in\lambda}a_{\lambda}(x),\\
    \sum_{x\in\lambda}h_{\lambda}(x)&=n(\lambda)+n(\lambda^{\rmT})+|\lambda|=\frac{1}{2}(||\lambda||^{2}+||\lambda^{\rmT}||^{2}).
\end{align}

\subsection{Box contents of the Young diagram}
\begin{figure}[t]
    \centering
    \includegraphics[width=8cm]{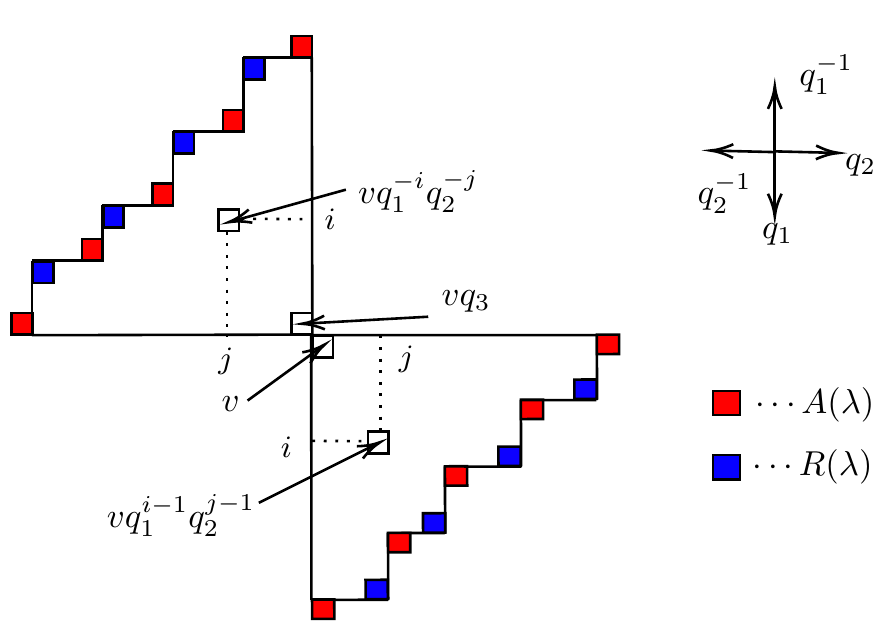}
    \caption{Box contents of two types of Young diagrams. Depending on the parity of the representation (i.e. level $(0,\pm1)$), there are two types of box contents $\chi_{x}^{\spm},\,x\in\lambda$, where $\chi_{x}^{\splus}=vq_{1}^{i-1}q_{2}^{j-1}$ and $\chi_{x}^{\sminus}=vq_{1}^{-i}q_{2}^{-j}$. The arm direction is defined in the horizontal direction, while the leg direction is defined in the vertical direction for both parities of Young diagrams. The sets $A(\lambda)$ and $R(\lambda)$ are sets of addable boxes and removable boxes of the Young diagram, respectively.}
    \label{fig:appendix-Youngcoord}
\end{figure}
In the main text, we used two types of box contents of the Young diagram, depending on the parity of the representation. For representations with $(0,\sigma)$\,$(\sigma=\pm 1)$, the box contents $\chi_{x}^{(\sigma)}$ are given as
\begin{align}
    \chi_{x}^{(\sigma)}=\begin{dcases}vq_{1}^{i-1}q_{2}^{j-1},\quad \sigma=+,\\ vq_{1}^{-i}q_{2}^{-j},\quad \sigma=-,\end{dcases}
\end{align}
where $x=(i,j)\in\lambda$. We illustrate the arm direction in the horizontal direction (with the coordinate $q_{2}$), and the leg direction in the vertical direction (with the coordinate $q_{1}$).  
They are illustrated as in Figure \ref{fig:appendix-Youngcoord}. We also denote the sets of addable and removable boxes of the Young diagram as $A(\lambda)$ and $R(\lambda)$, respectively.

\section{Special functions}\label{sec:appendix-specialfunc}
We denote the q-analog of the double gamma function as 
\begin{align}
    \mathcal{G}(z)=\mathcal{G}(z;q_{1},q_{2})=\exp\left(-\sum_{n=1}^{\infty}\frac{z^{n}}{n}\frac{1}{(1-q_{1}^{n})(1-q_{2}^{n})}\right),\quad |z|<1.
\end{align}
We can rewrite this using infinite products
\begin{align}
    \mathcal{G}(z)=\begin{dcases}\prod_{k,k'=0}^{\infty}(1-zq_{1}^{k}q_{2}^{k'})\quad (|z|,|q_{1}|,|q_{2}|<1)\\
    \prod_{k,k'=0}^{\infty}(1-zq_{1}^{-1}q_{1}^{-k}q_{2}^{k'})^{-1}\quad (|z|<1,|q_{1}|>1,|q_{2}|<1)\\
    \prod_{k,k'=0}^{\infty}(1-zq_{1}^{-k-1}q_{2}^{-k'-1})\quad (|z|<1,\,|q_{1}|,|q_{2}|>1).
    \end{dcases}
\end{align}
For the region $|z|>1$, we need to analytic continuate using the following formula
\begin{align}
    \mathcal{G}(z^{-1})=\mathcal{G}(q_{1}q_{2}z)\Gamma(z^{-1};q_{1},q_{2})^{-1},
\end{align}
where $\Gamma(z;q_{1},q_{2})$ is the elliptic gamma function
\begin{align}
    \Gamma(z;p,q)=\prod_{m,n=0}^{\infty}\frac{1-p^{n+1}q^{m+1}z^{-1}}{1-p^{n}q^{m}z}.
\end{align}
Using this property, for example, the perturbative part of the vector multiplet part is converted as
\begin{align}
    \prod_{i,j=1}^{n}\mathcal{G}(v_{i}/v_{j})\rightarrow \prod_{i<j}\mathcal{G}(q_{1}q_{2}v_{i}/v_{j})\mathcal{G}(v_{i}/v_{j}),\quad |v_{i}/v_{j}|<1\,\,\text{for}\,\,i<j.
\end{align}
Namely, we regularize the perturbative part by omitting the contact term comming from $\mathcal{G}(1)$ and use the elliptic gamma function to do the analytic continuation. We also omit the elliptic gamma function from the partition function. We will finally see that this form reproduces the partition function obtained from the intertwiner formalism.

\section{Formulas for Nekrasov factors}\label{sec:appendix-nekrasov}
\subsection{Equivalent expressions of Nekrasov factors}\label{sec:appendix-equiv-nekrasov}
In this section, we consider equivalent expressions of the Nekrasov factors appearing in supergroup gauge theories. For the ordinary group gauge theories, similar formulas have been derived in the past literature \cite{Awata:2008ed,Nekrasov:2002qd,Nakajima:2003uh,Nakajimalecturebook}. For the supergroup case, see \cite{Kimura-Pestun:supergroup,Kimura:instantoncounting}.

There are four types of Nekrasov factors which were defined as
\begin{align}
    N_{\sigma\sigma'}(v_{1},\lambda\,|\,v_{2},\nu)&=\mathbb{I}[\ch\bfN_{1}^{\sigma\vee}\ch\bfK^{\sigma'}_{2}+q_{3}\ch\bfK_{1}^{\sigma\vee}\ch\bfN^{\sigma'}_{2}-\ch\swedge\bfQ^{\vee}\ch\bfK_{1}^{\sigma\vee}\ch\bfK^{\sigma'}_{2}]\\
    &=\prod_{x\in\lambda}\left(1-\frac{\chi_{x}^{(\sigma)}}{q_{3}v_{2}}\right)\prod_{x\in\nu}\left(1-\frac{v_{1}}{\chi_{x}^{(\sigma')}}\right)\prod_{\substack{x\in\lambda\\y\in\nu}}S\left(\frac{\chi_{x}^{(\sigma)}}{\chi_{y}^{(\sigma')}}\right)\\
    &=\prod_{x\in\lambda}\left(1-Q\frac{q_{x}^{(\sigma)}}{q_{3}}\right)\prod_{x\in\nu}\left(1-Q\frac{1}{q_{x}^{(\sigma')}}\right)\prod_{\substack{x\in\lambda\\y\in\nu}}S\left(Q\frac{q_{x}^{(\sigma)}}{q_{y}^{(\sigma')}}\right),
    \label{eq:Nekrasovdef}
\end{align}
where we used 
\begin{align}
    \chi_{x}^{(\sigma)}=v_{1}q_{x}^{(\sigma)},\quad \chi_{x}^{(\sigma')}=v_{2}q_{x}^{(\sigma')},\quad Q=v_{1}/v_{2}.
\end{align}
Obviously, they depend only on the ratio $Q$, and so we also denote the Nekrasov factors as $N_{\lambda\nu}^{\sigma\sigma'}(Q;q_{1},q_{2})$. The $q$-coordinates $q_{x}^{(\sigma)}$ are the coordinates omitting the spectral parameter. Using the equivariant index, we can further derive different appearing but equivalent expressions. 

For each gauge node, we can introduce the so-called $x$-variables \cite{Kimura:2015rgi,Kimura-Pestun:supergroup,Kimura:instantoncounting}:
\begin{align}
\begin{split}
    \mathcal{X}^{\sigma}_{i}&=\left\{x^{\sigma}_{i,k}=v_{i}^{(\sigma)}q_{1}^{\sigma(k-1)}q_{2}^{\sigma\lambda^{\sigma}_{i,k}}\middle|\,k=1,\ldots,\infty\right\},\quad \mathring{\mathcal{X}}^{\sigma}_{i}=\left\{x^{\sigma}_{i,k}=v_{i}^{(\sigma)}q_{1}^{\sigma(k-1)}\middle|\,k=1,\ldots,\infty\right\},\\
    \check{\mathcal{X}}^{\sigma}_{i}&=\left\{\check{x}^{\sigma}_{i,k}=v_{i}^{(\sigma)}q_{2}^{\sigma(k-1)}q_{1}^{\sigma\lambda^{\rmT,\sigma}_{i,k}}\,|\,k=1,\ldots,\infty\right\},\quad \mathring{\check{\mathcal{X}}}^{\sigma}_{i}=\left\{x^{\sigma}_{i,k}=v_{i}^{(\sigma)}q_{2}^{\sigma(k-1)}\middle|\,k=1,\ldots,\infty\right\}.
\end{split}\label{eq:appendix-xvariables}
\end{align}
For simplicity, we only consider when the rank is one for each node and only consider two nodes. When we do not need to consider bifundamental contributions, we omit the subindex $i$ denoting the node of the quiver. 
We introduce bundles $\bfX_{i},\bfX^{\rmT}_{i}$ and set the supercharacter as 
\begin{align}
\sch\bfX_{i}&=\sum_{\sigma=\pm}\sigma\,\ch\bfX_{i,\sigma},\quad
    \ch \bfX_{i,\sigma}=\sum_{x\in\mathcal{X}^{\sigma}_{i}}x,\\
\sch\bfX^{\rmT}_{i}&=\sum_{\sigma=\pm}\sigma\,\ch\bfX^{\rmT}_{i,\sigma},\quad \ch\bfX^{\rmT}_{i,\sigma}=\sum_{x\in\check{\mathcal{X}}_{i}^{\sigma}}x.
\end{align}

\bclaim
We have the following relation
\begin{align}
\begin{split}
    \ch\bfY^{\sigma}&\coloneqq\ch\bfN^{\sigma}-\ch\swedge\bfQ\,\ch\bfK^{\sigma}\\
    &=(1-q_{1}^{\sigma})\,\ch\bfX_{\sigma}\\
    &=(1-q_{2}^{\sigma})\,\ch\bfX^{\rmT}_{\sigma}.
\end{split}\label{eq:appendix-XYrelation}
\end{align}
Generalizing this, it is obvious to see
\begin{align}
    \begin{split}
        \mathcal{Y}_{\lambda}^{\splus}(z)&=(1-v/z)\prod_{x\in\lambda}S\left(\chi_{x}^{\splus}/z\right)\\
        &=\frac{\prod_{x\in A(\lambda)}(1-\chi_{x}^{\splus}/z)}{\prod_{x\in R(\lambda)}(1-q_{3}^{-1}\chi_{x}^{\splus}/z)}\\
        &=\prod_{x\in\mathcal{X}^{+}}\frac{1-x/z}{1-q_{1}x/z}=\prod_{x\in\check{\mathcal{X}}^{+}}\frac{1-x/z}{1-q_{2}x/z},
    \end{split}\\
    \begin{split}
        \mathcal{Y}^{\sminus}_{\lambda}(z)&=(1-v/z)\prod_{x\in\lambda}S(\chi_{x}^{\sminus}/z)\\
        &=\frac{\prod_{x\in A(\lambda)}(1-q_{3}^{-1}\chi_{x}^{\sminus}/z)}{\prod_{x\in R(\lambda)}(1-\chi_{x}^{\sminus}/z)}\\
        &=\prod_{x\in\mathcal{X}^{-}}\frac{1-x/z}{1-q_{1}^{-1}x/z}=\prod_{x\in\check{\mathcal{X}}^{-}}\frac{1-x/z}{1-q_{2}^{-1}x/z}.
    \end{split}
\end{align}
\eclaim
\begin{proof}
Using 
\begin{align}
(1-q_{1})(1-q_{2})\sum_{\sAbox\,\in\lambda}\chi_{\sAbox}^{\splus}&=
\adjustbox{valign=c}{\includegraphics[width=3.5cm]{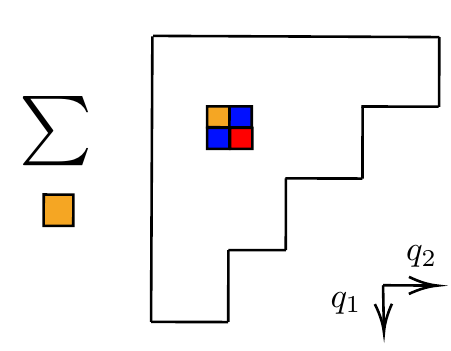}}\,=\,\adjustbox{valign=c}{\includegraphics[width=2.5cm]{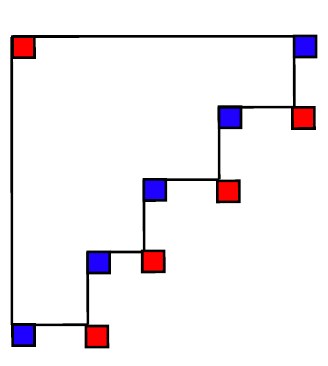}},\label{eq:appendix-chY-plus}\\
(1-q_{1})(1-q_{2})\sum_{\sAbox\in\lambda}\chi_{\sAbox}^{\sminus}&=\adjustbox{valign=c}{\includegraphics[width=3.5cm]{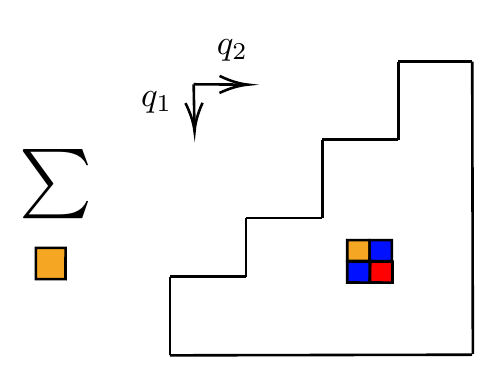}}\,=\adjustbox{valign=c}{\includegraphics[width=2.5cm]{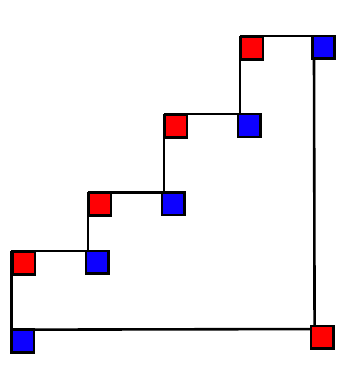}}
,\label{eq:appendix-chY-minus}
\end{align} we have
\begin{align}
\ch\bfY^{\sigma}&=v-(1-q_{1})(1-q_{2})\sum_{\sAbox\in\lambda}\chi_{\sAbox}^{(\sigma)}
=\begin{dcases}
\sum_{\sAbox\in A(\lambda)}\chi^{\splus}_{\sAbox}-q_{1}q_{2}\sum_{\sAbox\in R(\lambda)}\chi^{\splus}_{\sAbox},\quad \sigma=+\\
\sum_{\sAbox\in A(\lambda)}q_{1}q_{2}\chi^{\sminus}_{\sAbox}-\sum_{\sAbox\in R(\lambda)}\chi_{\sAbox}^{\sminus},\quad \sigma=-
\end{dcases}.
\end{align}
The boxes in the Young diagram are drawn with three colors: yellow, red, and blue. Red and yellow boxes are understood as terms with positive signs while blue boxes are understood as terms with negative signs. For example, for a box $\sAbox\in\lambda$, we have
\begin{align}
    (1-q_{1})(1-q_{2})\chi^{\splus}_{\sAbox}=\chi_{\sAbox}^{\splus}-q_{1}\chi_{\sAbox}^{\splus}-q_{2}\chi_{\sAbox}^{\splus}+q_{1}q_{2}\chi_{\sAbox}^{\splus}
\end{align}
and thus, we have a yellow box, two blue boxes next to it (in the $q_{1},q_{2}$ directions), and a red box in the diagonal direction ($q_{1}q_{2}$ direction).
In the first equality of (\ref{eq:appendix-chY-plus}) and (\ref{eq:appendix-chY-minus}), we took the sum with respect to the yellow box, which is a box in the Young diagram. After taking the sum, terms with positive signs and negative signs cancel with each other and we obtain the second equality. 

Similarly, using 
\begin{align}
(1-q_{1})\ch\bfX_{+}&=\adjustbox{valign=c}{\includegraphics[width=2.5cm]{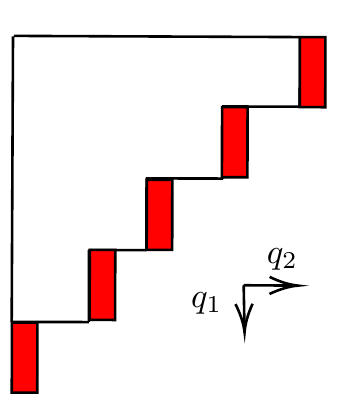}}-\quad
\adjustbox{valign=c}{\includegraphics[width=2.5cm]{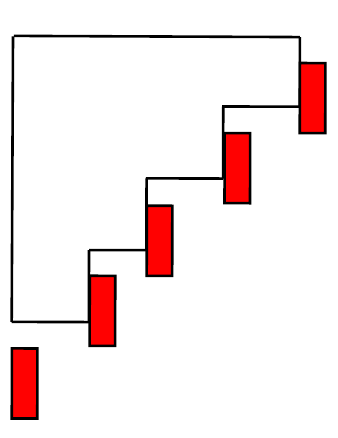}}=\quad
\adjustbox{valign=c}{\includegraphics[width=2.6cm]{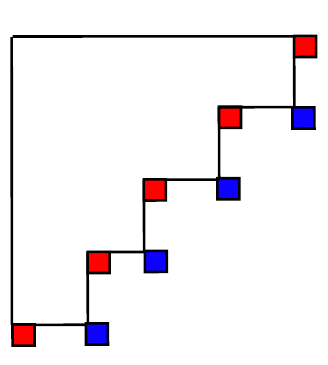}},\\
(1-q_{1}^{-1})\ch\bfX_{-}&=\adjustbox{valign=c}{\includegraphics[width=2.5cm]{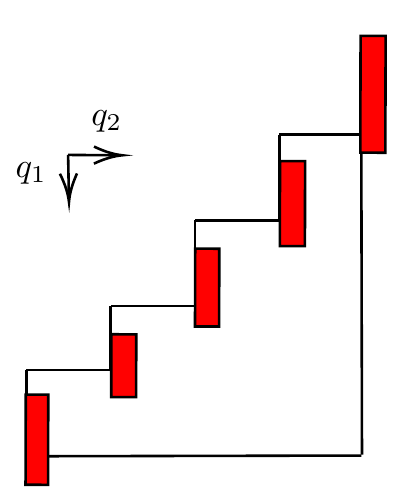}}
-\adjustbox{valign=c}{\includegraphics[width=2.3cm]{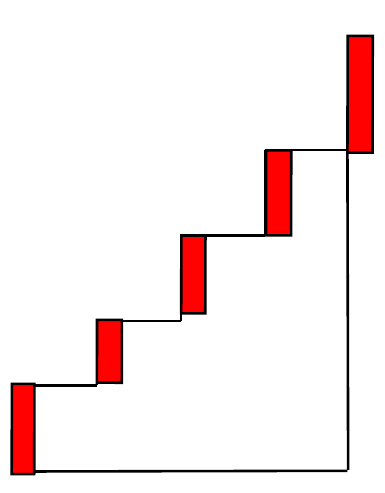}}=
\adjustbox{valign=c}{\includegraphics[width=2.6cm]{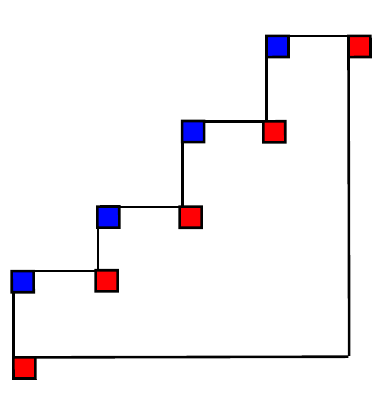}}
\end{align}

we have
\begin{align}
    (1-q_{1}^{\sigma})\ch\bfX_{\sigma}=\begin{dcases}
\sum_{\sAbox\in A(\lambda)}\chi^{\splus}_{\sAbox}-q_{1}q_{2}\sum_{\sAbox\in R(\lambda)}\chi^{\splus}_{\sAbox},\quad \sigma=+\\
\sum_{\sAbox\in A(\lambda)}q_{1}q_{2}\chi^{\sminus}_{\sAbox}-\sum_{\sAbox\in R(\lambda)}\chi_{\sAbox}^{\sminus},\quad \sigma=-
\end{dcases}.
\end{align}

For the transpose, we have
\begin{align}
(1-q_{2})\ch\bfX_{+}^{\rmT}&=
\adjustbox{valign=c}{\includegraphics[width=2.7cm]{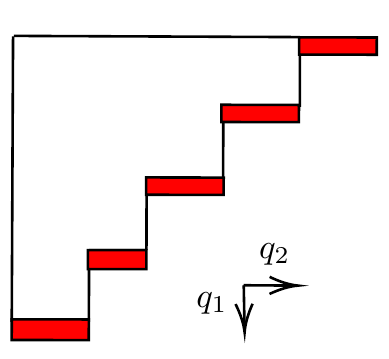}}-
\adjustbox{valign=c}{\includegraphics[width=2.7cm]{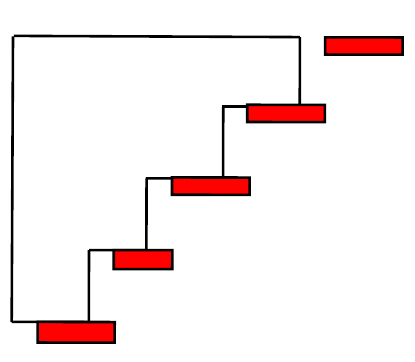}}=\adjustbox{valign=c}{\includegraphics[width=2.5cm]{figures/Young_diagrams/7.pdf}},\\
(1-q_{2}^{-1})\ch\bfX^{\rmT}_{-}&=
\adjustbox{valign=c}{\includegraphics[width=3cm]{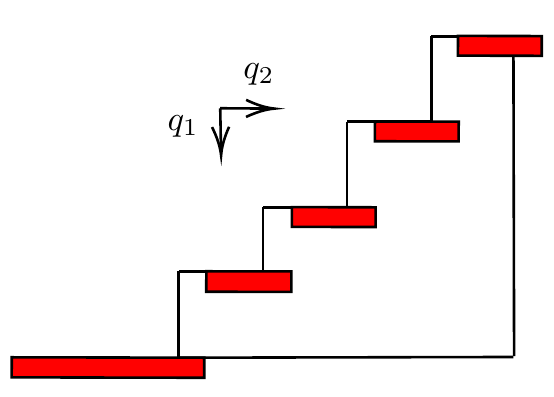}}-
\adjustbox{valign=c}{\includegraphics[width=3cm]{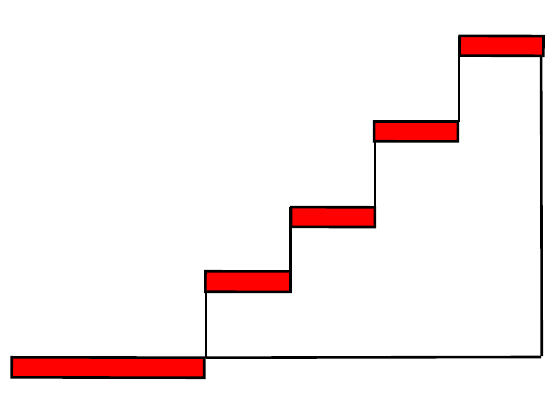}}=\adjustbox{valign=c}{\includegraphics[width=2.5cm]{figures/Young_diagrams/10.pdf}}\end{align}
\end{proof}

\bclaim
We have the following equivalent expressions of the Nekrasov factors.
\begin{align}
    N_{\lambda\nu}^{\sigma\sigma'}(Q;q_{1},q_{2})&=\prod_{x\in\lambda}\left(1-\frac{\chi_{x}^{(\sigma)}}{q_{3}v_{2}}\right)\prod_{x\in\nu}\left(1-\frac{v_{1}}{\chi_{x}^{(\sigma')}}\right)\prod_{\substack{x\in\lambda\\y\in\nu}}S\left(\frac{\chi_{x}^{(\sigma)}}{\chi_{y}^{(\sigma')}}\right),\\
    \begin{split}
    N_{\lambda\nu}^{\sigma\sigma'}(Q;q_{1},q_{2})&=\prod_{i,j=1}^{\infty}\left(\frac{1-Qq_{1}^{\sigma i-\frac{\sigma-1}{2}-\sigma'\nu_{j}^{\rmT}}q_{2}^{-\sigma'j+\frac{\sigma'+1}{2}+\sigma\lambda_{i}}}{1-Qq_{1}^{\sigma i-\frac{\sigma-1}{2}}q_{2}^{-\sigma' j+\frac{1+\sigma'}{2}}}\right)^{\sigma\sigma'}\\
    &=\prod_{i,j=1}^{\infty}\left(\frac{1-Qq_{1}^{-\sigma'j+\frac{\sigma'+1}{2}+\sigma\lambda_{i}^{\rmT}}q_{2}^{\sigma i-\frac{\sigma-1}{2}-\sigma'\nu_{j}}}{1-Qq_{1}^{-\sigma'j+\frac{\sigma'+1}{2}}q_{2}^{\sigma i-\frac{\sigma-1}{2}}}\right)^{\sigma\sigma'},
    \end{split}\\
    \begin{split}
    N_{\lambda\nu}^{\sigma\sigma'}(Q;q_{1},q_{2})&=\prod_{(x,x')\in\mathcal{X}^{\sigma}_{\lambda}\times\mathcal{X}^{\sigma'}_{\nu}}\left(\frac{(q_{2}q_{1}^{\frac{1+\sigma}{2}-\sigma'}x/x';q_{2})_{\infty}}{(q_{2}q_{1}^{\frac{1+\sigma}{2}}x/x';q_{2})_{\infty}}\right)^{\sigma}\\
    &\quad\times\prod_{(x,x')\in\mathring{\mathcal{X}}^{\sigma}_{\lambda}\times\mathring{\mathcal{X}}^{\sigma'}_{\nu}}\left(\frac{(q_{2}q_{1}^{\frac{1+\sigma}{2}}x/x';q_{2})_{\infty}}{(q_{2}q_{1}^{\frac{1+\sigma}{2}-\sigma'}x/x';q_{2})_{\infty}}\right)^{\sigma}\\
    &=\prod_{(x,x')\in\check{\mathcal{X}}^{\sigma}_{\lambda}\times\check{\mathcal{X}}^{\sigma'}_{\nu}}\left(\frac{(q_{1}q_{2}^{\frac{1+\sigma}{2}-\sigma'}x/x';q_{1})_{\infty}}{(q_{1}q_{2}^{\frac{1+\sigma}{2}}x/x';q_{1})_{\infty}}\right)^{\sigma}\\
    &\quad\times\prod_{(x,x')\in\mathring{\check{\mathcal{X}}}^{\sigma}_{\lambda}\times\mathring{\check{\mathcal{X}}}^{\sigma'}_{\nu}}\left(\frac{(q_{1}q_{2}^{\frac{1+\sigma}{2}}x/x';q_{1})_{\infty}}{(q_{1}q_{2}^{\frac{1+\sigma}{2}-\sigma'}x/x';q_{1})_{\infty}}\right)^{\sigma},
    \end{split}
\end{align}
where 
\begin{align}
    \begin{split}
        \mathcal{X}_{\lambda}^{\sigma}=\{v_{1}q_{1}^{\sigma(k-1)}q_{2}^{\sigma\lambda_{k}}\}_{k=1}^{\infty},&\quad \mathcal{X}_{\lambda}^{\sigma'}=\{v_{2}q_{1}^{\sigma'(k-1)}q_{2}^{\sigma'\nu_{k}}\}_{k=1}^{\infty},\\
        \check{\mathcal{X}}^{\sigma}_{\lambda}=\{v_{1}q_{2}^{\sigma(k-1)}q_{1}^{\sigma\lambda_{k}^{\rmT}}\}_{k=1}^{\infty},&\quad \check{\mathcal{X}}^{\sigma'}_{\nu}=\{v_{2}q_{2}^{\sigma'(k-1)}q_{1}^{\sigma'\nu_{k}^{\rmT}}\}_{k=1}^{\infty},\\
        \mathring{\mathcal{X}}^{\sigma}_{\lambda}=\{v_{1}q_{1}^{\sigma(k-1)}\}_{k=1}^{\infty},&\quad\mathring{\mathcal{X}}_{\nu}^{\sigma'}=\{v_{2}q_{1}^{\sigma'(k-1)}\}_{k=1}^{\infty},\\
        \mathring{\check{\mathcal{X}}}^{\sigma}_{\lambda}=\{v_{1}q_{2}^{\sigma(k-1)}\}_{k=1}^{\infty},&\quad\mathring{\check{\mathcal{X}}}_{\nu}^{\sigma'}=\{v_{2}q_{2}^{\sigma'(k-1)}\}_{k=1}^{\infty}.
    \end{split}
\end{align}
\eclaim
\begin{proof}
We can write the Nekrasov factors in two ways 
\begin{align}
    N_{\sigma\sigma'}(v_{1},\lambda\,|\,v_{2},\nu)&=\mathbb{I}[\ch\bfN_{1}^{\sigma\vee}\ch\bfK^{\sigma'}_{2}+q_{3}\ch\bfK_{1}^{\sigma\vee}\ch\bfN^{\sigma'}_{2}-\ch\swedge\bfQ^{\vee}\ch\bfK_{1}^{\sigma\vee}\ch\bfK^{\sigma'}_{2}]\\
    &=\mathbb{I}\left[-\frac{\ch\bfY^{\sigma\vee}_{1}[\lambda]\ch\bfY^{\sigma'}_{2}[\nu]-\ch\bfY^{\sigma\vee}_{1}[\emptyset]\ch\bfY^{\sigma'}_{2}[\emptyset]}{\ch\swedge\bfQ}\right],
\end{align}
where we explicitly wrote the Young diagram dependence in the second equation. Namely, the Nekrasov factor is obtained by excluding the Young diagram independent part. We can rewrite the character $\ch\bfY^{\sigma}[\lambda]$ as
\begin{align}
    \ch\bfY^{\sigma}[\lambda]&=v(1-q_{1}^{\sigma})\sum_{i=1}^{\infty}q_{1}^{\sigma(i-1)}q_{2}^{\sigma\lambda_{i}}\label{eq:appendix-chY-1}\\
    &=v(1-q_{2}^{\sigma})\sum_{j=1}^{\infty}q_{2}^{\sigma(j-1)}q_{1}^{\sigma\lambda_{j}^{\rmT}}\label{eq:appendix-chY-2}.
\end{align}
Then, 
\begin{align}
\begin{split}
    -\frac{\ch\bfY^{\sigma\vee}_{1}[\lambda]\ch\bfY^{\sigma'}_{2}[\nu]}{\ch\swedge\bfQ}&=-\frac{v_{2}}{v_{1}}\frac{(1-q_{1}^{-\sigma})(1-q_{2}^{\sigma'})}{(1-q_{1})(1-q_{2})}\sum_{i,j=1}^{\infty}q_{1}^{-\sigma(i-1)+\sigma'\nu_{j}^{\rmT}}q_{2}^{\sigma'(j-1)-\sigma\lambda_{i}}\\
    &=-\frac{v_{2}}{v_{1}}(-q_{1})^{-\frac{1+\sigma}{2}}(-q_{2})^{-\frac{1-\sigma'}{2}}\sum_{i,j=1}^{\infty}q_{1}^{-\sigma(i-1)+\sigma'\nu_{j}^{\rmT}}q_{2}^{\sigma'(j-1)-\sigma\lambda_{i}}\\
    &=\sigma\sigma'\frac{v_{2}}{v_{1}}\sum_{i,j=1}^{\infty}q_{1}^{-\sigma i+\frac{\sigma-1}{2}+\sigma'\nu_{j}^{\rmT}}q_{2}^{\sigma'j-\frac{\sigma'+1}{2}-\sigma\lambda_{i}},
\end{split}\label{eq:appendix-Nekrasov-chYrepresentation1}
\end{align}
where we used 
\begin{align}
    \frac{(1-q_{1}^{-\sigma})(1-q_{2}^{\sigma'})}{(1-q_{1})(1-q_{2})}&=(-q_{1})^{-\frac{1+\sigma}{2}}(-q_{2})^{-\frac{1-\sigma'}{2}},\quad (-1)^{\frac{\sigma'-\sigma}{2}}=\sigma\sigma'.
\end{align}
Here we used (\ref{eq:appendix-chY-1}) for $\lambda$ and (\ref{eq:appendix-chY-2}) for $\nu$. Using (\ref{eq:appendix-chY-1}) and (\ref{eq:appendix-chY-2}) oppositely, we have
\begin{align}
    -\frac{\ch\bfY^{\sigma\vee}_{1}[\lambda]\ch\bfY^{\sigma'}_{2}[\nu]}{\ch\swedge\bfQ}&=\sigma\sigma'\frac{v_{2}}{v_{1}}\sum_{i,j=1}^{\infty}q_{1}^{\sigma'j-\frac{\sigma'+1}{2}-\sigma\lambda_{i}^{\rmT}}q_{2}^{-\sigma i+\frac{\sigma-1}{2}+\sigma'\nu_{j}}.\label{eq:appendix-Nekrasov-chYrepresentation2}
\end{align}
In fact, permuting $q_{1}\leftrightarrow q_{2}$ and taking the transpose $\lambda_{i}\rightarrow \lambda^{\rmT}_{i}$, $\nu_{j}\rightarrow \nu_{j}^{\rmT}$ of (\ref{eq:appendix-Nekrasov-chYrepresentation1}), we obtain (\ref{eq:appendix-Nekrasov-chYrepresentation2})

Taking the index, the Nekrasov factors will be written as 
\begin{align}
    N_{\sigma\sigma'}(v_{1},\lambda\,|\,v_{2},\nu)\coloneqq N_{\lambda\nu}^{\sigma\sigma'}(Q;q_{1},q_{2})&=\prod_{i,j=1}^{\infty}\left(\frac{1-Qq_{1}^{\sigma i-\frac{\sigma-1}{2}-\sigma'\nu_{j}^{\rmT}}q_{2}^{-\sigma'j+\frac{\sigma'+1}{2}+\sigma\lambda_{i}}}{1-Qq_{1}^{\sigma i-\frac{\sigma-1}{2}}q_{2}^{-\sigma' j+\frac{1+\sigma'}{2}}}\right)^{\sigma\sigma'}\\
    &=\prod_{i,j=1}^{\infty}\left(\frac{1-Qq_{1}^{-\sigma'j+\frac{\sigma'+1}{2}+\sigma\lambda_{i}^{\rmT}}q_{2}^{\sigma i-\frac{\sigma-1}{2}-\sigma'\nu_{j}}}{1-Qq_{1}^{-\sigma'j+\frac{\sigma'+1}{2}}q_{2}^{\sigma i-\frac{\sigma-1}{2}}}\right)^{\sigma\sigma'},
\end{align}
where $Q=v_{1}/v_{2}$.
Explicitly we have
\begin{align}
\begin{split}
    N_{\lambda\nu}^{++}(Q;q_{1},q_{2})&=\prod_{i,j=1}^{\infty}\frac{1-Qq_{1}^{i-\nu_{j}^{\rmT}}q_{2}^{-j+1+\lambda_{i}}}{1-Qq_{1}^{i}q_{2}^{-j+1}}=\prod_{i,j=1}^{\infty}\frac{1-Qq_{2}^{i-\nu_{j}}q_{1}^{-j+1+\lambda^{\rmT}_{i}}}{1-Qq_{2}^{i}q_{1}^{-j+1}},\\
    N_{\lambda\nu}^{+-}(Q;q_{1},q_{2})&=\prod_{i,j=1}^{\infty}\frac{1-Qq_{1}^{i}q_{2}^{j}}{1-Qq_{1}^{i+\nu_{j}^{\rmT}}q_{2}^{j+\lambda_{i}}}=\prod_{i,j=1}^{\infty}\frac{1-Qq_{2}^{i}q_{1}^{j}}{1-Qq_{2}^{i+\nu_{j}}q_{1}^{j+\lambda^{\rmT}_{i}}},\\
    N^{-+}_{\lambda\nu}(Q;q_{1},q_{2})&=\prod_{i,j=1}^{\infty}\frac{1-Qq_{1}^{-i+1}q_{2}^{-j+1}}{1-Qq_{1}^{-i+1-\nu_{j}^{\rmT}}q_{2}^{-j+1-\lambda_{i}}}=\prod_{i,j=1}^{\infty}\frac{1-Qq_{2}^{-i+1}q_{1}^{-j+1}}{1-Qq_{2}^{-i+1-\nu_{j}}q_{1}^{-j+1-\lambda^{\rmT}_{i}}},\\ N^{--}_{\lambda\nu}(Q;q_{1},q_{2})&=\prod_{i,j=1}^{\infty}\frac{1-Qq_{1}^{-i+1+\nu^{\rmT}_{j}}q_{1}^{j-\lambda_{i}}}{1-Qq_{1}^{-i+1}q_{2}^{j}}=\prod_{i,j=1}^{\infty}\frac{1-Qq_{2}^{-i+1+\nu_{j}}q_{1}^{j-\lambda^{\rmT}_{i}}}{1-Qq_{2}^{-i+1}q_{1}^{j}}.
\end{split}
\end{align}

We can also obtain the expressions using the $x$-variables:
\begin{align}
\begin{split}
    -\frac{\ch\bfY^{\sigma\vee}_{1}[\lambda]\ch\bfY^{\sigma'}_{2}[\nu]}{\ch\swedge\bfQ}&=-\frac{(1-q_{1}^{-\sigma})(1-q_{1}^{\sigma'})}{(1-q_{1})(1-q_{2})}\sum_{(x,x')\in\mathcal{X}^{\sigma}_{\lambda}\times\mathcal{X}^{\sigma'}_{\nu}}\frac{x'}{x}\\
    &=-\sigma(1-q_{1}^{\sigma'})q_{1}^{-\frac{1+\sigma}{2}}\sum_{(x,x')\in\mathcal{X}^{\sigma}_{\lambda}\times\mathcal{X}_{\nu}^{\sigma'}}\sum_{j=0}^{\infty}q_{2}^{-1-j}\frac{x'}{x}.
\end{split}
\end{align}
Taking the index we obtain the above formula. Explicitly, they are written as 
\begin{align}
    \mathbb{I}\left[-\frac{\ch\bfY^{\sigma\vee}_{1}[\lambda]\ch\bfY^{\sigma'}_{2}[\nu]}{\ch\swedge\bfQ}\right]&=\begin{dcases}
    \prod_{(x,x')\in\mathcal{X}_{\lambda}^{+}\times\mathcal{X}^{+}_{\nu}}\frac{(q_{2}x/x';q_{2})_{\infty}}{(q_{1}q_{2}x/x';q_{2})_{\infty}},\quad \sigma=\sigma'=+,\\
    \prod_{(x,x')\in\mathcal{X}^{+}_{\lambda}\times\mathcal{X}^{-}_{\nu}}\frac{(q_{1}q_{2}x/x';q_{2})_{\infty}}{(q_{1}^{2}q_{2}x/x';q_{2})_{\infty}},\quad \sigma=+,\,\sigma'=-,\\
    \prod_{(x,x')\in\mathcal{X}^{-}_{\lambda}\times\mathcal{X}^{+}_{\nu}}\frac{(q_{2}x/x';q_{2})_{\infty}}{(q_{1}^{-1}q_{2}x/x';q_{2})_{\infty}},\quad \sigma=-,\,\sigma'=+,\\
    \prod_{(x,x')\in\mathcal{X}^{-}_{\lambda}\times\mathcal{X}^{-}_{\nu}}\frac{(q_{2}x/x';q_{2})_{\infty}}{(q_{1}q_{2}x/x';q_{2})_{\infty}},\quad \sigma=\sigma'=-.
    \end{dcases}
\end{align}
For the transpose, we just need to change $q_{1}\leftrightarrow q_{2}$, $\lambda\rightarrow \lambda^{\rmT}$, $\nu\rightarrow \nu^{\rmT}$, and $\mathcal{X}^{\sigma}_{\lambda}\rightarrow \check{\mathcal{X}}^{\sigma}_{\lambda}$.
\end{proof}

\bclaim
For the diagonal Nekrasov factors $N^{++}_{\lambda\nu}(Q;q_{1},q_{2}), N^{--}_{\lambda\nu}(Q;q_{1},q_{2})$, we have another description using the arm and leg length
\begin{align}
\begin{split}
    N^{++}_{\lambda\nu}(Q;q_{1},q_{2})&=\prod_{\sAbox\,\in\lambda}\left(1-Qq_{1}^{l_{\lambda}(\sAbox)+1}q_{2}^{-a_{\nu}(\sAbox)}\right)\prod_{\sAbox\,\in\nu}\left(1-Qq_{1}^{-l_{\nu}(\sAbox)}q_{2}^{a_{\lambda}(\sAbox)+1}\right)\\
    &=\prod_{\sAbox\,\in\nu}\left(1-Qq_{1}^{l_{\lambda}(\sAbox)+1}q_{2}^{-a_{\nu}(\sAbox)}\right)\prod_{\sAbox\,\in\lambda}\left(1-Qq_{1}^{-l_{\nu}(\sAbox)}q_{2}^{a_{\lambda}(\sAbox)+1}\right),
\end{split}\label{eq:appendix-armleg-Nek}\\
N^{--}_{\lambda\nu}(Q;q_{1},q_{2})&=N^{++}_{\nu\lambda}(Q;q_{1},q_{2})\label{eq:appendix-diagonalsymm}.
\end{align}
In particular, for $\lambda=\nu, Q=1$, we have
\begin{align}
    N^{++}_{\lambda\lambda}(1;q_{1},q_{2})=N^{--}_{\lambda\lambda}(1;q_{1},q_{2}),
\end{align}
which means the vector multiplet contributions of $U(1\,|\,0)$ and $U(0\,|\,1)$ gauge theories are the same.
\eclaim
\begin{proof}
See for example \cite{Nakajimalecturebook, Flume:2002az,Kimura:instantoncounting} for the derivation of the first equation (\ref{eq:appendix-armleg-Nek}). 

The second equation (\ref{eq:appendix-diagonalsymm}) comes from 
\begin{align}
\begin{split}
    N_{--}(v_{1},\lambda\,|\,v_{2},\nu)&=\prod_{\sAbox\,\in\lambda}\left(1-\frac{\chi_{\sAbox}^{\sminus}}{q_{3}v_{2}}\right)\prod_{\sAbox\,\in\nu}\left(1-\frac{v_{1}}{\chi_{\sAbox}^{\sminus}}\right)\prod_{\substack{\sAbox\,\in\lambda\\\sAboxF\,\in\nu}}S\left(\frac{\chi_{\sAbox}^{\sminus}}{\chi_{\sAboxF}^{\sminus}}\right)\\
    &=\prod_{\sAbox\,\in\lambda}\left(1-\frac{v_{1}}{v_{2}q_{\sAbox}^{\splus}}\right)\prod_{\sAbox\,\in\nu}\left(1-\frac{v_{1}q_{\sAbox}^{\splus}}{q_{3}v_{2}}\right)\prod_{\substack{\sAbox\,\in\lambda\\\sAboxF\,\in\nu}}S\left(\frac{v_{1}q_{\sAboxF}^{\splus}}{v_{2}q_{\sAbox}^{\splus}}\right)\\
    &=N_{++}(v_{1},\nu\,|\,v_{2},\lambda)
\end{split}
\end{align}
where we used 
\begin{align}
    \chi_{\sAbox}^{\sminus}=v_{1}q_{\sAbox}^{\sminus}=v_{1}q_{3}/q_{\sAbox}^{\splus},\quad \Abox\in\lambda,\quad\chi_{\sAbox}^{\sminus}=v_{2}q_{\sAbox}^{\sminus}=v_{2}q_{3}/q_{\sAbox}^{\splus},\quad \Abox\in\nu.
\end{align}
\end{proof}

\subsection{Symmetries of Nekrasov factors}\label{sec:appendix-nekrasov-symm}
Let us study the symmetries of the Nekrasov factors.
\bclaim
We have the following symmetries of the Nekrasov factors:
\begin{align}
\begin{split}
    N^{\sigma\sigma'}_{\lambda\nu}(Q;q_{1},q_{2})&=(-Q)^{|\lambda|+|\nu|}q_{3}^{-\frac{1+\sigma}{2}|\lambda|-\frac{1-\sigma'}{2}|\nu|}q_{1}^{\sigma n(\lambda)-\sigma' n(\nu)}q_{2}^{\sigma n(\lambda^{\rmT})-\sigma' n(\nu^{\rmT})}\\
    &\qquad\times N^{\sigma'\sigma}_{\nu\lambda}(q_{3}Q^{-1};q_{1},q_{2}),
\end{split} \label{eq:Nekrasovsymm1}\\
    N^{\sigma\sigma'}_{\lambda\nu}(Q;q_{1}^{-1},q_{2}^{-1})&=N^{\sigma'\sigma}_{\nu\lambda}(q_{3}Q;q_{1},q_{2})\label{eq:Nekrasovsymm2}
\end{align}
\eclaim
\begin{proof}
Using (\ref{eq:Nekrasovdef}), we have
\begin{align}
    N^{\sigma\sigma'}_{\lambda\nu}(Q;q_{1},q_{2})=\prod_{x\in\lambda}\left(-\frac{\chi_{x}^{(\sigma)}}{q_{3}v_{2}}\right)\prod_{x\in\nu}\left(-\frac{v_{1}}{\chi_{x}^{(\sigma')}}\right)N^{\sigma'\sigma}_{\nu\lambda}(q_{3}Q^{-1};q_{1},q_{2}).
\end{align}
Applying the identities
\begin{align}
    \prod_{x\in\lambda}\left(-\frac{\chi_{x}^{(\sigma)}}{q_{3}v_{2}}\right)&=(-Q)^{|\lambda|}q_{3}^{-\frac{1+\sigma}{2}|\lambda|}q_{1}^{\sigma n(\lambda)}q_{2}^{\sigma n(\lambda^{\rmT})},\\
    \prod_{x\in\nu}\left(-\frac{v_{1}}{\chi_{x}^{(\sigma')}}\right)&=(-Q)^{|\nu|}q_{3}^{-\frac{1-\sigma'}{2}}q_{1}^{-\sigma'n(\nu)}q_{2}^{-\sigma' n(\nu^{\rmT})},\\
    \prod_{x\in\lambda}\chi_{x}^{(\sigma)}&=v^{|\lambda|}q_{3}^{\frac{1-\sigma}{2}|\lambda|}q_{1}^{\sigma n(\lambda)}q_{2}^{\sigma n(\lambda^{\rmT})}
\end{align}
we obtain (\ref{eq:Nekrasovsymm1}).

The second indentity (\ref{eq:Nekrasovsymm2}) is obtained as 
\begin{align}
\begin{split}
    N^{\sigma\sigma'}_{\lambda\nu}(Q;q_{1}^{-1},q_{2}^{-1})&=\prod_{x\in\lambda}\left(1-Q\frac{q_{3}}{q_{x}^{(\sigma)}(q_{1},q_{2})}\right)\prod_{x\in\nu}\left(1-Q q_{x}^{(\sigma')}(q_{1},q_{2})\right)\prod_{\substack{x\in\lambda\\ y\in\nu}}S\left(Q\frac{q_{y}^{(\sigma')}(q_{1},q_{2})}{q_{x}^{(\sigma)}(q_{1},q_{2})};q_{1}^{-1},q_{2}^{-1}\right)\\
    &=\prod_{x\in\lambda}\left(1-Q\frac{q_{3}}{q_{x}^{(\sigma)}(q_{1},q_{2})}\right)\prod_{x\in\nu}\left(1-Q q_{x}^{(\sigma')}(q_{1},q_{2})\right)\prod_{\substack{x\in\lambda\\ y\in\nu}}S\left(Qq_{3}\frac{q_{y}^{(\sigma')}(q_{1},q_{2})}{q_{x}^{(\sigma)}(q_{1},q_{2})};q_{1},q_{2}\right)\\
    &=N^{\sigma'\sigma}_{\nu\lambda}(Qq_{3};q_{1},q_{2})
\end{split}
\end{align}
where we explicitly wrote the $q_{1},q_{2}$ dependence of the $q$-coordinates and the scattering function $S(z)$ and used
\begin{align}
    &q_{x}^{(\sigma)}(q_{1}^{-1},q_{2}^{-1})=1/q_{x}^{(\sigma)}(q_{1},q_{2}),\\
    &S(z;q_{1}^{-1},q_{2}^{-1})=S(z^{-1};q_{1},q_{2})=S(q_{3}z;q_{1},q_{2}).
\end{align}
\end{proof}

\subsection{Formulas for refined topological vertex}\label{sec:appendix-formula-topver}
We summarize the formulas we used to derive the identification of intertwiners and topological vertices. We set the deformation parameters as $q_{1}=q,\,q_{2}=t^{-1}$.
\bclaim
By direct calculation, we have
\begin{align}
\begin{split}
    (v\gamma)^{-|\lambda|}\prod_{x\in\lambda}\chi_{x}^{\splus}&=q^{\frac{1}{2}||\lambda^{\rmT}||^{2}}t^{-\frac{1}{2}||\lambda||^{2}},\\
    (q_{3}v\gamma)^{-|\lambda|}\prod_{x\in\lambda}\chi_{x}^{\sminus}&=q_{3}^{-|\lambda|}q^{-\frac{1}{2}||\lambda^{\rmT}||^{2}}t^{\frac{1}{2}||\lambda||^{2}},
\end{split}\\
\begin{split}
    N_{++}(v,\lambda\,|\,v,\lambda)&=(-\gamma)^{-|\lambda|}q^{-\frac{1}{2}||\lambda^{\rmT}||^{2}}t^{-\frac{1}{2}||\lambda||^{2}}\wt{Z}^{-1}_{\lambda}(q,t)\wt{Z}^{-1}_{\lambda^{\rmT}}(t,q),\\
    N_{--}(v,\lambda\,|\,v,\lambda)
    &=(-\gamma)^{-|\lambda|}q^{\frac{1}{2}||\lambda^{\rmT}||^{2}}t^{\frac{1}{2}||\lambda||^{2}}\wt{Z}^{-1}_{\lambda^{\rmT}}(t^{-1},q^{-1})\wt{Z}^{-1}_{\lambda}(q^{-1},t^{-1})\\
    &=N_{++}(v,\lambda\,|\,v,\lambda),
\end{split}\\
\begin{split}
    a_{\lambda}^{\splus}&=(-\gamma)^{|\lambda|}q^{||\lambda^{\rmT}||^{2}}\wt{Z}_{\lambda}(q,t)\wt{Z}_{\lambda^{\rmT}}(t,q),\\
    a_{\lambda}^{\sminus}&=(-\gamma)^{-|\lambda|}q^{-||\lambda^{\rmT}||^{2}}\wt{Z}_{\lambda^{\rmT}}(t^{-1},q^{-1})\wt{Z}_{\lambda}(q^{-1},t^{-1}),
\end{split}
\end{align}
where 
\begin{align}
    \wt{Z}_{\lambda}(t,q)=\prod_{(i,j)\in\lambda}(1-t^{l_{\lambda}(i,j)+1}q^{a_{\lambda}(i,j)})^{-1}.
\end{align}
\eclaim

\section{Derivation of level \texorpdfstring{$(0,-1)$}{(0,-1)} representation}\label{sec:appendix-level(0,-1)rep}
We first give the following ansatz
\begin{align}
\begin{split}
    x^{+}(z)\dket{v,\lambda}&=\sum_{x\in R(\lambda)}\delta\left(z/\chi_{x}^{\sminus}\right)R^{\sminus}_{\lambda}(x)\dket{v,\lambda-x},\\
    x^{-}(z)\dket{v,\lambda}&=\sum_{x\in A(\lambda)}\delta\left(z/\chi_{x}^{\sminus}\right)A_{\lambda}^{\sminus}(x)\dket{v,\lambda+x},\\
    \psi^{\pm}(z)\dket{v,\lambda}&=\left[\Psi^{\sminus}_{\lambda}(z)\right]_{\pm}\dket{v,\lambda}.
\end{split}
\end{align}
Let us first derive the eigenvalue $\Psi_{\lambda}^{\sminus}(z)$. Inserting these equations to the definition of the algebra, we have the following recursion formula
\begin{align}
    \frac{\Psi_{\lambda}^{\sminus}(z)}{\Psi_{\lambda-x}^{\sminus}(z)}=g\left(\frac{\chi_{x}^{\sminus}}{z}\right)=\frac{S(\chi_{x}^{\sminus}/z)}{S(q_{3}\chi_{x}^{\sminus}/z)}
\end{align}
and obtain
\begin{align}
    \Psi^{\sminus}_{\lambda}(z)=\Psi^{\sminus}_{\emptyset}(z)\prod_{x\in\lambda}g\left(\frac{\chi_{x}^{\sminus}}{z}\right).
\end{align}
The vacuum charge function is defined so that the there is a pole in the origin of the Young diagram with coordinate $q_{3}v$ and a zero in $v$, which terminate the growing of the crystal in the third direction. This gives 
\begin{align}
    \Psi_{\lambda}^{\sminus}(z)=\gamma\frac{(1-v/z)\prod_{x\in\lambda}S(\chi_{x}^{\sminus}/z)}{(1-q_{3}v/z)\prod_{x\in\lambda}S(q_{3}\chi_{x}^{\sminus}/z)}=\gamma\frac{\mathcal{Y}^{\sminus}_{\lambda}(z)}{\mathcal{Y}^{\sminus}_{\lambda}(q_{3}^{-1}z)}.
\end{align}
The next thing we need to do is determine the coefficients $A^{\sminus}_{\lambda}(x),R^{\sminus}_{\lambda}(x)$. Using the $x^{\pm}(z)$ relation and the poles structure of $\Psi_{\lambda}^{\sminus}(z)$, we have
\begin{align}
\begin{split}
    A^{\sminus}_{\lambda}(x)R^{\sminus}_{\lambda+x}(x)&=\gamma\frac{(1-q_{1})(1-q_{2})}{(1-q_{3}^{-1})}\underset{z=\chi_{x}^{\ssminus}}{\Res}z^{-1}\mathcal{Y}_{\lambda}^{\sminus}(q_{3}^{-1}z)^{-1}\,\mathcal{Y}_{\lambda}^{\sminus}(\chi_{x}^{\sminus}),\quad x\in A(\lambda),\\
    R^{\sminus}_{\lambda}(x)A^{\sminus}_{\lambda-x}(x)&=-\gamma\frac{(1-q_{1})(1-q_{2})}{(1-q_{3}^{-1})}\underset{z=\chi_{x}^{\ssminus}}{\Res}z^{-1}\mathcal{Y}_{\lambda}^{\sminus}(z)\,\mathcal{Y}_{\lambda}^{\sminus}(q_{3}^{-1}\chi_{x}^{\sminus})^{-1},\quad x\in R(\lambda).
\end{split}
\end{align}
An answer for these relations is 
\begin{align}
    A^{\sminus}_{\lambda}(x)=\underset{z=\chi_{x}^{\sminus}}{\Res}z^{-1}\mathcal{Y}_{\lambda}^{\sminus}(q_{3}^{-1}z)^{-1},\quad R^{\sminus}_{\lambda}(x)=\gamma\underset{z=\chi_{x}^{\sminus}}{\Res}z^{-1}\mathcal{Y}_{\lambda}^{\sminus}(z).
\end{align}
One can check this indeed satisfy the above relations.

\section{Contraction formulas}\label{sec:appendix-contraction}
\subsection{Vertex operator representations}\label{sec:appendix-vertexcontr}
The normal ordering formulas are 
\begin{align}
\varphi^{+}(z)\varphi^{-}(w)&=\frac{g(\gamma z/w)}{g(\gamma^{-1}z/w)}:\varphi^{-}(w)\varphi^{+}(z):,\\
    \eta(z)\eta(w)&=\frac{(1-w/z)(1-q_{3}^{-1}w/z)}{(1-q_{1}w/z)(1-q_{2}w/z)}:\eta(z)\eta(w):=S(w/z)^{-1}:\eta(z)\eta(w):,\\
    \xi(z)\xi(w)&=\frac{(1-q_{3}w/z)(1-w/z)}{(1-q_{1}^{-1}w/z)(1-q_{2}^{-1}w/z)}:\xi(z)\xi(w):=S(q_{3}w/z)^{-1}:\xi(z)\xi(w):,\\
    \eta(z)\xi(w)&=\frac{(1-q_{1}\gamma w/z)(1-q_{2}\gamma w/z)}{(1-\gamma w/z)(1-\gamma^{-1}w/z)}:\eta(z)\xi(w):=S(\gamma w/z):\eta(z)\xi(w):,\\
    \xi(z)\eta(w)&=\frac{(1-q_{1}\gamma w/z)(1-q_{2}\gamma w/z)}{(1-\gamma w/z)(1-\gamma^{-1} w/z)}:\xi(z)\eta(w):=S(\gamma w/z):\xi(z)\eta(w):.
\end{align}
\begin{align}
    \eta(z)\varphi^{+}(w)&=:\eta(z)\varphi^{+}(w):,\\
    \varphi^{+}(z)\eta(w)&=g\bl(\gamma^{-1/2}w/z\br)^{-1}:\varphi^{+}(z)\eta(w):=\frac{S(\gamma^{3/2}w/z)}{S(\gamma^{-1/2}w/z)}:\varphi^{+}(z)\eta(w):,\\
    \varphi^{-}(z)\eta(w)&=:\varphi^{-}(z)\eta(w):,\\
    \eta(z)\varphi^{-}(w)&=g\bl(\gamma^{-1/2}w/z\br)^{-1}:\varphi^{-}(w)\eta(z):=\frac{S(\gamma^{3/2}w/z)}{S(\gamma^{-1/2}w/z)}:\eta(z)\varphi^{-}(w):,\\
    \xi(z)\varphi^{+}(w)&=:\xi(z)\varphi^{+}(w):,\\
    \varphi^{+}(z)\xi(w)&=g\bl(\gamma^{1/2}w/z\br):\varphi^{+}(z)\xi(w):=\frac{S(\gamma^{1/2}w/z)}{S(\gamma^{5/2}w/z)}:\varphi^{+}(z)\xi(w):,\\
    \varphi^{-}(z)\xi(w)&=:\varphi^{-}(z)\xi(w):,\\
    \xi(z)\varphi^{-}(w)&=g\bl(\gamma^{\frac{1}{2}}w/z\br):\xi(z)\varphi^{-}(w):=\frac{S(\gamma^{1/2}w/z)}{S(\gamma^{5/2}w/z)}:\xi(z)\varphi^{-}(w):,\\
    \varphi^{+}(z)&=:\eta(z\gamma^{1/2})\xi(z\gamma^{-1/2}):,\\
    \varphi^{-}(z)&=:\eta(z\gamma^{-1/2})\xi(z\gamma^{1/2}):.,\\
    \gamma^{5/2}&=q_{3}\gamma^{1/2},\\
    \gamma^{2}&=q_{3},\\
    \gamma^{3/2}&=q_{3}\gamma^{-1/2}.
\end{align}

\subsection{Contractions of intertwiners}\label{sec:appendix-contraction-intertwiner}
The normal ordering formulas for the Drinfeld currents and the intertwiners are
\begin{align}
\eta(z)\Phi_{\sigma,\emptyset}[v]&=\left(1-v/z\right)^{-\sigma}:\eta(z)\Phi_{\sigma,\emptyset}[v]:,\\
\Phi_{\sigma,\emptyset}[v]\eta(z)&=\left(1-q_{3}^{-1}z/v\right)^{-\sigma}:\Phi_{\sigma,\emptyset}[v]\eta(z):,\\
\xi(z)\Phi_{\sigma,\emptyset}[v]&=\bl(1-\gamma v/z\br)^{\sigma}:\xi(z)\Phi_{\sigma,\emptyset}[v]:,\\
\Phi_{\sigma,\emptyset}[v]\xi(z)&=\bl(1-\gamma^{-1}z/v\br)^{\sigma}:\Phi_{\sigma,\emptyset}[v]\xi(z):,\\
\varphi^{-}(z)\Phi_{\sigma,\emptyset}[v]&=:\varphi^{-}(z)\Phi_{\sigma,\emptyset}[v]:,\\
\Phi_{\sigma,\emptyset}[v]\varphi^{-}(z)&=\left(\frac{1-\gamma^{-1/2}z/v}{1-\gamma^{-5/2}z/v}\right)^{\sigma}:\Phi_{\sigma,\emptyset}[v]\varphi^{-}(z):,\\
\varphi^{+}(z)\Phi_{\sigma,\emptyset}[v]&=\left(\frac{1-\gamma^{3/2}v/z}{1-\gamma^{-1/2}v/z}\right)^{\sigma}:\varphi^{+}(z)\Phi_{\sigma,\emptyset}[v]:,\\
\Phi_{\sigma,\emptyset}[v]\varphi^{+}(z)&=:\Phi_{\sigma,\emptyset}[v]\varphi^{+}(z):,
\end{align}
and 
\begin{align}
    \eta(z)\Phi^{\ast}_{\sigma,\emptyset}[v]&=\bl(1-\gamma v/z\br)^{\sigma}:\eta(z)\Phi^{\ast}_{\sigma,\emptyset}[v]:,\\
    \Phi^{\ast}_{\sigma,\emptyset}[v]\eta(z)&=\bl(1-\gamma^{-1}z/v\br)^{\sigma}:\Phi^{\ast}_{\sigma,\emptyset}[v]\eta(z):,\\
    \xi(z)\Phi^{\ast}_{\sigma,\emptyset}[v]&=\bl(1-q_{3}v/z\br)^{-\sigma}:\Phi^{\ast}_{\sigma,\emptyset}[v]\xi(z):,\\
    \Phi^{\ast}_{\sigma,\emptyset}[v]\xi(z)&=\bl(1-z/v\br)^{-\sigma}:\Phi^{\ast}_{\sigma,\emptyset}[v]\xi(z):,\\
    \varphi^{-}(z)\Phi^{\ast}_{\sigma,\emptyset}[v]&=:\varphi^{-}(z)\Phi^{\ast}_{\sigma,\emptyset}[v]:,\\
    \Phi^{\ast}_{\sigma,\emptyset}[v]\varphi^{-}(z)&=\frac{1-\gamma^{-3/2}z/v}{1-\gamma^{1/2}z/v}:\Phi^{\ast}_{\sigma,\emptyset}[v]\varphi^{-}(z):,\\
    \varphi^{+}(z)\Phi^{\ast}_{\sigma,\emptyset}[v]&=\frac{1-\gamma^{1/2}v/z}{1-\gamma^{5/2}v/z}:\varphi^{+}(z)\Phi^{\ast}_{\sigma,\emptyset}[v]:,\\
    \Phi^{\ast}_{\sigma,\emptyset}[v]\varphi^{+}(z)&=:\Phi^{\ast}_{\sigma,\emptyset}[v]\varphi^{+}(z):.
\end{align}
We also have the following contraction formulas
\begin{align}
    \eta(z)\Phi_{\sigma,\lambda}[u,v]&=\left[\mathcal{Y}^{(\sigma)}_{\lambda}(z)^{-\sigma}\right]_{+}:\eta(z)\Phi_{\sigma,\lambda}[u,v]:,\\
    \Phi_{\sigma,\lambda}[u,v]\eta(z)&=\left[\left(-\frac{vq_{3}}{z}\frac{1}{\mathcal{Y}^{(\sigma)}_{\lambda}(q_{3}^{-1}z)}\right)^{\sigma}\right]_{-}:\Phi_{\sigma,\lambda}[u,v]\eta(z):,\\
    \xi(z)\Phi_{\sigma,\lambda}[u,v]&=\left[\mathcal{Y}^{(\sigma)}_{\lambda}(\gamma^{-1}z)^{\sigma}\right]_{+}:\xi(z)\Phi_{\sigma,\lambda}[u,v]:,\\
    \Phi_{\sigma,\lambda}[u,v]\xi(z)&=\left[\left(-\frac{\gamma^{-1}z}{v}\mathcal{Y}^{(\sigma)}_{\lambda}(\gamma^{-1}z)\right)^{\sigma}\right]_{-}:\Phi_{\sigma,\lambda}[u,v]\xi(z):,\\
    \varphi^{-}(z)\Phi_{\sigma,\lambda}[u,v]&=:\varphi^{-}(z)\Phi_{\sigma,\lambda}[u,v]:,\\
   \begin{split} \Phi_{\sigma,\lambda}[u,v]\varphi^{-}(z)&=\left[\left(\gamma^{2}\frac{\mathcal{Y}^{(\sigma)}_{\lambda}(\gamma^{-1/2}z)}{\mathcal{Y}^{(\sigma)}_{\lambda}(\gamma^{-5/2}z)}\right)^{\sigma}\right]_{-}:\Phi_{\sigma,\lambda}[u,v]\varphi^{-}(z):\\
    &=\left[\gamma^{\sigma} {\Psi^{(\sigma)}_{\lambda}(\gamma^{-1/2}z)}^{-\sigma}\right]_{-}:\Phi_{\sigma,\lambda}[u,v]\varphi^{-}(z):,
    \end{split}\\
    \begin{split}
    \varphi^{+}(z)\Phi_{\sigma,\lambda}[u,v]&=\left[\left(\frac{\mathcal{Y}^{(\sigma)}_{\lambda}(\gamma^{-3/2}z)}{\mathcal{Y}^{(\sigma)}_{\lambda}(\gamma^{1/2}z)}\right)^{\sigma}\right]_{+}:\varphi^{+}(z)\Phi_{\sigma,\lambda}[u,v]:\\
    &=\left[\gamma^{\sigma}\Psi^{(\sigma)}_{\lambda}(\gamma^{1/2}z)^{\sigma}\right]_{+}:\varphi^{+}(z)\Phi_{\sigma,\lambda}[u,v]:,
    \end{split}\\
    \Phi_{\sigma,\lambda}[u,v]\varphi^{+}(z)&=:\Phi_{\sigma,\lambda}[u,v]\varphi^{+}(z):
\end{align}
and 
\begin{align}
\eta(z)\Phi_{\sigma,\lambda}^{(n)\ast}[u,v]&=\left[\mathcal{Y}^{(\sigma)}_{\lambda}(\gamma^{-1}z)^{\sigma}\right]_{+}:\eta(z)\Phi_{\sigma,\lambda}^{(n)\ast}[u,v]:,\\
\Phi_{\sigma,\lambda}^{(n)\ast}[u,v]\eta(z)&=\left[\left(-\frac{\gamma^{-1}z}{v}\mathcal{Y}^{(\sigma)}_{\lambda}(\gamma^{-1}z)\right)^{\sigma}\right]_{-}:\Phi_{\sigma,\lambda}^{(n)\ast}[u,v]\eta(z):,\\
\xi(z)\Phi_{\sigma,\lambda}^{(n)\ast}[u,v]&=\left[\mathcal{Y}^{(\sigma)}_{\lambda}(q_{3}^{-1}z)^{-\sigma}\right]_{+}:\xi(z)\Phi_{\sigma,\lambda}^{(n)\ast}[u,v]:,\\
\Phi_{\sigma,\lambda}^{(n)\ast}[u,v]\xi(z)&=\left[\left(-\frac{v}{z}\frac{1}{\mathcal{Y}^{(\sigma)}_{\lambda}(z)}\right)^{\sigma}\right]_{-}:\Phi_{\sigma,\lambda}^{(n)\ast}[u,v]\xi(z):,\\
\varphi^{-}(z)\Phi_{\sigma,\lambda}^{(n)\ast}[u,v]&=:\varphi^{-}(z)\Phi_{\sigma,\lambda}^{(n)\ast}[u,v]:,\\
\begin{split}
\Phi_{\sigma,\lambda}^{(n)\ast}[u,v]\varphi^{-}(z)&=\left[\left(\gamma^{-2}\frac{\mathcal{Y}^{(\sigma)}_{\lambda}(\gamma^{-3/2}z)}{\mathcal{Y}^{(\sigma)}_{\lambda}(\gamma^{1/2}z)}\right)^{\sigma}\right]_{-}:\Phi_{\sigma,\lambda}^{(n)\ast}[u,v]\varphi^{-}(z):\\
&=\left[\gamma^{-\sigma}\Psi^{(\sigma)}_{\lambda}(\gamma^{1/2}z)^{\sigma}\right]_{-}:\Phi_{\sigma,\lambda}^{(n)\ast}[u,v]\varphi^{-}(z):,
\end{split}\\
\begin{split}
\varphi^{+}(z)\Phi_{\sigma,\lambda}^{(n)\ast}[u,v]&=\left[\left(\frac{\mathcal{Y}^{(\sigma)}_{\lambda}(\gamma^{-1/2}z)}{\mathcal{Y}^{(\sigma)}_{\lambda}(\gamma^{-5/2}z)}\right)^{\sigma}\right]_{+}:\varphi^{+}(z)\Phi_{\sigma,\lambda}^{(n)\ast}[u,v]:\\
&=\left[\gamma^{-\sigma}{\Psi^{(\sigma)}_{\lambda}(\gamma^{-1/2}z)}^{-\sigma}\right]_{+}:\varphi^{+}(z)\Phi_{\sigma,\lambda}^{(n)\ast}[u,v]:,
\end{split}\\
\Phi_{\sigma,\lambda}^{(n)\ast}[u,v]\varphi^{+}(z)&=:\Phi_{\sigma,\lambda}^{(n)\ast}[u,v]\varphi^{+}(z):.
\end{align}

The contraction of the intertwiners are 
\begin{align}
    \contraction{}{\Phi_{\sigma',\emptyset}[v_{2}]}{}{\Phi_{\sigma,\emptyset}[v_{1}]}\Phi_{\sigma',\emptyset}[v_{2}]\Phi_{\sigma,\emptyset}[v_{1}]&=\bl(\cG(q_{3}^{-1}v_{1}/v_{2})\br)^{\sigma\sigma'}:\Phi_{\sigma',\emptyset}[v_{2}]\Phi_{\sigma,\emptyset}[v_{1}]:,\\
    \contraction{}{\Phi^{\ast}_{\sigma',\emptyset}[v_{2}]}{}{\Phi^{\ast}_{\sigma,\emptyset}[v_{1}]}\Phi^{\ast}_{\sigma',\emptyset}[v_{2}]\Phi^{\ast}_{\sigma,\emptyset}[v_{1}]&=\bl(\cG(v_{1}/v_{2})\br)^{\sigma\sigma'}:\Phi^{\ast}_{\sigma',\emptyset}[v_{2}]\Phi^{\ast}_{\sigma,\emptyset}[v_{1}]:,\\
    \contraction{}{\Phi_{\sigma',\emptyset}[v_{2}]}{}{\Phi^{\ast}_{\sigma,\emptyset}[v_{1}]}\Phi_{\sigma',\emptyset}[v_{2}]\Phi^{\ast}_{\sigma,\emptyset}[v_{1}]&=\bl(\cG(\gamma^{-1}v_{1}/v_{2})\br)^{-\sigma\sigma'}:\Phi_{\sigma',\emptyset}[v_{2}]\Phi^{\ast}_{\sigma,\emptyset}[v_{1}]:,\\
    \contraction{}{\Phi^{\ast}_{\sigma',\emptyset}[v_{2}]}{}{\Phi_{\sigma,\emptyset}[v_{1}]}\Phi^{\ast}_{\sigma',\emptyset}[v_{2}]\Phi_{\sigma,\emptyset}[v_{1}]&=\bl(\cG(\gamma^{-1}v_{1}/v_{2})\br)^{-\sigma\sigma'}:\Phi^{\ast}_{\sigma',\emptyset}[v_{2}]\Phi_{\sigma,\emptyset}[v_{1}]:
\end{align}
and 
\begin{align}
    \contraction{}{\Phi_{\sigma',\mu}^{(n_{2})}[u_{2},v_{2}]}{}{\Phi^{(n_{1})}_{\sigma,\lambda}[u_{1},v_{1}]}\Phi_{\sigma',\mu}^{(n_{2})}[u_{2},v_{2}]\Phi^{(n_{1})}_{\sigma,\lambda}[u_{1},v_{1}]&=\bl(\cG(q_{3}^{-1}v_{1}/v_{2})\br)^{\sigma\sigma'}N_{\sigma\sigma'}(v_{1},\lambda\,|\,v_{2},\mu)^{-\sigma\sigma'}:\Phi_{\sigma',\mu}^{(n_{2})}[u_{2},v_{2}]\Phi^{(n_{1})}_{\sigma,\lambda}[u_{1},v_{1}]:,\\
      \contraction{}{\Phi_{\sigma',\mu}^{(n_{2})\ast}[u_{2},v_{2}]}{}{\Phi^{(n_{1})\ast}_{\sigma,\lambda}[u_{1},v_{1}]}\Phi_{\sigma',\mu}^{(n_{2})\ast}[u_{2},v_{2}]\Phi^{(n_{1})\ast}_{\sigma,\lambda}[u_{1},v_{1}]&=\bl(\cG(v_{1}/v_{2})\br)^{\sigma\sigma'}N_{\sigma \sigma'}(q_{3}v_{1},\lambda\,|\,v_{2},\mu)^{-\sigma\sigma'}:\Phi_{\sigma',\mu}^{(n_{2})\ast}[u_{2},v_{2}]\Phi^{(n_{1)\ast}}_{\sigma,\lambda}[u_{1},v_{1}]:,\\
       \contraction{}{\Phi_{\sigma',\mu}^{(n_{2})\ast}[u_{2},v_{2}]}{}{\Phi^{(n_{1})}_{\sigma,\lambda}[u_{1},v_{1}]}\Phi_{\sigma',\mu}^{(n_{2})\ast}[u_{2},v_{2}]\Phi^{(n_{1})}_{\sigma,\lambda}[u_{1},v_{1}]&=\bl(\cG(\gamma^{-1}v_{1}/v_{2})\br)^{-\sigma\sigma'}N_{\sigma\sigma'}(\gamma v_{1},\lambda\,|\,v_{2},\mu)^{\sigma\sigma'}:\Phi_{\sigma',\mu}^{(n_{2})\ast}[u_{2},v_{2}]\Phi^{(n_{1})}_{\sigma,\lambda}[u_{1},v_{1}]:,\\
        \contraction{}{\Phi_{\sigma',\mu}^{(n_{2})}[u_{2},v_{2}]}{}{\Phi^{(n_{1})\ast}_{\sigma,\lambda}[u_{1},v_{1}]}\Phi_{\sigma',\mu}^{(n_{2})}[u_{2},v_{2}]\Phi^{(n_{1})\ast}_{\sigma,\lambda}[u_{1},v_{1}]&=\bl(\cG(\gamma^{-1}v_{1}/v_{2})\br)^{-\sigma\sigma'}N_{\sigma\sigma'}(\gamma v_{1},\lambda\,|\,v_{2},\mu)^{\sigma\sigma'}:\Phi_{\sigma',\mu}^{(n_{2})}[u_{2},v_{2}]\Phi^{(n_{1})\ast}_{\sigma,\lambda}[u_{1},v_{1}]:,
\end{align}
where $\sigma,\sigma'=\pm$.
\section{Intertwiner property (AFS property)}\label{sec:appendix-AFSproperty}
We illustrate the AFS properties of the positive and negative intertwiners using diagrams. For the intertwiner, we have the following three properties:
\begin{enumerate}
    \item $x^{+}(z)$
    \begin{align}
    \adjustbox{valign=c}{
        \begin{tikzpicture}[]
            \begin{scope}[scale=1.5]
                \node [above, scale=0.8] at (-1,0.8){$ $};
                \node [below,scale=0.8] at (-1,-0.8){$x^{+}(z)$};
                \node [right,scale=0.8] at (0,0){$ $};
                \draw[postaction={segment={mid arrow}}] (0,0) -- (-1,0);
		        \draw[postaction={segment={mid arrow}}] (-1,0.8) -- (-1,0)--(-1,-0.8);   
            \end{scope}
        \end{tikzpicture}}\quad=\quad\adjustbox{valign=c}{\begin{tikzpicture}[]
            \begin{scope}[scale=1.5]
                \node [above, scale=0.8] at (-1,0.8){$ $};
                \node [below,scale=0.8] at (-1,-0.8){$ $};
                \node [right,scale=0.8] at (0,0){$x^{+}(z)$};
                \draw[postaction={segment={mid arrow}}] (0,0) -- (-1,0);
		        \draw[postaction={segment={mid arrow}}] (-1,0.8) -- (-1,0)--(-1,-0.8);   
            \end{scope}
        \end{tikzpicture}}\quad+\quad \adjustbox{valign=c}{\begin{tikzpicture}[]
            \begin{scope}[scale=1.5]
                \node [above, scale=0.8] at (-1,0.8){$x^{+}(z)$};
                \node [below,scale=0.8] at (-1,-0.8){$ $};
                \node [right,scale=0.8] at (0,0){$\psi^{-}(z)$};
                \draw[postaction={segment={mid arrow}}] (0,0) -- (-1,0);
		        \draw[postaction={segment={mid arrow}}] (-1,0.8) -- (-1,0)--(-1,-0.8);   
            \end{scope}
        \end{tikzpicture}}
    \end{align}
    \item $x^{-}(z)$
    \begin{align}
    \adjustbox{valign=c}{
        \begin{tikzpicture}[]
            \begin{scope}[scale=1.5]
                \node [above, scale=0.8] at (-1,0.8){$ $};
                \node [below,scale=0.8] at (-1,-0.8){$x^{-}(z)$};
                \node [right,scale=0.8] at (0,0){$ $};
                \draw[postaction={segment={mid arrow}}] (0,0) -- (-1,0);
		        \draw[postaction={segment={mid arrow}}] (-1,0.8) -- (-1,0)--(-1,-0.8);   
            \end{scope}
        \end{tikzpicture}}\quad=\quad\adjustbox{valign=c}{\begin{tikzpicture}[]
            \begin{scope}[scale=1.5]
                \node [above, scale=0.8] at (-1,0.8){$\psi^{+}(\gamma^{1/2}z) $};
                \node [below,scale=0.8] at (-1,-0.8){$ $};
                \node [right,scale=0.8] at (0,0){$x^{-}(\gamma z)$};
                \draw[postaction={segment={mid arrow}}] (0,0) -- (-1,0);
		        \draw[postaction={segment={mid arrow}}] (-1,0.8) -- (-1,0)--(-1,-0.8);   
            \end{scope}
        \end{tikzpicture}}\quad+\quad \adjustbox{valign=c}{\begin{tikzpicture}[]
            \begin{scope}[scale=1.5]
                \node [above, scale=0.8] at (-1,0.8){$x^{-}(z)$};
                \node [below,scale=0.8] at (-1,-0.8){$ $};
                \node [right,scale=0.8] at (0,0){$ $};
                \draw[postaction={segment={mid arrow}}] (0,0) -- (-1,0);
		        \draw[postaction={segment={mid arrow}}] (-1,0.8) -- (-1,0)--(-1,-0.8);   
            \end{scope}
        \end{tikzpicture}}
    \end{align}
    \item $\psi^{\pm}(z)$
    \begin{align}
    \adjustbox{valign=c}{
        \begin{tikzpicture}[]
            \begin{scope}[scale=1.5]
                \node [above, scale=0.8] at (-1,0.8){$ $};
                \node [below,scale=0.8] at (-1,-0.8){$\psi^{\pm}(z)$};
                \node [right,scale=0.8] at (0,0){$ $};
                \draw[postaction={segment={mid arrow}}] (0,0) -- (-1,0);
		        \draw[postaction={segment={mid arrow}}] (-1,0.8) -- (-1,0)--(-1,-0.8);   
            \end{scope}
        \end{tikzpicture}}\quad=\quad\adjustbox{valign=c}{\begin{tikzpicture}[]
            \begin{scope}[scale=1.5]
                \node [above, scale=0.8] at (-1,0.8){$\psi^{\pm}(z)$};
                \node [below,scale=0.8] at (-1,-0.8){$ $};
                \node [right,scale=0.8] at (0,0){$ \psi^{\pm}(\gamma^{\pm 1/2}z)$};
                \draw[postaction={segment={mid arrow}}] (0,0) -- (-1,0);
		        \draw[postaction={segment={mid arrow}}] (-1,0.8) -- (-1,0)--(-1,-0.8);   
            \end{scope}
        \end{tikzpicture}}
    \end{align}
\end{enumerate}

For the dual intertwiner, we have 
\begin{enumerate}
    \item $x^{-}(z)$
    \begin{align}
        \adjustbox{valign=c}{
        \begin{tikzpicture}[]
            \begin{scope}[scale=1.5]
                \node [above, scale=0.8] at (-1,0.8){$x^{+}(z)$};
                \node [below,scale=0.8] at (-1,-0.8){$ $};
                \node [left,scale=0.8] at (-2,0){$ $};
                \draw[postaction={segment={mid arrow}}] (-1,0) -- (-2,0);
		        \draw[postaction={segment={mid arrow}}] (-1,0.8) -- (-1,0)--(-1,-0.8);   
            \end{scope}
        \end{tikzpicture}}\quad =\quad \adjustbox{valign=c}{
        \begin{tikzpicture}[]
            \begin{scope}[scale=1.5]
                \node [above, scale=0.8] at (-1,0.8){$ $};
                \node [below,scale=0.8] at (-1,-0.8){$x^{+}(z)$};
                \node [left,scale=0.8] at (-2,0){$ $};
                \draw[postaction={segment={mid arrow}}] (-1,0) -- (-2,0);
		        \draw[postaction={segment={mid arrow}}] (-1,0.8) -- (-1,0)--(-1,-0.8);   
            \end{scope}
        \end{tikzpicture}}\quad +\quad \adjustbox{valign=c}{
        \begin{tikzpicture}[]
            \begin{scope}[scale=1.5]
                \node [above, scale=0.8] at (-1,0.8){$ $};
                \node [below,scale=0.8] at (-1,-0.8){$\psi^{-}(\gamma^{1/2}z)$};
                \node [left,scale=0.8] at (-2,0){$x^{+}(\gamma z)$};
                \draw[postaction={segment={mid arrow}}] (-1,0) -- (-2,0);
		        \draw[postaction={segment={mid arrow}}] (-1,0.8) -- (-1,0)--(-1,-0.8);   
            \end{scope}
        \end{tikzpicture}}
    \end{align}
    \item $x^{-}(z)$
    \begin{align}
        \adjustbox{valign=c}{
        \begin{tikzpicture}[]
            \begin{scope}[scale=1.5]
                \node [above, scale=0.8] at (-1,0.8){$x^{-}(z)$};
                \node [below,scale=0.8] at (-1,-0.8){$ $};
                \node [left,scale=0.8] at (-2,0){$ $};
                \draw[postaction={segment={mid arrow}}] (-1,0) -- (-2,0);
		        \draw[postaction={segment={mid arrow}}] (-1,0.8) -- (-1,0)--(-1,-0.8);   
            \end{scope}
        \end{tikzpicture}}\quad=\quad\adjustbox{valign=c}{
        \begin{tikzpicture}[]
            \begin{scope}[scale=1.5]
                \node [above, scale=0.8] at (-1,0.8){$ $};
                \node [below,scale=0.8] at (-1,-0.8){$x^{-}(z)$};
                \node [left,scale=0.8] at (-2,0){$\psi^{+}(z)$};
                \draw[postaction={segment={mid arrow}}] (-1,0) -- (-2,0);
		        \draw[postaction={segment={mid arrow}}] (-1,0.8) -- (-1,0)--(-1,-0.8);   
            \end{scope}
        \end{tikzpicture}}\quad +\quad \adjustbox{valign=c}{
        \begin{tikzpicture}[]
            \begin{scope}[scale=1.5]
                \node [above, scale=0.8] at (-1,0.8){$ $};
                \node [below,scale=0.8] at (-1,-0.8){$ $};
                \node [left,scale=0.8] at (-2,0){$x^{-}(z)$};
                \draw[postaction={segment={mid arrow}}] (-1,0) -- (-2,0);
		        \draw[postaction={segment={mid arrow}}] (-1,0.8) -- (-1,0)--(-1,-0.8);   
            \end{scope}
        \end{tikzpicture}}
    \end{align}
    \item $\psi^{\pm}(z)$
    \begin{align}
        \adjustbox{valign=c}{
        \begin{tikzpicture}[]
            \begin{scope}[scale=1.5]
                \node [above, scale=0.8] at (-1,0.8){$\psi^{\pm}(z)$};
                \node [below,scale=0.8] at (-1,-0.8){$ $};
                \node [left,scale=0.8] at (-2,0){$ $};
                \draw[postaction={segment={mid arrow}}] (-1,0) -- (-2,0);
		        \draw[postaction={segment={mid arrow}}] (-1,0.8) -- (-1,0)--(-1,-0.8);   
            \end{scope}
        \end{tikzpicture}}\quad=\quad\adjustbox{valign=c}{
        \begin{tikzpicture}[]
            \begin{scope}[scale=1.5]
                \node [above, scale=0.8] at (-1,0.8){$ $};
                \node [below,scale=0.8] at (-1,-0.8){$\psi^{\pm}(z)$};
                \node [left,scale=0.8] at (-2,0){$\psi^{\pm}(\gamma^{\mp1/2}z)$};
                \draw[postaction={segment={mid arrow}}] (-1,0) -- (-2,0);
		        \draw[postaction={segment={mid arrow}}] (-1,0.8) -- (-1,0)--(-1,-0.8);   
            \end{scope}
        \end{tikzpicture}}
        \end{align}
\end{enumerate}
Note that the graphical look of the intertwiner properties does not depend on the parities of the crystal representations because their central charges are $\hat{\gamma}=1$, which means they do not change the variables inside the Drinfeld currents.

Using the ket and bra representations in section~\ref{sec:crystalrep}, we also have the following property for $g(z)\in\mathcal{E}$
\begin{align}
    \adjustbox{valign=c}{
        \begin{tikzpicture}[]
            \begin{scope}[scale=1.5]
                \node [above, scale=0.8] at (-1,0.8){$ $};
                \node [below,scale=0.8] at (-1,-0.8){$ $};
                \node [above right,scale=0.8] at (-2.2,0){$g(z)$};
                \draw[postaction={segment={mid arrow}}] (-1,0) -- (-2.3,0);
		        \draw[postaction={segment={mid arrow}}] (-1,0.8) -- (-1,0)--(-1,-0.8);  
		        \draw[postaction={segment={mid arrow}}] (-2.3,0.8) -- (-2.3,0)--(-2.3,-0.8);  
            \end{scope}
        \end{tikzpicture}}\qquad=\qquad\adjustbox{valign=c}{
        \begin{tikzpicture}[]
            \begin{scope}[scale=1.5]
                \node [above, scale=0.8] at (-1,0.8){$ $};
                \node [below,scale=0.8] at (-1,-0.8){$ $};
                \node [above left,scale=0.8] at (-1.1,0){$g(z)$};
                \draw[postaction={segment={mid arrow}}] (-1,0) -- (-2.3,0);
		        \draw[postaction={segment={mid arrow}}] (-1,0.8) -- (-1,0)--(-1,-0.8);  
		        \draw[postaction={segment={mid arrow}}] (-2.3,0.8) -- (-2.3,0)--(-2.3,-0.8);  
            \end{scope}
        \end{tikzpicture}}
\end{align}
where this is understood as $\Phi_{\sigma} g(z)\cdot\Phi_{\sigma}^{\ast}=\Phi_{\sigma}\cdot g(z)\Phi_{\sigma}^{\ast}$. Namely, after inserting $\mathbbm{1}=\sum_{\lambda}a_{\lambda}^{(\sigma)}\dket{v,\lambda}\dbra{v,\lambda}$, we have this identity. 

We also have similar properties for the gluings in the vertical direction such as
\begin{align}
    \adjustbox{valign=c}{\begin{tikzpicture}
            \begin{scope}[scale=1.2]
            \node[below right] at (-1,0.1){$g(z)$};
               \draw[postaction={segment={mid arrow}}] (-1,0.2) -- (-2,0.2);
              \draw[postaction={segment={mid arrow}}] (0,-1.2)--(-1,-1.2);
		        \draw[postaction={segment={mid arrow}}] (-1,1) -- (-1,0)--(-1,-1)--(-1,-2);
            \end{scope}
        \end{tikzpicture}}\qquad=\qquad\adjustbox{valign=c}{\begin{tikzpicture}
            \begin{scope}[scale=1.2]
            \node[above right] at (-1,-1.1){$g(z)$};
               \draw[postaction={segment={mid arrow}}] (-1,0.2) -- (-2,0.2);
              \draw[postaction={segment={mid arrow}}] (0,-1.2)--(-1,-1.2);
		        \draw[postaction={segment={mid arrow}}] (-1,1) -- (-1,0)--(-1,-1)--(-1,-2);
            \end{scope}
        \end{tikzpicture}}
\end{align}

\section{Schur functions and free field realization}\label{sec:appendix-Schurfunc}
\paragraph{Boson-fermion correspondence}
We use the following notations for free bosons and fermions:
\begin{align}
\begin{split}
    [a_{n},\hat{Q}]=\delta_{n,0},\quad [a_{m},a_{n}]=m\delta_{m+n,0},\quad\{\psi_{r},\psi^{\ast}_{s}\}=\delta_{r+s,0},\quad \{\psi_{r},\psi_{s}\}=\{\psi^{\ast}_{r},\psi^{\ast}_{s}\}=0,\\
    \phi(z)=\hat{Q}+a_{0}\log z-\sum_{n\neq 0}\frac{a_{n}}{n}z^{-n},\quad \psi(z)=\sum_{r\in\mathbb{Z}+1/2}\frac{\psi_{r}}{z^{r+1/2}},\quad \psi^{\ast}(z)=\sum_{r\in\mathbb{Z}+1/2}\frac{\psi^{\ast}_{r}}{z^{r+1/2}}.
\end{split}
\end{align}
We also define the $U(1)$ current $J(z)$ as 
\begin{align}
    J(z)=\partial \phi(z)=\sum_{n\in\mathbb{Z}}\frac{a_{n}}{z^{n+1}}.
\end{align}
Then, the boson-fermion correspondence is 
\begin{align}
    J(z)=:\psi^{\ast}(z)\psi(z):,\quad \psi(z)=:e^{-\phi(z)}:,\quad \psi^{\ast}(z)=:e^{\phi(z)}:
\end{align}
Using the above correspondence, we can construct the Fock basis of the free boson using the fermion representation as
\begin{align}
\begin{split}
    \ket{\lambda}&=(-1)^{\sum_{i}(n_{i}-\frac{1}{2})}\psi^{\ast}_{-m_{1}}\cdots\psi^{\ast}_{-m_{s}}\psi_{-n_{s}}\cdots\psi_{-n_{1}}\ket{0},\\
    \bra{\lambda}&=(-1)^{\sum_{i}(n_{i}-\frac{1}{2})}\bra{0}\psi^{\ast}_{n_{1}}\cdots\psi_{n_{s}}^{\ast}\psi_{m_{s}}\cdots\psi_{m_{1}},\\
    \braket{\lambda\,|\,\mu}&=\delta_{\lambda,\mu},\quad 1=\sum_{\lambda}\ket{\lambda}\hspace{-0.1cm}\bra{\lambda}.
\end{split}\label{eq:appendix-Fockbasis}
\end{align}
where $m_{1}>m_{2}>\cdots m_{s}$ and $n_{1}>n_{2}>\cdots>n_{s}$ are the Frobenius coordinates.

\paragraph{Free field realization of Schur functions}
We introduce the following vertex operators
\begin{align}
    V_{\pm}(x)=\exp\left(\sum_{n=1}^{\infty}\frac{p_{n}(x)}{n}a_{\pm n}\right),&\quad \widetilde{V}_{\pm}(x)=\exp\left(\sum_{n=1}^{\infty}-\frac{(-1)^{n}p_{n}(x)}{n}a_{\pm n}\right),\label{eq:appendix-Schurvertexop-def}
\end{align}
where $p_{n}(x)=\sum_{i=1}^{\infty}x_{i}^{n}$ for $x=(x_{1},x_{2},x_{3},\ldots
)$. Note we are using the multivariable notation. The contraction formulas are
\begin{align}
    V_{+}(x)V_{-}(y)&=\prod_{i,j}(1-x_{i}y_{j})^{-1}V_{-}(y)V_{+}(x),\\
    \wt{V}_{+}(x)\wt{V}_{-}(y)&=\prod_{i,j}(1-x_{i}y_{j})^{-1}\wt{V}_{-}(y)\wt{V}_{+}(x),\\
    V_{+}(x)\wt{V}_{-}(y)&=\prod_{i,j}(1+x_{i}y_{j})\wt{V}_{-}(y)V_{+}(x),\\
    \wt{V}_{+}(x)V_{-}(y)&=\prod_{i,j}(1+x_{i}y_{j})V_{-}(y)\wt{V}_{+}(x).
\end{align}
Free field realizations of skew Schur functions are 
\begin{align}
    s_{\lambda/\mu}(x)&=\bra{\mu}V_{+}(x)\ket{\lambda}=\bra{\lambda}V_{-}(x)\ket{\mu},\label{eq:appendix-Schurfreefield1}\\
    s_{\lambda^{\rmT}/\mu^{\rmT}}(x)&=\bra{\mu}\wt{V}_{+}(x)\ket{\lambda}=\bra{\lambda}\wt{V}_{-}(x)\ket{\mu}\label{eq:appendix-Schurfreefield2},
\end{align}
where $\lambda^{\rmT}$ is the transpose of $\lambda$. We also have the following formulas
\begin{align}
    \sum_{\lambda}s_{\lambda/\mu}(x)s_{\lambda/\nu}(y)&=\prod_{i,j}(1-x_{i}y_{j})^{-1}\sum_{\eta}s_{\nu/\eta}(x)s_{\mu/\eta}(y),\\
    \sum_{\lambda}s_{\lambda/\mu^{\rmT}}(x)s_{\lambda^{\rmT}/\nu}(y)&=\prod_{i,j}(1+x_{i}y_{j})\sum_{\eta}s_{\nu^{\rmT}/\eta}(x)s_{\mu/\eta^{\rmT}}(y),
\end{align}
and 
\begin{align}
    s_{\lambda/\mu}(\alpha x)=\alpha^{|\lambda|-|\mu|}s_{\lambda/\mu}(x),\quad
    s_{\alpha}(q^{\rho+\beta})=(-1)^{|\alpha|}s_{\alpha^{\rmT}}(q^{-\rho-\beta^{\rmT}}).\label{eq:appendix-Schurproperty}
\end{align}
Note also that 
\begin{align}
    \ket{\lambda^{\rmT}}=(-1)^{|\lambda|}\ket{\lambda}.
\end{align}

\bibliography{supergroupAGT}

\providecommand{\href}[2]{#2}\begingroup\raggedright\begin{thebibliography}{100}

\bibitem{Nekrasov:2002qd}
N.~A. Nekrasov, ``{Seiberg-Witten prepotential from instanton counting},''
  \href{http://dx.doi.org/10.4310/ATMP.2003.v7.n5.a4}{{\em Adv. Theor. Math.
  Phys.} {\bfseries 7} no.~5, (2003) 831--864},
\href{http://arxiv.org/abs/hep-th/0206161}{{\ttfamily arXiv:hep-th/0206161
  [hep-th]}}.

\bibitem{Nekrasov:2003rj}
N.~Nekrasov and A.~Okounkov, ``{Seiberg-Witten theory and random partitions},''
  \href{http://dx.doi.org/10.1007/0-8176-4467-9_15}{{\em Prog. Math.}
  {\bfseries 244} (2006) 525--596},
  \href{http://arxiv.org/abs/hep-th/0306238}{{\ttfamily arXiv:hep-th/0306238}}.

\bibitem{Nekrasov2004}
N.~A. Nekrasov, ``Seiberg-Witten prepotential from instanton counting,''
  \href{http://dx.doi.org/10.4310/ATMP.2003.v7.n5.a4}{{\em
  Adv.Theor.Math.Phys.} {\bfseries 7} (2004) 831--864},
\href{http://arxiv.org/abs/hep-th/0206161}{{\ttfamily arXiv:hep-th/0206161
  [hep-th]}}.

\bibitem{Nakajima:2003uh}
H.~Nakajima and K.~Yoshioka, ``{Lectures on instanton counting},'' in {\em {CRM
  Workshop on Algebraic Structures and Moduli Spaces}}.
\newblock 11, 2003.
\newblock \href{http://arxiv.org/abs/math/0311058}{{\ttfamily
  arXiv:math/0311058}}.

\bibitem{Nakajima:2005fg}
H.~Nakajima and K.~Yoshioka, ``{Instanton counting on blowup. II. K-theoretic
  partition function},'' \href{http://arxiv.org/abs/math/0505553}{{\ttfamily
  arXiv:math/0505553}}.

\bibitem{Nakajima2005}
H.~Nakajima and K.~Yoshioka, ``Instanton counting on blowup. 1.,''
  \href{http://dx.doi.org/10.1007/s00222-005-0444-1}{{\em Invent.Math.}
  {\bfseries 162} (2005) 313--355},
\href{http://arxiv.org/abs/math/0306198}{{\ttfamily arXiv:math/0306198
  [math-ag]}}.

\bibitem{Nakajimalecturebook}
H.~Nakajima, ``Lectures on Hilbert Schemes of Points on Surfaces,''.

\bibitem{Pestun:2016zxk}
V.~Pestun {\em et~al.}, ``{Localization techniques in quantum field
  theories},'' \href{http://dx.doi.org/10.1088/1751-8121/aa63c1}{{\em J. Phys.
  A} {\bfseries 50} no.~44, (2017) 440301},
  \href{http://arxiv.org/abs/1608.02952}{{\ttfamily arXiv:1608.02952
  [hep-th]}}.

\bibitem{Nakajima:1994nid}
H.~Nakajima, ``{Instantons on ALE spaces, quiver varieties, and Kac-Moody
  algebras},'' \href{http://dx.doi.org/10.1215/S0012-7094-94-07613-8}{{\em Duke
  Math. J.} {\bfseries 76} no.~2, (1994) 365--416}.

\bibitem{schiffmann2012cherednik}
O.~Schiffmann and E.~Vasserot, ``{Cherednik algebras, W algebras and the
  equivariant cohomology of the moduli space of instantons on $A^{2}$},''
  \href{http://arxiv.org/abs/1202.2756}{{\ttfamily arXiv:1202.2756 [math.QA]}}.

\bibitem{Nekrasov:2015wsu}
N.~Nekrasov, ``{BPS/CFT correspondence: non-perturbative Dyson-Schwinger
  equations and qq-characters},''
  \href{http://dx.doi.org/10.1007/JHEP03(2016)181}{{\em JHEP} {\bfseries 03}
  (2016) 181}, \href{http://arxiv.org/abs/1512.05388}{{\ttfamily
  arXiv:1512.05388 [hep-th]}}.

\bibitem{Nekrasov:2016gud}
N.~Nekrasov and N.~S. Prabhakar, ``{Spiked Instantons from Intersecting
  D-branes},'' \href{http://dx.doi.org/10.1016/j.nuclphysb.2016.11.014}{{\em
  Nucl. Phys. B} {\bfseries 914} (2017) 257--300},
  \href{http://arxiv.org/abs/1611.03478}{{\ttfamily arXiv:1611.03478
  [hep-th]}}.

\bibitem{Nekrasov:2016qym}
N.~Nekrasov, ``{BPS/CFT CORRESPONDENCE II: INSTANTONS AT CROSSROADS, MODULI AND
  COMPACTNESS THEOREM},''
  \href{http://dx.doi.org/10.4310/ATMP.2017.v21.n2.a4}{{\em Adv. Theor. Math.
  Phys.} {\bfseries 21} (2017) 503--583},
  \href{http://arxiv.org/abs/1608.07272}{{\ttfamily arXiv:1608.07272
  [hep-th]}}.

\bibitem{Nekrasov:2016ydq}
N.~Nekrasov, ``{BPS/CFT Correspondence III: Gauge Origami partition function
  and qq-characters},'' \href{http://dx.doi.org/10.1007/s00220-017-3057-9}{{\em
  Commun. Math. Phys.} {\bfseries 358} no.~3, (2018) 863--894},
  \href{http://arxiv.org/abs/1701.00189}{{\ttfamily arXiv:1701.00189
  [hep-th]}}.

\bibitem{Nekrasov:2017gzb}
N.~Nekrasov, ``{BPS/CFT correspondence V: BPZ and KZ equations from
  qq-characters},'' \href{http://arxiv.org/abs/1711.11582}{{\ttfamily
  arXiv:1711.11582 [hep-th]}}.

\bibitem{Nekrasov:2017rqy}
N.~Nekrasov, ``{BPS/CFT correspondence IV: sigma models and defects in gauge
  theory},'' \href{http://dx.doi.org/10.1007/s11005-018-1115-7}{{\em Lett.
  Math. Phys.} {\bfseries 109} no.~3, (2019) 579--622},
  \href{http://arxiv.org/abs/1711.11011}{{\ttfamily arXiv:1711.11011
  [hep-th]}}.

\bibitem{Alday2010}
L.~F. Alday, D.~Gaiotto, and Y.~Tachikawa, ``Liouville Correlation Functions
  from Four-dimensional Gauge Theories,''
  \href{http://dx.doi.org/10.1007/s11005-010-0369-5}{{\em Lett.Math.Phys.}
  {\bfseries 91} (2010) 167--197},
\href{http://arxiv.org/abs/0906.3219}{{\ttfamily arXiv:0906.3219 [hep-th]}}.

\bibitem{Wyllard2009}
N.~Wyllard, ``{$A_{N-1}$ conformal Toda field theory correlation functions from
  conformal N = 2 SU(N) quiver gauge theories},''
  \href{http://dx.doi.org/10.1088/1126-6708/2009/11/002}{{\em JHEP} {\bfseries
  0911} (2009) 002},
\href{http://arxiv.org/abs/0907.2189}{{\ttfamily arXiv:0907.2189 [hep-th]}}.

\bibitem{Gaiotto:2009we}
D.~Gaiotto, ``{N=2 dualities},''
  \href{http://dx.doi.org/10.1007/JHEP08(2012)034}{{\em JHEP} {\bfseries 08}
  (2012) 034},
\href{http://arxiv.org/abs/0904.2715}{{\ttfamily arXiv:0904.2715 [hep-th]}}.

\bibitem{Gaiotto:2009ma}
D.~Gaiotto, ``{Asymptotically free N=2 theories and irregular conformal
  blocks},''
\href{http://arxiv.org/abs/0908.0307}{{\ttfamily arXiv:0908.0307 [hep-th]}}.

\bibitem{LeFloch:2020uop}
B.~Le~Floch, ``{A slow review of the AGT correspondence},''
  \href{http://dx.doi.org/10.1088/1751-8121/ac5945}{{\em J. Phys. A} {\bfseries
  55} no.~35, (2022) 353002}, \href{http://arxiv.org/abs/2006.14025}{{\ttfamily
  arXiv:2006.14025 [hep-th]}}.

\bibitem{awata2010five}
H.~Awata and Y.~Yamada, ``{Five-dimensional AGT conjecture and the deformed
  Virasoro algebra},'' {\em Journal of High Energy Physics} {\bfseries 2010}
  no.~1, (2010) 1--11, \href{http://arxiv.org/abs/0910.4431}{{\ttfamily
  arXiv:0910.4431 [hep-th]}}.

\bibitem{Awata_2010}
H.~Awata and Y.~Yamada, ``{Five-Dimensional AGT Relation and the Deformed
  $\beta$-Ensemble},'' \href{http://dx.doi.org/10.1143/ptp.124.227}{{\em
  Progress of Theoretical Physics} {\bfseries 124} no.~2, (Aug, 2010)
  227–262}. \url{http://dx.doi.org/10.1143/PTP.124.227}.

\bibitem{Yanagida_2010}
S.~Yanagida, ``{Five-dimensional SU(2) AGT conjecture and recursive formula of
  deformed Gaiotto state},'' \href{http://dx.doi.org/10.1063/1.3505826}{{\em
  Journal of Mathematical Physics} {\bfseries 51} no.~12, (Dec, 2010) 123506}.
  \url{http://dx.doi.org/10.1063/1.3505826}.

\bibitem{awata2011notes}
H.~Awata, B.~Feigin, A.~Hoshino, M.~Kanai, J.~Shiraishi, and S.~Yanagida,
  ``{Notes on Ding-Iohara algebra and AGT conjecture},''
  \href{http://arxiv.org/abs/1106.4088}{{\ttfamily arXiv:1106.4088 [math-ph]}}.

\bibitem{ohkubo2016crystallization}
Y.~Ohkubo, H.~Awata, and H.~Fujino, ``{Crystallization of deformed Virasoro
  algebra, Ding-Iohara-Miki algebra and 5D AGT correspondence},''
  \href{http://arxiv.org/abs/1512.08016}{{\ttfamily arXiv:1512.08016
  [math-ph]}}.

\bibitem{Vafa:2001qf}
C.~Vafa, ``{Brane / anti-brane systems and U(N|M) supergroup},''
  \href{http://arxiv.org/abs/hep-th/0101218}{{\ttfamily arXiv:hep-th/0101218}}.

\bibitem{Okuda:2006fb}
T.~Okuda and T.~Takayanagi, ``{Ghost D-branes},''
  \href{http://dx.doi.org/10.1088/1126-6708/2006/03/062}{{\em JHEP} {\bfseries
  03} (2006) 062}, \href{http://arxiv.org/abs/hep-th/0601024}{{\ttfamily
  arXiv:hep-th/0601024}}.

\bibitem{Mikhaylov:2014aoa}
V.~Mikhaylov and E.~Witten, ``{Branes And Supergroups},''
  \href{http://dx.doi.org/10.1007/s00220-015-2449-y}{{\em Commun. Math. Phys.}
  {\bfseries 340} no.~2, (2015) 699--832},
  \href{http://arxiv.org/abs/1410.1175}{{\ttfamily arXiv:1410.1175 [hep-th]}}.

\bibitem{Dijkgraaf:2016lym}
R.~Dijkgraaf, B.~Heidenreich, P.~Jefferson, and C.~Vafa, ``{Negative Branes,
  Supergroups and the Signature of Spacetime},''
  \href{http://dx.doi.org/10.1007/JHEP02(2018)050}{{\em JHEP} {\bfseries 02}
  (2018) 050}, \href{http://arxiv.org/abs/1603.05665}{{\ttfamily
  arXiv:1603.05665 [hep-th]}}.

\bibitem{Hanany:1996ie}
A.~Hanany and E.~Witten, ``{Type IIB superstrings, BPS monopoles, and
  three-dimensional gauge dynamics},''
  \href{http://dx.doi.org/10.1016/S0550-3213(97)00157-0}{{\em Nucl. Phys. B}
  {\bfseries 492} (1997) 152--190},
  \href{http://arxiv.org/abs/hep-th/9611230}{{\ttfamily arXiv:hep-th/9611230}}.

\bibitem{Kimura-Sugimoto:antivertex}
T.~Kimura and Y.~Sugimoto, ``{Topological Vertex/anti-Vertex and Supergroup
  Gauge Theory},'' \href{http://dx.doi.org/10.1007/JHEP04(2020)081}{{\em JHEP}
  {\bfseries 04} (2020) 081}, \href{http://arxiv.org/abs/2001.05735}{{\ttfamily
  arXiv:2001.05735 [hep-th]}}.

\bibitem{Kimura-Pestun:supergroup}
T.~Kimura and V.~Pestun, ``{Super instanton counting and localization},''
  \href{http://arxiv.org/abs/1905.01513}{{\ttfamily arXiv:1905.01513
  [hep-th]}}.

\bibitem{Iqbal:2007ii}
A.~Iqbal, C.~Kozcaz, and C.~Vafa, ``{The Refined topological vertex},''
  \href{http://dx.doi.org/10.1088/1126-6708/2009/10/069}{{\em JHEP} {\bfseries
  10} (2009) 069},
\href{http://arxiv.org/abs/hep-th/0701156}{{\ttfamily arXiv:hep-th/0701156
  [hep-th]}}.

\bibitem{Awata:2005fa}
H.~Awata and H.~Kanno, ``{Instanton counting, Macdonald functions and the
  moduli space of D-branes},''
  \href{http://dx.doi.org/10.1088/1126-6708/2005/05/039}{{\em JHEP} {\bfseries
  05} (2005) 039}, \href{http://arxiv.org/abs/hep-th/0502061}{{\ttfamily
  arXiv:hep-th/0502061}}.

\bibitem{Awata:2008ed}
H.~Awata and H.~Kanno, ``{Refined BPS state counting from Nekrasov's formula
  and Macdonald functions},''
  \href{http://dx.doi.org/10.1142/S0217751X09043006}{{\em Int. J. Mod. Phys.}
  {\bfseries A24} (2009) 2253--2306},
\href{http://arxiv.org/abs/0805.0191}{{\ttfamily arXiv:0805.0191 [hep-th]}}.

\bibitem{Aganagic:2003db}
M.~Aganagic, A.~Klemm, M.~Marino, and C.~Vafa, ``{The Topological vertex},''
  \href{http://dx.doi.org/10.1007/s00220-004-1162-z}{{\em Commun. Math. Phys.}
  {\bfseries 254} (2005) 425--478},
\href{http://arxiv.org/abs/hep-th/0305132}{{\ttfamily arXiv:hep-th/0305132
  [hep-th]}}.

\bibitem{Kimura-Nieri:defects}
T.~Kimura and F.~Nieri, ``{Intersecting defects and supergroup gauge theory},''
  \href{http://dx.doi.org/10.1088/1751-8121/ac2716}{{\em J. Phys. A} {\bfseries
  54} no.~43, (2021) 435401}, \href{http://arxiv.org/abs/2105.02776}{{\ttfamily
  arXiv:2105.02776 [hep-th]}}.

\bibitem{Nieri:2021xpe}
F.~Nieri, ``{Defects at the Intersection: the Supergroup Side},'' in {\em {14th
  International Workshop on Lie Theory and Its Applications in Physics}}.
\newblock 12, 2021.
\newblock \href{http://arxiv.org/abs/2112.07603}{{\ttfamily arXiv:2112.07603
  [hep-th]}}.

\bibitem{Chen:2020rxu}
H.-Y. Chen, T.~Kimura, and N.~Lee, ``{Quantum Integrable Systems from
  Supergroup Gauge Theories},''
  \href{http://dx.doi.org/10.1007/JHEP09(2020)104}{{\em JHEP} {\bfseries 09}
  (2020) 104}, \href{http://arxiv.org/abs/2003.13514}{{\ttfamily
  arXiv:2003.13514 [hep-th]}}.

\bibitem{Aharony:1997bh}
O.~Aharony, A.~Hanany, and B.~Kol, ``{Webs of (p,q) five-branes,
  five-dimensional field theories and grid diagrams},''
  \href{http://dx.doi.org/10.1088/1126-6708/1998/01/002}{{\em JHEP} {\bfseries
  01} (1998) 002}, \href{http://arxiv.org/abs/hep-th/9710116}{{\ttfamily
  arXiv:hep-th/9710116}}.

\bibitem{Katz:1996fh}
S.~H. Katz, A.~Klemm, and C.~Vafa, ``{Geometric engineering of quantum field
  theories},'' \href{http://dx.doi.org/10.1016/S0550-3213(97)00282-4}{{\em
  Nucl. Phys. B} {\bfseries 497} (1997) 173--195},
  \href{http://arxiv.org/abs/hep-th/9609239}{{\ttfamily arXiv:hep-th/9609239}}.

\bibitem{Miki2007}
K.~Miki, ``{A (q, $\gamma$) analog of the $W_{1+\infty}$ algebra},''
  \href{http://dx.doi.org/10.1063/1.2823979}{{\em Journal of Mathematical
  Physics} {\bfseries 48} no.~12, (2007) 3520}.
  \url{http://scitation.aip.org/content/aip/journal/jmp/48/12/10.1063/1.2823979}.

\bibitem{Ding:1996mq}
J.-t. Ding and K.~Iohara, ``{Generalization and deformation of Drinfeld quantum
  affine algebras},''
\href{http://dx.doi.org/10.1023/A:1007341410987}{{\em Lett. Math. Phys.}
  {\bfseries 41} (1997) 181--193}.

\bibitem{Feigin2011}
B.~Feigin, E.~Feigin, M.~Jimbo, T.~Miwa, and E.~Mukhin, ``{Quantum continuous
  $gl(\infty)$ : Tensor products of Fock modules and $W_n$ characters},'' {\em
  Kyoto Journal of Mathematics} {\bfseries 51} no.~2, (2011) 365--392,
  \href{http://arxiv.org/abs/1002.3113}{{\ttfamily arXiv:1002.3113 [math.QA]}}.

\bibitem{feigin2011quantum}
B.~Feigin, E.~Feigin, M.~Jimbo, T.~Miwa, E.~Mukhin, {\em et~al.}, ``Quantum
  continuous $gl(\infty)$: Semiinfinite construction of representations,'' {\em
  Kyoto Journal of Mathematics} {\bfseries 51} no.~2, (2011) 337--364,
  \href{http://arxiv.org/abs/1002.3100}{{\ttfamily arXiv:1002.3100 [math.QA]}}.

\bibitem{feigin2012quantum}
B.~Feigin, M.~Jimbo, T.~Miwa, E.~Mukhin, {\em et~al.}, ``Quantum toroidal
  $gl_1$-algebra: Plane partitions,'' {\em Kyoto Journal of Mathematics}
  {\bfseries 52} no.~3, (2012) 621--659,
  \href{http://arxiv.org/abs/1110.5310}{{\ttfamily arXiv:1110.5310}}.

\bibitem{Awata:2011ce}
H.~Awata, B.~Feigin, and J.~Shiraishi, ``{Quantum Algebraic Approach to Refined
  Topological Vertex},'' \href{http://dx.doi.org/10.1007/JHEP03(2012)041}{{\em
  JHEP} {\bfseries 03} (2012) 041},
\href{http://arxiv.org/abs/1112.6074}{{\ttfamily arXiv:1112.6074 [hep-th]}}.

\bibitem{Awata:2016riz}
H.~Awata, H.~Kanno, T.~Matsumoto, A.~Mironov, A.~Morozov, A.~Morozov,
  Y.~Ohkubo, and Y.~Zenkevich, ``{Explicit examples of DIM constraints for
  network matrix models},''
  \href{http://dx.doi.org/10.1007/JHEP07(2016)103}{{\em JHEP} {\bfseries 07}
  (2016) 103}, \href{http://arxiv.org/abs/1604.08366}{{\ttfamily
  arXiv:1604.08366 [hep-th]}}.

\bibitem{Mironov:2016yue}
A.~Mironov, A.~Morozov, and Y.~Zenkevich,
  ``{Ding\textendash{}Iohara\textendash{}Miki symmetry of network matrix
  models},'' \href{http://dx.doi.org/10.1016/j.physletb.2016.09.033}{{\em Phys.
  Lett. B} {\bfseries 762} (2016) 196--208},
  \href{http://arxiv.org/abs/1603.05467}{{\ttfamily arXiv:1603.05467
  [hep-th]}}.

\bibitem{Awata:2016bdm}
H.~Awata, H.~Kanno, A.~Mironov, A.~Morozov, A.~Morozov, Y.~Ohkubo, and
  Y.~Zenkevich, ``{Anomaly in RTT relation for DIM algebra and network matrix
  models},'' \href{http://dx.doi.org/10.1016/j.nuclphysb.2017.03.003}{{\em
  Nucl. Phys. B} {\bfseries 918} (2017) 358--385},
  \href{http://arxiv.org/abs/1611.07304}{{\ttfamily arXiv:1611.07304
  [hep-th]}}.

\bibitem{Awata:2016mxc}
H.~Awata, H.~Kanno, A.~Mironov, A.~Morozov, A.~Morozov, Y.~Ohkubo, and
  Y.~Zenkevich, ``{Toric Calabi-Yau threefolds as quantum integrable systems. $
  \mathrm{\mathcal{R}} $ -matrix and $
  \mathrm{\mathcal{R}}\mathcal{T}\mathcal{T} $ relations},''
  \href{http://dx.doi.org/10.1007/JHEP10(2016)047}{{\em JHEP} {\bfseries 10}
  (2016) 047}, \href{http://arxiv.org/abs/1608.05351}{{\ttfamily
  arXiv:1608.05351 [hep-th]}}.

\bibitem{Awata_2017}
H.~Awata, H.~Kanno, A.~Mironov, A.~Morozov, A.~Morozov, Y.~Ohkubo, and
  Y.~Zenkevich, ``{Generalized Knizhnik-Zamolodchikov equation for
  Ding-Iohara-Miki algebra},''
  \href{http://dx.doi.org/10.1103/PhysRevD.96.026021}{{\em Phys. Rev. D}
  {\bfseries 96} no.~2, (2017) 026021},
  \href{http://arxiv.org/abs/1703.06084}{{\ttfamily arXiv:1703.06084
  [hep-th]}}.

\bibitem{Awata_2018}
H.~Awata, H.~Kanno, A.~Mironov, A.~Morozov, K.~Suetake, and Y.~Zenkevich,
  ``{$(q,t)$-KZ equations for quantum toroidal algebra and Nekrasov partition
  functions on ALE spaces},''
  \href{http://dx.doi.org/10.1007/JHEP03(2018)192}{{\em JHEP} {\bfseries 03}
  (2018) 192}, \href{http://arxiv.org/abs/1712.08016}{{\ttfamily
  arXiv:1712.08016 [hep-th]}}.

\bibitem{Zenkevich:2018fzl}
Y.~Zenkevich, ``{Higgsed network calculus},''
  \href{http://dx.doi.org/10.1007/JHEP08(2021)149}{{\em JHEP} {\bfseries 08}
  (2021) 149}, \href{http://arxiv.org/abs/1812.11961}{{\ttfamily
  arXiv:1812.11961 [hep-th]}}.

\bibitem{zenkevich2019mathfrakgln}
Y.~Zenkevich, ``{$\mathfrak{gl}_N$ Higgsed networks},''
  \href{http://arxiv.org/abs/1912.13372}{{\ttfamily arXiv:1912.13372
  [hep-th]}}.

\bibitem{zenkevich2020mixed}
Y.~Zenkevich, ``{Mixed network calculus},''
  \href{http://arxiv.org/abs/2012.15563}{{\ttfamily arXiv:2012.15563
  [hep-th]}}.

\bibitem{Ghoneim:2020sqi}
M.~Ghoneim, C.~Kozcaz, K.~Kursun, and Y.~Zenkevich, ``{4d higgsed network
  calculus and elliptic DIM algebra},''
  \href{http://arxiv.org/abs/2012.15352}{{\ttfamily arXiv:2012.15352
  [hep-th]}}.

\bibitem{Bourgine:2017jsi}
J.-E. Bourgine, M.~Fukuda, K.~Harada, Y.~Matsuo, and R.-D. Zhu, ``{(p,q)-webs
  of DIM representations, 5d $ \mathcal{N}=1 $ instanton partition functions
  and qq-characters},'' \href{http://dx.doi.org/10.1007/JHEP11(2017)034}{{\em
  JHEP} {\bfseries 11} (2017) 034},
\href{http://arxiv.org/abs/1703.10759}{{\ttfamily arXiv:1703.10759 [hep-th]}}.

\bibitem{Bourgine:2017rik}
J.-E. Bourgine, M.~Fukuda, Y.~Matsuo, and R.-D. Zhu, ``{Reflection states in
  Ding-Iohara-Miki algebra and brane-web for D-type quiver},''
  \href{http://dx.doi.org/10.1007/JHEP12(2017)015}{{\em JHEP} {\bfseries 12}
  (2017) 015}, \href{http://arxiv.org/abs/1709.01954}{{\ttfamily
  arXiv:1709.01954 [hep-th]}}.

\bibitem{Zhu:2017ysu}
R.-D. Zhu, ``{An Elliptic Vertex of Awata-Feigin-Shiraishi type for
  M-strings},'' \href{http://dx.doi.org/10.1007/JHEP08(2018)050}{{\em JHEP}
  {\bfseries 08} (2018) 050}, \href{http://arxiv.org/abs/1712.10255}{{\ttfamily
  arXiv:1712.10255 [hep-th]}}.

\bibitem{Bourgine:2018uod}
J.~E. Bourgine and K.~Zhang, ``{A note on the algebraic engineering of 4D
  $\mathcal{N}=2$ super Yang-Mills theories},''
  \href{http://dx.doi.org/10.1016/j.physletb.2018.11.066}{{\em Phys. Lett. B}
  {\bfseries 789} (2019) 610--619},
  \href{http://arxiv.org/abs/1809.08861}{{\ttfamily arXiv:1809.08861
  [hep-th]}}.

\bibitem{Bourgine:2019phm}
J.-E. Bourgine and S.~Jeong, ``{New quantum toroidal algebras from 5D $
  \mathcal{N} $ = 1 instantons on orbifolds},''
  \href{http://dx.doi.org/10.1007/JHEP05(2020)127}{{\em JHEP} {\bfseries 05}
  (2020) 127}, \href{http://arxiv.org/abs/1906.01625}{{\ttfamily
  arXiv:1906.01625 [hep-th]}}.

\bibitem{Nekrasov:2018gne}
N.~Nekrasov, ``{Superspin chains and supersymmetric gauge theories},''
  \href{http://dx.doi.org/10.1007/JHEP03(2019)102}{{\em JHEP} {\bfseries 03}
  (2019) 102}, \href{http://arxiv.org/abs/1811.04278}{{\ttfamily
  arXiv:1811.04278 [hep-th]}}.

\bibitem{Ishtiaque:2021jan}
N.~Ishtiaque, S.~F. Moosavian, S.~Raghavendran, and J.~Yagi, ``{Superspin
  chains from superstring theory},''
  \href{http://arxiv.org/abs/2110.15112}{{\ttfamily arXiv:2110.15112
  [hep-th]}}.

\bibitem{Orlando:2010uu}
D.~Orlando and S.~Reffert, ``{Relating Gauge Theories via Gauge/Bethe
  Correspondence},'' \href{http://dx.doi.org/10.1007/JHEP10(2010)071}{{\em
  JHEP} {\bfseries 10} (2010) 071},
  \href{http://arxiv.org/abs/1005.4445}{{\ttfamily arXiv:1005.4445 [hep-th]}}.

\bibitem{Kimura:instantoncounting}
T.~Kimura, \href{http://dx.doi.org/10.1007/978-3-030-76190-5}{{\em {Instanton
  Counting, Quantum Geometry and Algebra}}}.
\newblock Springer, 7, 2021.
\newblock \href{http://arxiv.org/abs/2012.11711}{{\ttfamily arXiv:2012.11711
  [hep-th]}}.

\bibitem{Nekrasov:1996cz}
N.~Nekrasov, ``{Five dimensional gauge theories and relativistic integrable
  systems},'' \href{http://dx.doi.org/10.1016/S0550-3213(98)00436-2}{{\em Nucl.
  Phys. B} {\bfseries 531} (1998) 323--344},
  \href{http://arxiv.org/abs/hep-th/9609219}{{\ttfamily arXiv:hep-th/9609219}}.

\bibitem{Tachikawa:2004ur}
Y.~Tachikawa, ``{Five-dimensional Chern-Simons terms and Nekrasov's instanton
  counting},'' \href{http://dx.doi.org/10.1088/1126-6708/2004/02/050}{{\em
  JHEP} {\bfseries 02} (2004) 050},
  \href{http://arxiv.org/abs/hep-th/0401184}{{\ttfamily arXiv:hep-th/0401184}}.

\bibitem{Aharony:2008ug}
O.~Aharony, O.~Bergman, D.~L. Jafferis, and J.~Maldacena, ``{N=6 superconformal
  Chern-Simons-matter theories, M2-branes and their gravity duals},''
  \href{http://dx.doi.org/10.1088/1126-6708/2008/10/091}{{\em JHEP} {\bfseries
  10} (2008) 091}, \href{http://arxiv.org/abs/0806.1218}{{\ttfamily
  arXiv:0806.1218 [hep-th]}}.

\bibitem{Marino:2009jd}
M.~Marino and P.~Putrov, ``{Exact Results in ABJM Theory from Topological
  Strings},'' \href{http://dx.doi.org/10.1007/JHEP06(2010)011}{{\em JHEP}
  {\bfseries 06} (2010) 011}, \href{http://arxiv.org/abs/0912.3074}{{\ttfamily
  arXiv:0912.3074 [hep-th]}}.

\bibitem{Drukker:2009hy}
N.~Drukker and D.~Trancanelli, ``{A Supermatrix model for N=6 super
  Chern-Simons-matter theory},''
  \href{http://dx.doi.org/10.1007/JHEP02(2010)058}{{\em JHEP} {\bfseries 02}
  (2010) 058}, \href{http://arxiv.org/abs/0912.3006}{{\ttfamily arXiv:0912.3006
  [hep-th]}}.

\bibitem{Poghossian:2010pn}
R.~Poghossian, ``{Deforming SW curve},''
  \href{http://dx.doi.org/10.1007/JHEP04(2011)033}{{\em JHEP} {\bfseries 04}
  (2011) 033}, \href{http://arxiv.org/abs/1006.4822}{{\ttfamily arXiv:1006.4822
  [hep-th]}}.

\bibitem{Fucito:2011pn}
F.~Fucito, J.~F. Morales, D.~R. Pacifici, and R.~Poghossian, ``{Gauge theories
  on $\Omega$-backgrounds from non commutative Seiberg-Witten curves},''
  \href{http://dx.doi.org/10.1007/JHEP05(2011)098}{{\em JHEP} {\bfseries 05}
  (2011) 098}, \href{http://arxiv.org/abs/1103.4495}{{\ttfamily arXiv:1103.4495
  [hep-th]}}.

\bibitem{Nekrasov:2013xda}
N.~Nekrasov, V.~Pestun, and S.~Shatashvili, ``{Quantum geometry and quiver
  gauge theories},'' \href{http://dx.doi.org/10.1007/s00220-017-3071-y}{{\em
  Commun. Math. Phys.} {\bfseries 357} no.~2, (2018) 519--567},
  \href{http://arxiv.org/abs/1312.6689}{{\ttfamily arXiv:1312.6689 [hep-th]}}.

\bibitem{Aharony:1997ju}
O.~Aharony and A.~Hanany, ``{Branes, superpotentials and superconformal fixed
  points},'' \href{http://dx.doi.org/10.1016/S0550-3213(97)00472-0}{{\em Nucl.
  Phys. B} {\bfseries 504} (1997) 239--271},
  \href{http://arxiv.org/abs/hep-th/9704170}{{\ttfamily arXiv:hep-th/9704170}}.

\bibitem{Katz:1997eq}
S.~Katz, P.~Mayr, and C.~Vafa, ``{Mirror symmetry and exact solution of 4-D N=2
  gauge theories: 1.},''
  \href{http://dx.doi.org/10.4310/ATMP.1997.v1.n1.a2}{{\em Adv. Theor. Math.
  Phys.} {\bfseries 1} (1998) 53--114},
  \href{http://arxiv.org/abs/hep-th/9706110}{{\ttfamily arXiv:hep-th/9706110}}.

\bibitem{Dijkgraaf:2002fc}
R.~Dijkgraaf and C.~Vafa, ``{Matrix models, topological strings, and
  supersymmetric gauge theories},''
  \href{http://dx.doi.org/10.1016/S0550-3213(02)00766-6}{{\em Nucl. Phys. B}
  {\bfseries 644} (2002) 3--20},
  \href{http://arxiv.org/abs/hep-th/0206255}{{\ttfamily arXiv:hep-th/0206255}}.

\bibitem{Hollowood:2003cv}
T.~J. Hollowood, A.~Iqbal, and C.~Vafa, ``{Matrix models, geometric engineering
  and elliptic genera},''
  \href{http://dx.doi.org/10.1088/1126-6708/2008/03/069}{{\em JHEP} {\bfseries
  03} (2008) 069}, \href{http://arxiv.org/abs/hep-th/0310272}{{\ttfamily
  arXiv:hep-th/0310272}}.

\bibitem{Leung:1997tw}
N.~C. Leung and C.~Vafa, ``{Branes and toric geometry},''
  \href{http://dx.doi.org/10.4310/ATMP.1998.v2.n1.a4}{{\em Adv. Theor. Math.
  Phys.} {\bfseries 2} (1998) 91--118},
  \href{http://arxiv.org/abs/hep-th/9711013}{{\ttfamily arXiv:hep-th/9711013}}.

\bibitem{Kapustin:1998fa}
A.~Kapustin, ``{D(n) quivers from branes},''
  \href{http://dx.doi.org/10.1088/1126-6708/1998/12/015}{{\em JHEP} {\bfseries
  12} (1998) 015}, \href{http://arxiv.org/abs/hep-th/9806238}{{\ttfamily
  arXiv:hep-th/9806238}}.

\bibitem{Hanany:1999sj}
A.~Hanany and A.~Zaffaroni, ``{Issues on orientifolds: On the brane
  construction of gauge theories with SO(2n) global symmetry},''
  \href{http://dx.doi.org/10.1088/1126-6708/1999/07/009}{{\em JHEP} {\bfseries
  07} (1999) 009}, \href{http://arxiv.org/abs/hep-th/9903242}{{\ttfamily
  arXiv:hep-th/9903242}}.

\bibitem{Hayashi:2015vhy}
H.~Hayashi, S.-S. Kim, K.~Lee, M.~Taki, and F.~Yagi, ``{More on 5d descriptions
  of 6d SCFTs},'' \href{http://dx.doi.org/10.1007/JHEP10(2016)126}{{\em JHEP}
  {\bfseries 10} (2016) 126}, \href{http://arxiv.org/abs/1512.08239}{{\ttfamily
  arXiv:1512.08239 [hep-th]}}.

\bibitem{Kimura:2019gon}
T.~Kimura and R.-D. Zhu, ``{Web Construction of ABCDEFG and Affine Quiver Gauge
  Theories},'' \href{http://dx.doi.org/10.1007/JHEP09(2019)025}{{\em JHEP}
  {\bfseries 09} (2019) 025}, \href{http://arxiv.org/abs/1907.02382}{{\ttfamily
  arXiv:1907.02382 [hep-th]}}.

\bibitem{Bourgine:2018fjy}
J.-E. Bourgine, ``{Fiber-base duality from the algebraic perspective},''
  \href{http://dx.doi.org/10.1007/JHEP03(2019)003}{{\em JHEP} {\bfseries 03}
  (2019) 003}, \href{http://arxiv.org/abs/1810.00301}{{\ttfamily
  arXiv:1810.00301 [hep-th]}}.

\bibitem{Sasa:2019rbk}
S.~Sasa, A.~Watanabe, and Y.~Matsuo, ``{A note on the S-dual basis in the free
  fermion system},'' \href{http://dx.doi.org/10.1093/ptep/ptz158}{{\em PTEP}
  {\bfseries 2020} no.~2, (2020) 023B02},
  \href{http://arxiv.org/abs/1904.04766}{{\ttfamily arXiv:1904.04766
  [hep-th]}}.

\bibitem{bershtein2018plane}
M.~Bershtein, B.~Feigin, and G.~Merzon, ``Plane partitions with a “pit”:
  generating functions and representation theory,'' {\em Selecta Mathematica}
  {\bfseries 24} no.~1, (2018) 21--62,
  \href{http://arxiv.org/abs/1512.08779}{{\ttfamily arXiv:1512.08779 [math]}}.

\bibitem{feigin2013representations}
B.~Feigin, M.~Jimbo, T.~Miwa, and E.~Mukhin, ``Representations of quantum
  toroidal gl(n),'' {\em Journal of Algebra} {\bfseries 380} (2013) 78--108,
  \href{http://arxiv.org/abs/1204.5378}{{\ttfamily 1204.5378}}.

\bibitem{Galakhov:2020vyb}
D.~Galakhov and M.~Yamazaki, ``{Quiver Yangian and Supersymmetric Quantum
  Mechanics},'' \href{http://arxiv.org/abs/2008.07006}{{\ttfamily
  arXiv:2008.07006 [hep-th]}}.

\bibitem{Li:2020rij}
W.~Li and M.~Yamazaki, ``{Quiver Yangian from Crystal Melting},''
  \href{http://dx.doi.org/10.1007/JHEP11(2020)035}{{\em JHEP} {\bfseries 11}
  (2020) 035}, \href{http://arxiv.org/abs/2003.08909}{{\ttfamily
  arXiv:2003.08909 [hep-th]}}.

\bibitem{Galakhov:2021vbo}
D.~Galakhov, W.~Li, and M.~Yamazaki, ``{Toroidal and elliptic quiver BPS
  algebras and beyond},'' \href{http://dx.doi.org/10.1007/JHEP02(2022)024}{{\em
  JHEP} {\bfseries 02} (2022) 024},
  \href{http://arxiv.org/abs/2108.10286}{{\ttfamily arXiv:2108.10286
  [hep-th]}}.

\bibitem{Galakhov:2021xum}
D.~Galakhov, W.~Li, and M.~Yamazaki, ``{Shifted quiver Yangians and
  representations from BPS crystals},''
  \href{http://dx.doi.org/10.1007/JHEP08(2021)146}{{\em JHEP} {\bfseries 08}
  (2021) 146}, \href{http://arxiv.org/abs/2106.01230}{{\ttfamily
  arXiv:2106.01230 [hep-th]}}.

\bibitem{Yamazaki:2022cdg}
M.~Yamazaki, ``{Quiver Yangians and Crystal Melting: A Concise Summary},'' in
  {\em {International Congress on Mathematical Physics}}.
\newblock 3, 2022.
\newblock \href{http://arxiv.org/abs/2203.14314}{{\ttfamily arXiv:2203.14314
  [hep-th]}}.

\bibitem{Noshita:2021dgj}
G.~Noshita and A.~Watanabe, ``{Shifted quiver quantum toroidal algebra and
  subcrystal representations},''
  \href{http://dx.doi.org/10.1007/JHEP05(2022)122}{{\em JHEP} {\bfseries 05}
  (2022) 122}, \href{http://arxiv.org/abs/2109.02045}{{\ttfamily
  arXiv:2109.02045 [hep-th]}}.

\bibitem{Noshita:2021ldl}
G.~Noshita and A.~Watanabe, ``{A note on quiver quantum toroidal algebra},''
  \href{http://dx.doi.org/10.1007/JHEP05(2022)011}{{\em JHEP} {\bfseries 05}
  (2022) 011}, \href{http://arxiv.org/abs/2108.07104}{{\ttfamily
  arXiv:2108.07104 [hep-th]}}.

\bibitem{Shiraishi:1995rp}
J.~Shiraishi, H.~Kubo, H.~Awata, and S.~Odake, ``{A Quantum deformation of the
  Virasoro algebra and the Macdonald symmetric functions},''
  \href{http://dx.doi.org/10.1007/BF00398297}{{\em Lett. Math. Phys.}
  {\bfseries 38} (1996) 33--51},
\href{http://arxiv.org/abs/q-alg/9507034}{{\ttfamily arXiv:q-alg/9507034
  [q-alg]}}.

\bibitem{Feigin:1995sf}
B.~Feigin and E.~Frenkel, ``{Quantum W algebras and elliptic algebras},''
  \href{http://dx.doi.org/10.1007/BF02108819}{{\em Commun. Math. Phys.}
  {\bfseries 178} (1996) 653--678},
  \href{http://arxiv.org/abs/q-alg/9508009}{{\ttfamily arXiv:q-alg/9508009}}.

\bibitem{Awata:1995zk}
H.~Awata, H.~Kubo, S.~Odake, and J.~Shiraishi, ``{Quantum W(N) algebras and
  Macdonald polynomials},'' \href{http://dx.doi.org/10.1007/BF02102595}{{\em
  Commun. Math. Phys.} {\bfseries 179} (1996) 401--416},
  \href{http://arxiv.org/abs/q-alg/9508011}{{\ttfamily arXiv:q-alg/9508011}}.

\bibitem{Awata:1996dx}
H.~Awata, H.~Kubo, S.~Odake, and J.~Shiraishi, ``{Quantum deformation of the
  W(N) algebra},'' in {\em {Extended and Quantum Algebras and their
  Applications to Physics Tianjin, China, August 19-24, 1996}}.
\newblock 1996.
\newblock
\href{http://arxiv.org/abs/q-alg/9612001}{{\ttfamily arXiv:q-alg/9612001
  [q-alg]}}.
\newblock

\bibitem{FHSSY:2010}
B.~Feigin, A.~Hoshino, J.~Shibahara, J.~Shiraishi, and S.~Yanagida, ``{Kernel
  function and quantum algebras},''
\href{http://arxiv.org/abs/1002.2485}{{\ttfamily arXiv:1002.2485 [math]}}.

\bibitem{Kojima2019}
T.~Kojima, ``Quadratic relations of the deformed $W$-superalgebra ${\cal W}_{q
  t}(\mathfrak{sl}(2|1))$,'' \href{http://arxiv.org/abs/1912.03096}{{\ttfamily
  arXiv:1912.03096 [math.QA]}}.

\bibitem{Kojima2021}
T.~Kojima, ``Quadratic relations of the deformed $W$-superalgebra
  $W_{qt}(A(M,N))$,'' \href{http://arxiv.org/abs/2101.01110}{{\ttfamily
  arXiv:2101.01110 [math.QA]}}.

\bibitem{Harada:2021xnm}
K.~Harada, Y.~Matsuo, G.~Noshita, and A.~Watanabe, ``{$q$-deformation of corner
  vertex operator algebras by Miura transformation},''
  \href{http://dx.doi.org/10.1007/JHEP04(2021)202}{{\em JHEP} {\bfseries 04}
  (2021) 202}, \href{http://arxiv.org/abs/2101.03953}{{\ttfamily
  arXiv:2101.03953 [hep-th]}}.

\bibitem{Bourgine:2021nyw}
J.-E. Bourgine, ``{Engineering 3D $\mathcal{N}=2$ theories using the quantum
  affine $\mathfrak{sl}(2)$ algebra},''
  \href{http://arxiv.org/abs/2107.10063}{{\ttfamily arXiv:2107.10063
  [hep-th]}}.

\bibitem{Bourgine:2022scz}
J.-E. Bourgine, ``{Shifted quantum groups and matter multiplets in
  supersymmetric gauge theories},''
  \href{http://arxiv.org/abs/2205.01309}{{\ttfamily arXiv:2205.01309
  [hep-th]}}.

\bibitem{Matsuo:2014rba}
Y.~Matsuo, C.~Rim, and H.~Zhang, ``{Construction of Gaiotto states with
  fundamental multiplets through Degenerate DAHA},''
  \href{http://dx.doi.org/10.1007/JHEP09(2014)028}{{\em JHEP} {\bfseries 09}
  (2014) 028}, \href{http://arxiv.org/abs/1405.3141}{{\ttfamily arXiv:1405.3141
  [hep-th]}}.

\bibitem{Bourgine:2015szm}
J.-E. Bourgine, Y.~Matsuo, and H.~Zhang, ``{Holomorphic field realization of
  SH$^{c}$ and quantum geometry of quiver gauge theories},''
  \href{http://dx.doi.org/10.1007/JHEP04(2016)167}{{\em JHEP} {\bfseries 04}
  (2016) 167}, \href{http://arxiv.org/abs/1512.02492}{{\ttfamily
  arXiv:1512.02492 [hep-th]}}.

\bibitem{Bourgine:2016vsq}
J.-E. Bourgine, M.~Fukuda, Y.~Matsuo, H.~Zhang, and R.-D. Zhu, ``{Coherent
  states in quantum $\mathcal{W}_{1+\infty}$ algebra and qq-character for 5d
  Super Yang-Mills},'' \href{http://dx.doi.org/10.1093/ptep/ptw165}{{\em PTEP}
  {\bfseries 2016} no.~12, (2016) 123B05},
  \href{http://arxiv.org/abs/1606.08020}{{\ttfamily arXiv:1606.08020
  [hep-th]}}.

\bibitem{Kimura:2015rgi}
T.~Kimura and V.~Pestun, ``{Quiver W-algebras},''
  \href{http://dx.doi.org/10.1007/s11005-018-1072-1}{{\em Lett. Math. Phys.}
  {\bfseries 108} no.~6, (2018) 1351--1381},
\href{http://arxiv.org/abs/1512.08533}{{\ttfamily arXiv:1512.08533 [hep-th]}}.

\bibitem{Taki:2007dh}
M.~Taki, ``{Refined Topological Vertex and Instanton Counting},''
  \href{http://dx.doi.org/10.1088/1126-6708/2008/03/048}{{\em JHEP} {\bfseries
  03} (2008) 048}, \href{http://arxiv.org/abs/0710.1776}{{\ttfamily
  arXiv:0710.1776 [hep-th]}}.

\bibitem{Bao:2013pwa}
L.~Bao, V.~Mitev, E.~Pomoni, M.~Taki, and F.~Yagi, ``{Non-Lagrangian Theories
  from Brane Junctions},''
  \href{http://dx.doi.org/10.1007/JHEP01(2014)175}{{\em JHEP} {\bfseries 01}
  (2014) 175}, \href{http://arxiv.org/abs/1310.3841}{{\ttfamily arXiv:1310.3841
  [hep-th]}}.

\bibitem{Foda:2018sce}
O.~Foda and R.-D. Zhu, ``{An elliptic topological vertex},''
  \href{http://dx.doi.org/10.1088/1751-8121/aae654}{{\em J. Phys. A} {\bfseries
  51} (2018) 465401}, \href{http://arxiv.org/abs/1805.12073}{{\ttfamily
  arXiv:1805.12073 [hep-th]}}.

\bibitem{Saito2013EllipticDA}
Y.~Saito, ``Elliptic Ding–Iohara Algebra and the Free Field Realization of
  the Elliptic Macdonald Operator,'' {\em Publications of The Research
  Institute for Mathematical Sciences} {\bfseries 50} (2013) 411--455.

\bibitem{bezerra2019quantum}
L.~Bezerra and E.~Mukhin, ``{Quantum toroidal algebra associated with
  $\mathfrak{gl}_{m|n}$},'' \href{http://arxiv.org/abs/1904.07297}{{\ttfamily
  arXiv:1904.07297 [math.QA]}}.

\bibitem{bezerra2021representations}
L.~Bezerra and E.~Mukhin, ``{Representations of quantum toroidal superalgebras
  and plane $\mathbf{s}$-partitions},''
  \href{http://arxiv.org/abs/2104.05841}{{\ttfamily arXiv:2104.05841
  [math.QA]}}.

\bibitem{feigin2021combinatorics}
B.~Feigin, M.~Jimbo, and E.~Mukhin, ``{Combinatorics of vertex operators and
  deformed $W$-algebra of type D$(2,1;\alpha)$},''
  \href{http://arxiv.org/abs/2103.15247}{{\ttfamily arXiv:2103.15247
  [math.QA]}}.

\bibitem{Litvinov:2016mgi}
A.~Litvinov and L.~Spodyneiko, ``{On W algebras commuting with a set of
  screenings},'' \href{http://dx.doi.org/10.1007/JHEP11(2016)138}{{\em JHEP}
  {\bfseries 11} (2016) 138}, \href{http://arxiv.org/abs/1609.06271}{{\ttfamily
  arXiv:1609.06271 [hep-th]}}.

\bibitem{Gaiotto:2017euk}
D.~Gaiotto and M.~Rap\v{c}\'{a}k, ``{Vertex Algebras at the Corner},''
  \href{http://dx.doi.org/10.1007/JHEP01(2019)160}{{\em JHEP} {\bfseries 01}
  (2019) 160},
\href{http://arxiv.org/abs/1703.00982}{{\ttfamily arXiv:1703.00982 [hep-th]}}.

\bibitem{Prochazka:2017qum}
T.~Proch\'{a}zka and M.~Rap\v{c}\'{a}k, ``{Webs of W-algebras},''
  \href{http://dx.doi.org/10.1007/JHEP11(2018)109}{{\em JHEP} {\bfseries 11}
  (2018) 109},
\href{http://arxiv.org/abs/1711.06888}{{\ttfamily arXiv:1711.06888 [hep-th]}}.

\bibitem{Prochazka:2018tlo}
T.~Proch\'{a}zka and M.~Rap\v{c}\'{a}k, ``{$ \mathcal{W} $ -algebra modules,
  free fields, and Gukov-Witten defects},''
  \href{http://dx.doi.org/10.1007/JHEP05(2019)159}{{\em JHEP} {\bfseries 05}
  (2019) 159},
\href{http://arxiv.org/abs/1808.08837}{{\ttfamily arXiv:1808.08837 [hep-th]}}.

\bibitem{Rapcak_2019}
M.~Rapcak, Y.~Soibelman, Y.~Yang, and G.~Zhao, ``{Cohomological Hall algebras,
  vertex algebras and instantons},''
  \href{http://dx.doi.org/10.1007/s00220-019-03575-5}{{\em Commun. Math. Phys.}
  {\bfseries 376} no.~3, (2019) 1803--1873},
  \href{http://arxiv.org/abs/1810.10402}{{\ttfamily arXiv:1810.10402
  [math.QA]}}.

\bibitem{rapcak2020cohomological}
M.~Rapcak, Y.~Soibelman, Y.~Yang, and G.~Zhao, ``{Cohomological Hall algebras
  and perverse coherent sheaves on toric Calabi-Yau 3-folds},''
  \href{http://arxiv.org/abs/2007.13365}{{\ttfamily arXiv:2007.13365
  [math.QA]}}.

\bibitem{NoshitaYITP}
G.~Noshita, ``{5D AGT correspondence of Supergroup Gauge Theories from
  Ding-Iohara-Miki algebra}.''.
  {\url{http://www2.yukawa.kyoto-u.ac.jp/~qft.web/2022/slides/noshita-go.pdf}}.

\bibitem{macdonald1998symmetric}
I.~Macdonald, {\em Symmetric Functions and Hall Polynomials}.
\newblock Oxford classic texts in the physical sciences. Clarendon Press, 1998.
\newblock \url{https://books.google.co.jp/books?id=srv90XiUbZoC}.

\bibitem{Flume:2002az}
R.~Flume and R.~Poghossian, ``{An Algorithm for the microscopic evaluation of
  the coefficients of the Seiberg-Witten prepotential},''
  \href{http://dx.doi.org/10.1142/S0217751X03013685}{{\em Int. J. Mod. Phys. A}
  {\bfseries 18} (2003) 2541},
  \href{http://arxiv.org/abs/hep-th/0208176}{{\ttfamily arXiv:hep-th/0208176}}.

\end{thebibliography}\endgroup
\bibliographystyle{utphys}
\end{document}